\documentclass[useAMS,usenatbib]{mn2e}
\usepackage{epsfig}          
\usepackage{graphicx,color}        
\usepackage{amssymb} 
\usepackage{color}           
\usepackage{url}
\usepackage{amsmath}

\usepackage{morefloats}

\usepackage{setspace}
\usepackage{rotating}
\usepackage{float}
\usepackage{textcomp}
\usepackage[english]{babel}
\usepackage{graphicx}
\usepackage{adjustbox}
\usepackage[mediumspace,mediumqspace,Grey,squaren]{SIunits}
\usepackage{tabularx}
\newcommand{\araa}{  {\it Annu. Rev. Astron. Astrophys.}}

\newcommand{\aap}{    {\it Astron. Astrophys.}}

\newcommand{\aj}{     {\it Astronom. J.}}
\newcommand{\apj}{    {\it Astrophys. J.}}
\newcommand{\apjl}{   {\it Astrophys. J. Lett.}}
\newcommand{\apss}{   {\it Astrophys. Spa. Sci.}}

\newcommand{\mnras}{  {\it Mon. Not. Roy. Astron. Soc.}}
\newcommand{\nat}{    {\it Nature}}

\newcommand{\apjs}{{\it Astrophys. J. Suppl. Series}}

\title{Near-Infrared Imaging of Barred Halo Dominated Low Surface Brightness Galaxies}

\author[Honey et al.]{ M. Honey$^{1,2}$\thanks{E-mail : mhoney@iiap.res.in},
M. Das$^1$\thanks{E-mail : mousumi@iiap.res.in (MD)}, J.P.~Ninan $^3$\thanks{E-mail : ninan@tifr.res.in)},
M.~Purvankara $^3$\thanks{E-mail : manoj.puravankara@tifr.res.in)}\\
$^1$Indian Institute of Astrophysics, Koramanagala, BangalorE-560034, India.\\
$^2$Pondicherry University, R. Venkataraman Nagar, Kalapet, 605014 Pondicherry, India.\\
$^3$Tata Institute of Fundamental Research, Mumbai 400005, India.\\
\date{Accepted 2016 June 22; Received 2016 June 19; in original form 2016 March 09}
\pagerange{\pageref{firstpage}--\pageref{lastpage}} \pubyear{2014}
}
\begin{document}
\maketitle

\label{firstpage}
\begin{abstract}
We present a near-infrared (NIR) imaging study of barred low surface brightness (LSB) galaxies using the TIFR near-infrared  Spectrometer and Imager (TIRSPEC). LSB galaxies are dark matter dominated, late type spirals that have low luminosity stellar disks but large neutral hydrogen (HI) gas disks. Using SDSS images of a very large sample of LSB galaxies derived from the literature, we found that the barred fraction is only 8.3\%. We imaged twenty five barred LSB galaxies in the J, H, K$_S$ wavebands and twenty nine in the K$_S$ band.
Most of the bars are much brighter than their stellar disks, which appear to be very diffuse. Our image analysis gives deprojected mean bar sizes of $R_{b}/R_{25}$~=~0.40 and
ellipticities e~$\approx$~0.45, which are similar to bars in high surface brightness galaxies. Thus, although bars are rare in LSB galaxies, they appear to be just as strong as 
bars found in normal galaxies. There is no correlation of $R_{b}/R_{25}$ or $e$ with the relative HI or stellar masses of the galaxies. In the (J-K$_S$) color images most of the
bars have no significant color gradient which indicates that their stellar population is uniformly distributed and confirms that they have low dust content. 
\end{abstract}

\begin{keywords} 
galaxies : spiral; galaxies: structure; galaxies: evolution; infrared : galaxies.
\end{keywords}

\section{\sc\bf Introduction }
Low Surface Brightness galaxies are extreme late type spiral galaxies that are optically dim and  
have a central disk surface brightness fainter than 22 magnitudes/arcsec$^{2}$ in the B band \citep{ImpeyBothun97}. They have diffuse stellar 
disks, that are low in metallicity \citep{McGaugh.1994} and dust content (\citealt{Rahman.etal.2007}; \citealt{Hinz.etal.2007}).
They are rich in neutral hydrogen (HI) gas \citep{O'Neil.etal.2004} but have low star formation rates 
(\citealt{Boissier.etal.2008}; \citealt{O'Neil.etal.2004}). They can be broadly classified into LSB spirals and LSB dwarf or irregular galaxies. 
Of the LSB spirals, a significant fraction have very large disks and HI gas masses; these galaxies are often referred to as giant LSB (GLSB) 
galaxies of which UGC~6614, Malin~1 and Malin~2 are very good examples (\citealt{Pickering.etal.1997}; \citealt{Sprayberry.etal.1995}). 
GLSB galaxies are usually seen in isolated environments \citep{Rosenbaum.etal.2009} but the smaller LSB dwarf and irregular galaxies
are found in both underdense regions \citep{Pustilink.etal.2011} as well as more crowded environments (\citealt{Merritt.etal.2014}; \citealt{Javanmardi.etal.2016}; \citealt{Davies.etal.2016}).
\par
One of the distinguishing features of LSB galaxies is their very large dark matter content \citep{deblok.etal.2001}. Their dominant dark matter
halos suppresses the formation of both global and local disk instabilities \citep{Mihos.etal.1997} thus hampering the
formation of bars and strong spiral arms in these galaxies \citep{Wadsley.etal.2004}. Thus it is not surprising that bars are
relatively rare in LSB galaxies and spiral arms are thin compared to those found in high surface brightness (HSB) galaxies. However, although barred
LSB galaxies are rare, they are the best systems in which to understand the formation and evolution of bars in dark matter dominated disks. 
Although there have been many simulation studies of bars in halo dominated disks (\citealt{long.etal.2014}; \citealt{saha.naab.2013}; \citealt{villa-vargas.etal.2010}), there are surprisingly no near-infrared (NIR) or optical studies of bars in halo dominated disk galaxies.
One of the main aims of this paper is to provide a deep NIR study of bars in LSB galaxies and see how they differ from normal bright galaxies.  
\par
Bars play an important role in disk evolution through several dynamical effects. First, they drive gas into the centers of galaxies resulting in 
the buildup of central mass concentrations that can lead to bulge growth (\citealt{norman.etal.1996}; \citealt{bournaud.etal.2005}; \citealt{fanali.etal.2015}) 
as well as disk star formation \citep{ellison.etal.2011}. During this process, gas may collect at the resonance radii in the disks - such as the 
corotation radii at the bar ends, circumnuclear rings within the bars or at resonance radii in the outer disks \citep{sellwood.wilkinson.1993}. If the 
gas surface density in these rings is large enough, local instabilities can result in the formation of bright star forming rings at the resonance radii 
\citep{Buta1986}. Such rings are clearly seen in the larger LSB galaxies such as UGC~6614 \citep{mapelli.etal.2008}. Secondly, bars can themselves also 
evolve into rounder bars or boxy bulges in disks. This change in bar morphology can occur rapidly due to bending instabilities in bars \citep{raha.etal.1991} 
or alternatively there maybe a slow dissolution of the bar structure caused by 
gas infall and the buildup of a central mass concentration (\citealt{das.etal.2003}; \citealt{das.etal.2008}). This slow internal evolution 
of bars is often referred to as the secular evolution of barred galaxies (\citealt{Kormendy.etal.2004}; \citealt{Combes.etal.1990}; \citealt{Combes.etal.1981}; \citealt{sheth.etal.2005}). LSB galaxies are
usually isolated and are hence ideal systems in which to study secular evolution of bars into boxier bulges. As our study shows, LSB bulges are not always 
classical bulges; a definite indication that bulges evolve even in the most isolated environments.
\par
In this paper we present 
near infrared (NIR) imaging in the J, H, K$_S$ bands of a sample of barred LSB galaxies using the TIFR NIR Spectrometer and Imager (TIRSPEC) that
is mounted on the Himalayan Chandra Telescope (HCT). Our aim is to study bar morphologies in LSB galaxies; their sizes, shapes, colors and 
correlation with other galaxy properties such as stellar and HI gas masses. We have also examined the K$_S$ band isophotes for signatures
of bar evolution (boxy shapes), interactions or nested bars (twisted isophotes). In the following sections we describe our sample selection and 
our estimate of bar fraction in LSB galaxies. We then describe our observations, results and discuss the implications of our study. 
\\

\section{ Sample selection}

Observational studies show that about one third of all disk galaxies are barred (\citealt{simmons.etal.2014}; \citealt{Menendez-Delmestre.etal.2007}; \citealt{Marinova.etal.2007}). To see if the fraction of bars is different for LSB galaxies we have visually examined the 
Sloan Digital Sky Survey(SDSS) images of a sample of 938 LSB galaxies that were collected from catalogues in the literature 
(\citealt{Impey.etal.1996}; \citealt{Schombert.etal.1988}; \citealt{Schombert.etal.1992}; \citealt{Schombert1998}). However, only 854 galaxies had SDSS images. We 
visually examined the SDSS (Data Release12) images of these galaxies and found that 69 are barred. We had previous HCT observations of
two more LSB galaxies UGC~2936 and UGC~11754 that are not covered in SDSS, but were clearly barred. This gives the barred fraction of
LSB galaxies to be 71/856 or \~ 8.3\%. However, we had taken those galaxies which are brighter enough to observe with TIRSPEC and excluded the very faint galaxies, since they are extremely difficult to image in NIR. We were also limited by sky conditions during our observations. Hence we 
were able to observe a total of 29 barred LSB galaxies. Of these, 25 were covered in the J, H and K$_S$ bands.
The galaxies LSBC~F570-01, IC~2423, 1442+0137 and LSBC~F675-01 have only K$_S$ band images. The sample details are given in Table~1.
 
 \begin{center}
\begin{table*}
\label{Table 1}
\caption{The barred LSB galaxies. The D$_{25}$ is the physical major axis of the galaxy at the 25 mag/arcsec{$^2$} in B band }
\begin{tabular}{l|c|c|c|c|c|c|c|c}

\hline

Galaxy &  RA \& DEC &  Redshift  &  morphology  &       D$_{25}$  & scale	& D$_{25}$ & Active \\
name   &  (J 2000)   & 		 & 	(NED)	&   	(arcsec)  & (kpc/arcsec)& (kpc)	   &	\\
\hline
CGCG 381-048 & 23h47m21.1s +01d56m01s   & 0.0176	& SBc		& ***		&  0.319	& ***   & **	\\
UGC 1920  & 02h27m51.8s +45d56m49s      & 0.0207	& (R')SB(s)ab	& 90.8		&  0.389        & 35.32 & **	\\
UGC 1455  & 01h58m48.0s +24d53m33s      & 0.0171	& SAB(rs)bc	& 165.30	&  0.316 	& 52.23 & **	\\
NGC 5905  & 15h15m23.3s +55d31m03s	& 0.0113	& SB(r)b	& 238.90	&  0.226	& 53.99	& yes-Sy1\\
UM163     & 23h30m32.3s -02d27m45s	& 0.0334	& SB(r)b pec    & 111.7		&  0.617	& 68.92 & yes-Sy1.2     \\
UGC 11754 & 21h29m31.5s +27d19m17s	& 0.0161	& SABcd		& 106.70	&  0.294	& 31.37	& ** \\
PGC 68495 & 22h17m13.1s +25d12m48s	& 0.0422	& SBc(r)	& 52.30		&  0.779	& 40.74 & ** \\
UGC 2936  & 04h02m48.2s +01d57m58s	& 0.0127	& SB(s)d        & 150.70	&  0.242	& 36.47 & yes-Sy2     \\
UGC 10405 & 16h28m54.1s +17d53m27s	& 0.0363	& Scd		& 95.10		&  0.695 	& 66.09 & ** \\
0223-0033 & 02h26m06.7s -00d19m55s	& 0.0214	& SB(rs)bc	& 109.20	&  0.400 	& 43.68	& yes-LINER \\
UGC 5035  & 09h27m10.2s +21d35m38s	& 0.0370	& (R)SB(s)a	& 65.80		&  0.722 	& 47.50	& ** \\
UGC 09087    & 14h12m16.8s +18d17m58s	& 0.0171	& S0		& 70.50		&  0.349	& 24.60 & **	\\
LSBC F568-08 & 10h34m08.7s +19d42m15s	& 0.0341	& Sb		& 57.09	(SDSS-r)&  0.671	& 38.30 & **  \\
LSBC F568-09 & 10h28m12.0s +18d36m23s	& 0.0269	& SBc-p-	& 44.30	(SDSS-r)&  0.539        & 23.88	& ** 	\\
UGC 9634     & 14h58m57.9s +20d03m10s   & 0.0428	& SB(s)b	& 62.80		&  0.820	& 51.50	& ** \\
IC 742       & 11h51m02.2s +20d47m59s   & 0.0214	& SBab		& 65.8		&  0.436	& 28.69 & ** \\
UGC 8794     & 13h52m57.7s +20d55m01s   & 0.0286	& Sb		& 73.8		&  0.566 	& 41.77	& ** \\
UGC 9927     & 15h36m27.8s +22d30m02s   & 0.0144	& SB0		& 54.7 		&  0.290	& 15.86 & ** \\
LSBC F584-01 & 16h05m36.8s +22d11m11s   & 0.0401	& S		& 52.3 		&  0.766 	& 40.06	& ** \\
LSBC F580-02 & 14h36m44.7s +21d04m22s   & 0.0181	& Sc(r)		& **		&  **		& **	& ** \\
UGC 3968     & 07h42m45.2s +66d15m30s	& 0.0226	& SB(r)c        & 84.8	 	&  0.441	& 37.40 & **  \\
1300+0144    & 13h03m16.0s +01d28m07s   & 0.0409	& Sc(f)		& 60.52(SDSS-r)	&  0.795	& 48.11	& ** \\
PGC 60365    & 17h28m54.6s +25d49m03s	& **		& SBb		& **		&  **		& **	&  ** \\
CGCG 006-023 & 09h16m13.7s +00d42m02s   & 0.0381	& Sbc           & 50.55 (SDSS-r)&  0.746        & 37.71 & **\\
1252+0230    & 12h55m25.9s +02d13m52s	& 0.0480	& Sc     	& 44.64 (SDSS-r)&  0.922        & 41.16	& ** \\
LSBC F570-01 & 11h17m57.9s +22d30m11s   & 0.0259	& SB0		& 69.27 (SDSS-r)&  0.521        & 36.08 & **  \\
1442+0137    & 14h45m00.2s +01d24m31s   & 0.0338	& Sc		& 32.42(SDSS-r)	&  0.661	& 21.43	& **\\
LSBC F675-01 & 22h40m35.3s +13d58m38s   & 0.0368	& S/ring	& 25.55(SDSS-r) &  0.678 	& 17.32 & **\\
IC 2423      & 08h54m47.1s +20d13m13s   & 0.0305	& SAB(s)b	& 62.8 (SDSS-r)	&  0.6		& 37.68 & ** \\

\hline

\end{tabular}
\end{table*}

\end{center}

\section{ Observations}

Our NIR and optical observations were done with the near-IR imager TIRSPEC which is mounted on the 2m Himalayan Chandra Telescope (HCT), which is part 
of the Indian Astronomical Observatory in Hanle in the Himalayas. The telescope is remotely operated from Centre for Research and Education in 
Science and Technology (CREST), Bangalore. TIRSPEC is a medium resolution NIR spectrometer and imager covering a wavelength range of 1 to 2.5 $\mu~m$.
It has a field of view of 307 $\times$ 307 arcsec$^{2}$ and a plate scale of 0.3 arcsec pixel$^{-1}$. The observations were done over the period
August 2014 to May 2015. We have taken the images in J, H and K$_S$ broad band filters. In each of the filters we have done a minimum of five 
dither positions and each dither position has three to twenty frames of 15 to 20 second exposures each according to the brightness of the galaxy.
Besides the target frames, the sky frames were also taken in J, H and K$_S$ bands immediately after the target frames in the different dither
positions with at least three frames in each position. The dark frames, twilight flats and morning flats were used for preprocessing the images.
The observation details are given in Table~2.

\par
For two galaxies, UGC~11754 and UM~163, we had to do separate R band observations in order to determine the position angle (PA) or major axis of 
the galaxy which is required for estimating the deprojected bar lengths and ellipticities. UGC~11754 is not covered in the SDSS and the 
galaxy UM~163 is on the edge of the frame, which made it difficult to obtain the position angle of the disk. Although we did obtain NIR images of these 
galaxies, it was extremely difficult to image the faint, diffuse outer parts of the stellar disks which are needed to derive the position angle of the 
galaxy. So we have carried out the optical R-band observation for these two galaxies with the Himalayan Faint Object Spectrograph Camera(HFOSC), that 
has a plate scale of 0.296 arcsec pixel$^{-1}$, and has a field of view of 10 $\times$ 10 arcmin$^{2}$. The details of optical observations are given in 
Table~3.
 
\begin{center}
\begin{table*}
\label{Table 2}
\caption{The Ks band observations of the sample galaxies.}

\begin{tabular}{l|l|c|c|c|c}

\hline

Galaxy	&	date&		number of        & number of frames   & Frame exposure & Total exposure time \\
	&	    &		dither positions & per dither position&	(s)	       &	(s)		\\
\hline

CGCG 381-048 &	22-08-2014	&	5	&	3	&	20	&	300	\\
UGC 1920     &	29-08-2014	&	5	&	3	&	20	&	300 	\\
UGC 1455     &	29-08-2014	&	5	&	6	&	15	&	450	\\
UGC 5905     &	17-09-2014	&	5	&	6	&	15	& 	450	\\
UM 163	     &	17-09-2014	&	5	&	6	&	20	&	600	\\
UGC 11754    &	23-09-2014	&	5	&	10	&	30	&	1500	\\
PGC 68495    &	23-09-2014	&	5	&	6	&	20	&	600	\\
UGC 2936     &	23-09-2014	&	5	&	6	&	20	&	600	\\
UGC 10405    &	12-10-2014	&	5	&	6	&	20	&	600	\\
0223-0033    &	10-11-2014	&	5	&	6	&	20	&	600	\\
UGC 5035     &	10-11-2014	&	5	&	6	&	20	&	600	\\
UGC 9087     &	02-12-2014	&	5	&	6	&	20	&	600	\\
LSBC F568-08 &	09-12-2014	&	5	&	6	&	20	&	600	\\
LSBC F568-09 &	09-01-2015	&	5	&	10	&	20	&	1000	\\
UGC 9634     &  20-03-2015	&	7	& 	10	&	20	&	1400	\\	
IC 742 	     &	18-05-2015	&	5	&	10	&	20	&	1000	\\
UGC 8794     &	18-05-2015	& 	5	&	10	&	20	&	1000	\\
UGC 9927     &	18-05-2015	& 	5	&	10	&	20	&	1000	\\
LSBC F584-01 &	28-05-2015	&	5	&	20	&	20	&	2000	\\
LSBC F580-02 &	08-02-2015	&	5	&	20	&	20	&	2000	\\
UGC3968      &	08-02-2015	&	10	&	10	&	20	&	2000	\\
1300+0144    &  21-05-2015	&	5	&	20	&	20	&	2000	\\
PGC60365     &	12-10-2014	&	5	&	6	&	20	&	600	\\
CGCG 006-023 &	09-12-2014	&	5	&	10	&	20	&	1000	\\
1252+0230    &	31-12-2014	&	5	&	15	&	20	&	1500	\\
LSBC F570-01 &  31-01-2015	&	5	&	10	&	20	&	1000	\\
1442+0137    &  21-05-2015	&	5	&	20	&	20	&	2000	\\
LSBC F675-01 &  21-05-2015	&	5	&	15	&	20	&	1500	\\
IC 2423      &  21-05-2015	&	5	&	10	&	20	&	1000	\\

\hline

\end{tabular}
\end{table*}

\end{center}

\begin{center}
\begin{table*}
\label{Table 3}
\caption{The optical observations of galaxies UGC~11754 and UM~163.}
\begin{tabular}{l|c|c}
\hline
Galaxy	&	date	&		Total exposure time \\
	&	   	&	  	    (s)	            \\
\hline	
UM~163	&    2014-07-07 &		300		\\
UGC~11754&   2014-07-07 &		300*3		\\
\hline

\end{tabular}
\end{table*}

\end{center}

\section{ Data reduction }
The NIR data reduction was carried out using the TIRSPEC semi automated reduction pipeline \citep{Ninan.etal.2014} written in python,
that allows the visual inspection of individual frames, so that the bad frames can be avoided. All the good NIR frames were selected and 
preprocessing steps were done which included bias subtraction and flat fielding. The flat fielded frames, which were taken in different dither 
positions were aligned  
and combined. For the astrometry, we have found the centres of the sources in the field of view of the target framess using IMEXAM  task in
IRAF\footnote{Image Reduction \& Analysis Facility Software } and compared those with the 2MASS images, which were retrieved from the data 
archive. We used the python program for incorporating the world coordinate system (WCS) coordinate information in the images. The J, H and 
K$_s$ images of each of the galaxies are shown in Figure~1. The J and H images of LSBC F584-01, the H images are UGC 3968 and 1300+0144 were 
not good since the sky conditions were bad due to the passing clouds. 
The K$_S$ band isophotes are overlaid on the K$_s$ band images.
The J-K$_S$ images were created by dividing the J band images with K$_s$ band images using task IMARITH , after the prepocessing steps such as 
dark subtraction, flat fielding, aligning and combining. The point spread function for J and K$_s$ bands are similar, so that we directly divided 
the images with out any smoothing. The J-K$_S$ images are shown in figure 2. 

\par 
 The optical images of galaxies UM~163 and UGC~11754 were reduced using standard tasks in IRAF. The target frames were bias subtracted and
 flat field corrected. Cosmic ray hits were removed using the task COSMICRAYS. The frames were aligned together using GEOMAP \& GEOTRAN tasks 
 and combined using IMCOMBINE. The optical images are shown in figure 3. 

\section{Image analysis}
The reduced image frames were analysed using the tasks in the package STSDAS\footnote{The Space Telescope Science Data Analysis System 
(STSDAS) is a software package, which is used for reducing and analyzing astronomical data.}. In the following section we describe how we 
fitted the ellipses and derived the deprojected bar lengths.
 
\subsection{Individual galaxy notes on J, H images \& K$_s$ band isophotes}
The bar is clearly visible in the NIR images and represents the red, old stellar population of the disk.  The K$_s$ band isophotes trace the 
stellar density. In most of the cases the LSB galaxy disks are found to host big, bright bulges; these have been discussed in earlier studies of 
bulges in LSB galaxies \citep{Galaz.etal.2002}. The diffuse stellar disk are barely visible in NIR for 
most of the sample. This can be due to two reasons : (i)~they have a predominantly younger population of stars in the disk which can not be traced very 
well in the NIR bands. However, most LSB disks show very little star formation. (ii)~Instead, it is more likely that the surface mass density of the 
stellar disks ($\Sigma_{*}$) are very low and hence show very little structure in either NIR or optical images.
 \par
Isophotal contours in K$_s$ do not show any significant twisting in the bar region for the sample galaxies. This suggests that the stellar orbits are 
aligned along the bar major axes (x$_1$ orbits) (\citealt{das.etal.2001}; \citealt{sormani.etal.2015}). If there are stellar orbits perpendicular to the bar length (x$_2$ orbits), 
they will be apparent in the K$_s$ contours, but most of our sample show simple, nested, contours aligned along the bar major axis. Just from visual inspection 
about half the sample of 29 galaxies in Figure~1 have bright, classical bulges. The remaining have either boxy bulges or are bulgeless. Several galaxies 
have short oval bars and boxy bulges as well. These systems may represent bars that are undergoing secular evolution and may finally evolve into large, 
boxy bulges.\citep{Kormendy.etal.2004}. The galaxies are individually described below.
\\
\\CGCG 381-048- It has a long thin bar which is clearly visible in all J, H and K$_s$ bands. Bulge is small but brighter. Disk is very
diffuse and hardly visible for short exposures.
\\UGC 1920- has a oval bar with a bright, classical bulge. Disk features are very faint.
\\UGC 1455- The galaxy has a roundish, disky bar. Classical bulge. Very bright and covers a large fraction of the bar. The disk is faint.
\\NGC 5905- Long thin bar and a bright, classical bulge. Traces of spiral arms and a corotation ring are visible in the NIR images.
\\UM 163-  Long thin bar and a large, classical bulge. But the disk is very faint.
\\UGC 11754- Has an oval shaped, short bar. No clear bulge. Disk features are visible.
\\PGC 68495- Disky, oval bar. Small bulge. Galactic disk is diffuse and faint.
\\UGC 2936- Highly inclined galaxy that has a short, oval bar and distinct spiral arms. The disk spiral arms are surprisingly clear.
\\UGC 10405- Very short bar with a bright bulge. Disk is poorly seen.
\\0223-0033-Bright bulge with a small, oval bar followed by the spiral arms. The galactic disk is clearly visible in all the three bands.
\\UGC 5035- Small, oval bar and a large, classical bulge.
\\UGC 9087- Long bar with bright classical bulge. Disk is very faint.
\\LSBC F568-08- Short, oval bar. No clear bulge. Maybe a bar evolving into a bulge.
\\LSBC F568-09- Short bar with bright, classical bulge. The bar isophotes are disky.
\\UGC 9634- Oval bar and a bright boxy bulge. Maybe undergoing secular evolution.   
\\IC 742- Faint but long bar with bright, oval bulge. Stellar disk is slightly seen.
\\UGC 8794- Boxy bulge and short bar. 
\\UGC 9927- Thick bar associated with a bright, classical bulge. Inner ring withing the bulge is visible.
\\LSBC F584-01- The weather conditions during the observation were not good which is reflected in the images.
The galaxy has a short, oval bar.
\\LSBC F580-02-This galaxy does not have a proper bulge. The isophotes show a thin bar formed in the diffuse stellar disk.
\\UGC 3968- The observations in the H band were interrupted by the clouds. The galaxy has a long thin bar and a bright oval bulge.
\\1300+0144- The sky conditions were bad during the H band observations. The galaxy is has a short bar but no bulge.
\\PGC60365- Bright, oval bulge and only a faint bar. May represent ongoing secular evolution.
\\CGCG 006-023- Classified as not barred in NED. But has a bright, classical bulge and a clear bar.
\\1252+0230- Short, oval bar which associated with a bright, oval bulge. May represent ongoing secular evolution.
\\LSBC F570-01 - Very large oval bulge or bar.
\\IC 2423 - Short bar and compact bulge.
\\1442+0137 - Small bar and small, compact bulge.
\\LSBC F675-01-Small bar with oval bulge. May represent ongoing secular evolution.
\subsection{Ellipse fit for NIR images}
We used the IRAF task ELLIPSE to find the bar length and ellipticities using the K$_S$ band images of the galaxies. The galaxy centers were 
first estimated using the task IMEXAM; the output obtained was given as the input for the task IMCNTR to determine the galaxy center more accurately. These values for the galaxy centers 
were taken as the input for the task ELLIPSE. An initial guess for the position angle was also made from visual inspection. During the fitting
procedure, nearby sources were masked in order to avoid a bias in fitting procedure. We compared the outputs with and with out fixing 
the center of the galaxy. We first did the fitting without fixing the center and analyzed the output tables with the plotting task ISOPALL. 
For galaxies that had good signal to noise, such as UGC~3968, UGC~9927, IC~2423, even without fixing the center gave good fitting results. We further checked its stop code values to ensure the quality of the fit. However, for galaxies with poorer signal,  the coordinates of the center 
can change from the central regions to the outer parts by more than two pixels. For these galaxies we redid the fit by fixing the center to the 
values obtained from the well fitted inner regions; we followed this method for the galaxies IC~742, UGC~8794, CGCG~006-023. 
\par
The bar semi major axis length and ellipticities were assumed to be the radii at which the surface
brightness, ellipticity, position angle and b4 parameters changed abruptly. An example is shown in Figure~4. In the surface brightness plots, 
the initial sudden drop in the surface brightness near the center represents the bugle regions. After that it remains flat, this represents the 
contribution from the bar. At further radii the decrease in surface brightness represents the disk region. In the ellipticity plots (where 
ellipticity is defined as 1-b/a, where b and a are the semi-minor and semi-major axes respectively) \citep{Jedrzejewski1987}, except for a 
few pixels in the centre, the  ellipticity initially increases, reaching a maximum value and remains at a constant value for a few pixels. 
Then the ellipticity decreases suddenly, this corresponds to the bar end. Similarly the position angle plot also has distinct regions for the 
different galaxy components. The b4 parameter reveals the diskyness or the boxyness of the fitted ellipse and by how much the 
fitted ellipse differ from the actual isophotes. For example the b4 parameter of UGC~3968 indicates the bar has a boxy shape.
\par
The increase in semi major axis at each level is 0.1 times the semi major axis of the  previous level. All the parameters obtained through the fitting procedure with their error mesurements are tabulated in Table~3.
 

\subsection{Deprojecting the bar length}
The observed image is the projected image of the galaxy on the sky. So the bar length and ellipticities that we observe are not the intrinsic
parameters of the bar. We deprojected the bar length and ellipcity using the following derived parameters - position angle of the galaxy, angle of
inclination of the galaxy and the parameters obtained from ellipse fit such as the bar semi major axis length, position angle of the bar with 
respect to the major axis of the galaxy and the bar ellipcity. The inclination and the position angle of the galaxies were taken from NASA Extragalactic Database (mainly from the RC3 catalogue and SDSS). Since the two surveys cover two different wavelength bands, the redder indicating the older population and the bluer band tracing the young stellar population, their values are connected with the intrinsic properties of the galaxy and give a good estimate of the size of the stellar disks. We selected the values corresponding to the 25~mag/arcsec{$^2$} isophote. However, for the galaxies which are covered in both the surveys, we selected the higher 25~mag/arcsec{$^2$} sizes to obtain an upper estimate of the galaxy
diameter. The angle of inclination can be found from the cos$^{-1}$(b/a) values, where b is the semi minor axis and a is the semi major axis of the galaxy as derived from the isophotes. 
\par
For galaxies LSBC~F580-02 and PGC~60365 the parameters were not listed in the literature, so we have taken their SDSS r-band images and fitted 
ellipses using the task ELLIPSE in IRAF. The position angle of the galaxy and the angle of inclination were determined from the fitted 
25~mag/arcsec{$^2$} isophote. For the galaxy UGC~11754 there was no literature value and no SDSS image; also for UM~163 the position angle is not 
in the literature and the SDSS image was on the edge of the frame so half of the galaxy image was cut out of the frame. Hence we have taken a 
r-band images of these galaxies, UGC~11754 and UM~163 using the HCT. We have the J, H, K$_S$ band images of both galaxies but the galaxy disk is poorly visible. We did the ELLIPSE fit, determined the position angle and the (b/a) value of the outer most isophote from the r-band images.
\par
If the galaxy is inclined at an angle $i$, $\alpha$ is the angle between the bar major axis and the galaxy axis in the sky plane and L$_{obs}$
is the observed bar semi major axis length in the sky plane, then the intrinsic barlength can be determined from the following relation 
(\citealt{Gadotti.etal.2007}; \citealt{Martin.1995})
\\
L$_{dep}$=L$_{obs}$(sin{$^2$}$\alpha$ sec{$^2$}$i$+cos{$^2$}$\alpha$){$^{1/2}$}

The L$_{dep}$ can be converted to the physical units L$_{bar}$  which is in kpc using the scale for the galaxy. However, this relation
is derived using a 1-D approximation, in which the bar is considered as a line. In actual case, the bar is not a 1-D line but has a 2-D shape. 
However, the relations will work well for $i$ less than 60  $^{\circ}$.
The uncertainties in deprojection will increase with increasing $i$ and all the deprojection methods behave badly when the inclination 
angle $i$ is larger than 60 $^{\circ}$ (\citealt{Gadotti.etal.2007}; \citealt{Zou.etal.2014}). Unfortunately some of our galaxies 0223-0033,
UGC~2936, UGC~9634, UGC~8794 and 1300+0144 have high inclination angles; hence, the results for these galaxies will have greater error.
So even though we calculated the bar parameters for these galaxies, we exclude them in the plots (Figures~5, 6, 7, 8).
The deprojected parameters are tabulated in Table~4 with their errors obtained using the propagation of errors in the observed parameters.

\subsection{Deprojecting the bar ellipticities}

Just like barlength, the observed ellipticities of the bars will also be different from the intrinsic values due to the projection effects. The deprojected
ellipticity can be calculated analytically from the following observed parameters - bar major axis length, ellipcity, angle between the galaxy major
axis and the bar major axis, the angle of inclination of the galaxy \citep{Gadotti.etal.2007}. In this method we consider the galaxy nodal axis
to lie along the x-axis or abscissae. The bar is an ellipse in the rotated co-ordinate system in which the bar major axis is the x-axis of the 
rotated system. The entire galactic co-ordinate plane is inclined at an angle $i$ with the sky plane. Using the general conic section equations for this 
system, we will get a quadratic equation. Solving this will give the semi-major and semi-minor axes values which can be used to get the deprojected 
ellipticities of the bars.

\section{Results}

\subsection{Distribution of barlengths and ellipticity of our sample}

The histogram of the deprojected bar length, deprojected ellipticities and the ratio of barlength to D$_{25}$ are shown in Figuere~5. 
Most of the galaxies have bar lengths of 8 to 17~kpc which is similar to the NIR barlengths seen in normal galaxies. For normal galaxies the
large scale bar sizes are in the range ~ 2-28 kpc with 50 \% of them clustered in the range of 4-10 kpc (\citealt{Marinova.etal.2007}; \citealt{Menendez-Delmestre.etal.2007}). Thus surprisingly our plot does not show any special trends for bars in LSB galaxies. Some LSB 
galaxies such as UM~163, UGC~5035, UGC~8794, PGC~60365 have very large bar lengths greater than 18~kpc. The distribution of the ratio
of barlength with the D$_{25}$ diameter ($D_{bar}/D_{25}$) peaks in the  0.4 to 0.6 range, which is also in good agreement with the bars 
observed in normal barred galaxies, for which this fraction is in the range of 0.1 to 0.5. It  has been observed that bars in early type galaxies 
have $D_{bar}/D_{25}$ twice as large compared to late type spirals \citep{Menendez-Delmestre.etal.2007}. However, although LSB galaxies are extreme late 
type spiral galaxies, we find that their bar lengths $D_{bar}/D_{25}$ are not confined to low values as expected for late type spirals. 
\par
These results are surprising since early studies have shown that bars are difficult to form in dark matter dominated galaxies 
(\citealt{ostriker.peebles.1973}; \citealt{hohl.1976}; \citealt{efstathiou.etal.1982}). The low stellar mass surface density may also be a reason for the lower
bar fraction in LSB galaxies \citep{Mihos.etal.1997}. Since the barlength is a measure of the global instability formed in a disk and the instability 
length scale, we would also expect bars in halo dominated galaxies such as LSB galaxies to be shorter. The presence of a range of barlengths suggests that 
it is not just the halo mass  that influences bar formation, but also perhaps the halo shape and spin plays an important role in bar formation in spiral
galaxies (\citealt{Athanassoula.etal.2013}; \citealt{saha.naab.2013}; \citealt{long.etal.2014}; \citealt{Heller.2007ApJ.671.1108}; \citealt{Heller.2007ApJ.226H}; 
\citealt{El-Zant.etal.2002}; \citealt{Martinez-Valpuesta.etal.2004}). Our study thus provides a test sample in which to 
study the effect of halo parameters on bar formation in disk galaxies.
\par
The ellipticities also show a similar trend. The histogram peaks in the ellipcity range 0.4 to 0.6 (Figure~5). There are very few cases 
that exhibit low ellipcity values (say the range of 0.2 to 0.3); some galaxies show very high values of ellipcity which is not expected for 
bars in late type, dark matter dominated galaxies. For normal barred galaxies  the ellipticity lies in the range 0.5 to 0.75
(\citealt{Marinova.etal.2007}; \citealt{Menendez-Delmestre.etal.2007}). All these suggest that the bars in LSB galaxies are similar to the bars
in normal galaxies. In figure 6, we have plotted the ellipcity against the ratio of barlength to D$_{25}$ ($D_{bar}/D_{25}$). The LSB galaxies here  
also show a scattered distribution similar to normal galaxies.

\subsection{Variation of bar properties with the gas fraction}
We have investigated the variation of bar properties with the baryonic masses of the galaxies. Since molecular gas is usually low or absent in LSB galaxies
\citep{das.etal.2006} we assume the baryonic mass to be the sum of the stellar mass and neutral hydrogen mass in LSB galaxies.
For stellar mass calculation we have used the model magnitudes and fluxes from the SDSS DR12 database. The SDSS (g-r) colors were converted into 
corresponding (B-V) values using the color relation B-V= 0.98$\times$ (g-r) + 0.22 \citep{Jester.etal.2005}. Using the (B-V) color
and the R-band flux values we have calculated the mass of the underlying stellar population using the color dependent mass to light ratio coefficient. 
Given for the closed box model\citep{Bell.etal.2001}. The (g-r) \& (B-V) colors and stellar masses are listed in the Table~5.
\par
We have taken the H1 fluxes from HYPERLEDA database and calculated the neutral hydrogen masses (M(HI)) using the relation 
M$_{HI}$= 2.36$\times$ 10$^5$D$^2$ $\int$ (Sd$\nu$) M$_{\odot}$ \citep{Roberts1975}, where D is the distance to the galaxy in Mpc and flux integral 
is in Jy Km s$^{-1}$. The HI mass for each of the galaxies are listed in Table~6. Since the LSB galaxies are poor in molecular hydrogen and dust, the 
total baryonic mass is calculated as the sum of stellar mass and the HI mass. Figure~7 (top) shows the plot of the relative bar length $D_{bar}/D_{25}$ 
against M(HI) normalized by the total baryonic mass (i.e. $M_{HI}$/M$_{HI}$+M$_{stellar}$). Figure~7 (middle) shows the bar ellipticity plotted against the 
$M_{HI}$/M$_{HI}$+M$_{stellar}$. Neither shows any corelation and there is a lot of scatter. The lower plot shows the variation of bar length $D_{bar}/D_{25}$
with the total baryonic mass. There is only a weak correlation. Simulations have shown that gas fraction
 can severely limit the secular bar growth \citep{villa-vargas.etal.2010}. However, we do not see any variation of bar properties 
with HI masses in our sample of LSB galaxies, even though the HI gas fraction of the baryonic mass is as large as  20 to 50\% . 
 
\subsection{The J-K$_s$ color plots}

Most of our J, H, K$_s$ images do not have much emission from the stellar disks, possibly because of the low stellar surface density (Figure~2). 
The exceptions are CGCG~381-048, NGC~5905 and UGC~3968, which show faint emission from their disk regions. Hence, most of our (J-K$_s$) color images 
(i.e. the J and K$_s$ flux ratios) mainly reveal the difference in stellar emission from the bulge and bar regions only. The (J-K$_s$) images in general 
show only a very narrow brightness range or color gradient, indicating that the stellar population and metallicity in the bar and bulge regions 
are very similar; or in other words the color profile is flat. It also shows there is not much dust content in these LSB galaxies which is expected from earlier
Spitzer observations of LSB galaxies (\citealt{Rahman.etal.2007}; \citealt{Hinz.etal.2007}). Although the gradient over the bar is very small in the color plots, 
in some galaxies the bulge is dark with respect to the bar. This indicates that the J and K$_s$ flux ratios are very low in those bulges which shows that are 
composed of an older stellar population than that of the bar. Below we briefly summarise the J-K$_s$ plots (Figure~2).

\noindent
(i)~For the galaxies CGCG~381-048, UGC~11754, PGC~68495, UGC~10405, UGC~5035, 0223-0033, UGC~9087, UGC~3968,  1300+0144, PGC~60365, CGCG~006-023, 1252+0230 the 
(J-K$_s$) color of their bulge and bar regions are almost the same i.e. they have a flat color gradient. We can conclude that the bulge and bar have almost 
similar stellar populations and low dust content.\\
(ii)~The galaxies UGC~1920,UGC~1455, NGC~5905, UM~163, UGC~2936, LSBC~F568-08, UGC~9634, IC~742, UGC~8794, UGC~9927, LSBC~F584-01, LSBC~F580-02 show  significant 
variations in color for the bar and central bulge regions, indicating the central bulge is made up of older stellar population.\\
(iii)~In some galaxies like CGCG~381-048, NGC~5905, UGC~11754, 0223-0033, UGC~3968, PGC 60365, CGCG~006-023 the disk and the bar show similar colors in their bars and 
bulges. In these galaxies bar might have helped the mixing of stellar population and metallicity.\\  
(iv)~In case of the galaxy  UGC~2936, even the spiral arms are visible in the disk representing the arm of older stars. this galaxy may have significant dust and 
ongoing star formation \citep{pickering.etal.1999}.

\subsection{Variation of J-K$_s$ with bar parameters}
We were not able to get the NIR standard star observations on the same day as our observation for most of our sample. Alternatively, we could have used NIR 
bright stars in our observing frames. But unfortunately most of our NIR frames did not have a good enough number of stars that did not show NIR variability 
for calibrating the NIR magnitudes. Hence to have uniform magnitudes we have used the J, and K$_s$ magnitudes from the 2MASS catalogue for extended sources 
for most of our sample galaxies (Table~6) \citep{skrutskie.etal.2006}; for some it was not available. The J and K$_s$ magnitudes with their errors are 
listed in table 6.
To see if whether the bar parameters are directly show any correlation with  the color of the galaxy, we have plotted the J-K$_s$ values with the bar deprojected
ellipticities and the ratio of barlength to the disk size $D_{bar}/D_{25}$. The plots are shown in Figure~8. The bar ellipticities show a weak relation with the 
J-K$_s$ color. But the relative bar lengths $D_{bar}/D_{25}$ show a  clear correlation with the J-K$_s$ colors. This is because a longer bar that extends 
further into the disk will trigger more star formation and thus make the galaxies bluer.
 
\subsection{J-K$_s$ variation with stellar and neutral hydrogen masses}

Figure~9 shows the J-K$_s$ color plotted against the ratio of  neutral hydrogen mass to stellar mass (M$_{HI}$/M$_{stellar}$). The J-K$_s$ color of the galaxies 
do not exhibit any strong correlation with the ratio M$_{HI}$/M$_{stellar}$. For normal galaxies those hosts strong bars, the bars can trigger disk star 
formation and gas infall (\citealt{ellison.etal.2011}; \citealt{renaud.etal.2015}). Hence, bluer colors maybe corelated with more gas depletion or lower gas mass 
factions M$_{HI}$/M$_{stellar}$. If that is true 
for our sample of LSB galaxies, Figure~9 should show decreasing J-K$_s$ values with increasing M$_{HI}$/M$_{stellar}$ ratios. However, we do not see any 
correlation at all, which suggests that the bars in LSB galaxies are not strong enough to trigger enough star formation to cause significant gas depletion
in their disks. However, star formation maybe episodic in nature \citep{mishra.etal.2015} and localized to small regions in the disk. 
    
\section{Comparing our observations with simulations in the literature}

Early simulations have shown that cold rotating stellar disk will become bar unstable \citep{ostriker.peebles.1973}. 
Later simulations showed that if the stellar disks 
are immersed in massive dark matter halos, the disc will be stable against bar formation. 
In these studies the halos were both non-rotatating and non responsive i.e 
they were fixed potentials (\citealt{hohl.1976}; \citealt{efstathiou.etal.1982}). Later simulations of halo dominated LSB galaxies   
have shown that they are usually stable against bar formation but if bars do form, they are smaller than those found in normal galaxies 
(\citealt{Mihos.etal.1997}; \citealt{Wadsley.etal.2004}). The lower fraction of barred LSB galaxies obtained in our sample selection (Section~2) supports these 
early simulations.

However, our observations clearly show that halo dominated galaxies can host strong bars (e.g. NGC~5905, IC~742, UGC~8794). This is where our observations differ
from early simulations and it is the most important result from our study. Our observations suggest that it is not just the relative disk to halo mass that determines 
whether a bar can form in a disk galaxy, other parameters such as the halo angular momentum and halo shape can influence bar formation and determine the bar strength and 
shape. Some simulations with live or responsive halos have shown that the halo concentration can have a pronounced effect on the bar formation\citep{Villa-vargas.etal.2009} 
and not the total mass of the halo. In the simulations, the spinning dark matter halos helps in the bar formation (\citealt{saha2013}; \citealt{long.etal.2014}). Besides these
factors the axes ratios or the shape also can influence the bar growth considerably even for massive halos
\citep{Athanassoula.etal.2013}. In these studies, the halo is composed of N-body particles 
just like the disk; there is angular momentum transfer between the stellar disk and halo (\citealt{Weinberg1985}; \citealt{Debattista.etal.1998}; \citealt{Villa-vargas.etal.2009}) which helps the disks 
form strong bars despite the large halo masses. In short the responsive halos can form strong bars (\citealt{Athanassoula2002}; \citealt{Athanassoula2003}). 
The bar-halo interaction can also result in the slow down of bar rotation in later evolutionary stages \citep{chemin.etal.2009}. In the secular
phase the bar growth triggered by buckling instabilities and vertical resonances will alter the bar shape, and transform the bar into peanut or boxy shapes (\citealt{Martinez-Valpuesta.etal.2006}
; \citealt{Combes.etal.1990}). LSB galaxies also show signatures of secular evolution since several galaxies in our sample appear to have boxy bulges (e.g. 1252+0230, PGC~60365, UGC~8794)
\citep{sheth.etal.2005}. We are exploring this in a separate study. Thus LSB galaxies are an important testbed for understanding the evolution of galaxy disks in halo 
dominated environments. 

\section{Conclusions}
{\bf 1.}~The fraction of bars in LSB galaxies is very low, as expected from the simulations of bar formation in halo dominated galaxies. We examined the SDSS images
of a sample of  856 LSB galaxies  and found that only 8.3\% have bars. This low fraction supports the simulations that show that massive dark matter halos suppress 
the formation of disk instabilities such as bars in spiral galaxies.  
\\
{\bf 2.~}About half our sample of 29 barred LSB galaxies host large, classical  bulges that are very bright in the NIR. The remaining other half have either boxy 
bulges or are bulgeless. The bars are also much brighter in NIR than the diffuse stellar disks. This indicates that both the bars and bulges have either an older stellar 
population and/or higher stellar surface density than the LSB disks. The K$_s$ isophotes do not show any twisting in the selected sample and are instead generally aligned 
along the bar major axes. 
\\
{\bf 3.~}Although the fraction of barred candidates in the dark matter dominated LSB galaxies is very small, the bar parameters such as
barlength and ellipticities have a range of values that are similar to those found in normal galaxies. Our results clearly show that halo dominated galaxies can host strong bars.
\\
{\bf 4.~}For more than half of the sample (25/29) the color parameter J-K$_s$ shows practically no variation between the bar and bulge regions. This indicates that they have 
similar stellar populations, metallicities and dust content. Some candidates have bulges that are significantly dimmer than the bar; these galaxies may have 
a older cold stellar population in the bulge.
\\
{\bf 5.~}The plots of J-K$_s$ with the bar length $D_{bar}/D_{25}$ shows a weak but significant correlation, which suggests that the bar may cause some local disk 
star formation which makes the J-K$_s$ bluer. But the plots of  J-K$_s$ with the ratios of the HI content and stellar mass (M$_{HI}$/M$_{stellar}$) do not 
show any correlation, which clearly shows that the star formation triggered by the bar is only local and not on a global disk scales.

{\bf \sc Acknowledgments}

The optical observations were done at the Indian Optical Observatory
(IAO) at Hanle. We thank the staff of IAO, Hanle and CREST, Hosakote, that made these
observations possible. The facilities at IAO and CREST are operated by the Indian Institute
of Astrophysics, Bangalore. This research has made use of the NASA/IPAC Extragalactic
Database (NED), which is operated by the Jet Propulsion Laboratory, California Institute
of Technology, under contract with the National Aeronautics and Space Administration.
We acknowledge the usage of the HyperLeda database\footnote{http://leda.univ-lyon1.fr/}\citep{Makarov2014}.
Our work has also used SDSS-III data. Funding for SDSS-III has been provided by
the Alfred P. Sloan Foundation, the Participating Institutions, the National Science
Foundation, and the U.S. Department of Energy Office of Science. The SDSS-III website
is http://www.sdss3.org/. SDSS-III is managed by the Astrophysical Research
Consortium for the Participating Institutions of the SDSS-III Collaboration including the
University of Arizona, the Brazilian Participation Group, Brookhaven National Laboratory,
Carnegie Mellon University, University of Florida, the French Participation Group, the
German Participation Group, Harvard University, the Instituto de Astrofisica de Canarias,
the Michigan State/Notre Dame/JINA Participation Group, Johns Hopkins University,
Lawrence Berkeley National Laboratory, Max Planck Institute for Astrophysics, Max
Planck Institute for Extraterrestrial Physics, New Mexico State University, New York
University, Ohio State University, Pennsylvania State University, University of Portsmouth,
Princeton University, the Spanish Participation Group, University of Tokyo, University of
Utah, Vanderbilt University, University of Virginia, University of Washington, and Yale
University. This publication makes use of data products from the Two Micron All Sky Survey, 
which is a joint project of the University of Massachusetts and the Infrared Processing and 
Analysis Center/California Institute of Technology, funded by the National Aeronautics and 
Space Administration and the National Science Foundation. This research made use of Montage.
 It is funded by the National Science Foundation under Grant Number ACI-1440620, and was previously
 funded by the National Aeronautics and Space Administration's Earth Science Technology Office,
 Computation Technologies Project, under Cooperative Agreement Number NCC5-626 between NASA and 
the California Institute of Technology.The color plots were generated using
the two-dimensional graphics environment Matplotlib \citep{Hunter2007}. We thank the anonymous referee for valuable comments and suggestions
on the draft.

%
%
%

\begin{figure*}
\centering

\caption{The J, H, K$_s$ band images of Low Surface Brightness galaxies. In all figures the north is up and east is in the left hand side. We have used 
the logarithmic scaling.The isophotal contours are overlaid on the K$_s$ band image.
Most of the isophotes are above 5$\sigma$ level except the galaxies IC 742, NGC 5905, PGC 68495, UM163, LSBC F568-08, UGC 5035,LSBC F568-09 have last contour level is 4$\sigma$ and for galaxies CGCG 381-048, 
UGC 1920, UGC 9927, UGC 9634 have last contour level is 3$\sigma$. For galaxies UGC 3968, LSBC F580-01, and 1300+0144 the H band observations were 
effected by the bad sky conditions.}
\includegraphics[trim = 10mm 17mm 20mm 8mm, clip,scale=0.37]{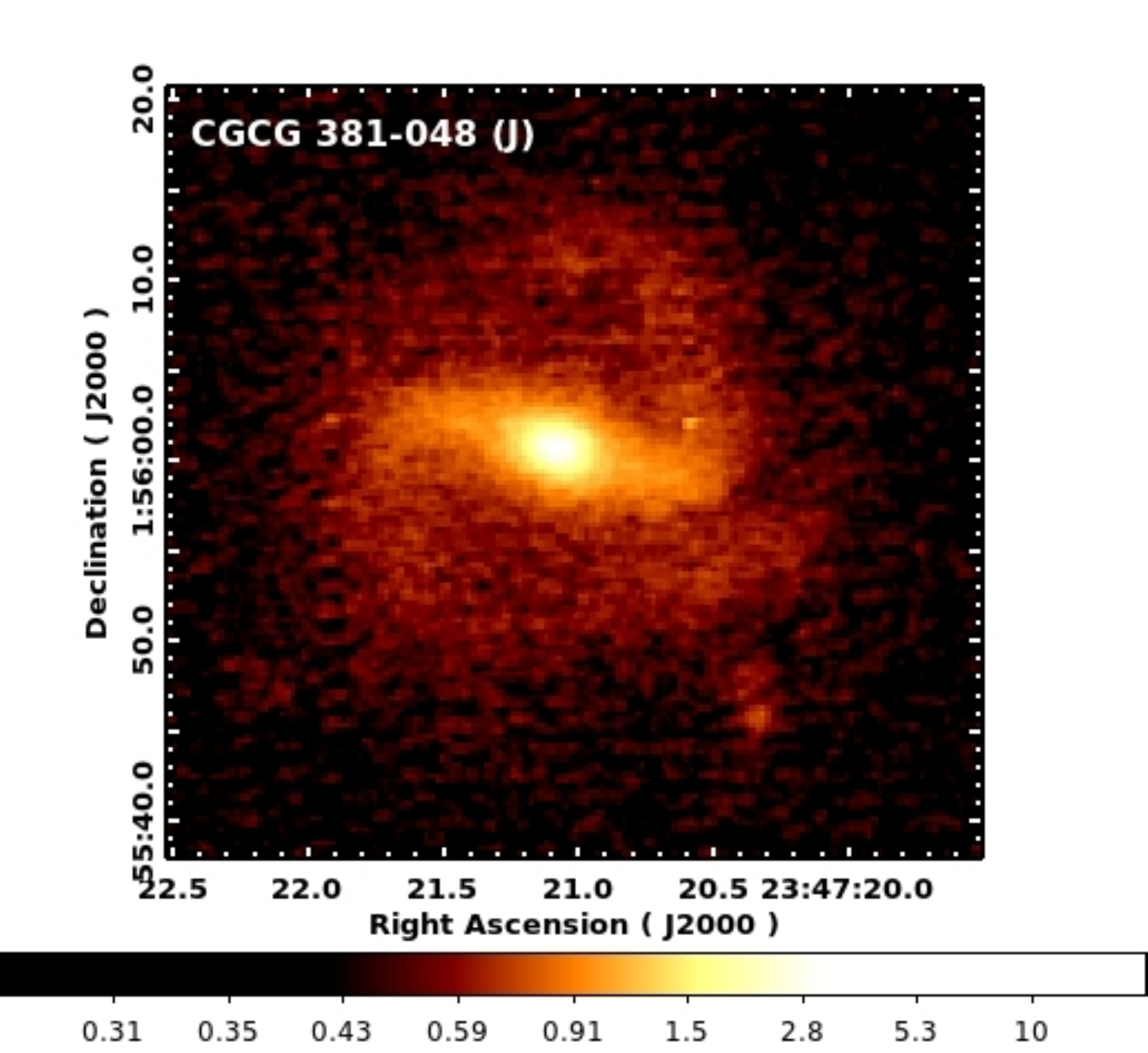}\includegraphics[trim = 10mm 17mm 20mm 8mm, clip,scale=0.37]{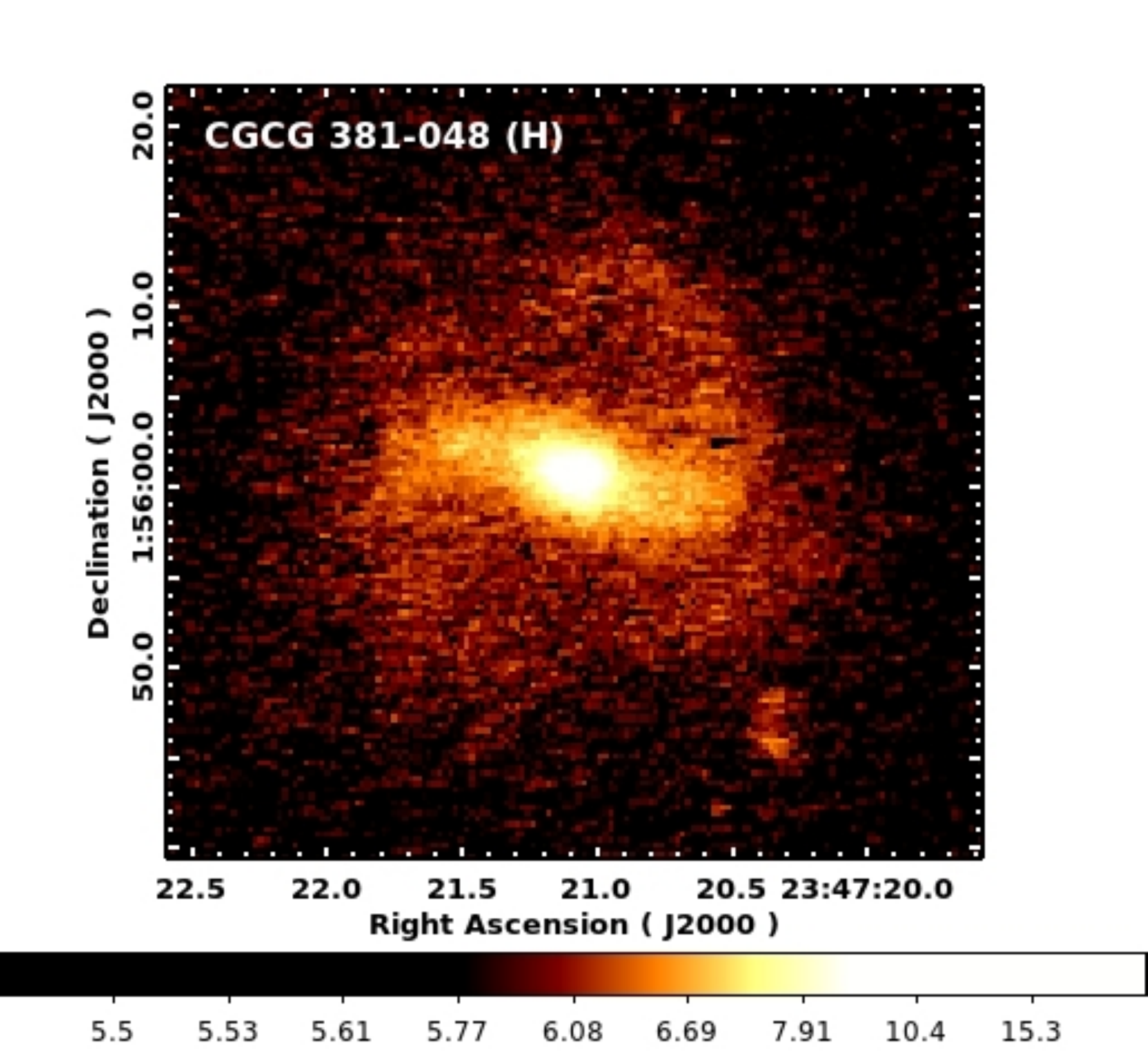}\includegraphics[trim = 10mm 17mm 20mm 8mm, clip,scale=0.37]{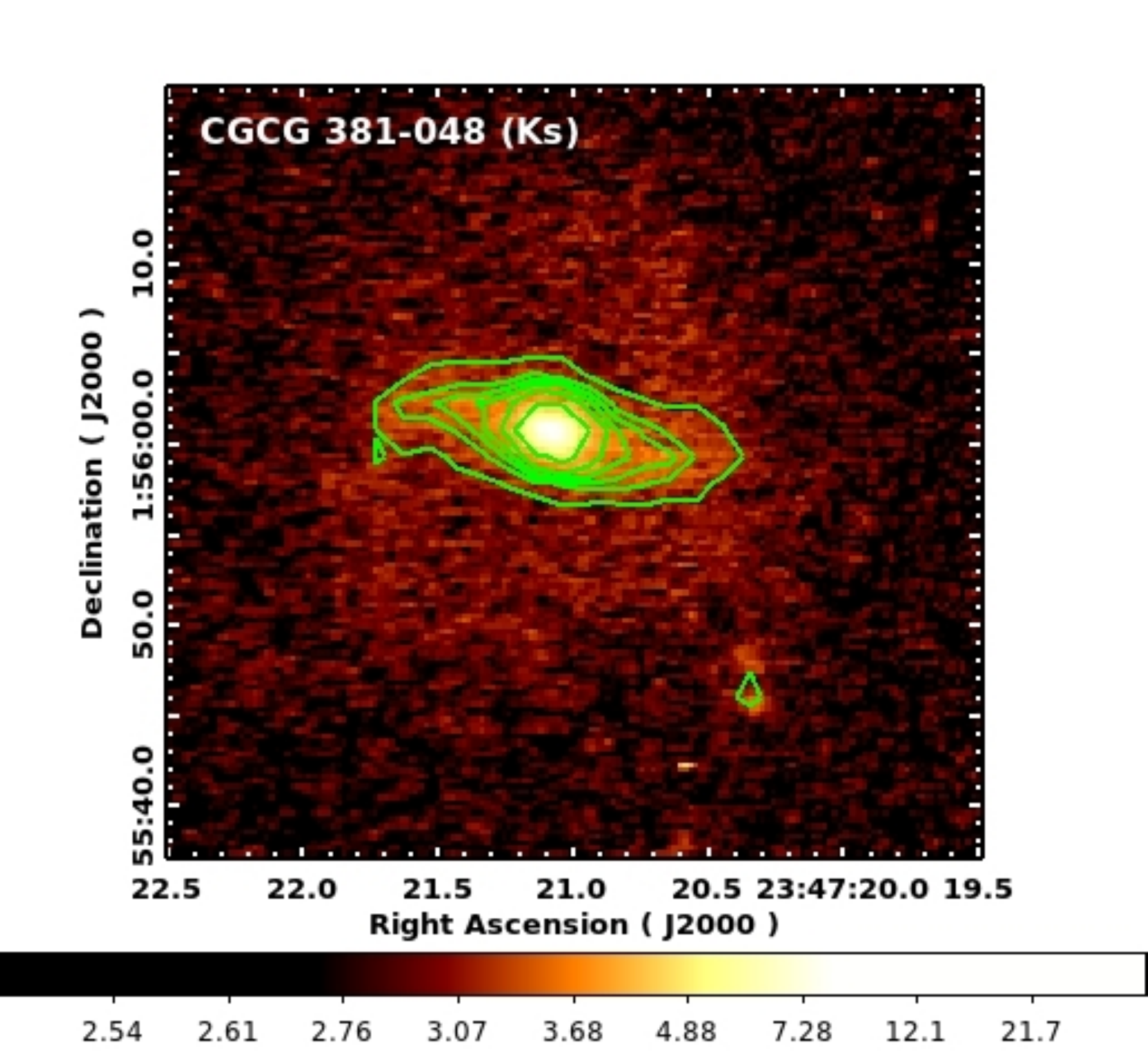}
\includegraphics[trim = 5mm 17mm 20mm 15mm, clip,scale=0.37]{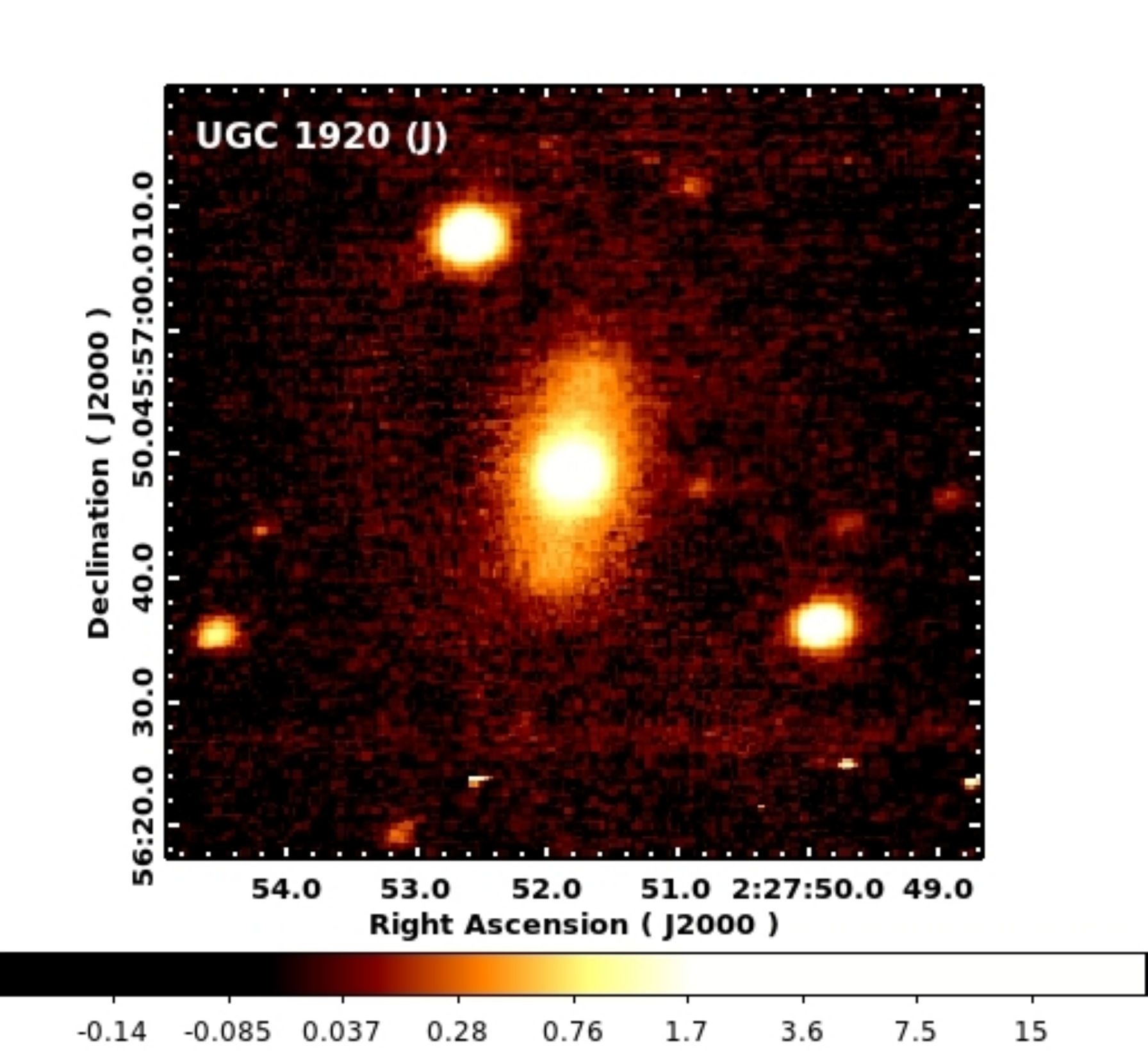}\includegraphics[trim = 10mm 17mm 20mm 15mm, clip,scale=0.37]{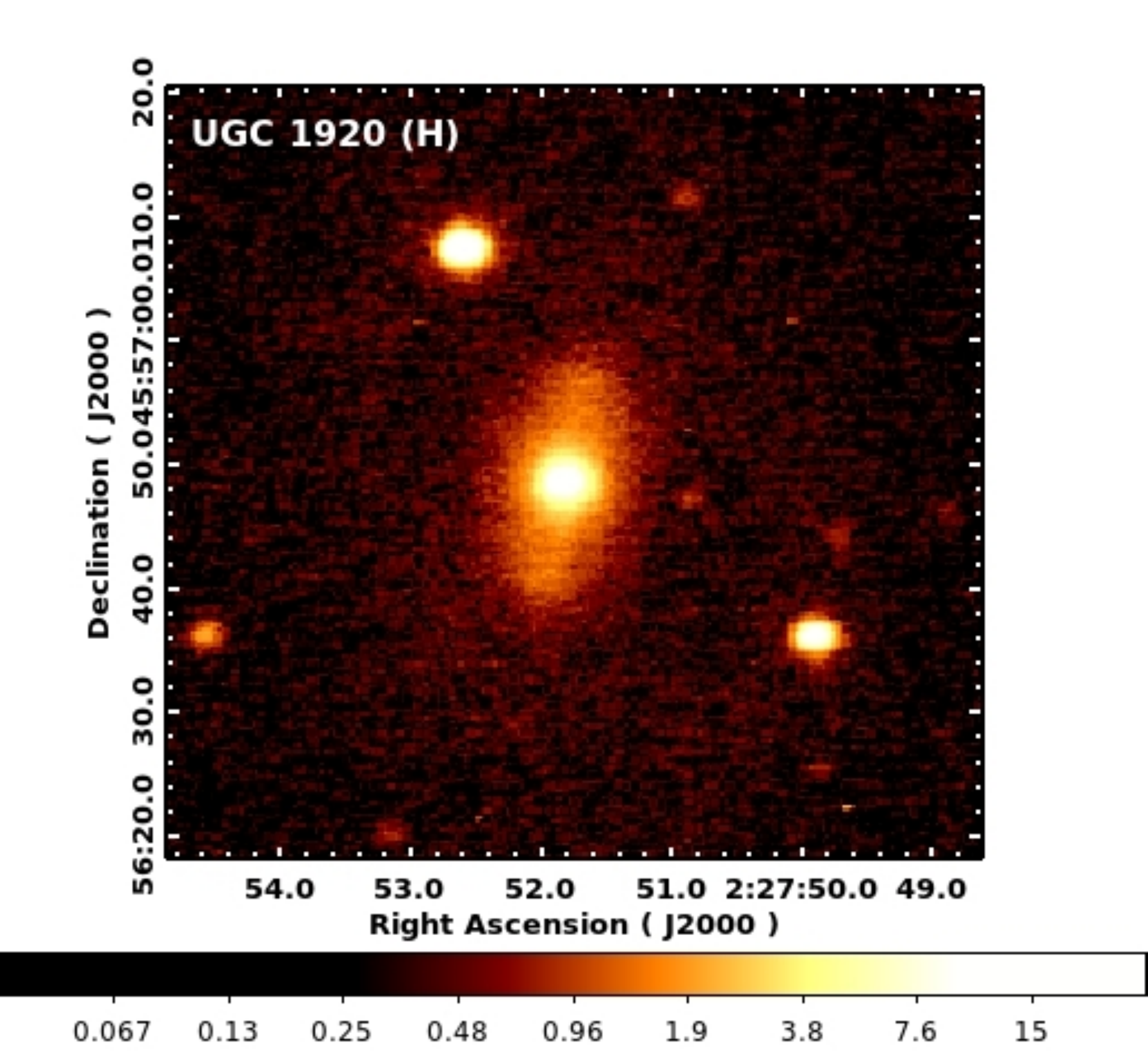}\includegraphics[trim = 25mm 23mm 20mm 15mm, clip,scale=0.45]{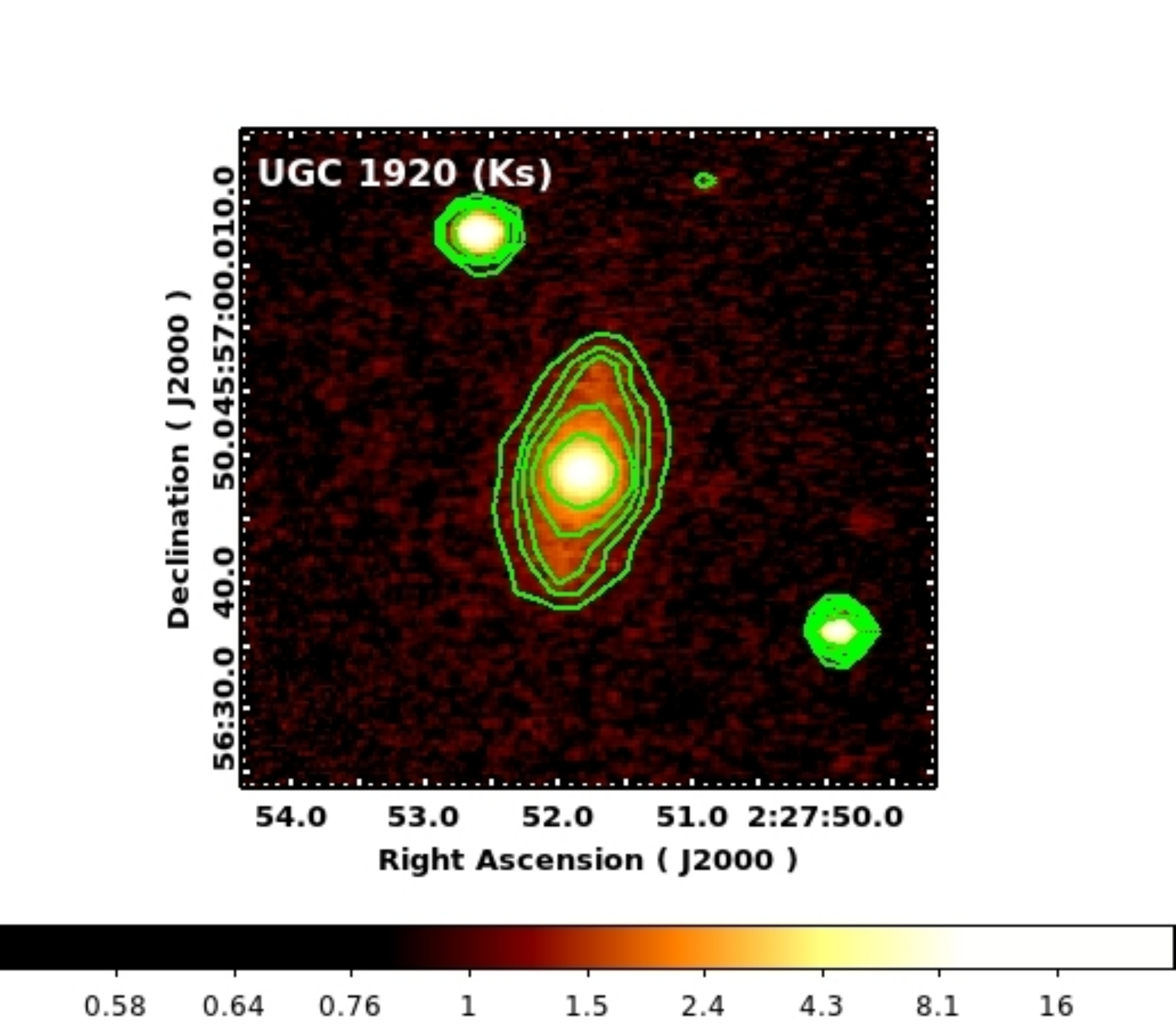}
\includegraphics[trim = 45mm 20mm 60mm 20mm, clip,width=5.5cm]{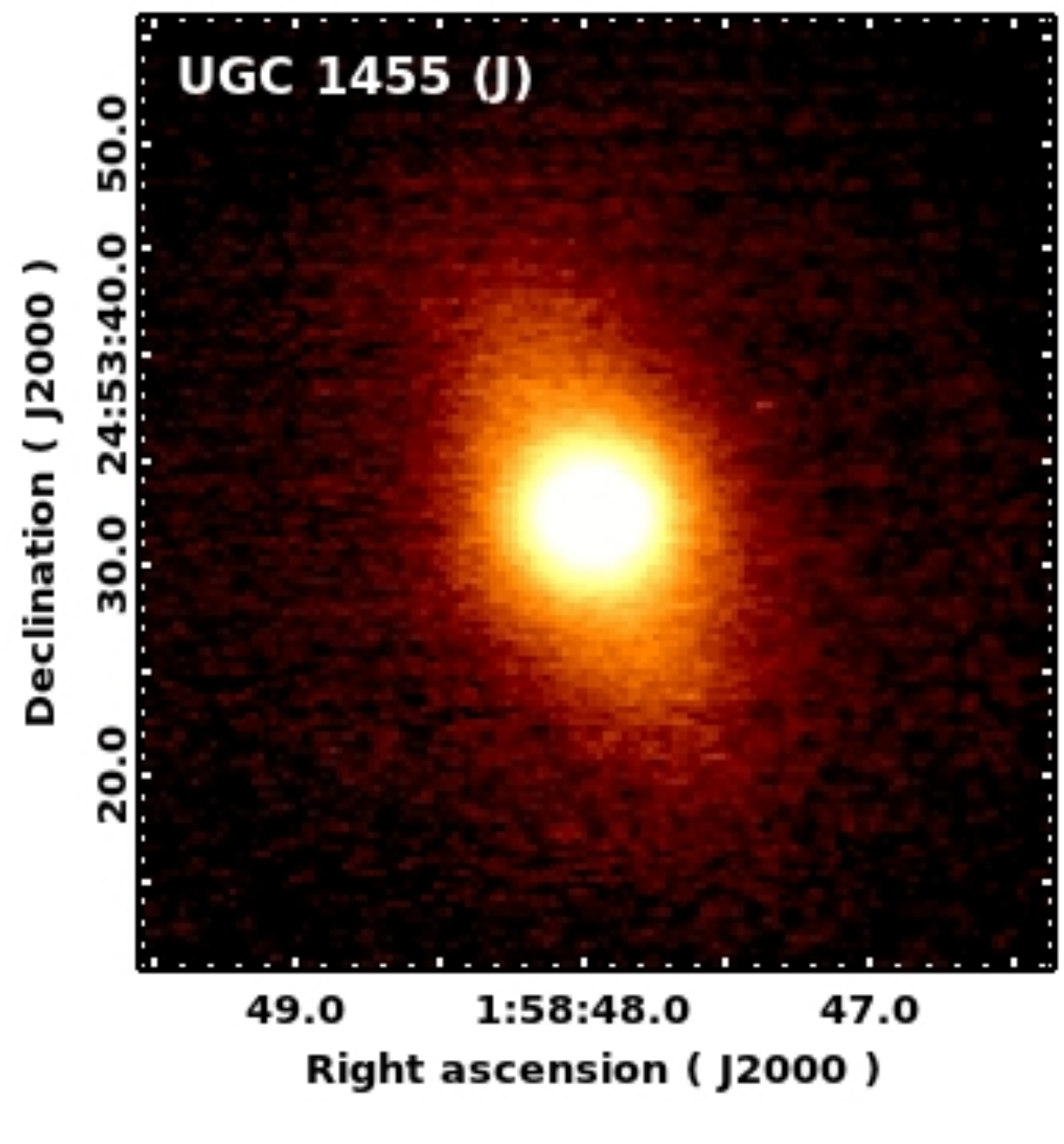}\includegraphics[trim = 45mm 20mm 60mm 20mm, clip,width=5.5cm]{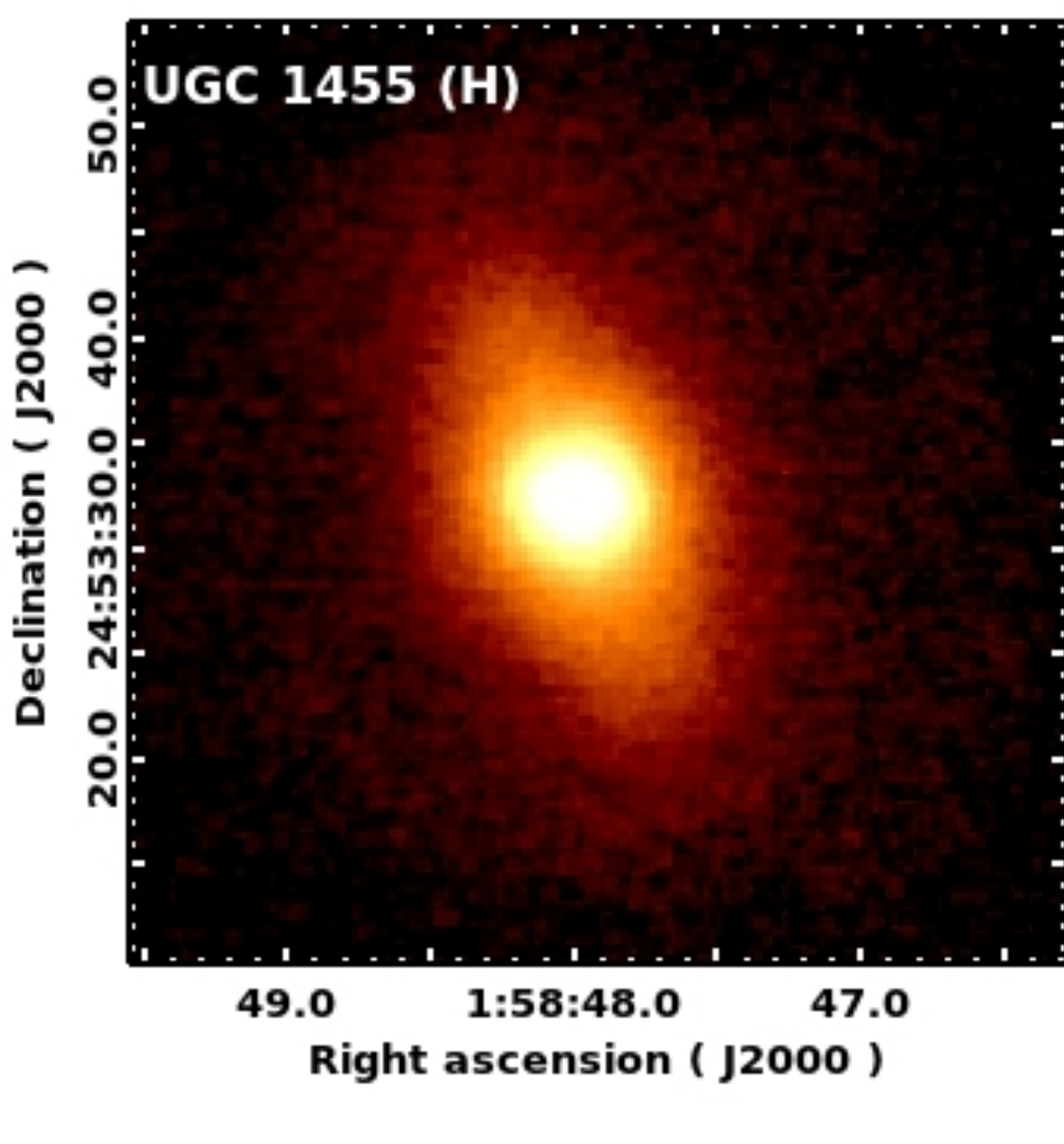}\includegraphics[trim = 50mm 20mm 60mm 20mm, clip,scale=0.46]{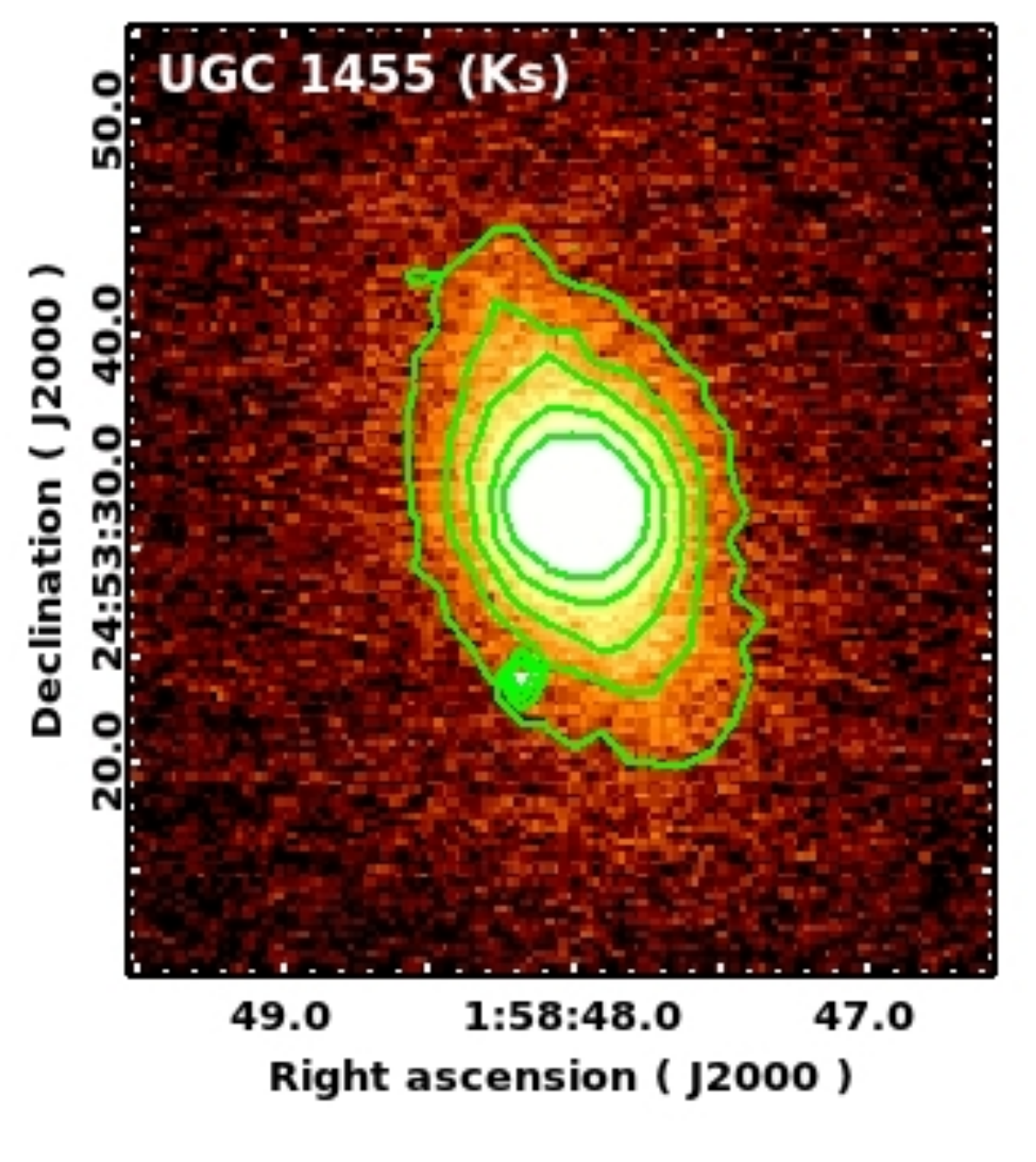}
\includegraphics[trim = 30mm 17mm 45mm 15mm, clip,scale=0.37]{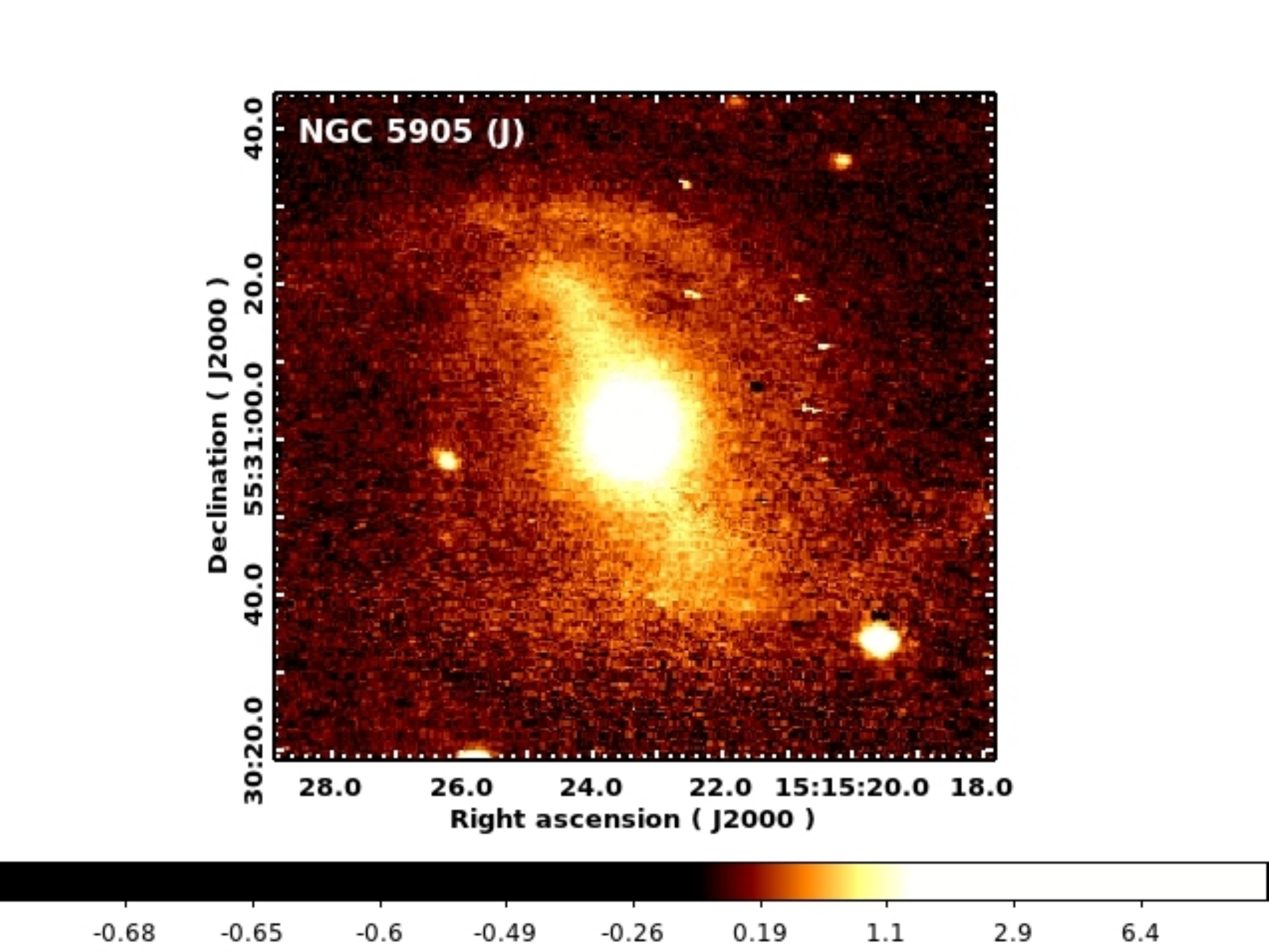}\includegraphics[trim = 30mm 17mm 45mm 15mm, clip,scale=0.37]{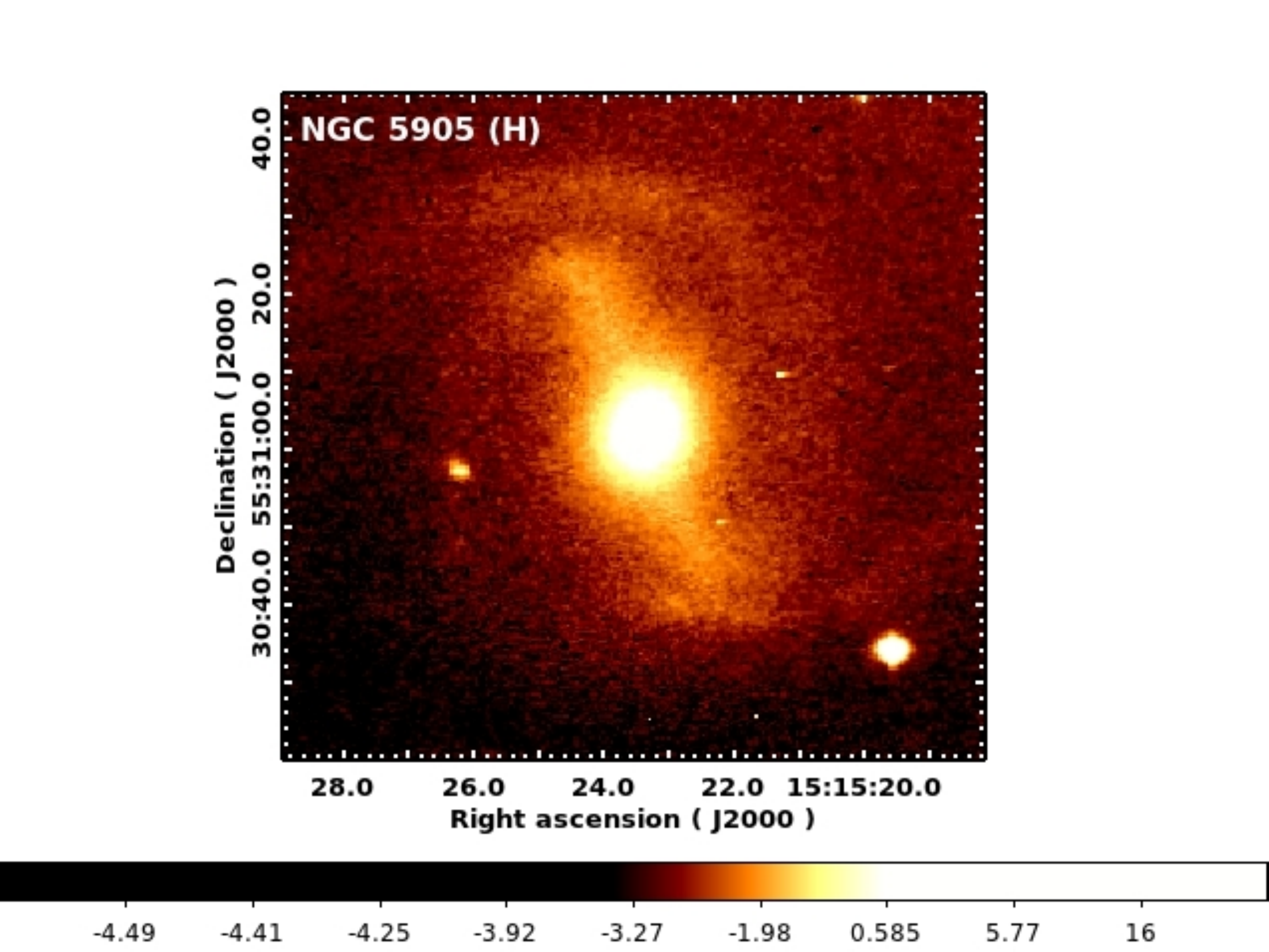}\includegraphics[trim = 30mm 17mm 45mm 15mm, clip,scale=0.37]{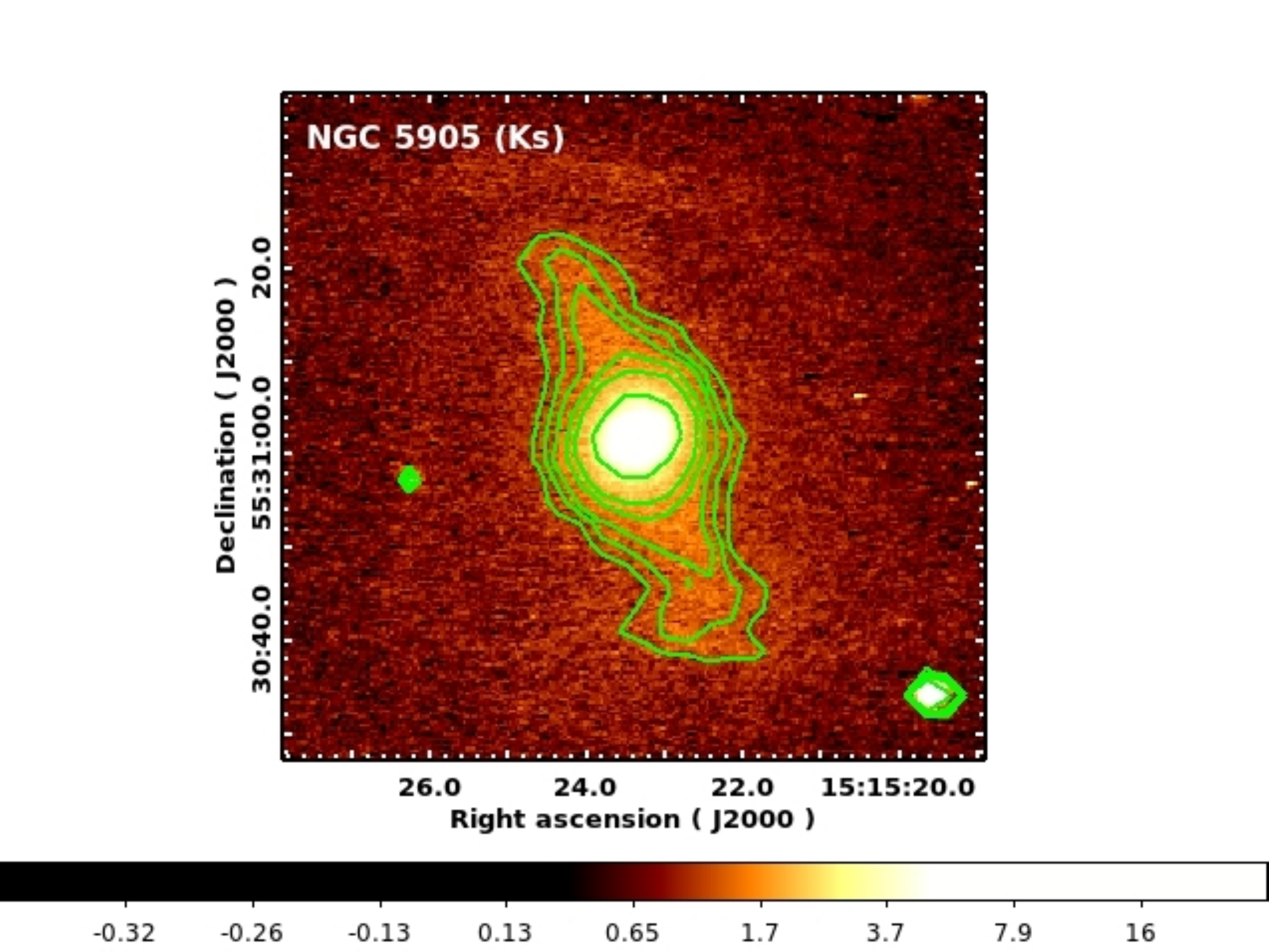}

 \end{figure*}
\begin{figure*}
\includegraphics[trim = 45mm 30mm 45mm 20mm, clip,scale=0.49]{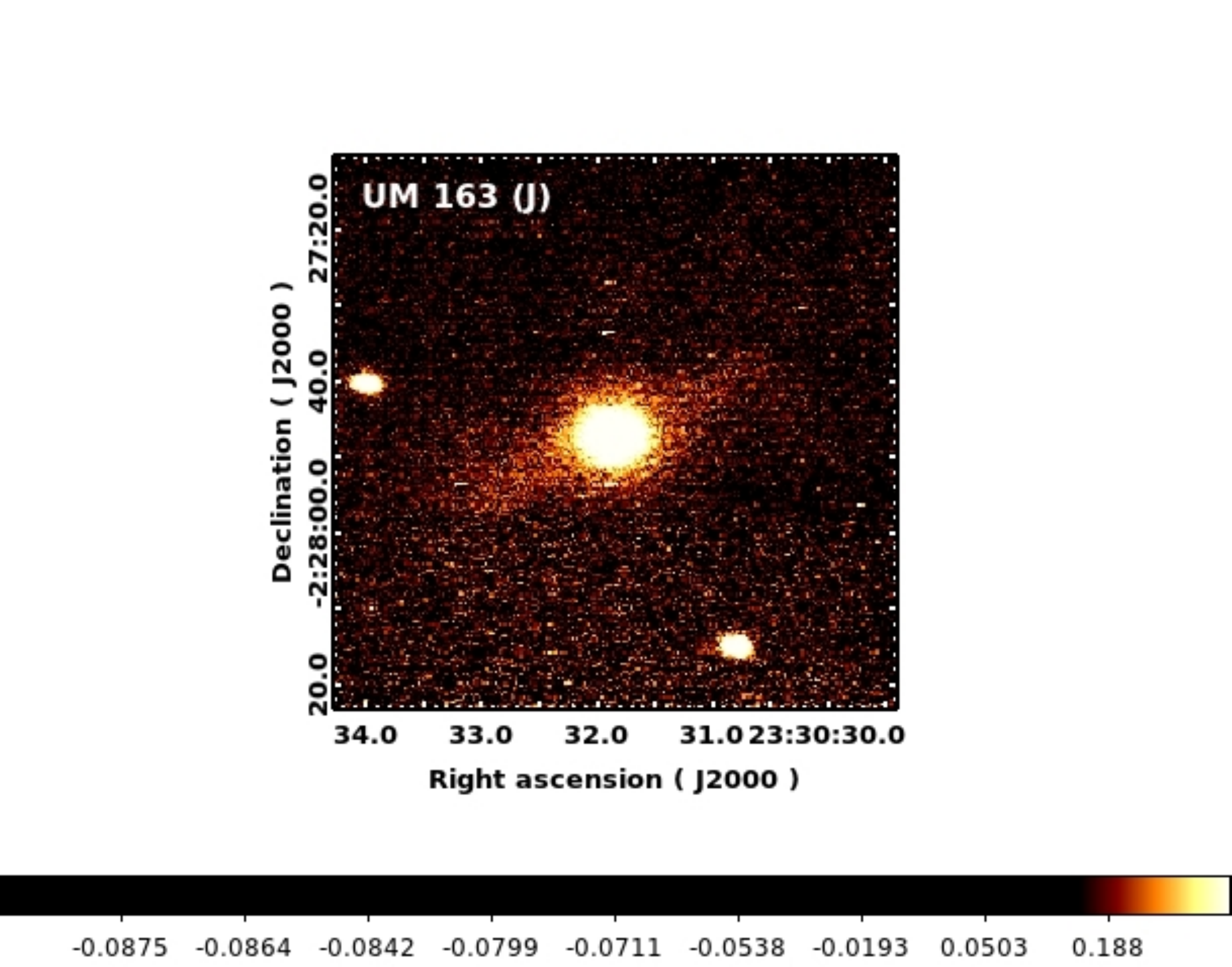}\includegraphics[trim = 30mm 25mm 30mm 8mm, clip,scale=0.49]{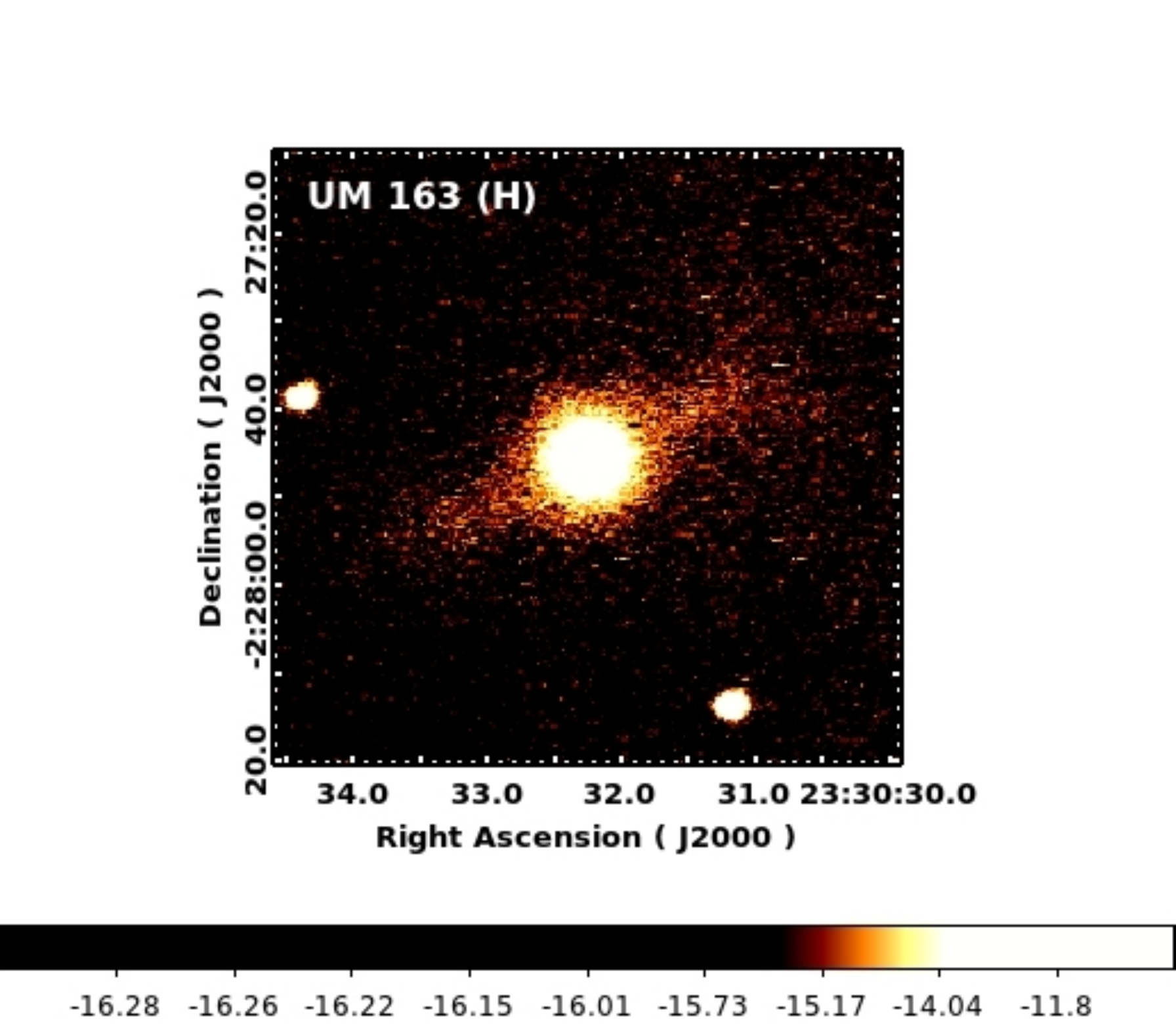}\includegraphics[trim = 35mm 17mm 20mm 8mm, clip,scale=0.405]{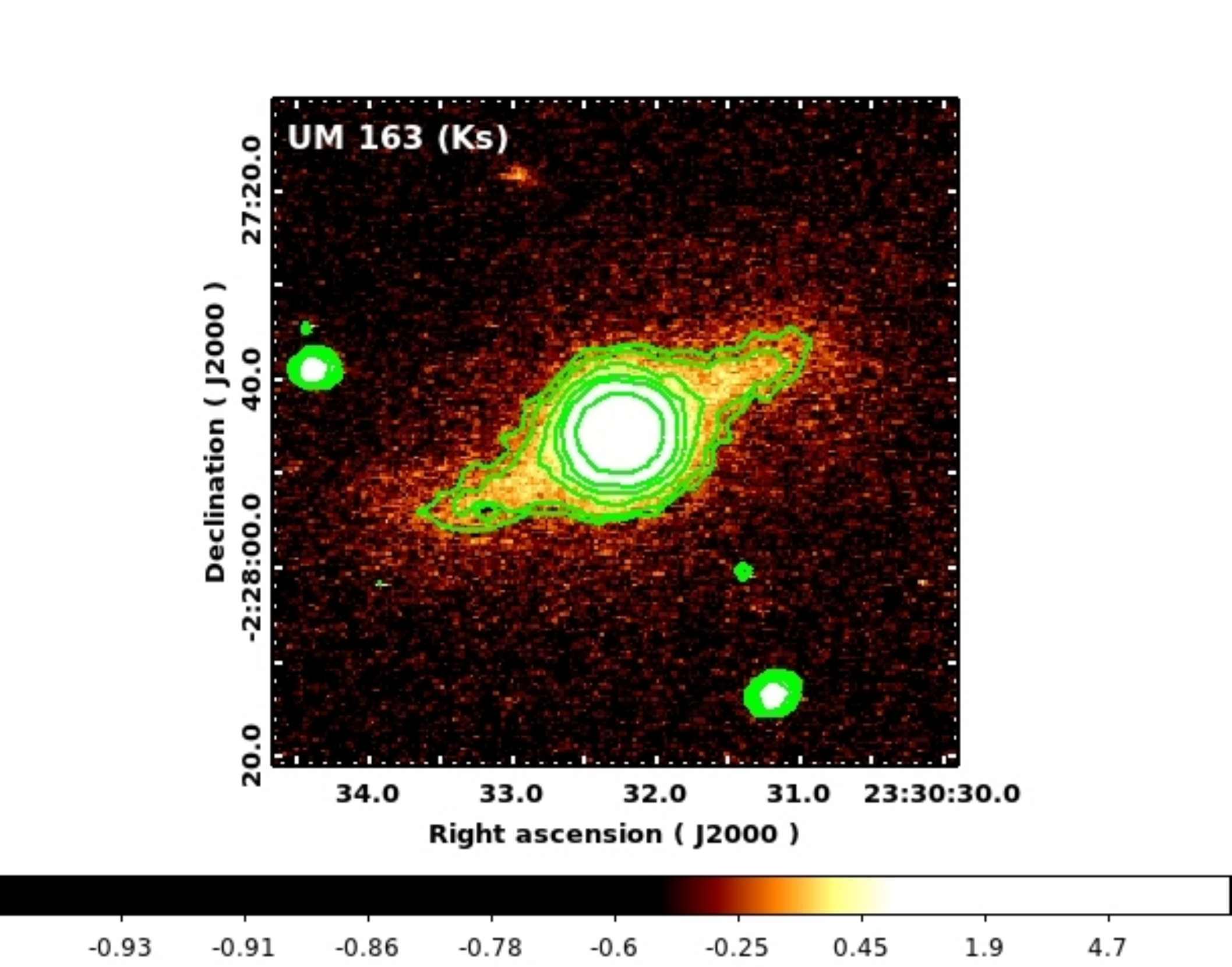}\\

\includegraphics[trim = 35mm 17mm 40mm 15mm, clip,scale=0.44]{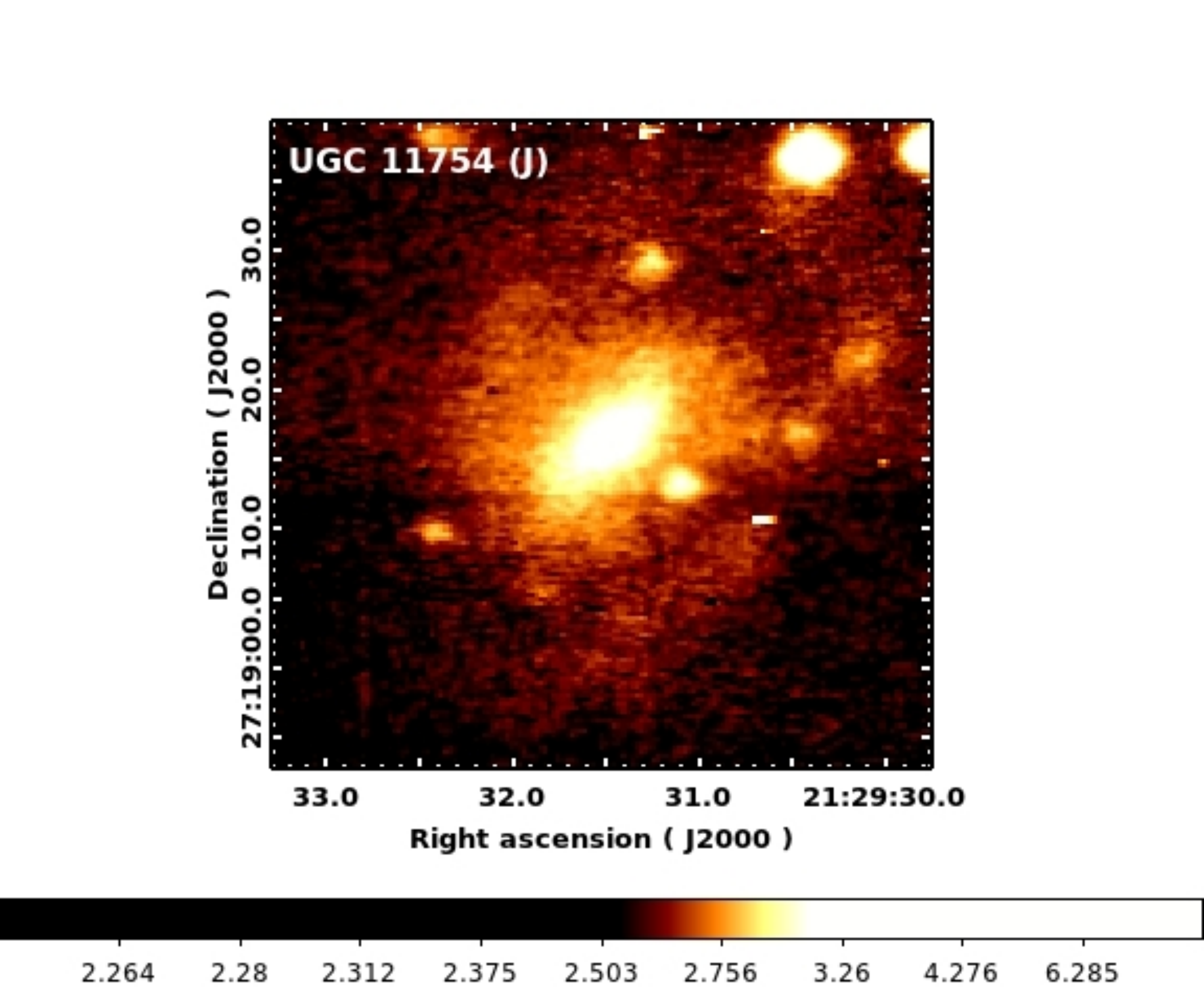}\includegraphics[trim = 40mm 25mm 50mm 20mm, clip,scale=0.49]{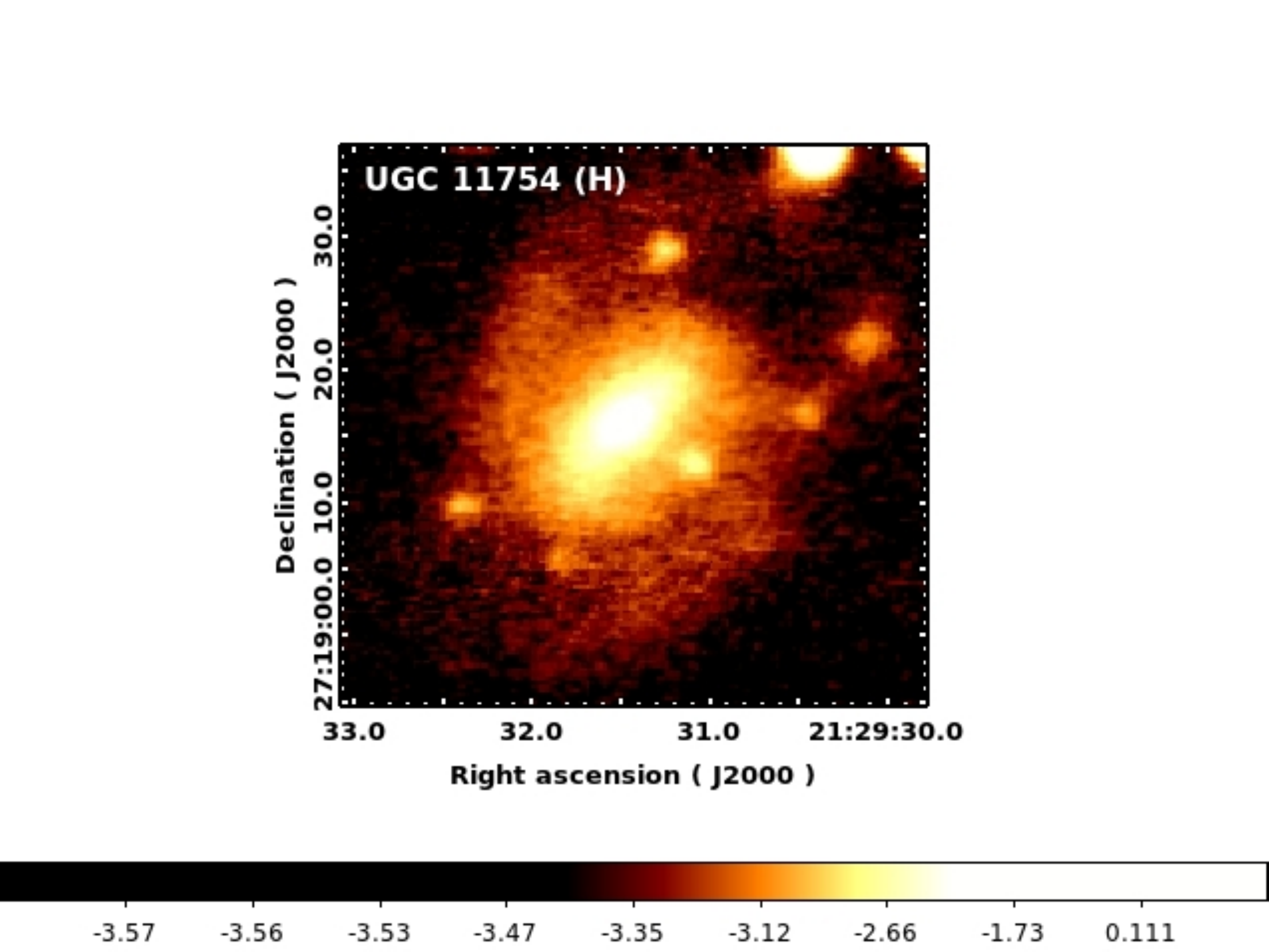}\includegraphics[trim = 35mm 17mm 35mm 15mm, clip,scale=0.44]{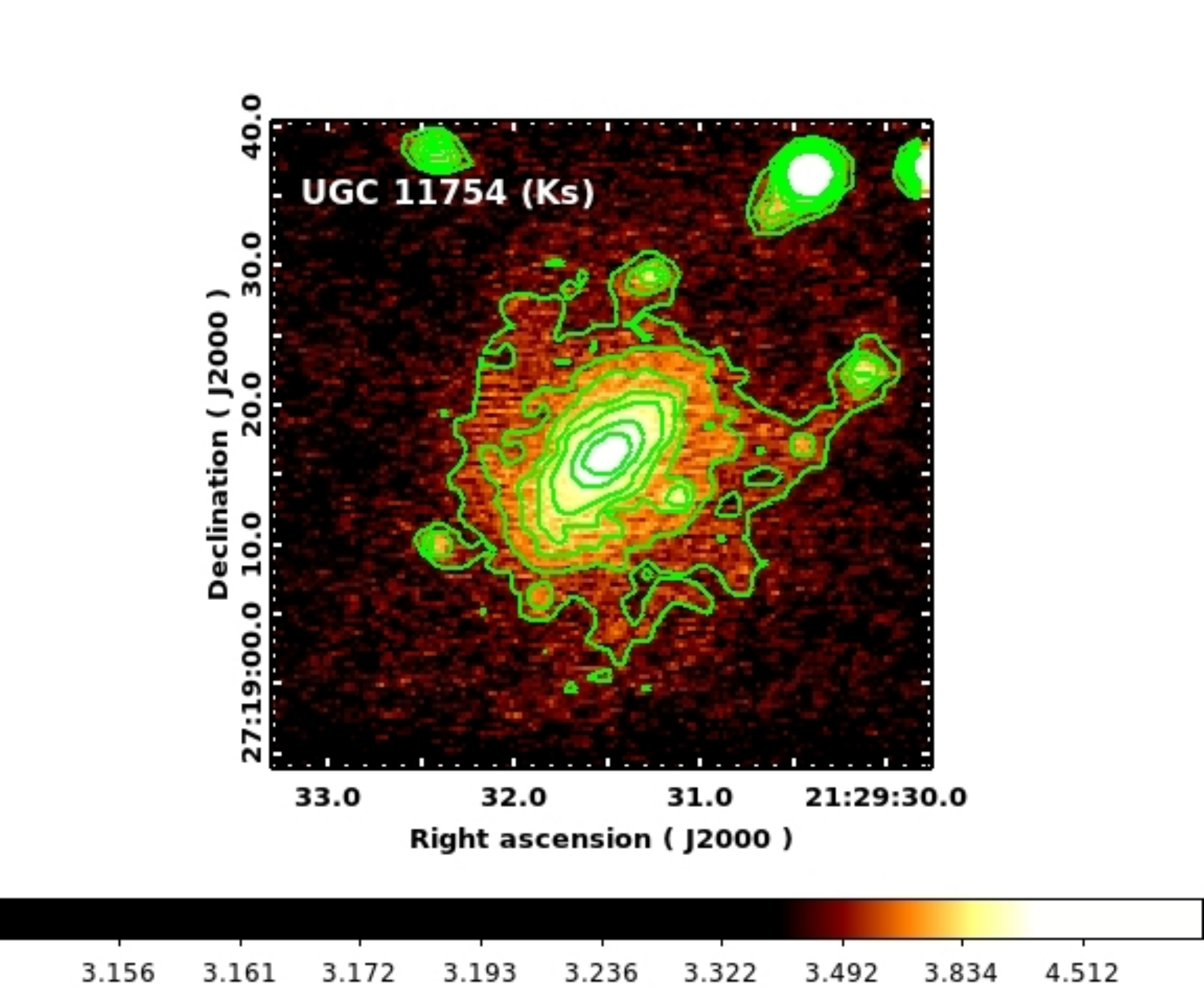}
\includegraphics[trim = 35mm 17mm 45mm 15mm, clip,scale=0.41]{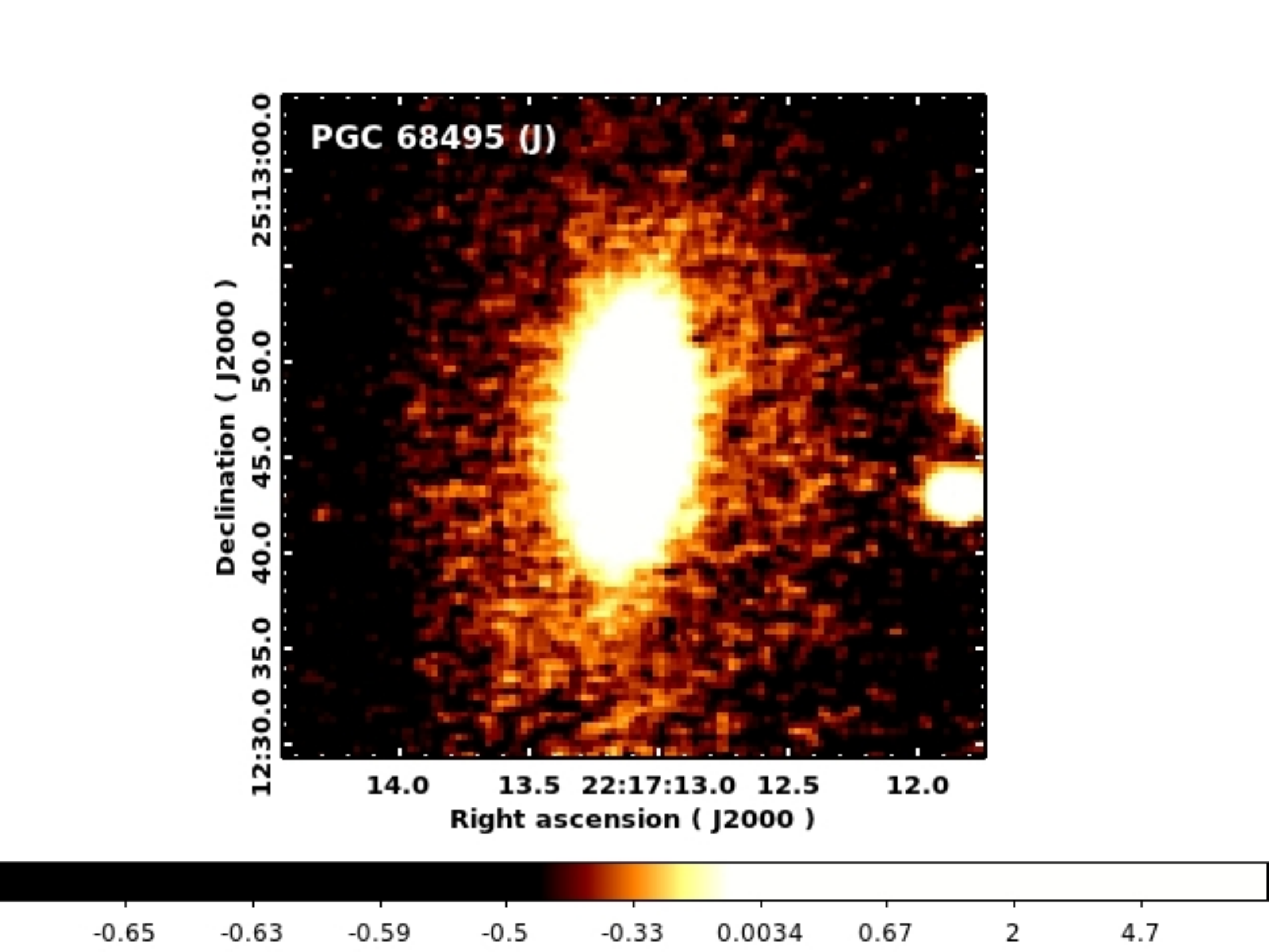}\includegraphics[trim = 35mm 17mm 45mm 15mm, clip,scale=0.41]{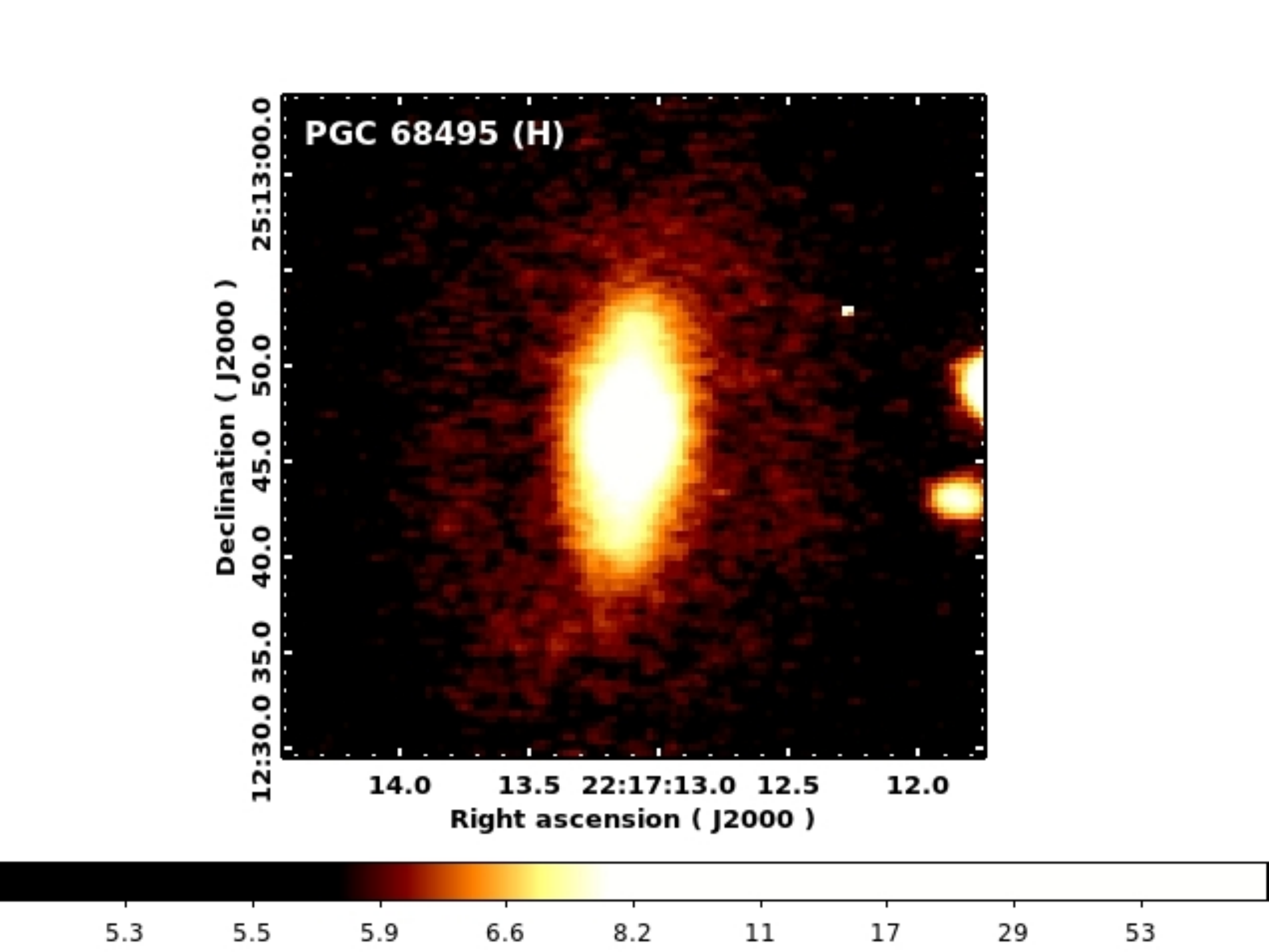}\includegraphics[trim = 35mm 17mm 45mm 15mm, clip,scale=0.41]{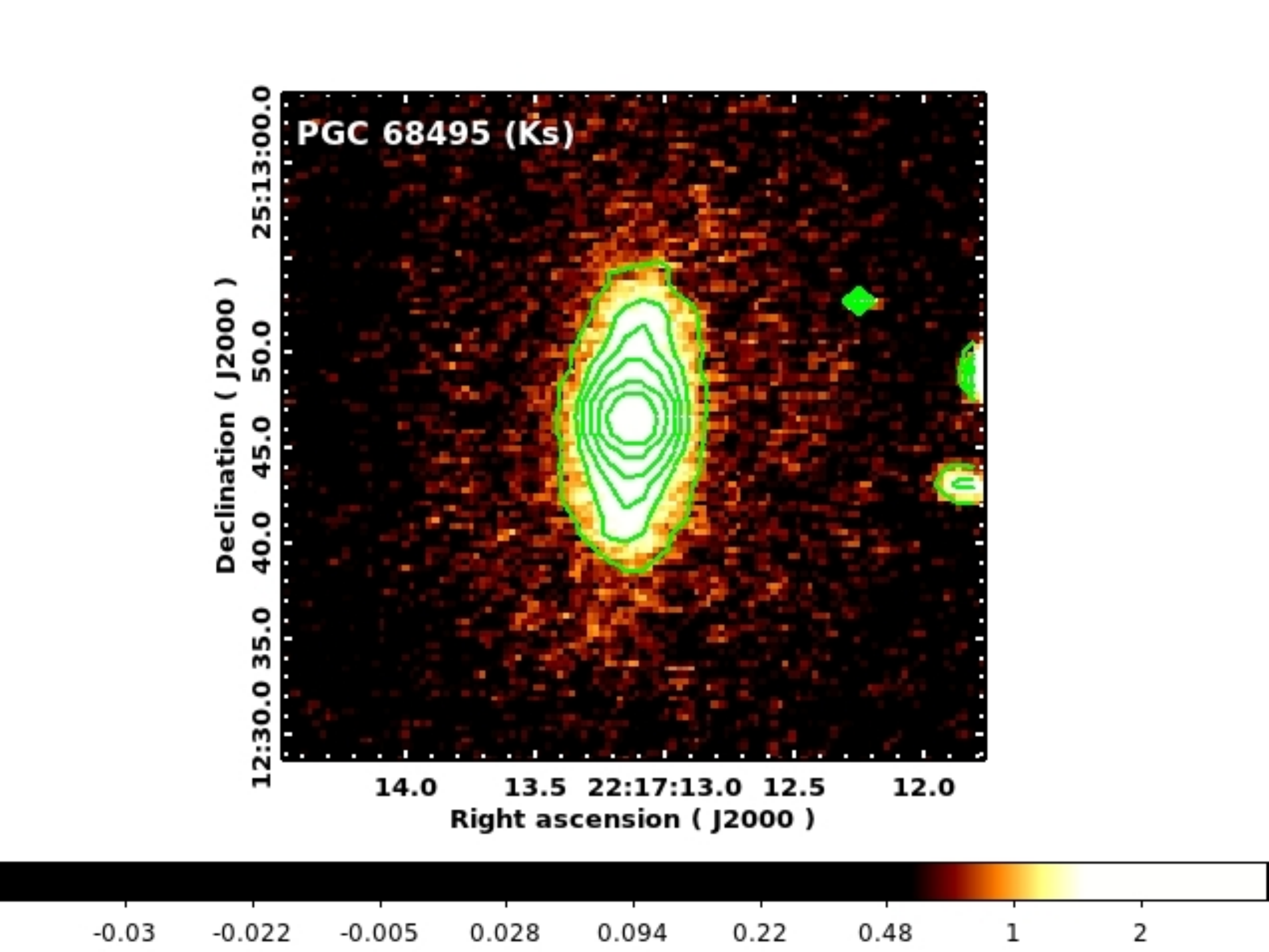}
\includegraphics[trim = 40mm 17mm 40mm 8mm, clip,scale=0.425]{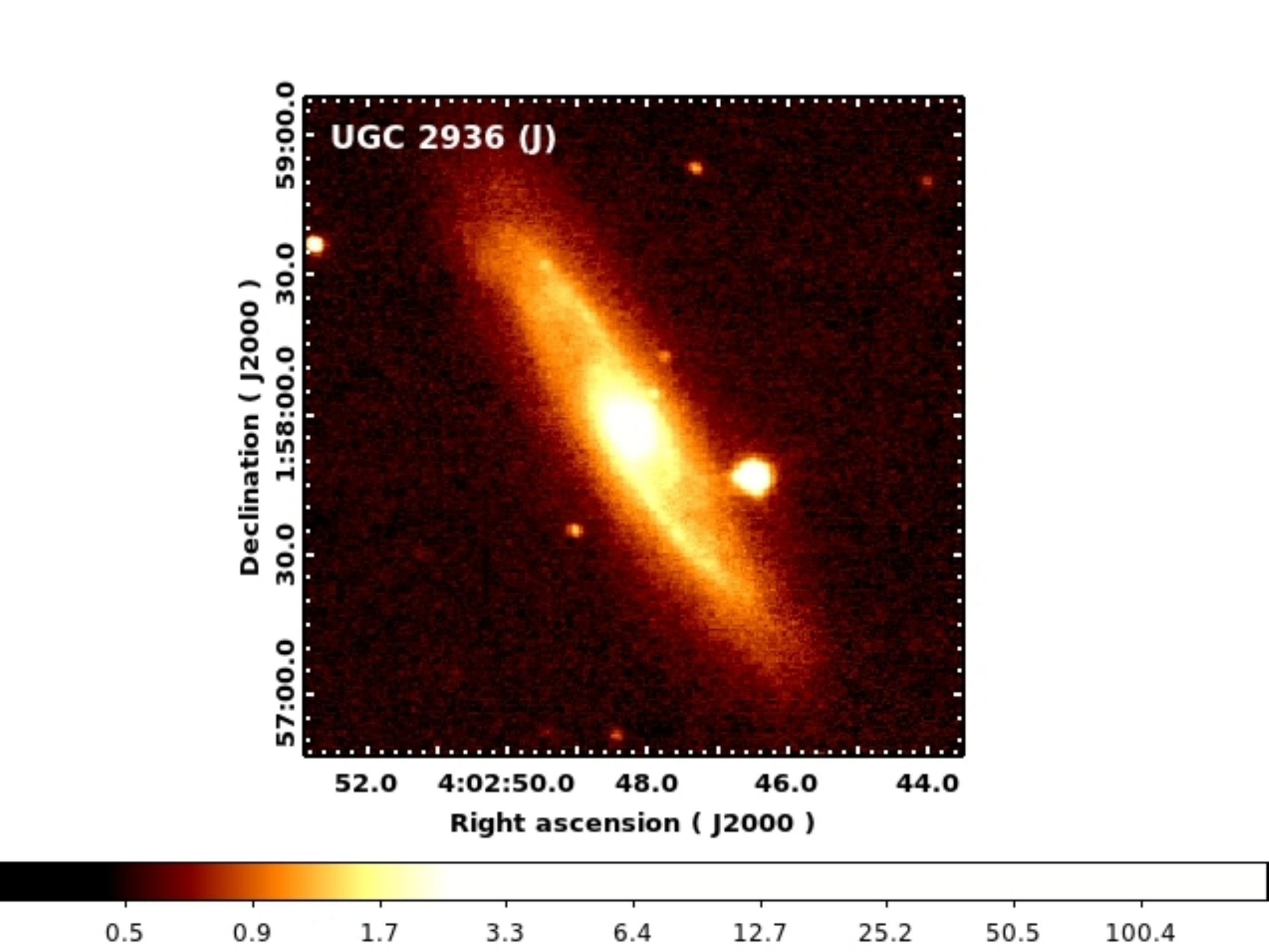}\includegraphics[trim = 40mm 17mm 40mm 8mm, clip,scale=0.425]{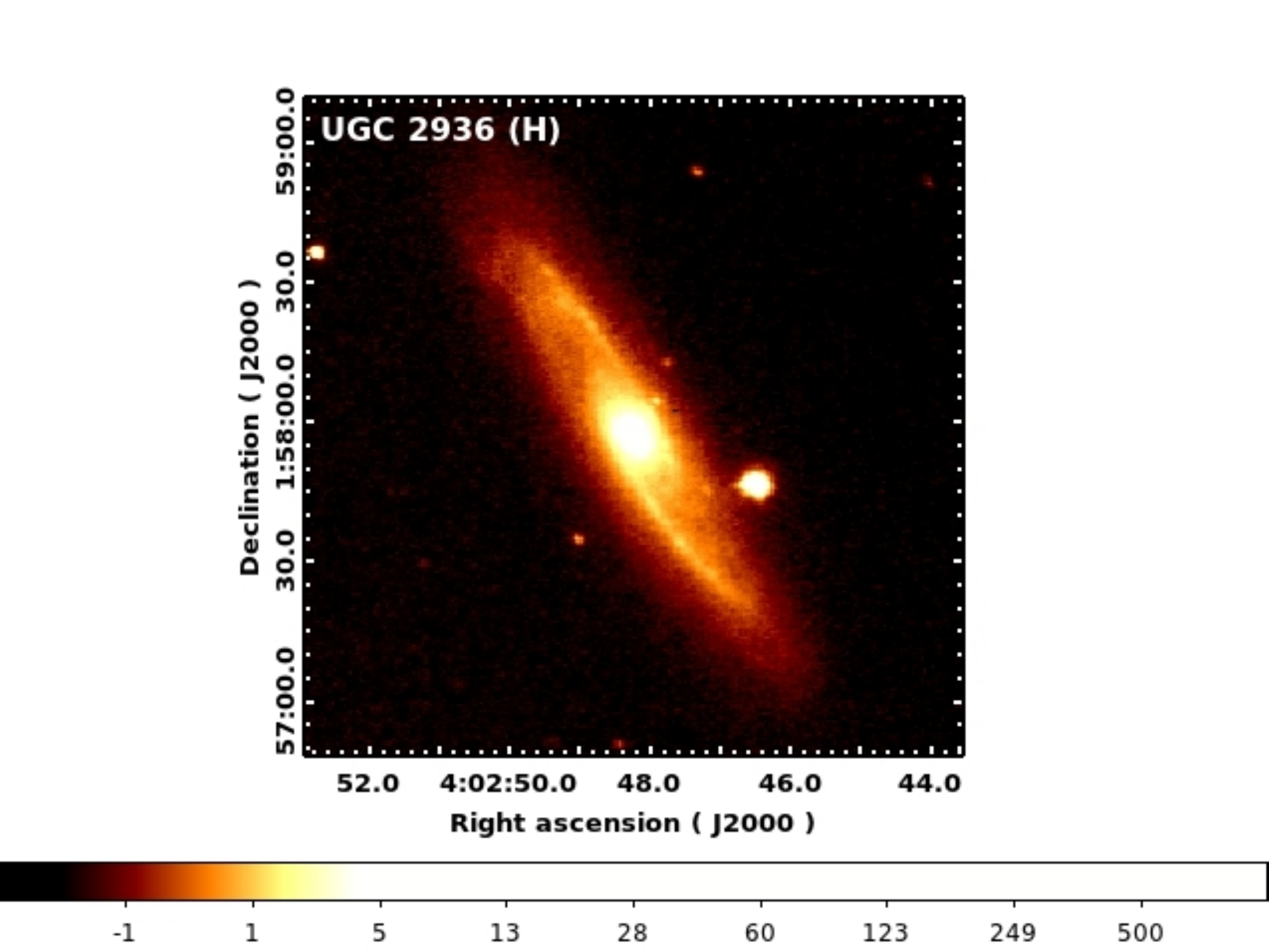}\includegraphics[trim = 40mm 17mm 40mm 8mm, clip,scale=0.425]{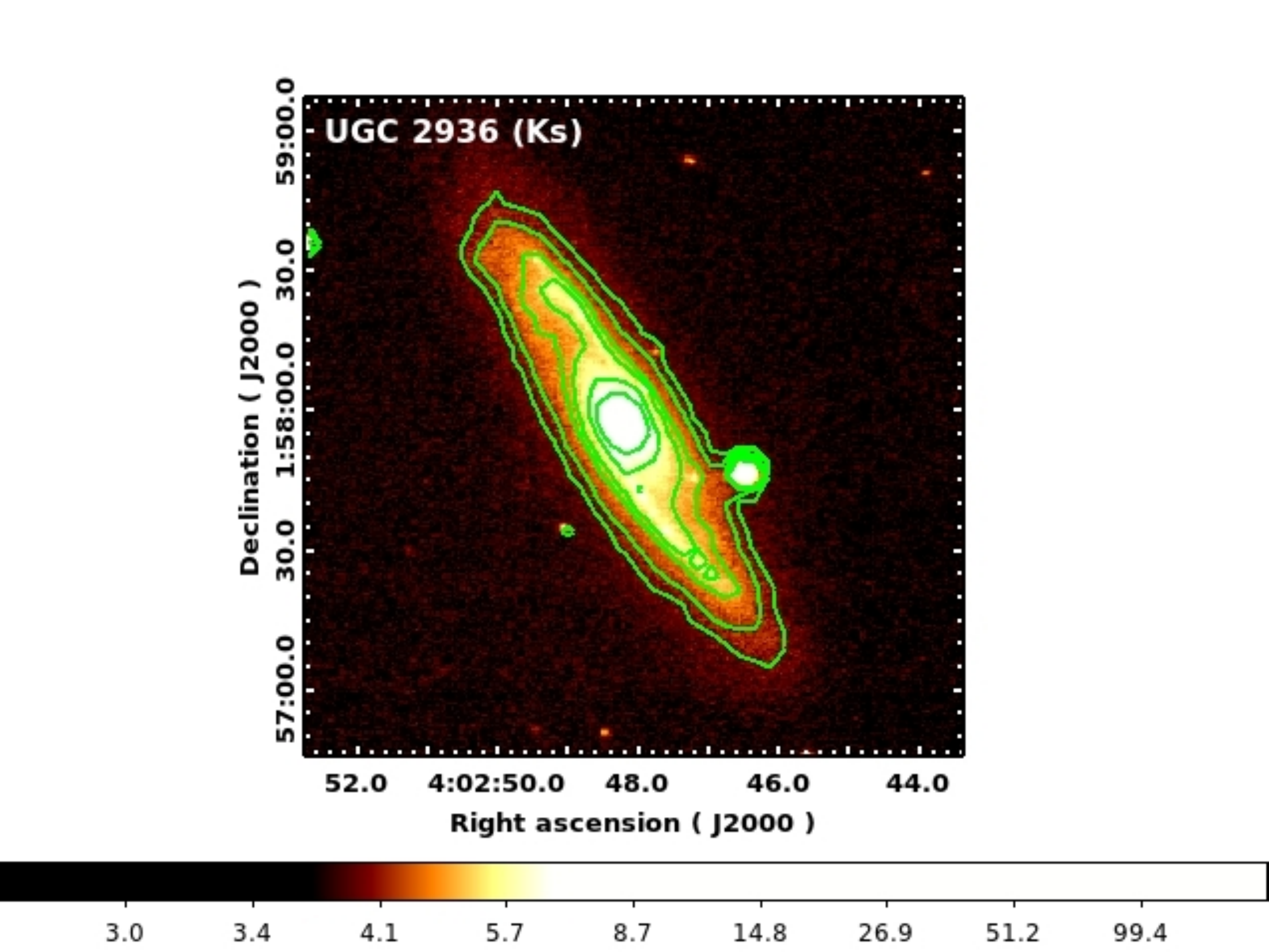}
\end{figure*}

\begin{figure*}
\includegraphics[trim = 40mm 17mm 40mm 8mm, clip,scale=0.44]{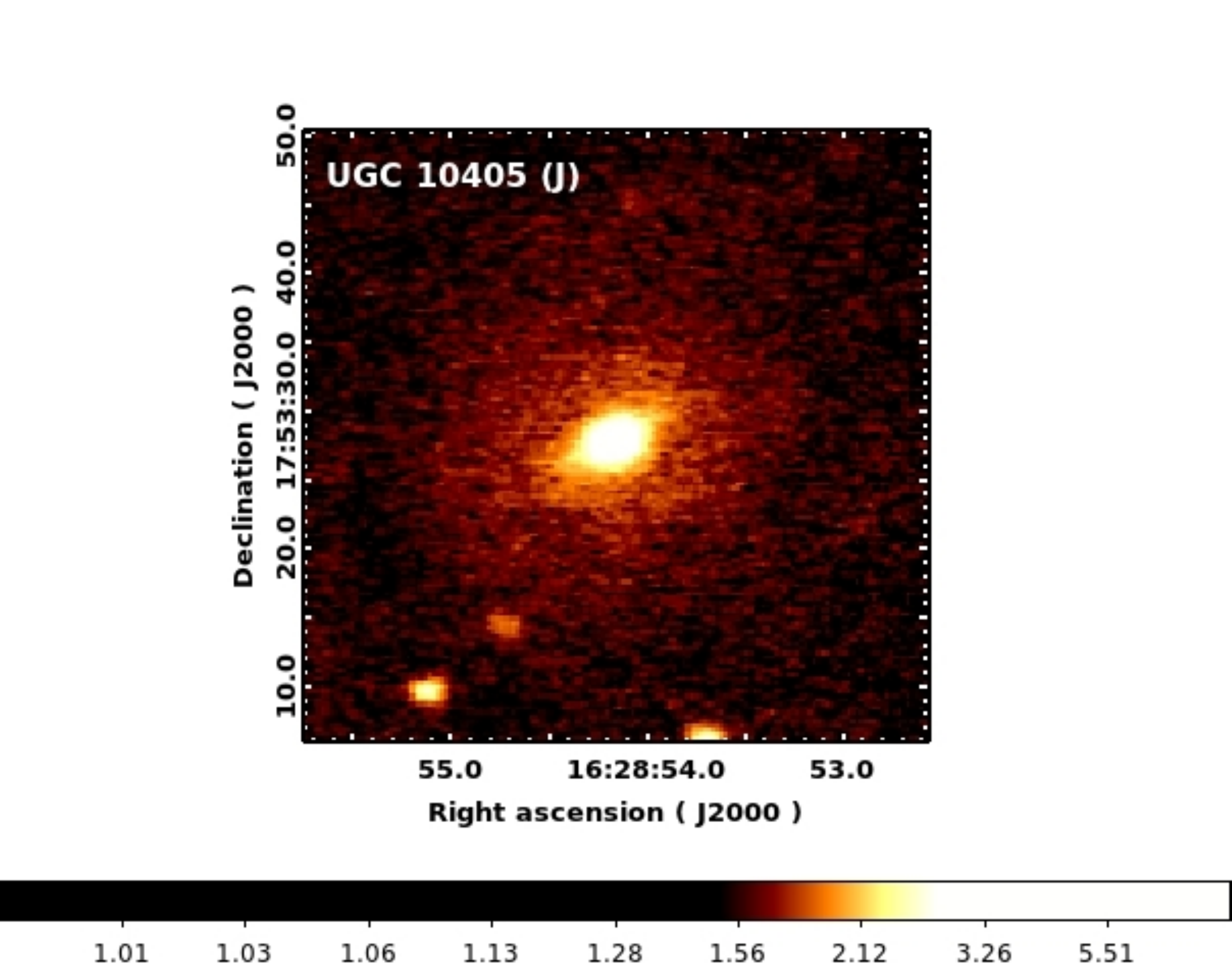}\includegraphics[trim = 40mm 17mm 40mm 8mm, clip,scale=0.44]{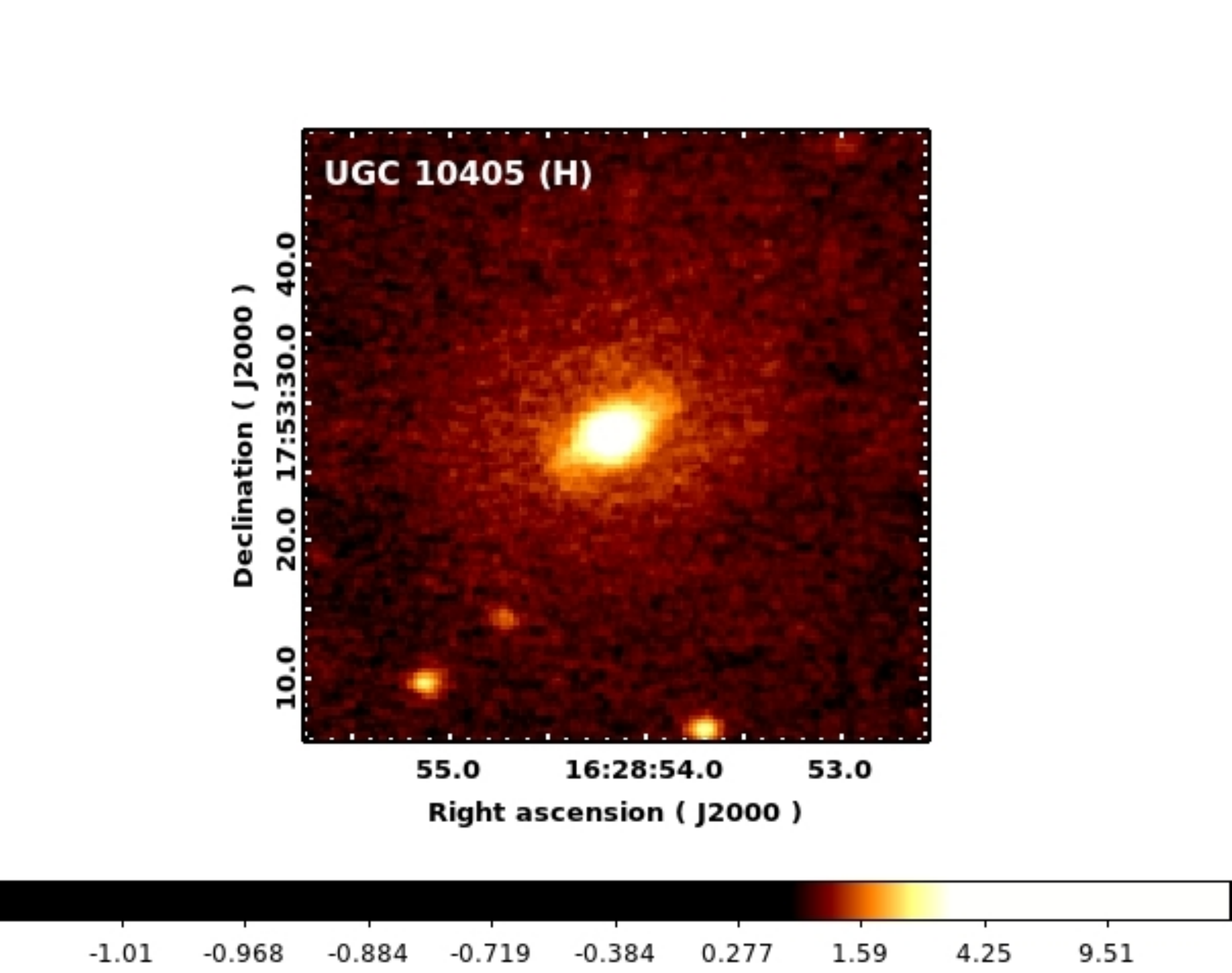}\includegraphics[trim = 40mm 17mm 40mm 8mm, clip,scale=0.44]{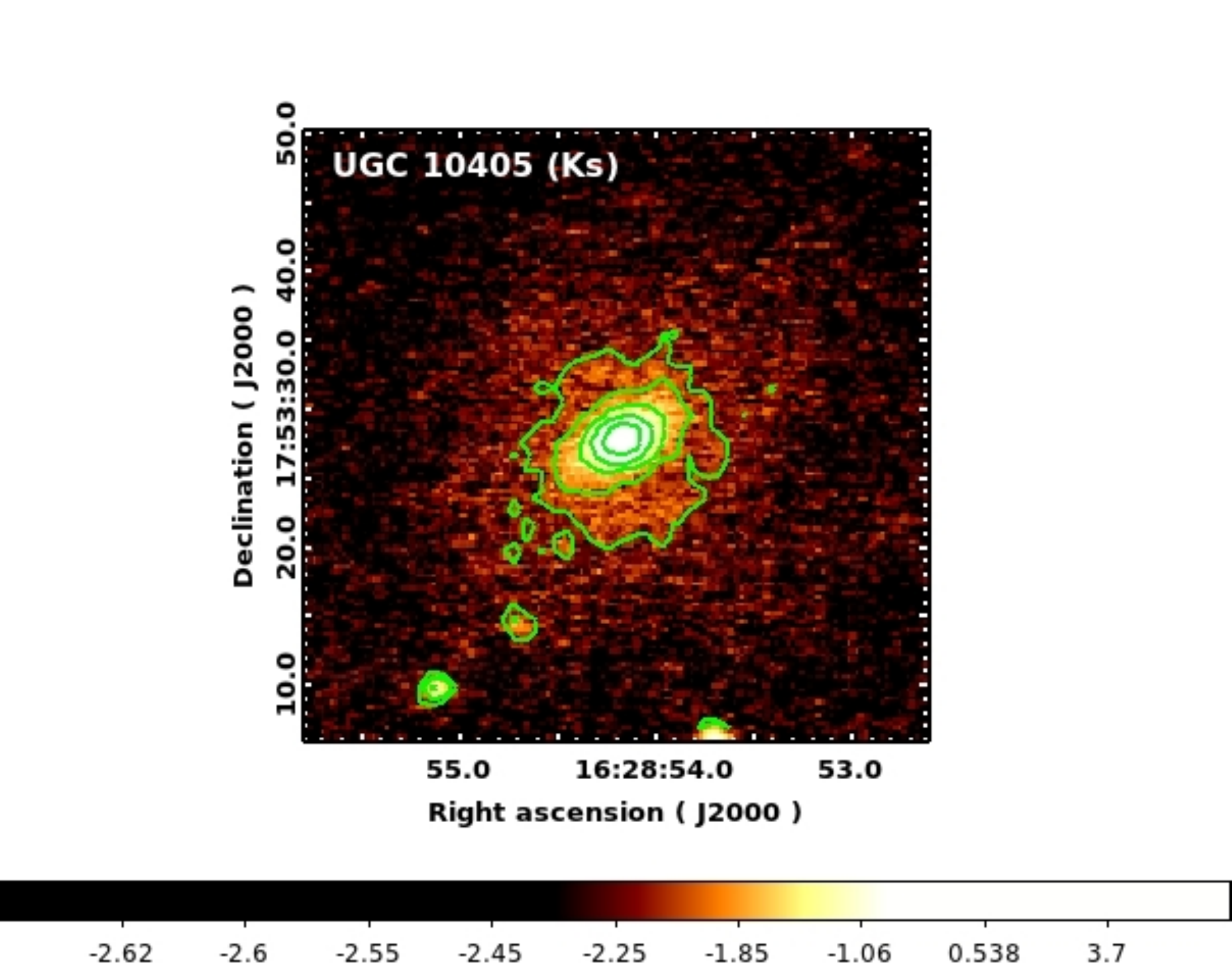}
\includegraphics[trim = 10mm 17mm 20mm 8mm, clip,scale=0.4]{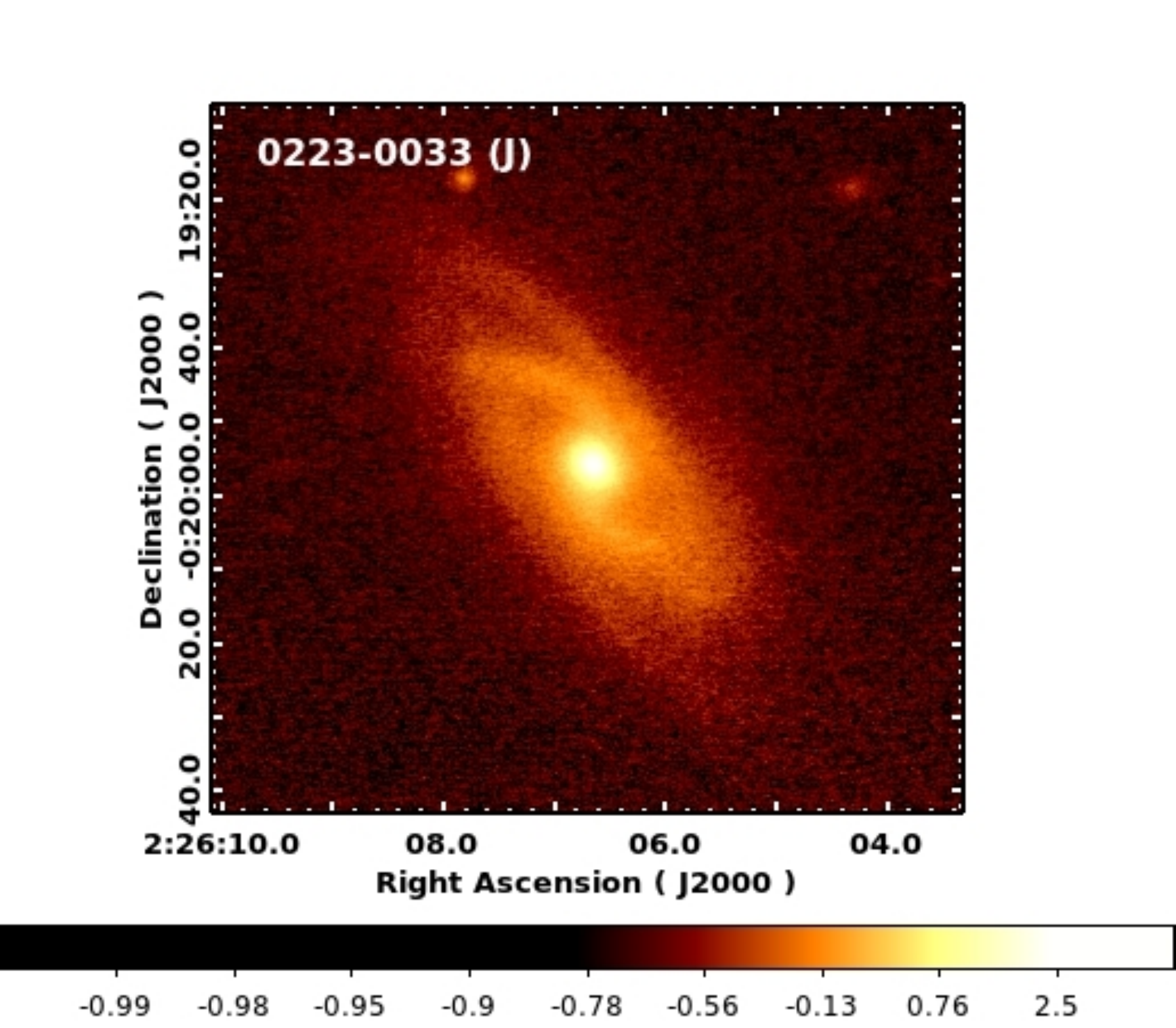}\includegraphics[trim = 17mm 17mm 20mm 8mm, clip,scale=0.4]{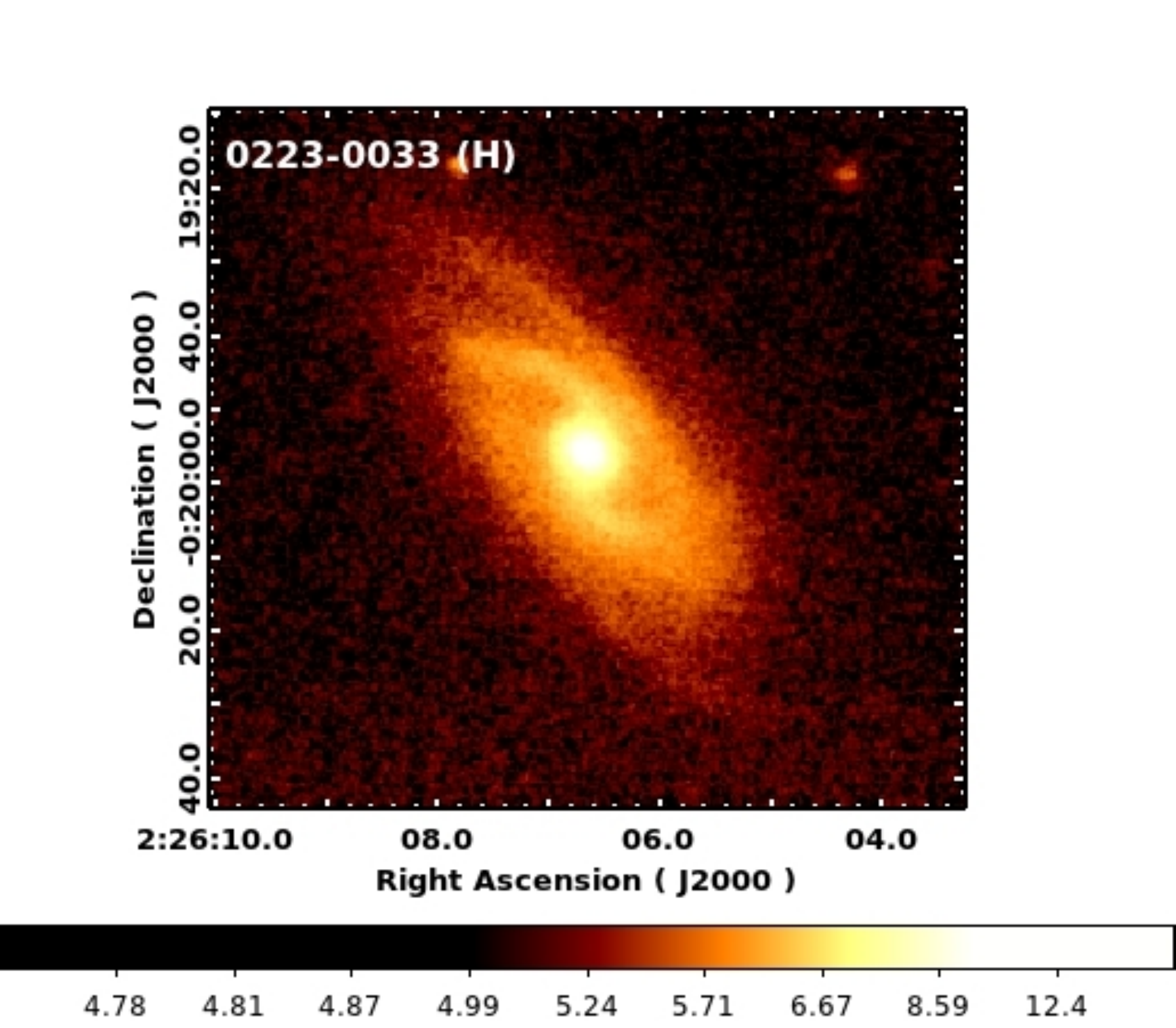}\includegraphics[trim = 17mm 17mm 20mm 8mm, clip,scale=0.4]{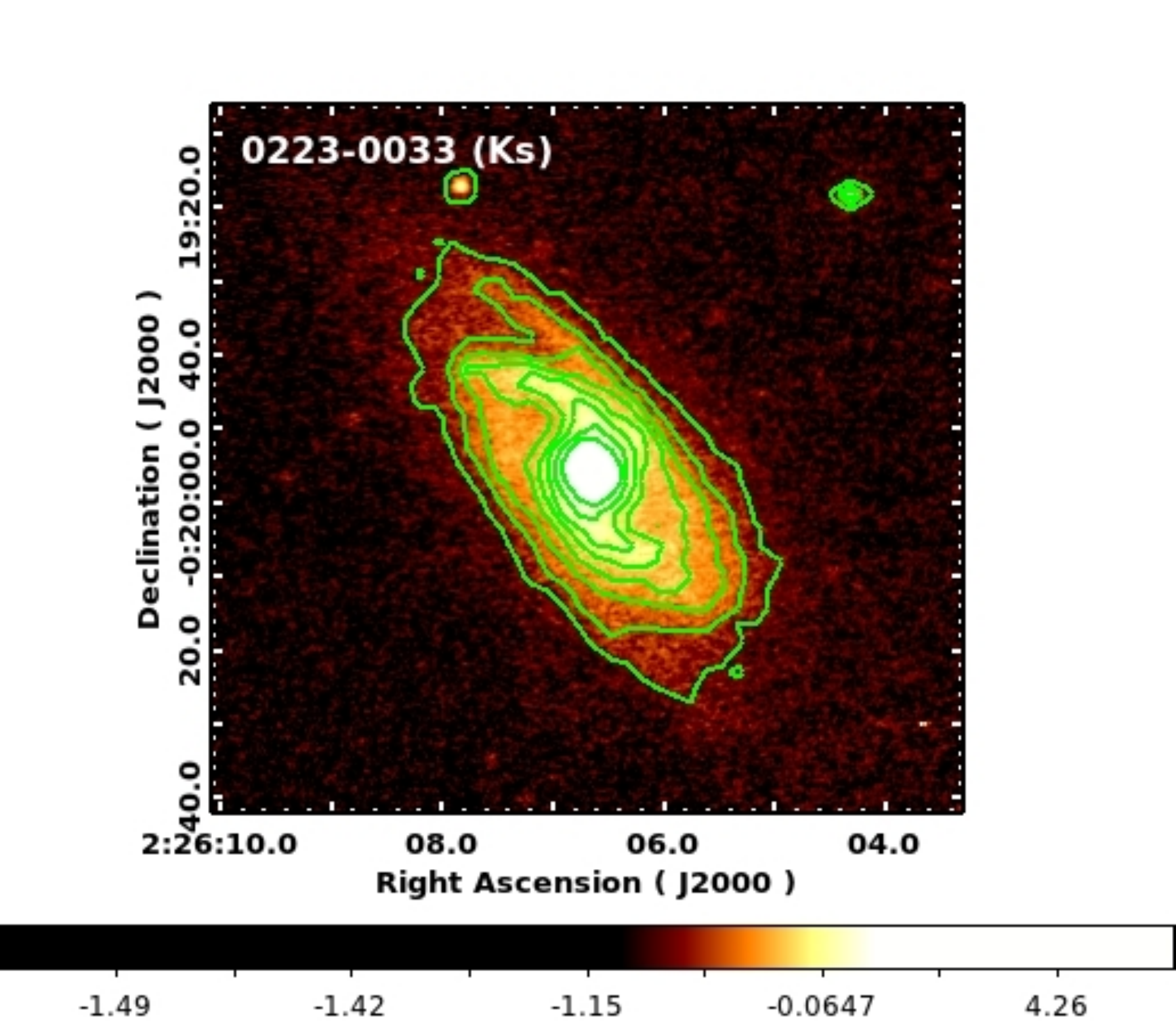}

\includegraphics[trim = 25mm 17mm 30mm 8mm, clip,scale=0.46]{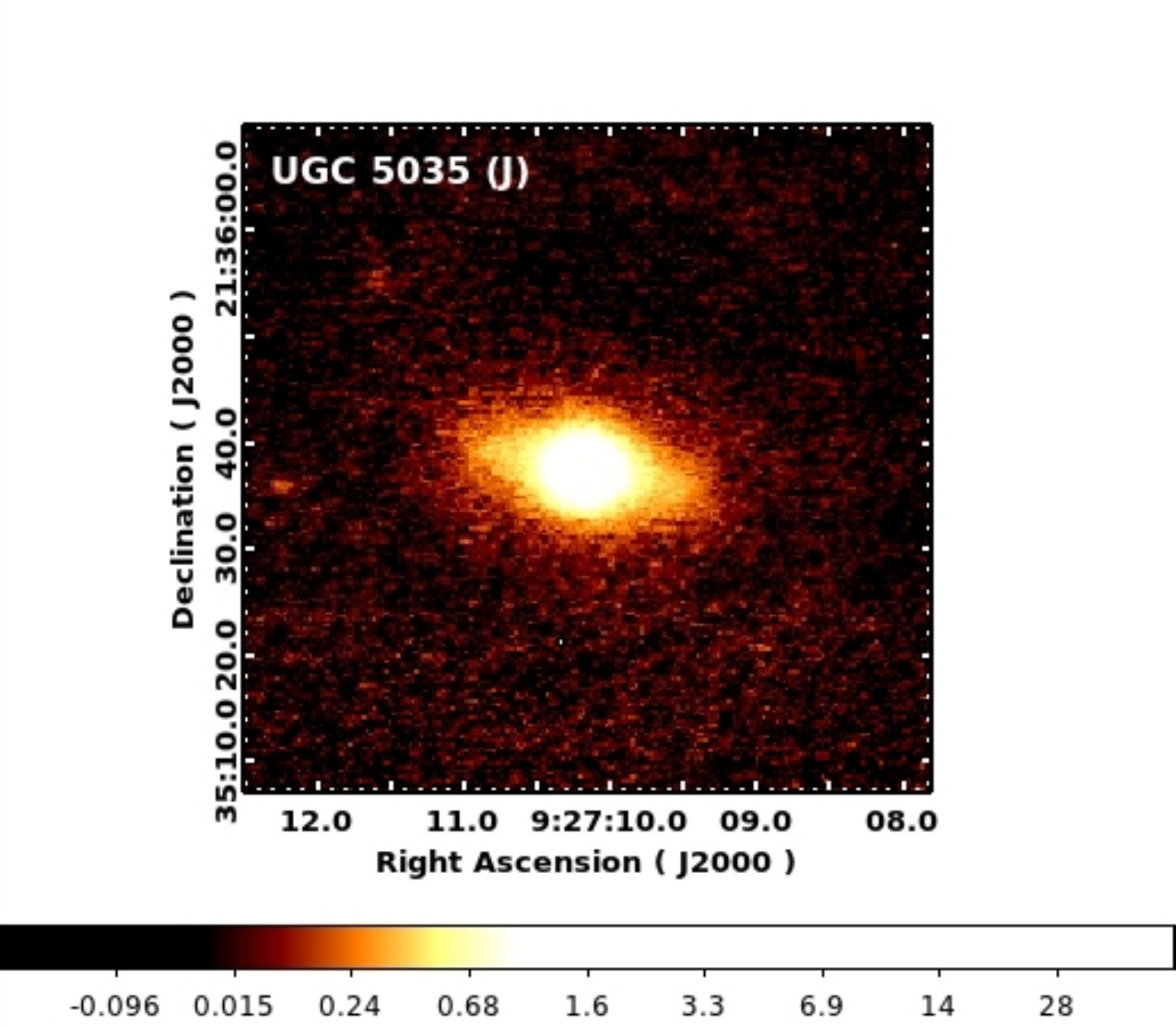}\includegraphics[trim = 25mm 17mm 30mm 8mm, clip,scale=0.46]{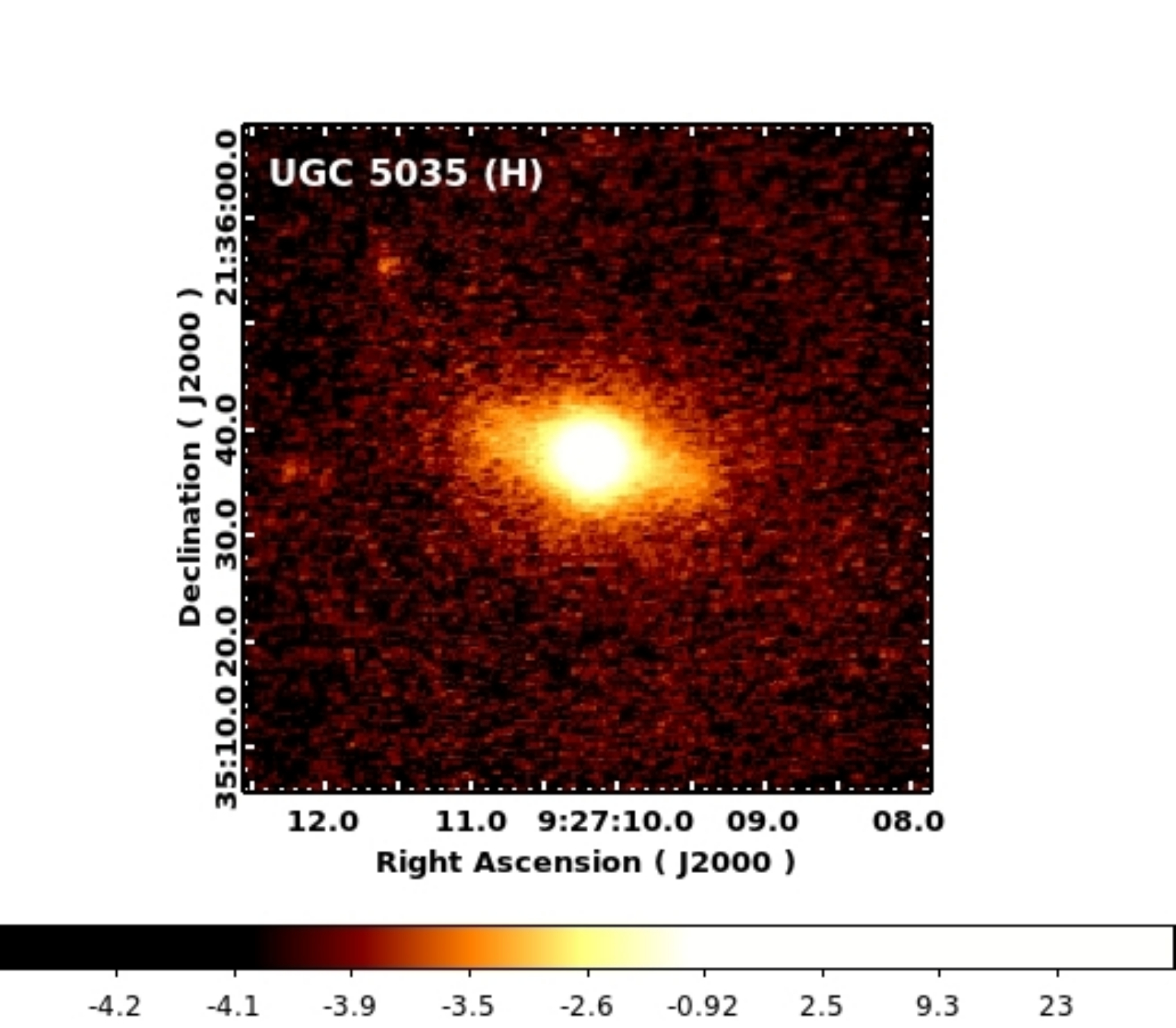}\includegraphics[trim = 25mm 17mm 30mm 8mm, clip,scale=0.46]{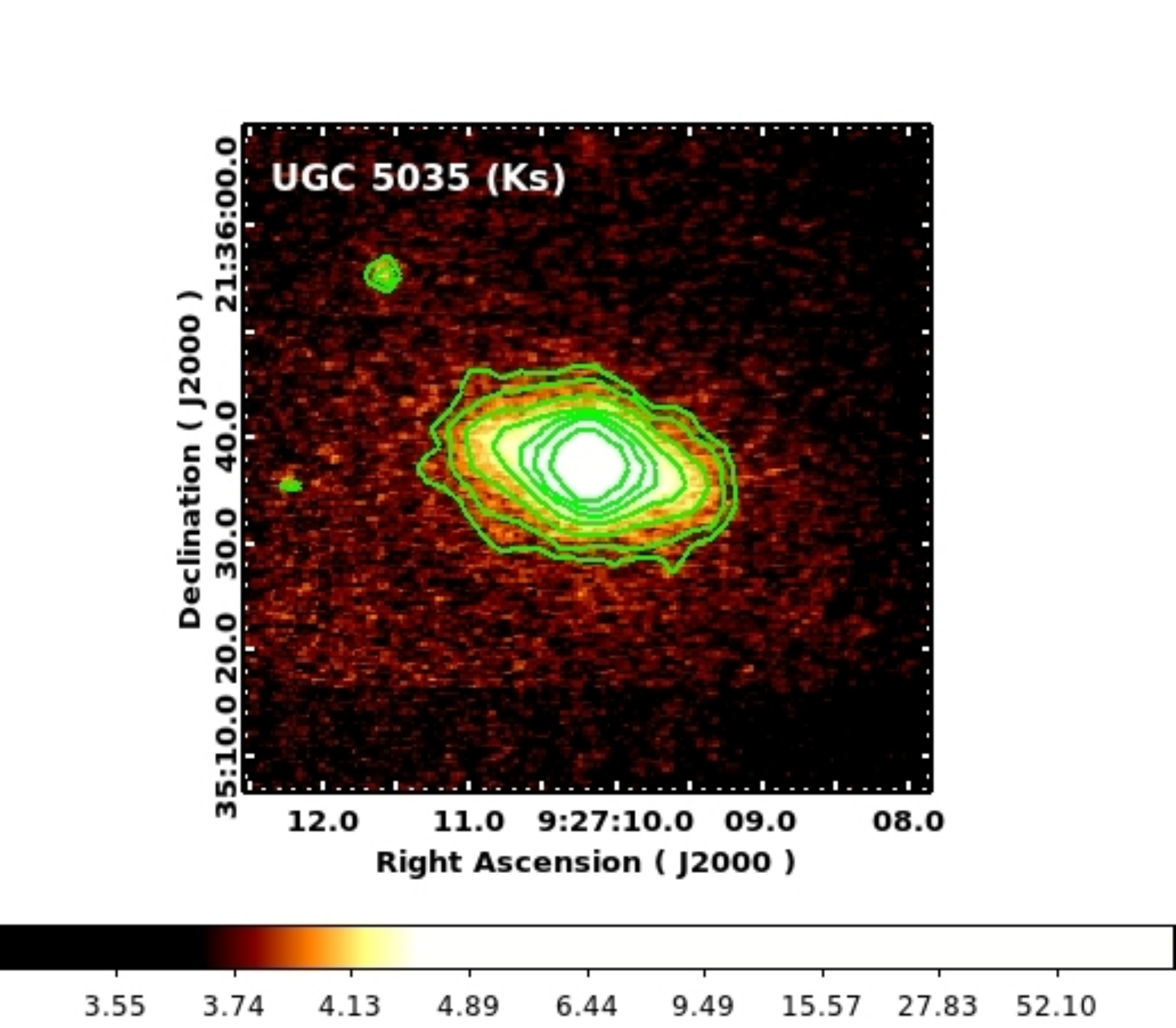}
\includegraphics[trim = 25mm 17mm 30mm 8mm, clip,scale=0.46]{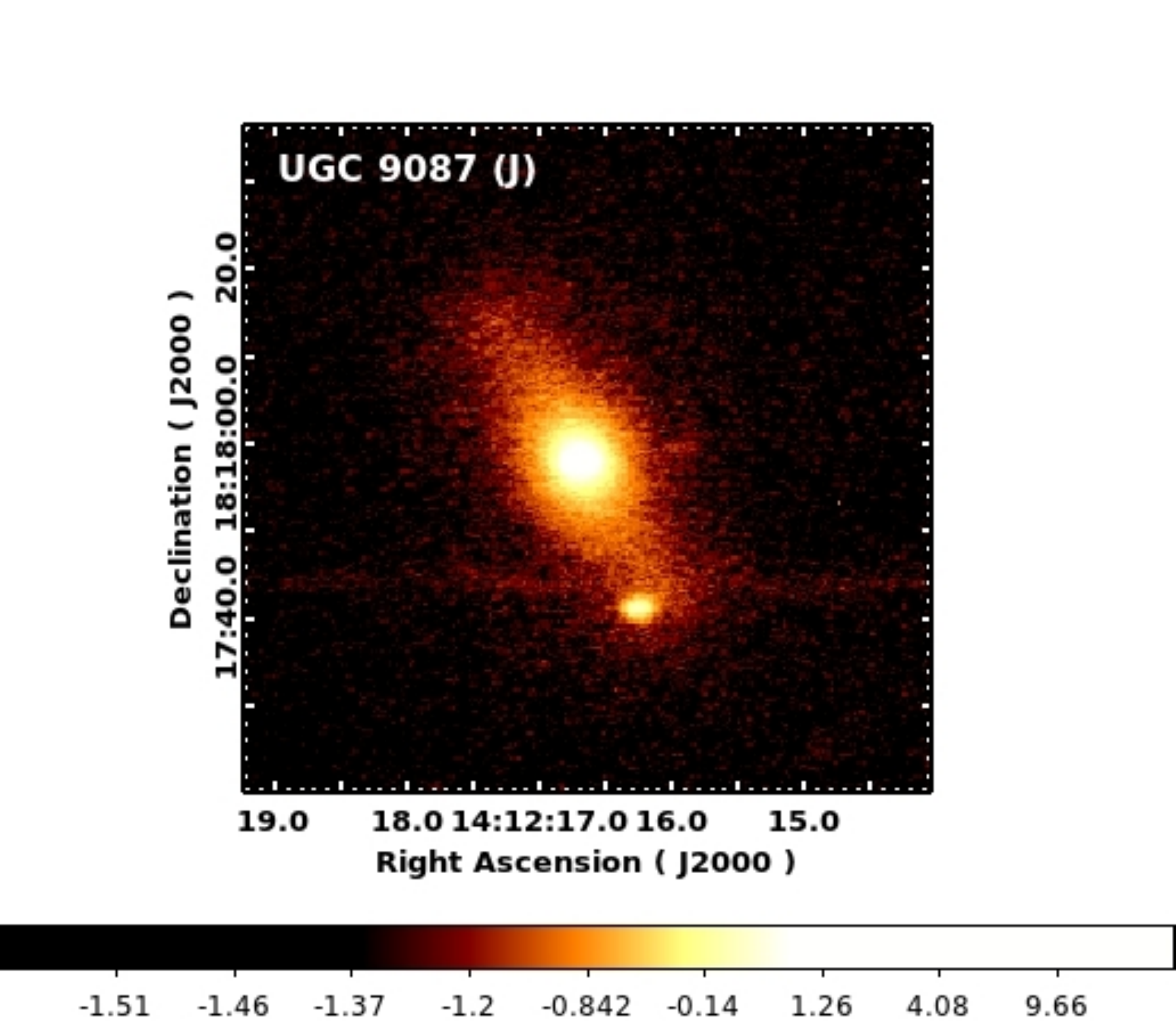}\includegraphics[trim = 25mm 17mm 30mm 8mm, clip,scale=0.46]{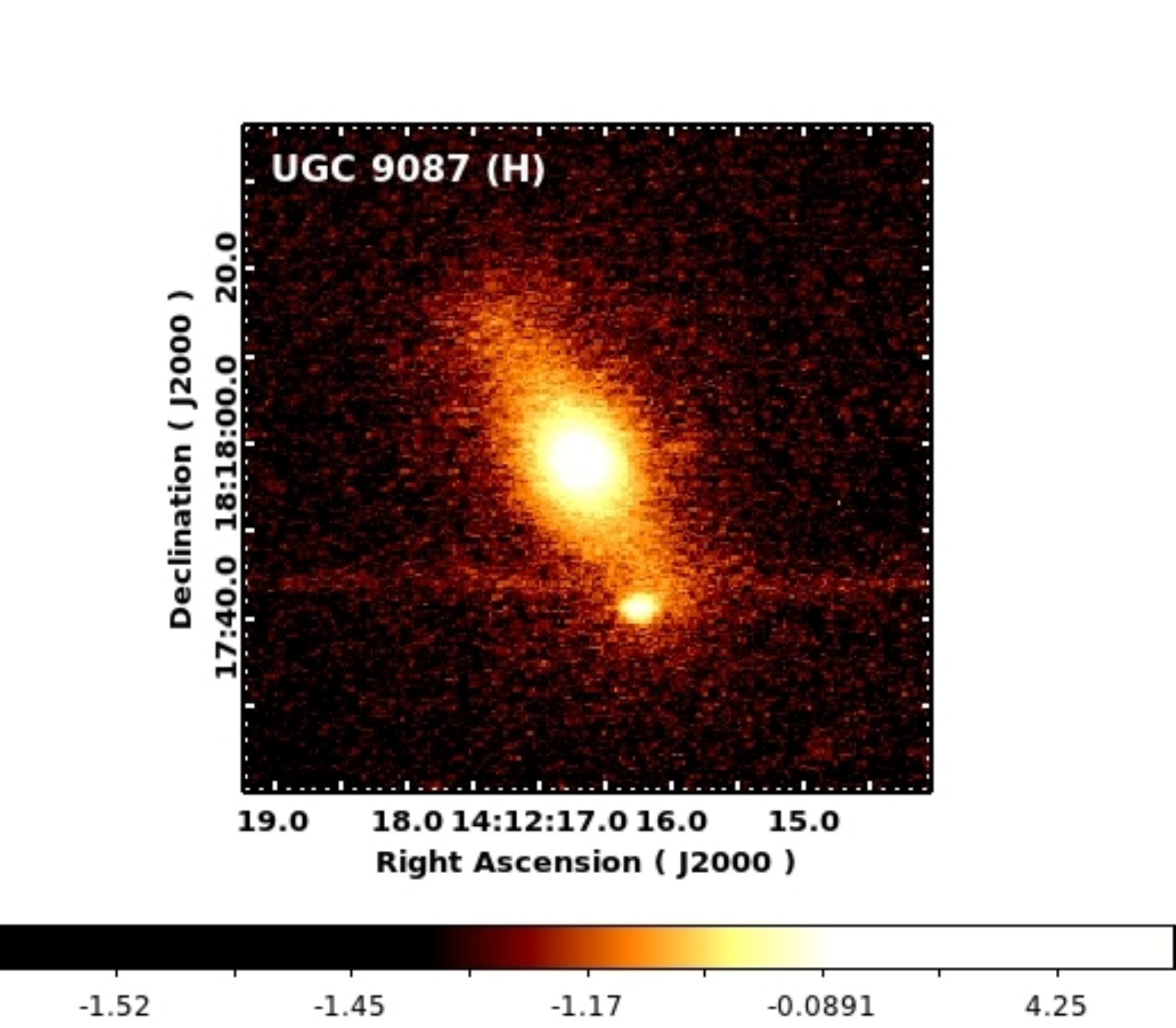}\includegraphics[trim = 25mm 17mm 30mm 8mm, clip,scale=0.46]{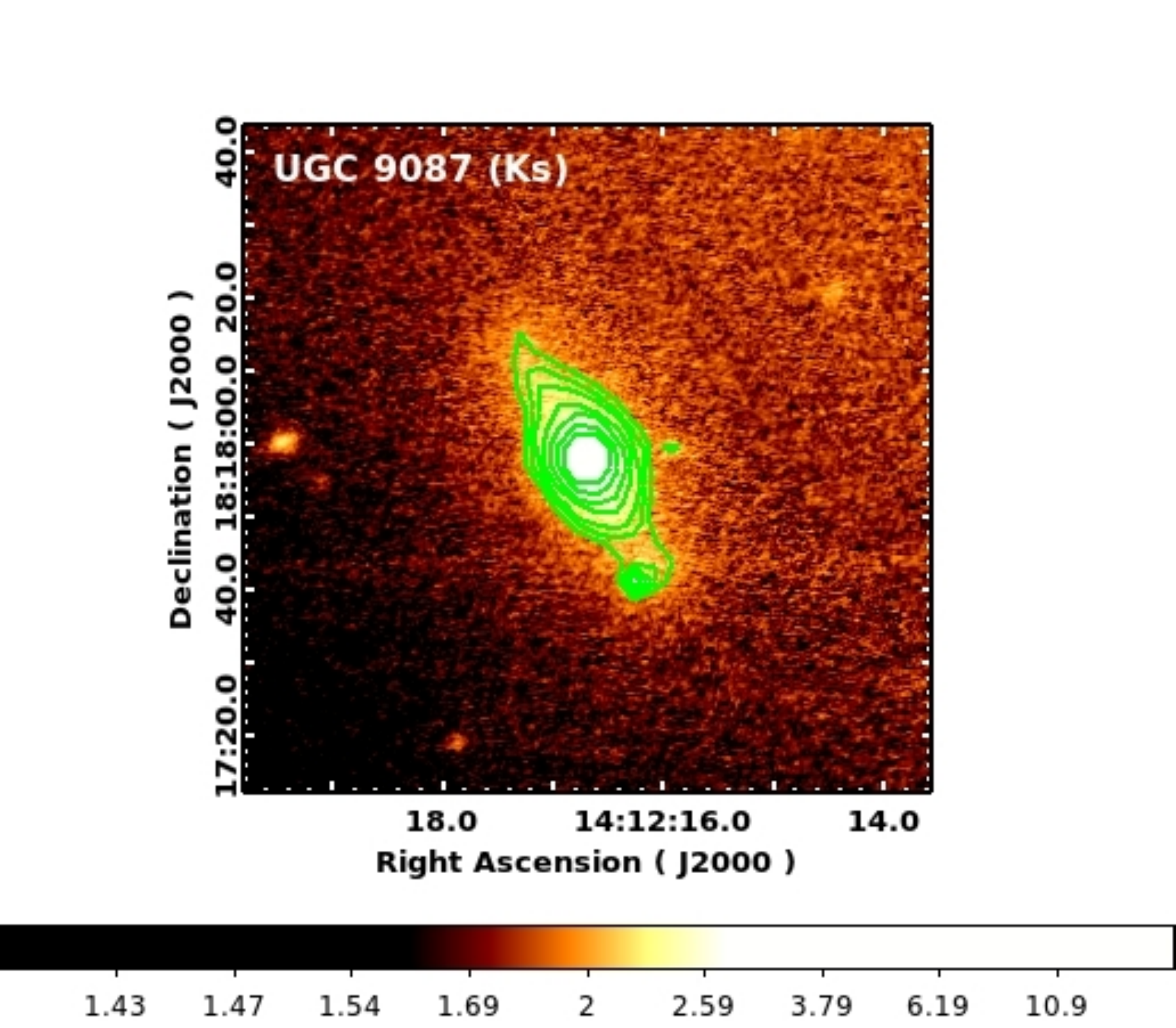}
\end{figure*}
\begin{figure*}
\includegraphics[trim = 25mm 17mm 30mm 8mm, clip,scale=0.44]{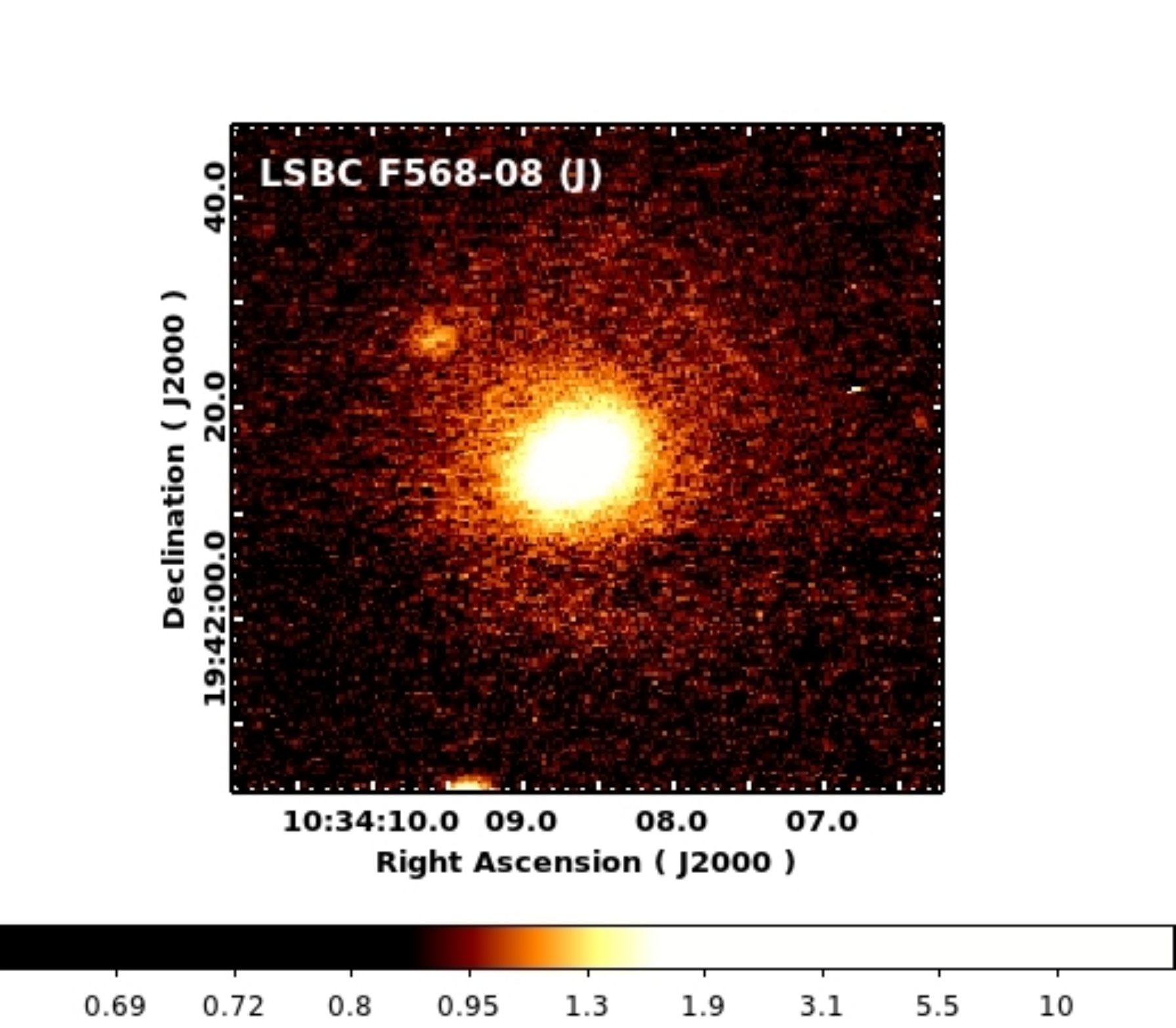}\includegraphics[trim = 25mm 22mm 30mm 8mm, clip,scale=0.45]{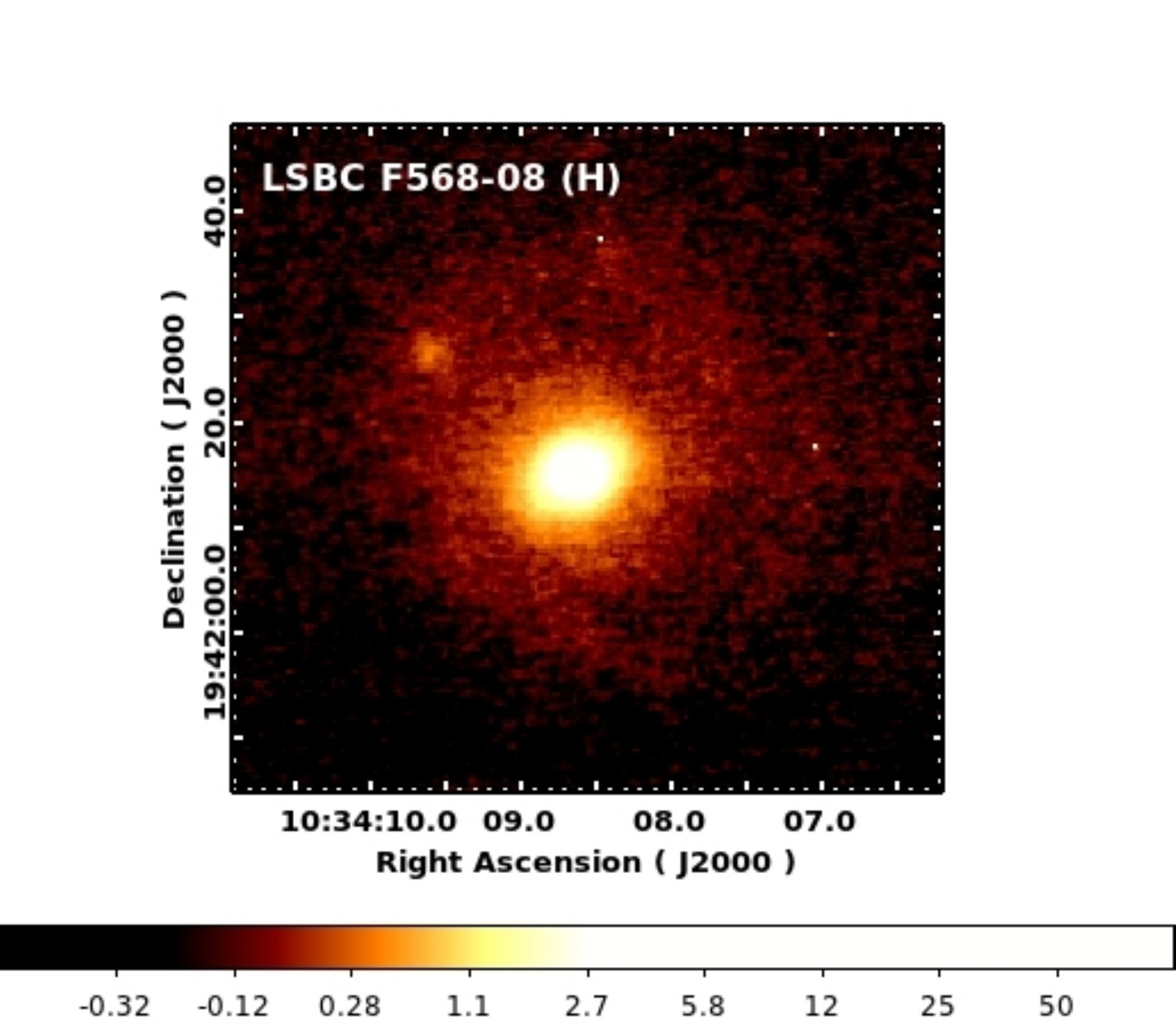}\includegraphics[trim = 20mm 17mm 30mm 8mm, clip,scale=0.44]{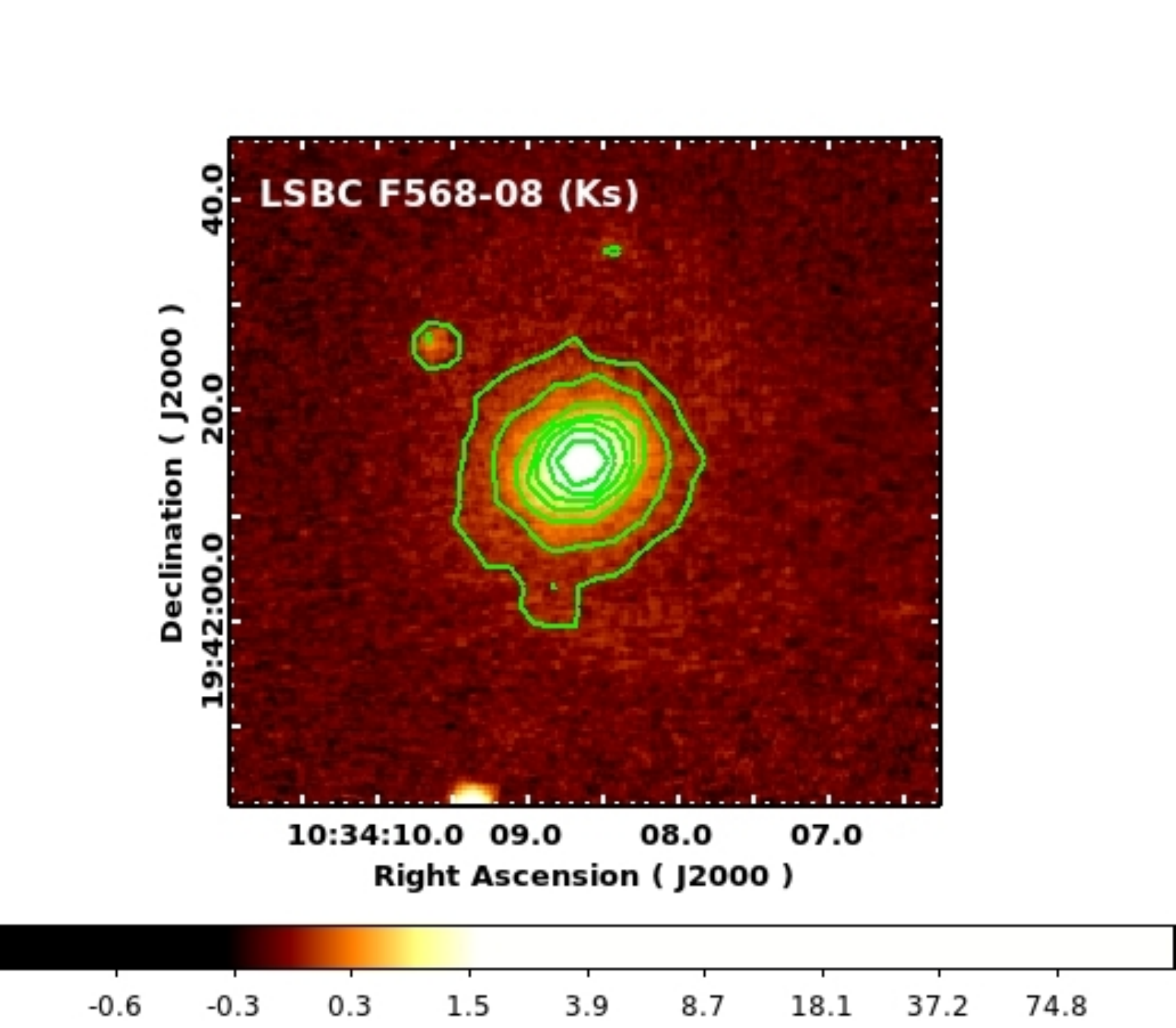}

\includegraphics[trim = 58mm 17mm 60mm 8mm, clip,scale=0.46]{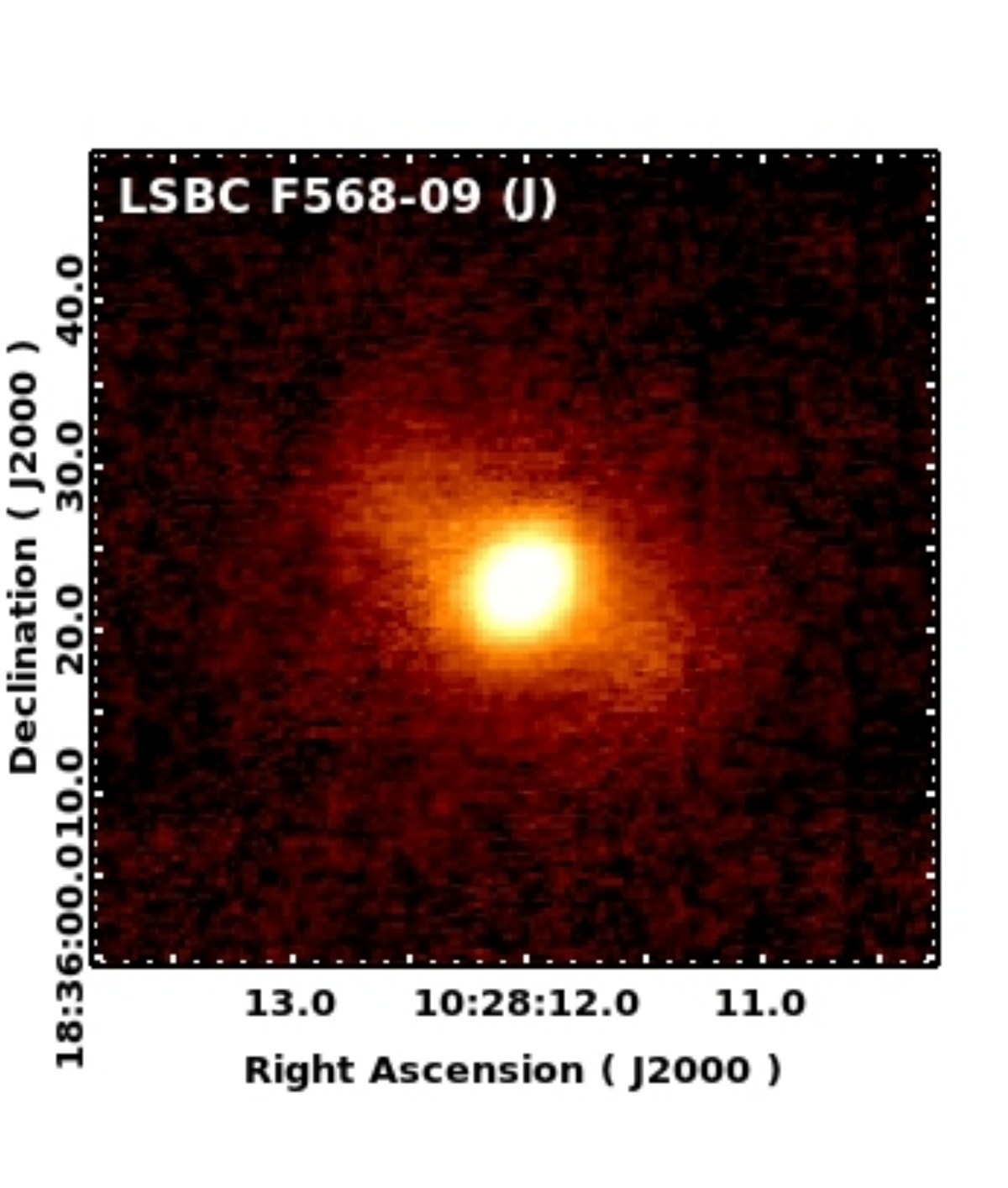}\includegraphics[trim = 0mm 0mm 0mm 0mm, clip,scale=0.39]{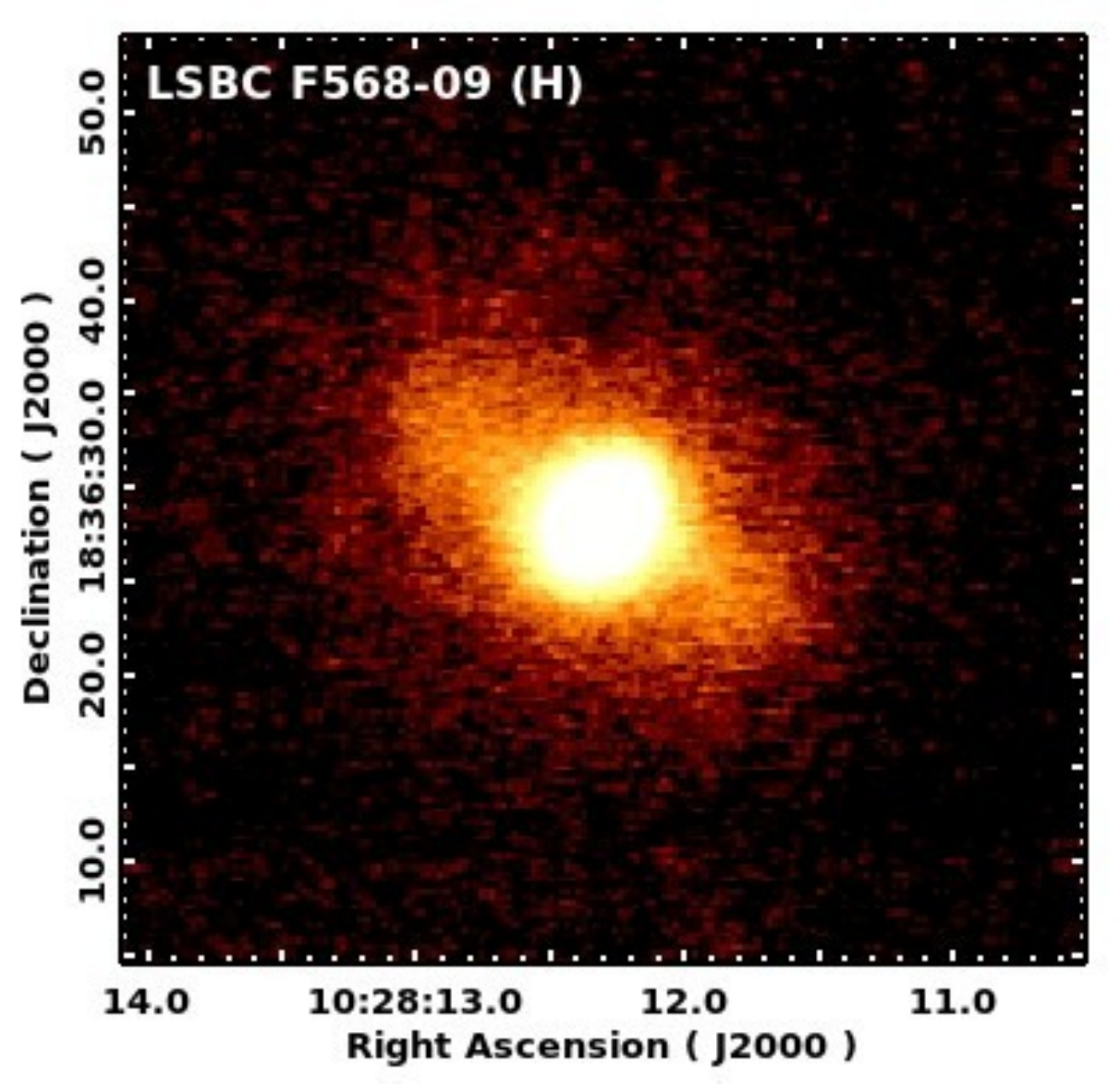}\includegraphics[trim = 48mm 17mm 40mm 8mm, clip,scale=0.46]{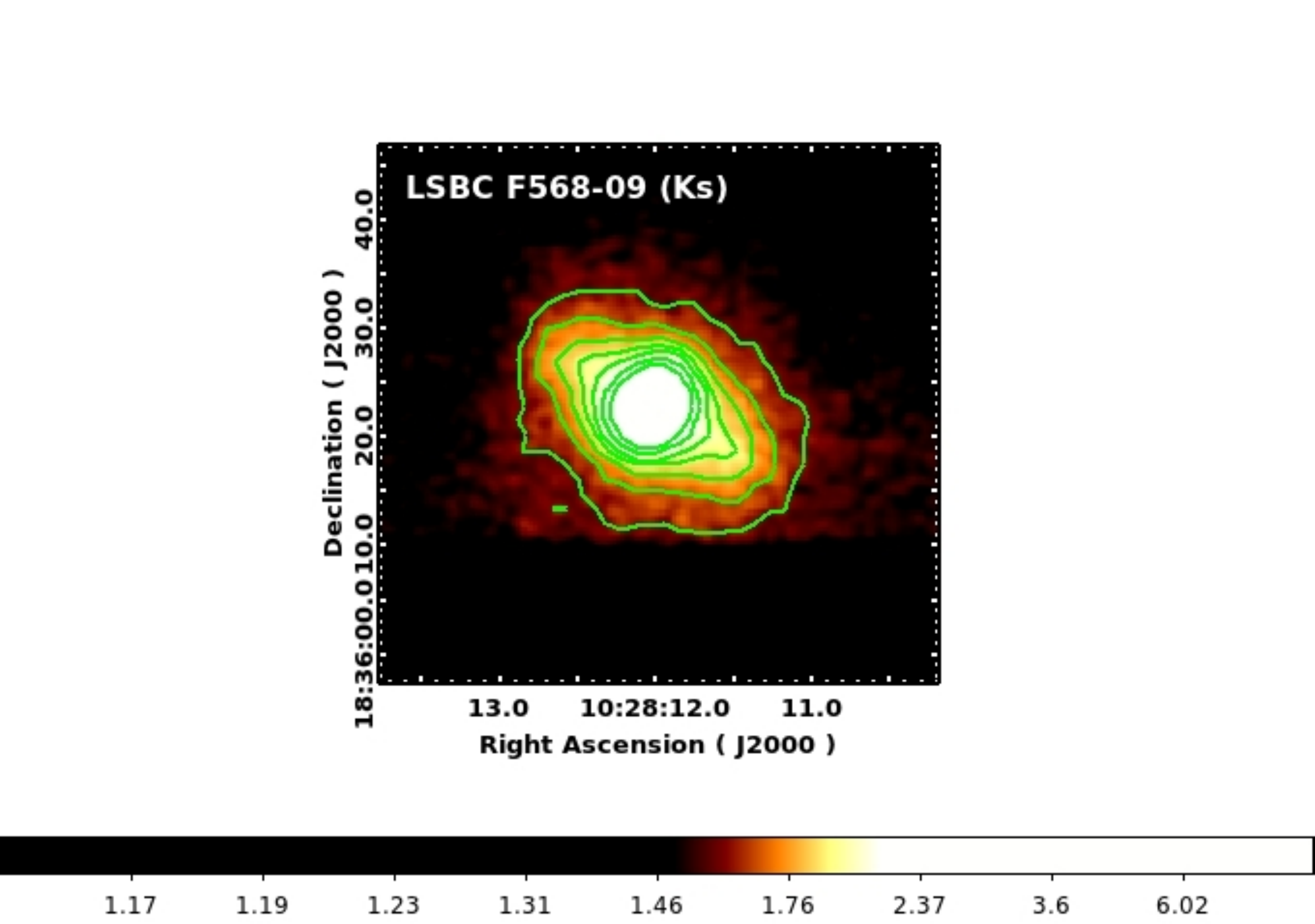}
\includegraphics[trim = 45mm 17mm 50mm 8mm, clip,scale=0.39]{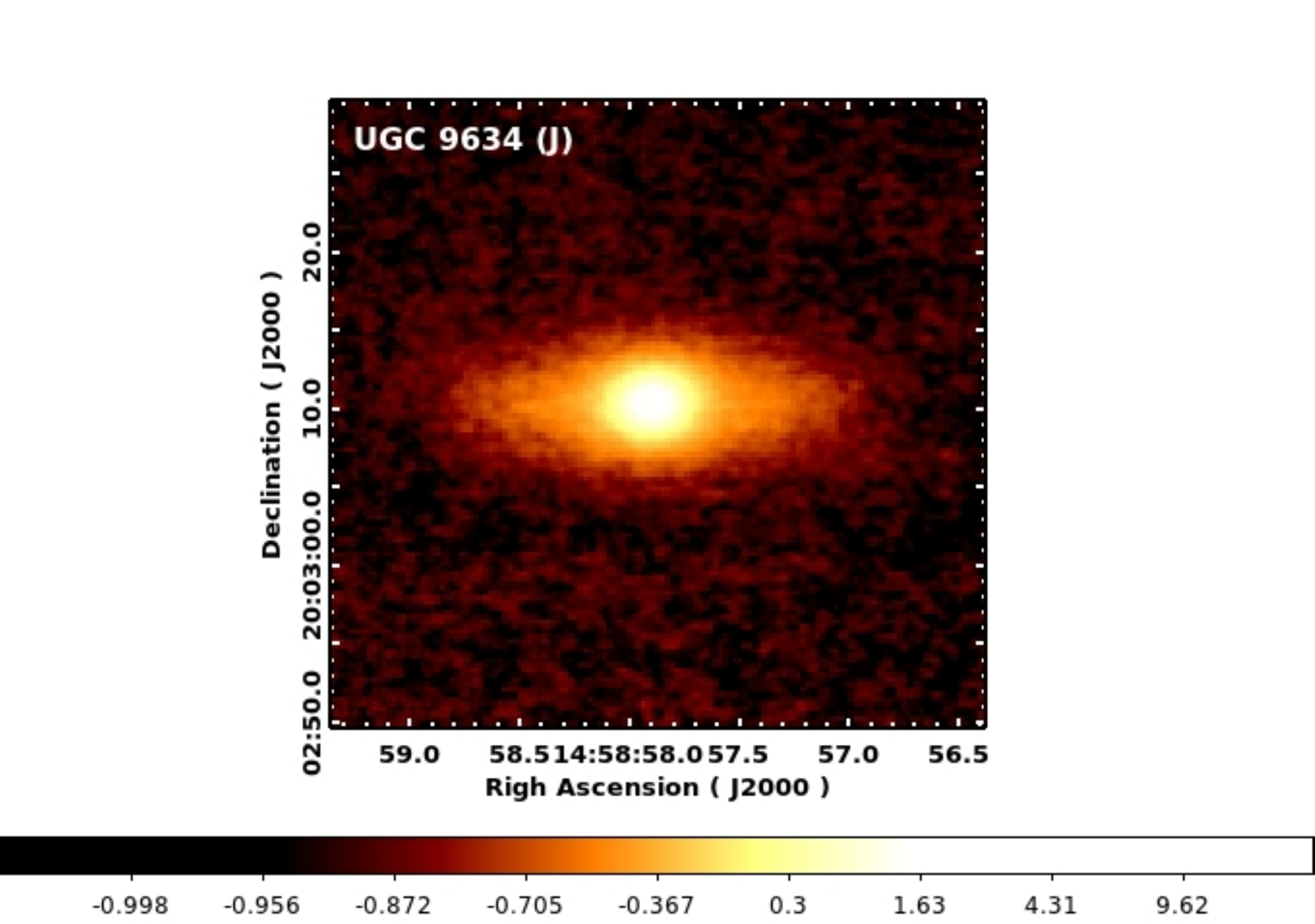}\includegraphics[trim = 45mm 17mm 50mm 8mm, clip,scale=0.39]{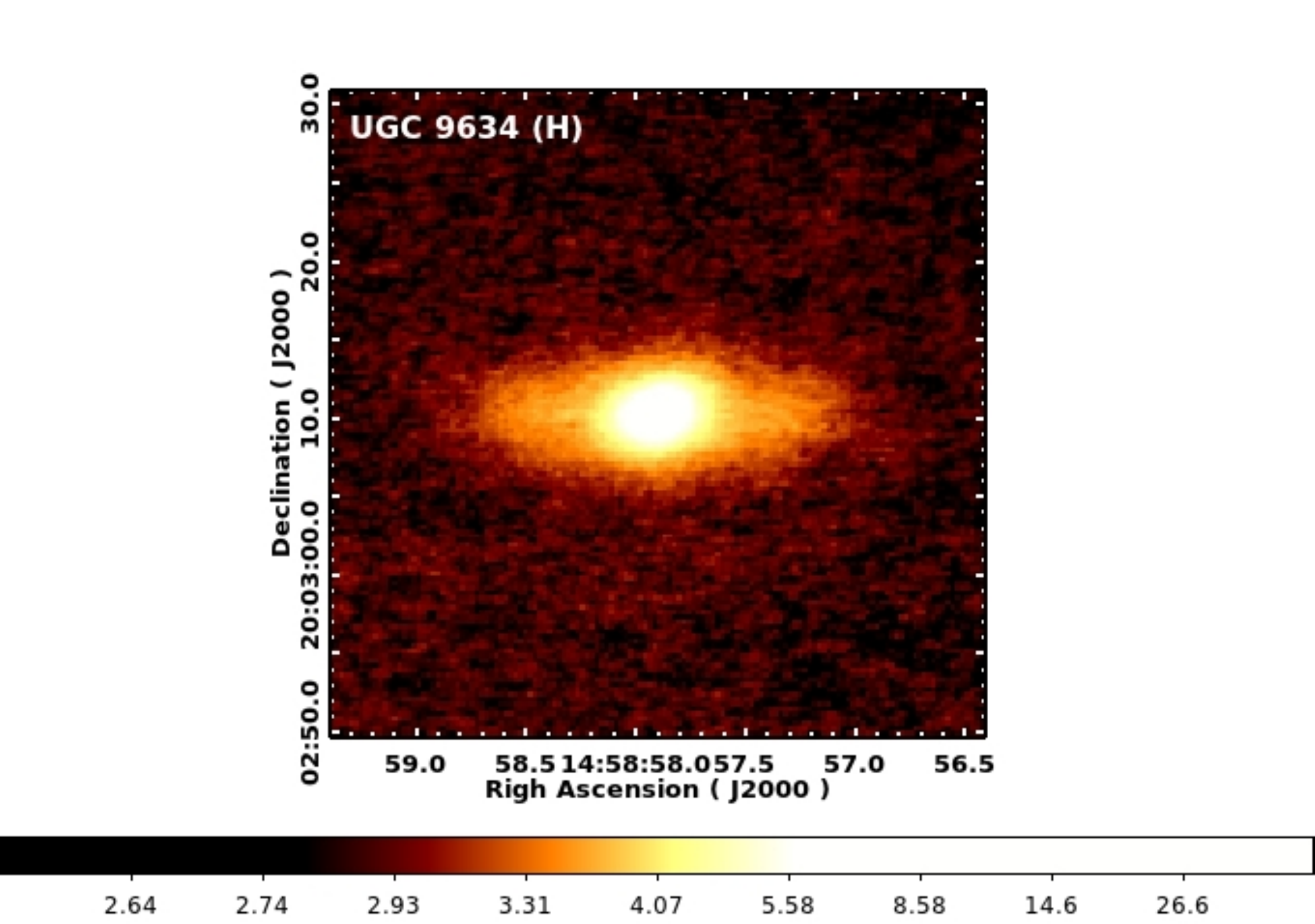}\includegraphics[trim = 45mm 17mm 50mm 8mm, clip,scale=0.39]{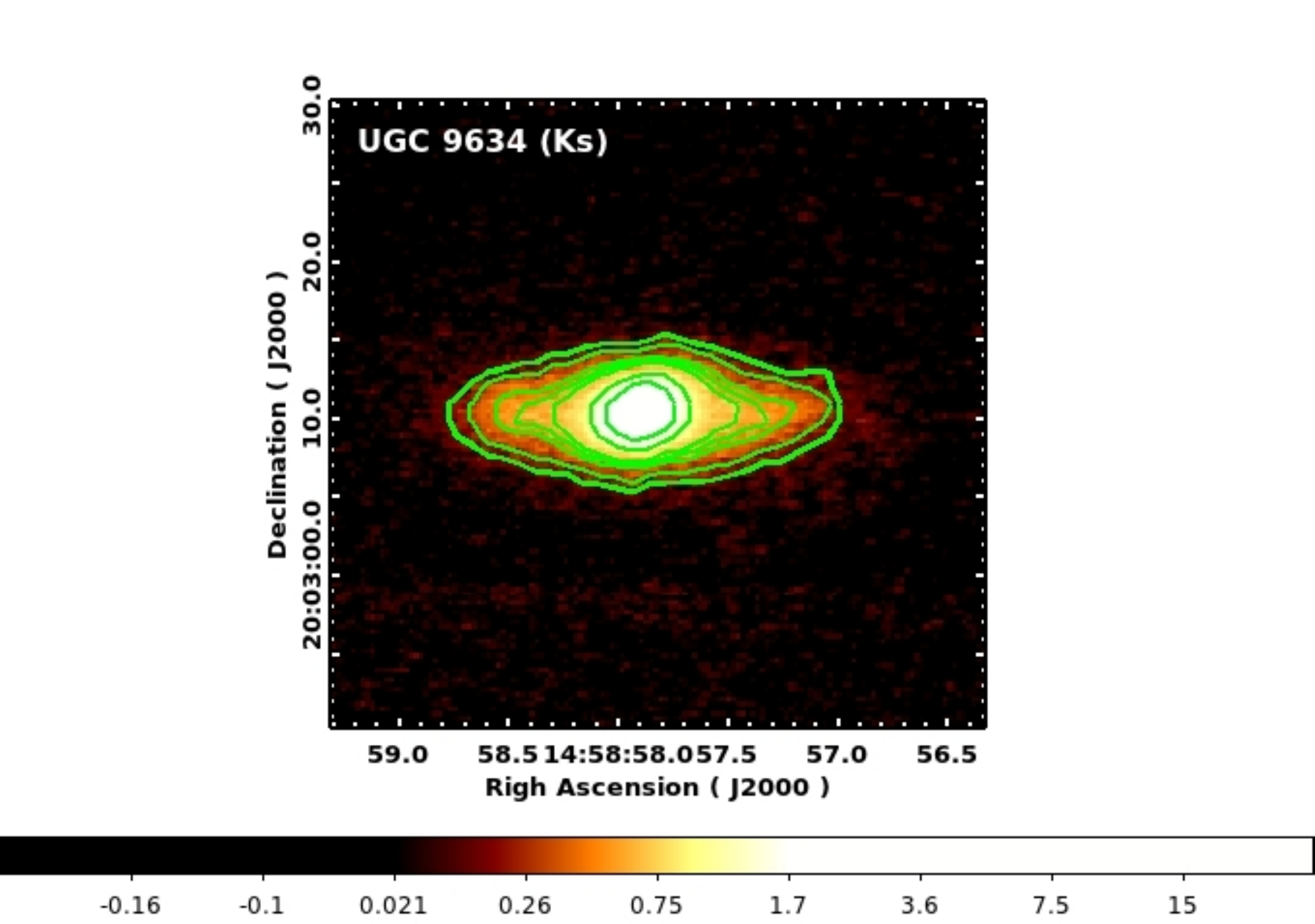}

\includegraphics[trim = 40mm 17mm 55mm 8mm, clip,scale=0.39]{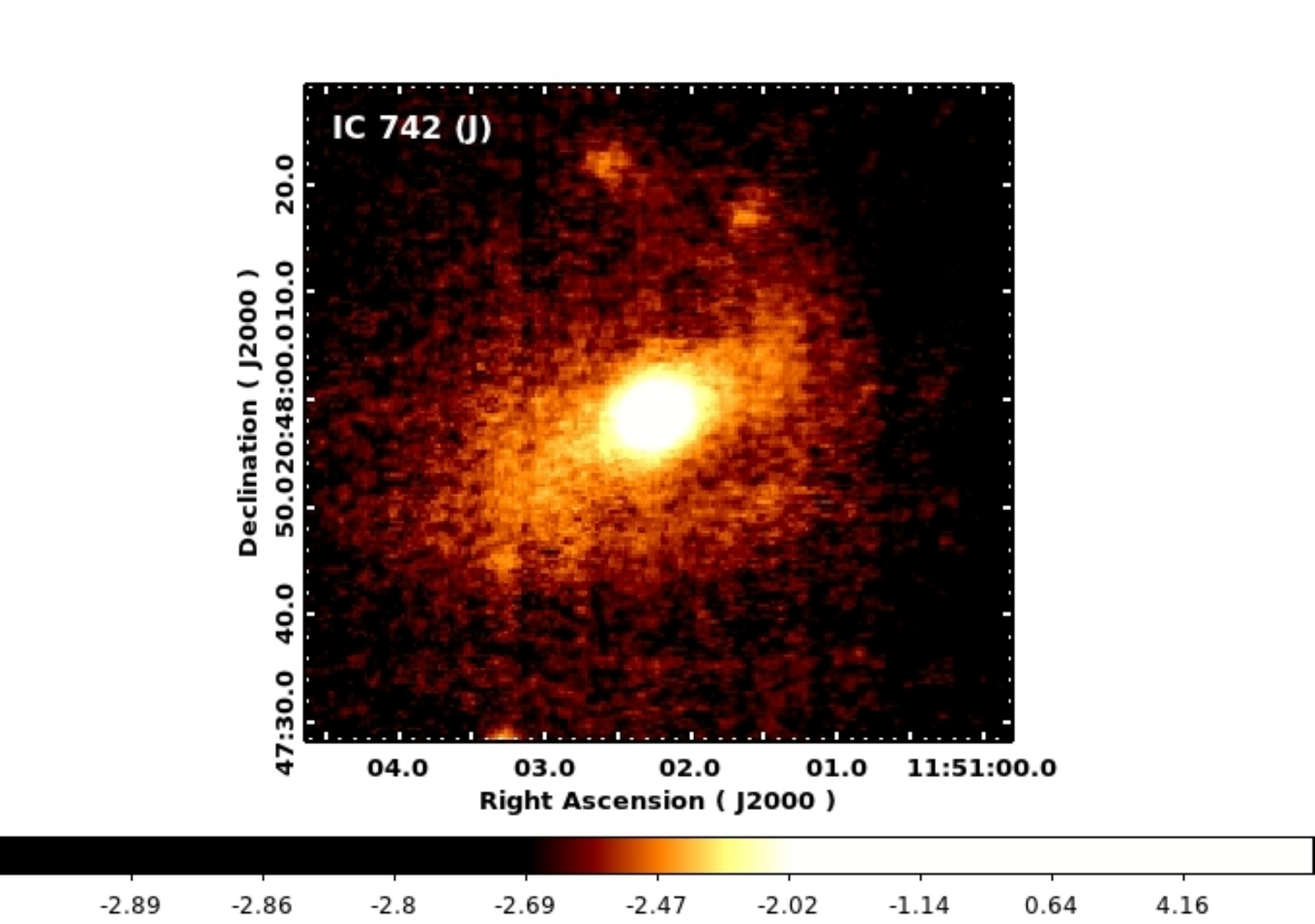}\includegraphics[trim = 40mm 17mm 55mm 8mm, clip,scale=0.39]{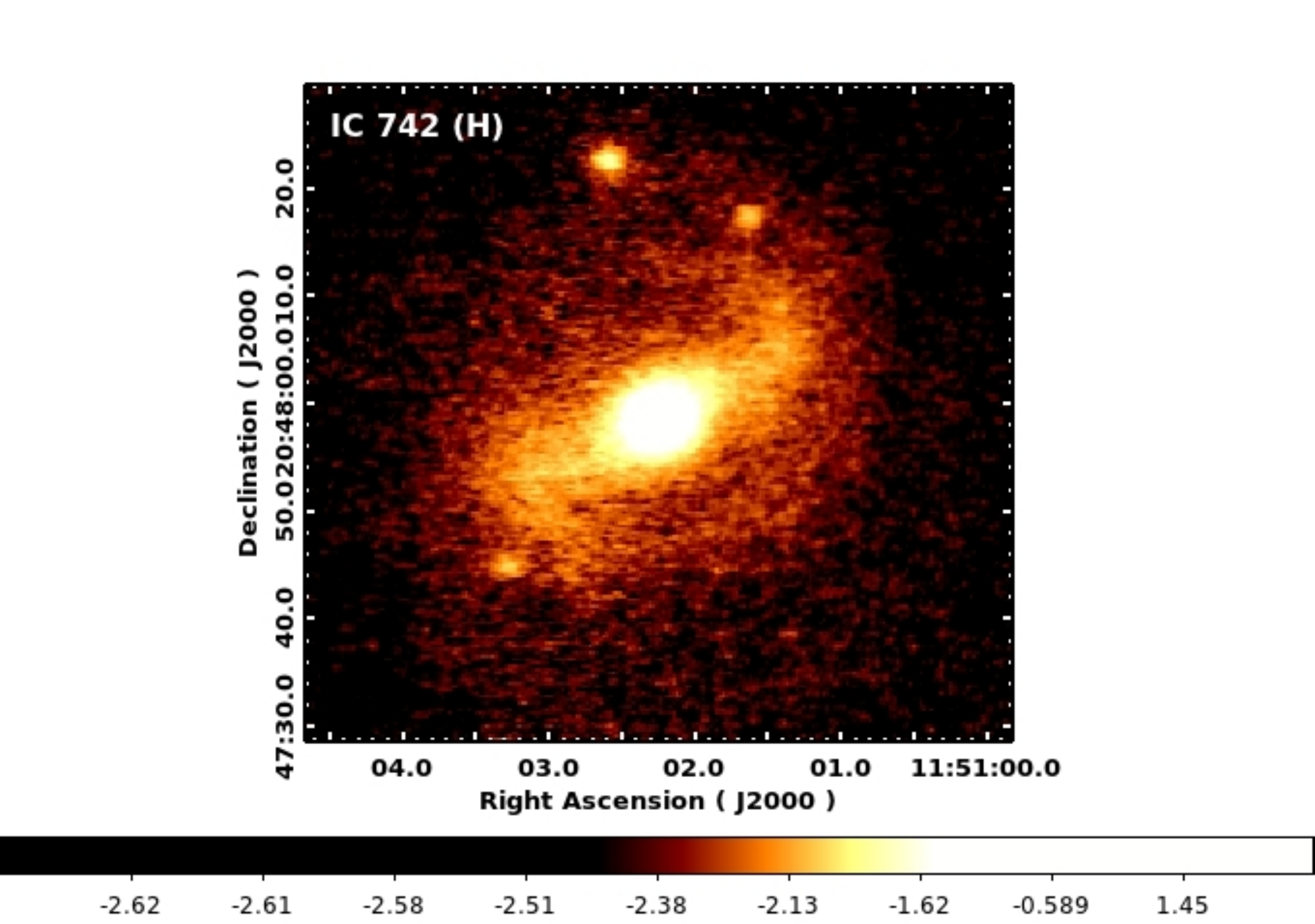}\includegraphics[trim = 40mm 17mm 55mm 8mm, clip,scale=0.39]{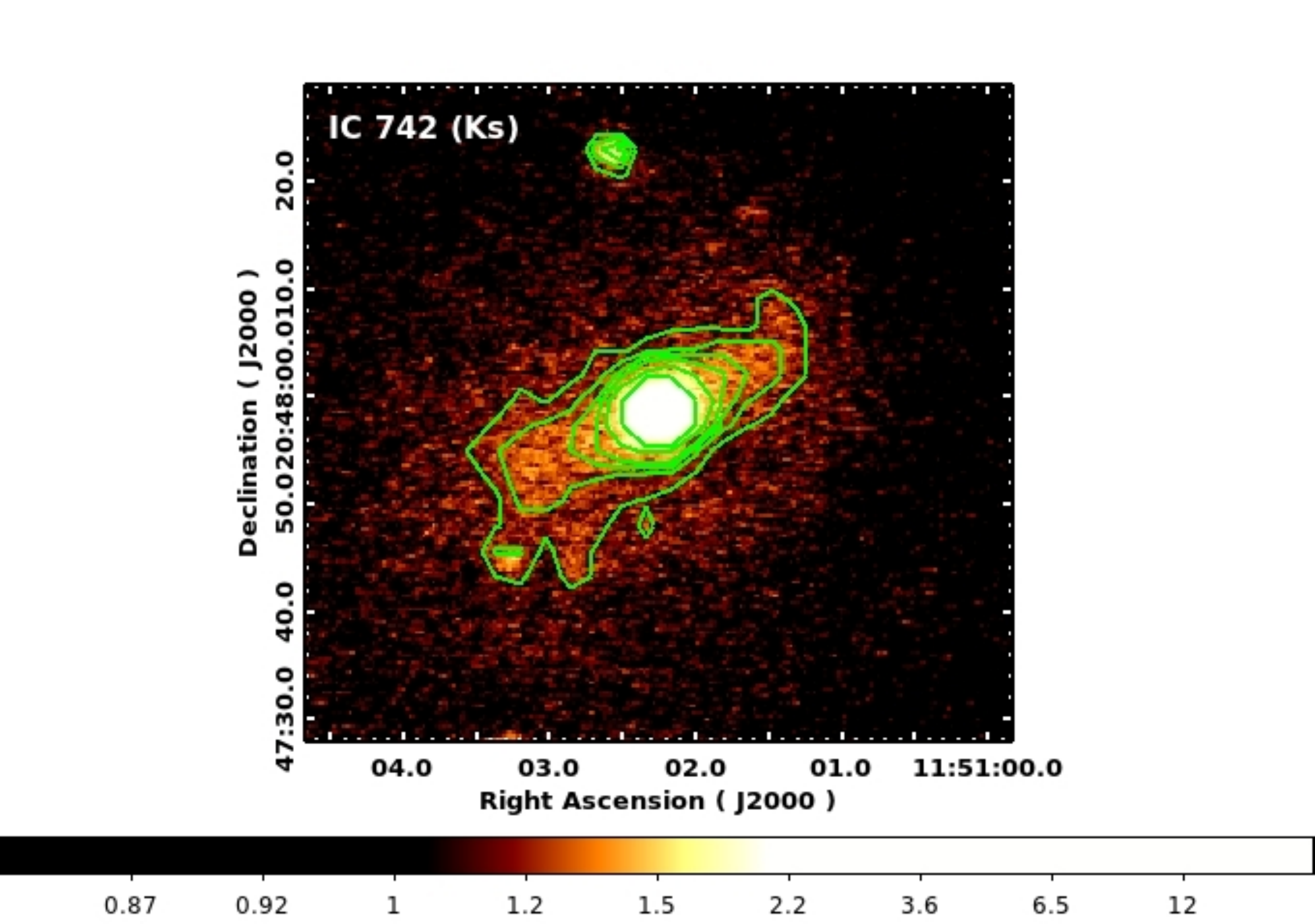}
\end{figure*}
\begin{figure*}
\includegraphics[trim = 45mm 17mm 65mm 8mm, clip,scale=0.42]{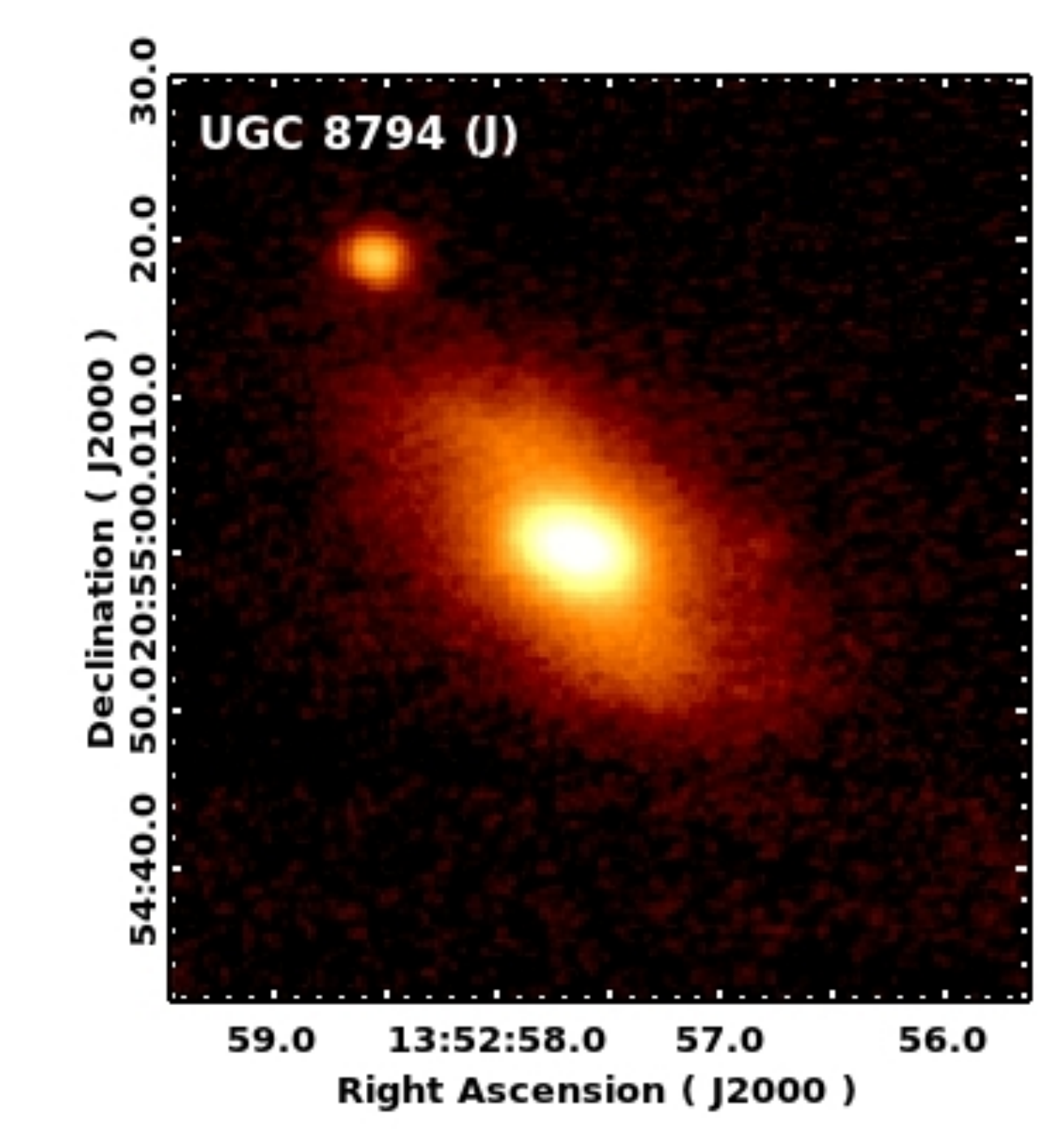}\includegraphics[trim = 48mm 17mm 60mm 8mm, clip,scale=0.42]{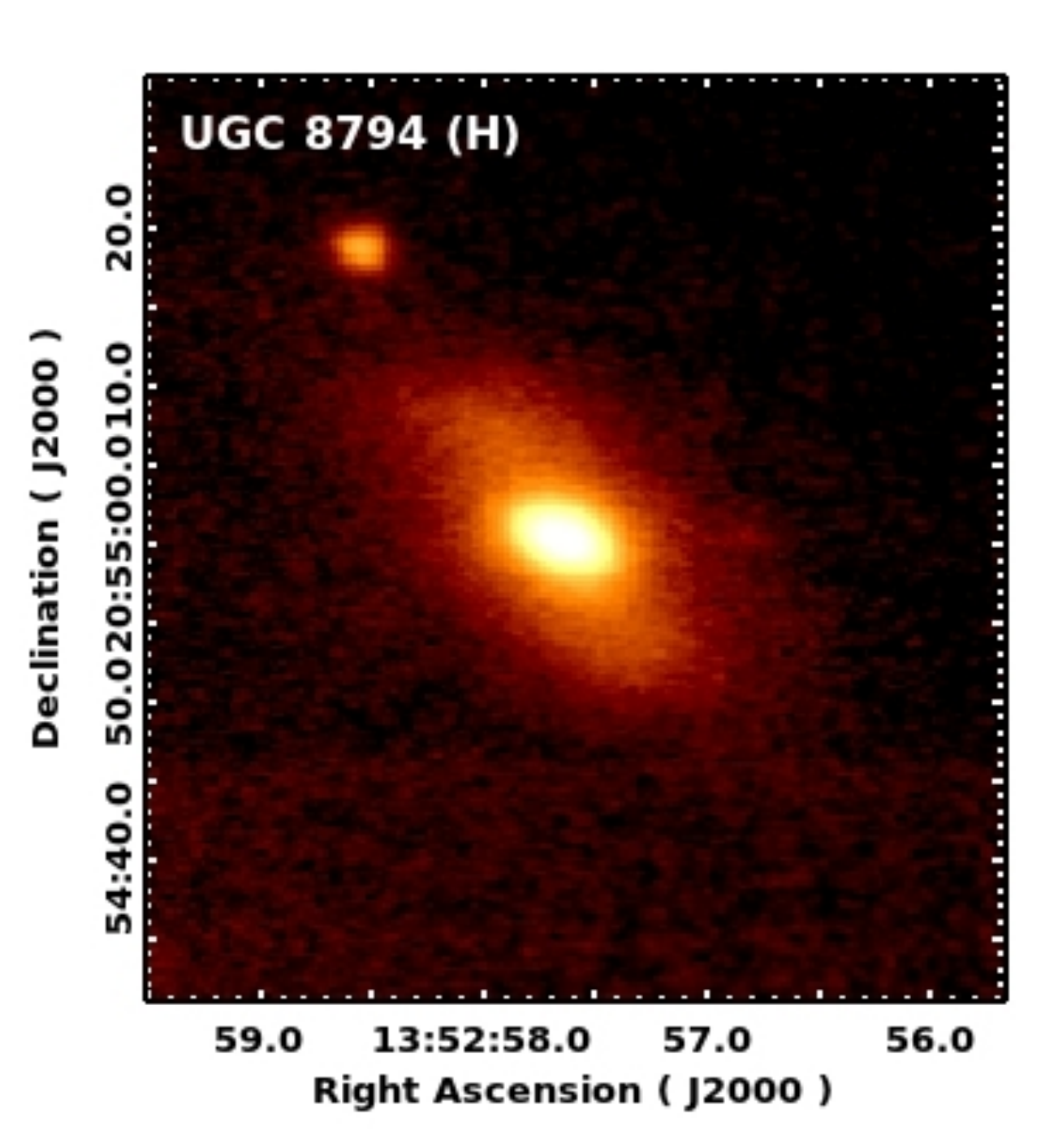}\includegraphics[trim = 55mm 17mm 60mm 8mm, clip,scale=0.42]{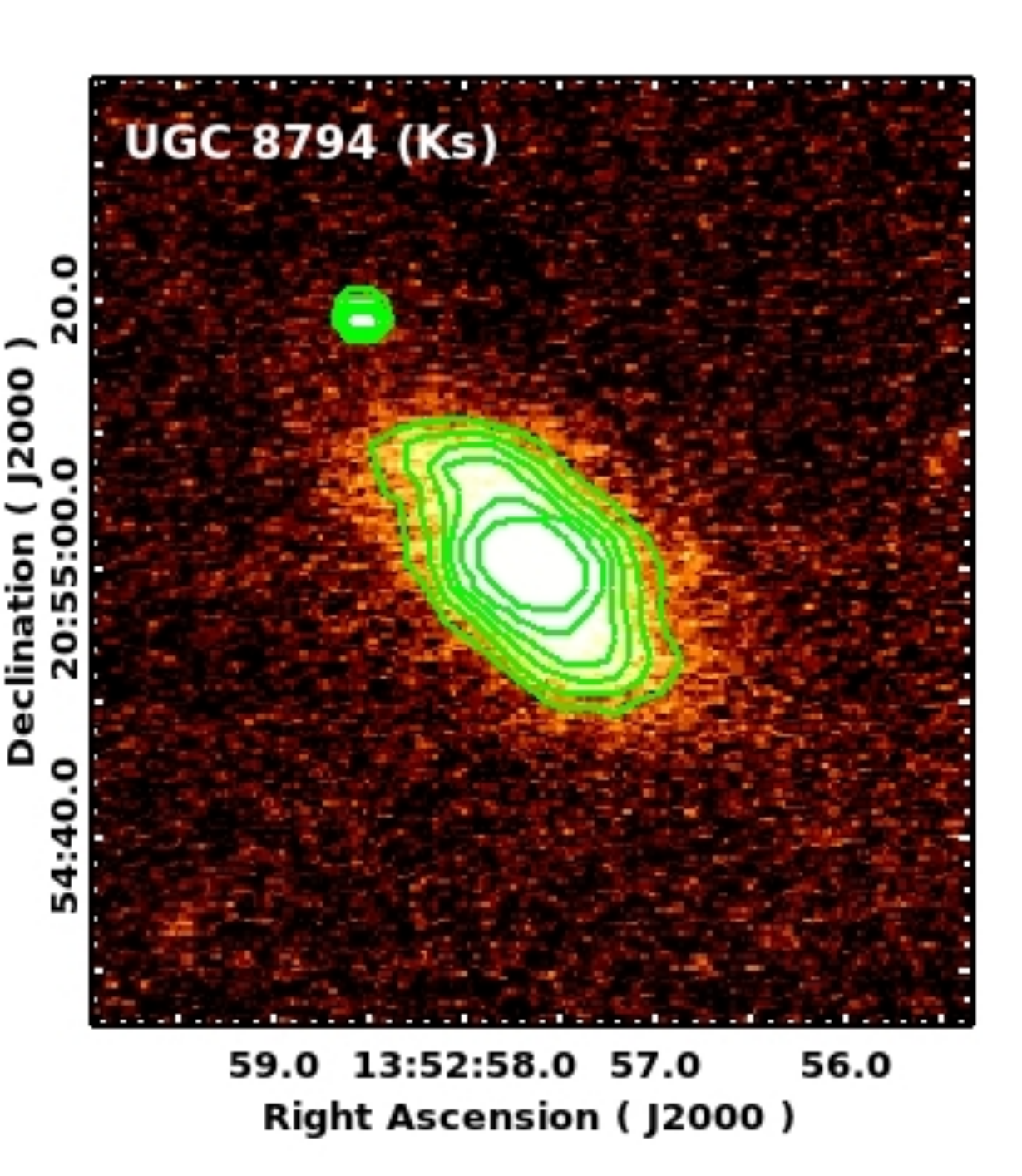}
\includegraphics[trim = 50mm 17mm 65mm 8mm, clip,scale=0.42]{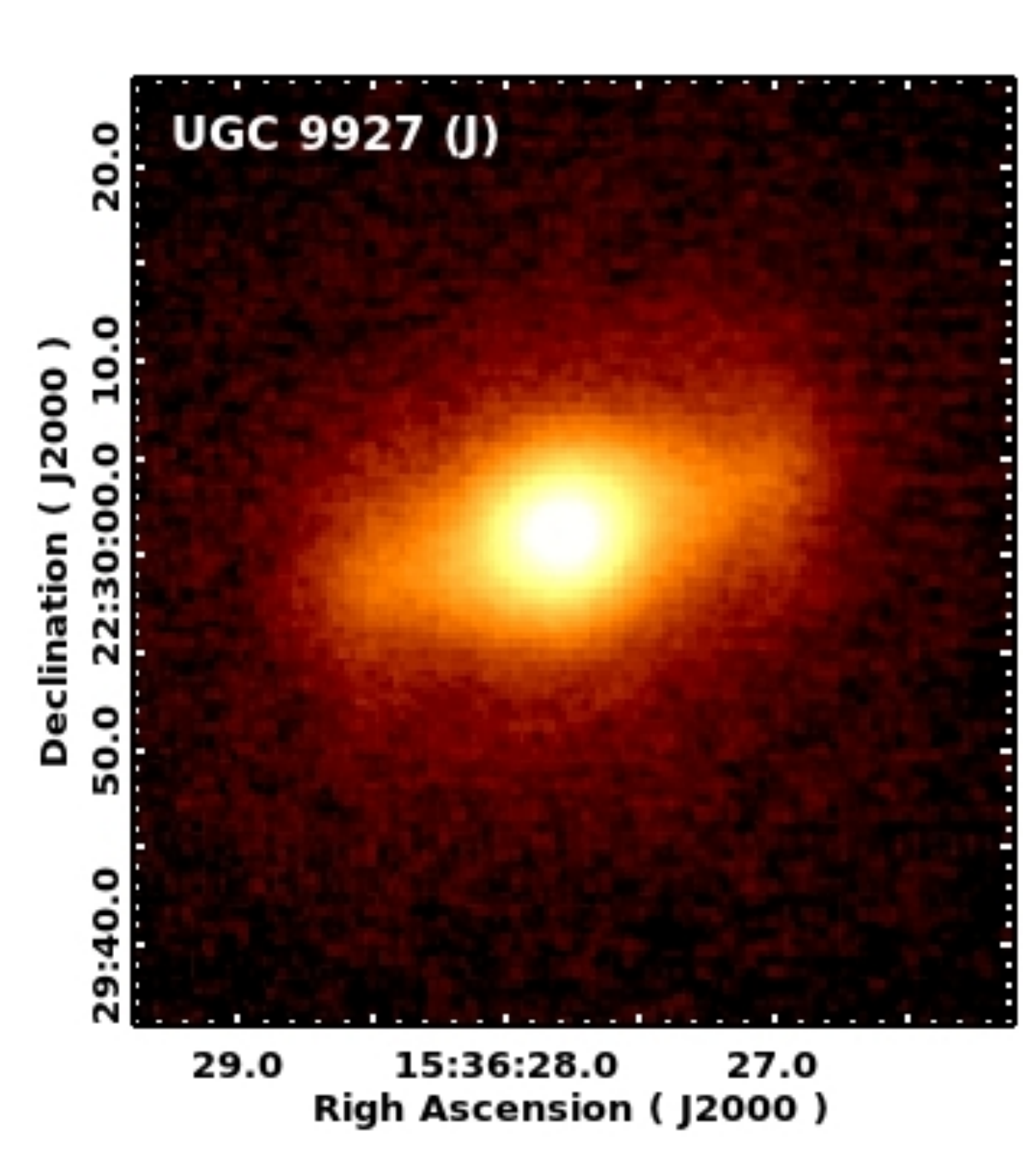}\includegraphics[trim = 50mm 17mm 65mm 8mm, clip,scale=0.42]{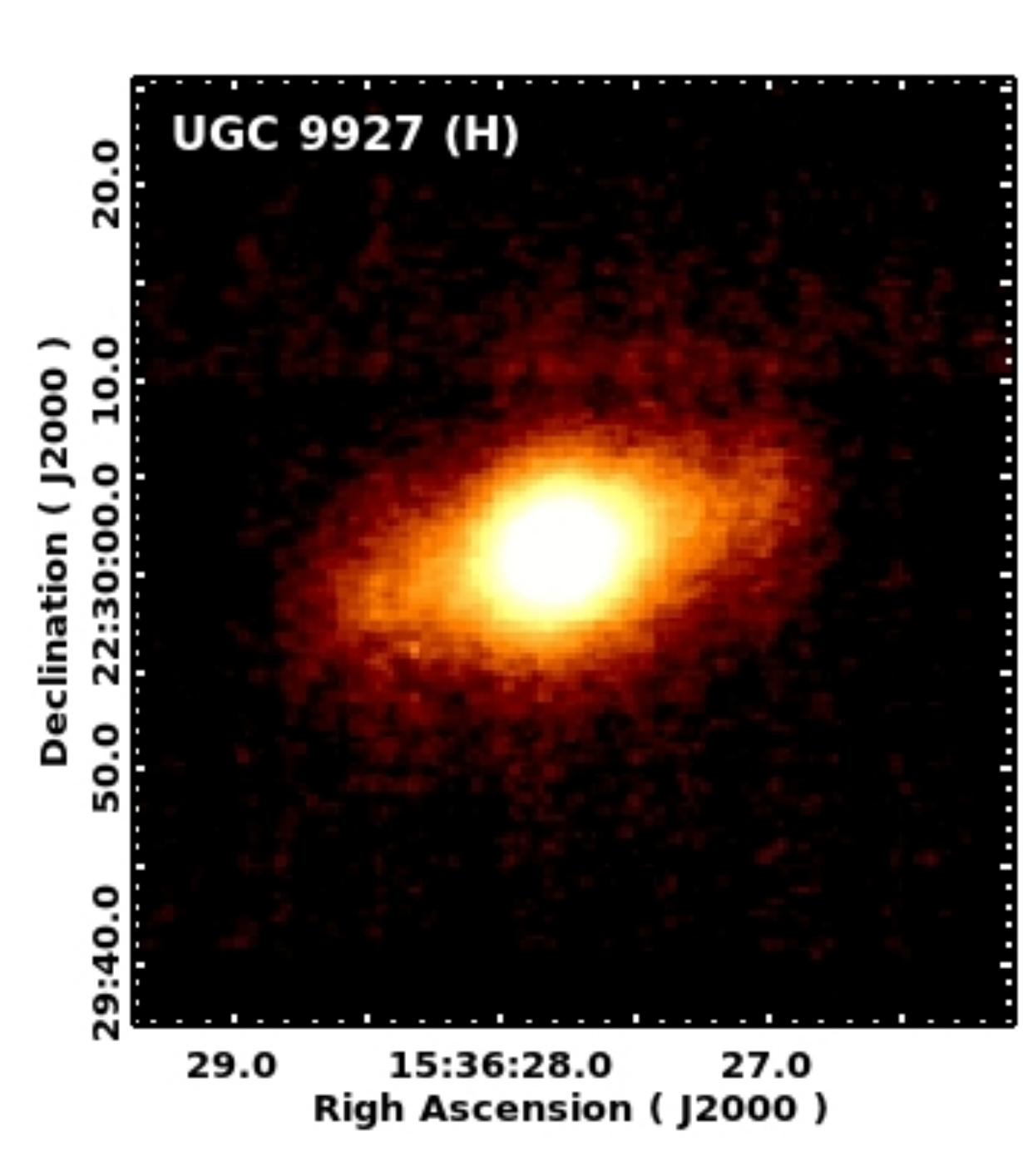}\includegraphics[trim = 50mm 17mm 65mm 8mm, clip,scale=0.42]{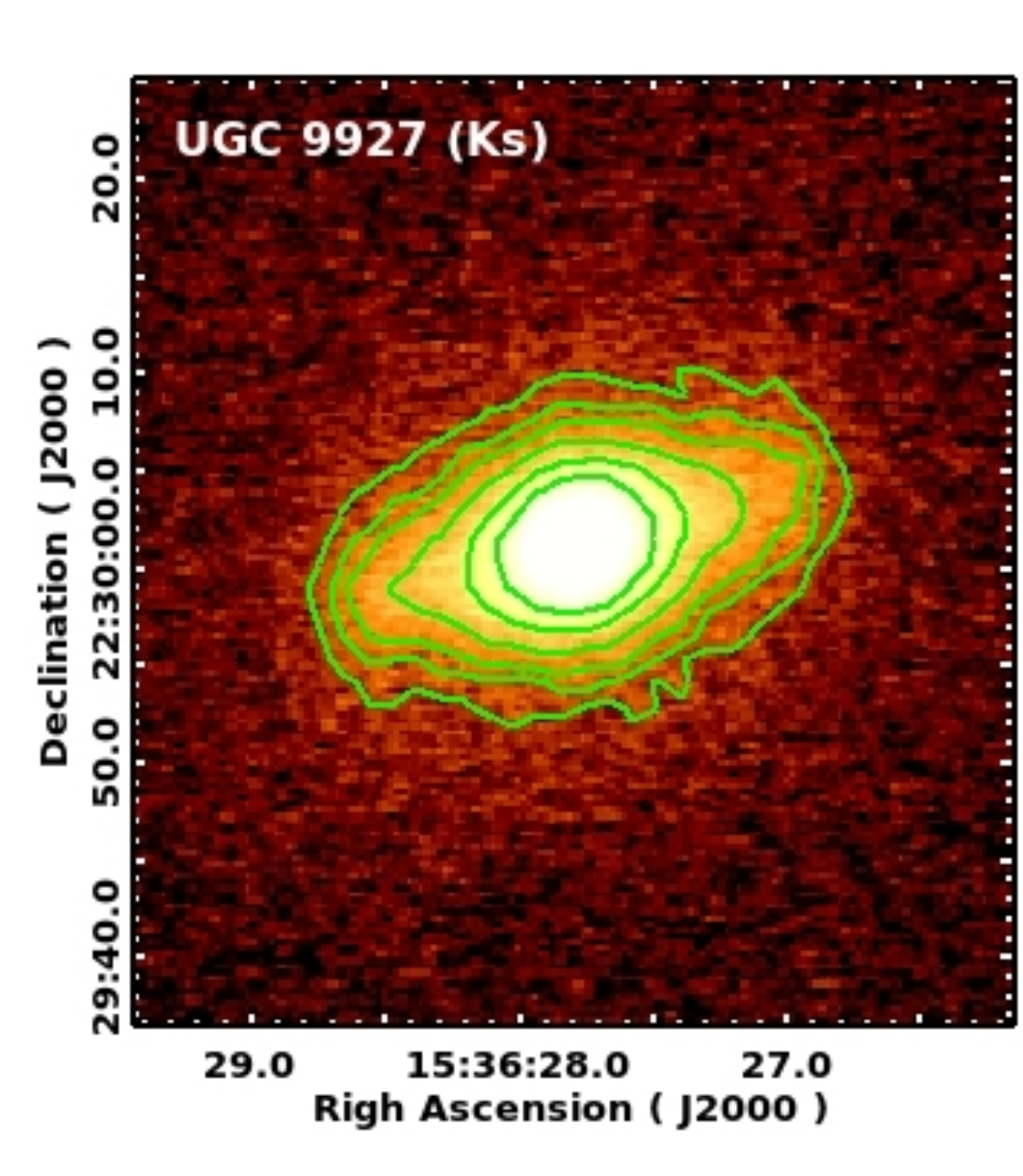}
\includegraphics[trim = 50mm 17mm 65mm 8mm, clip,scale=0.42]{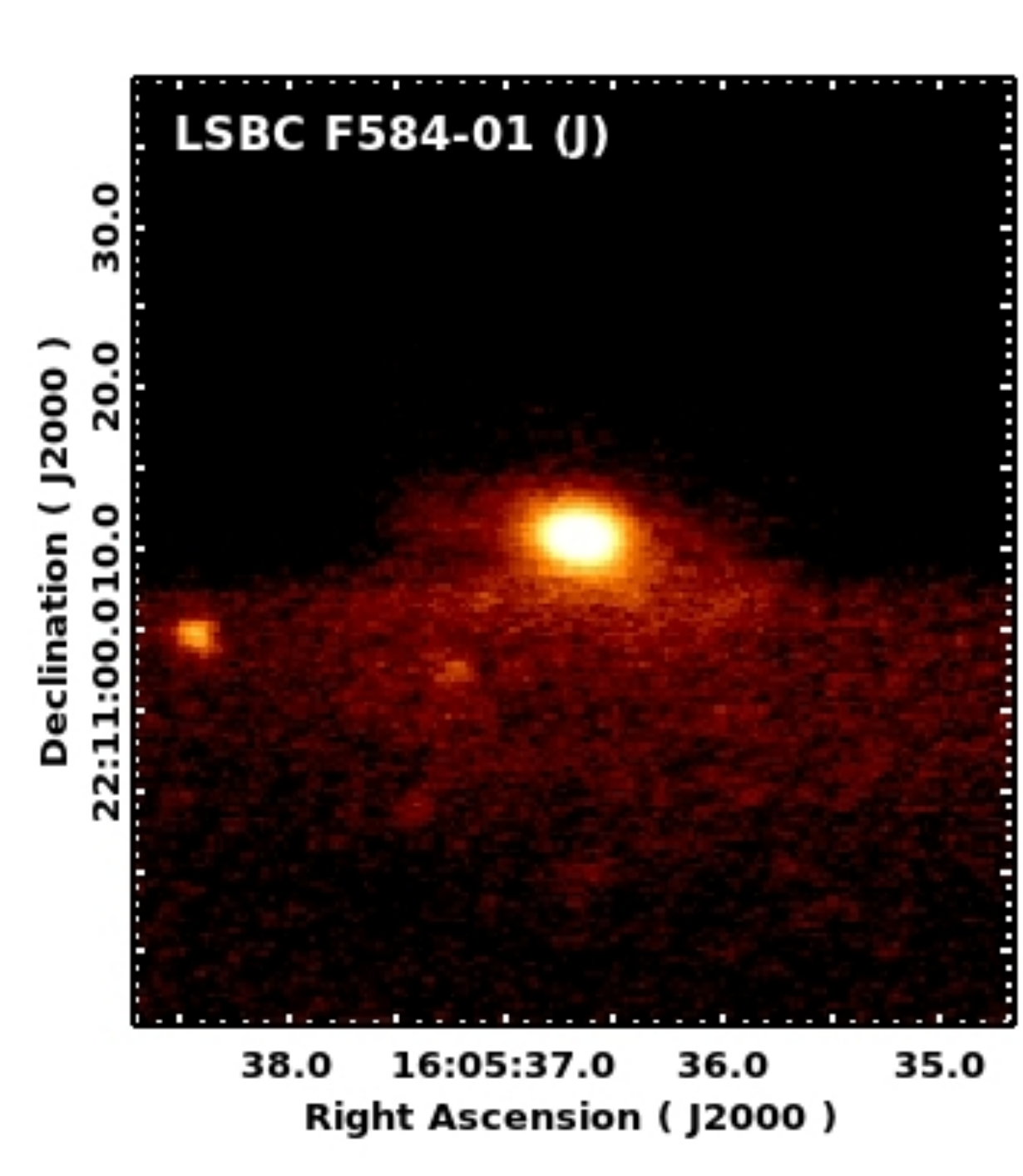}\includegraphics[trim = 50mm 17mm 65mm 8mm, clip,scale=0.42]{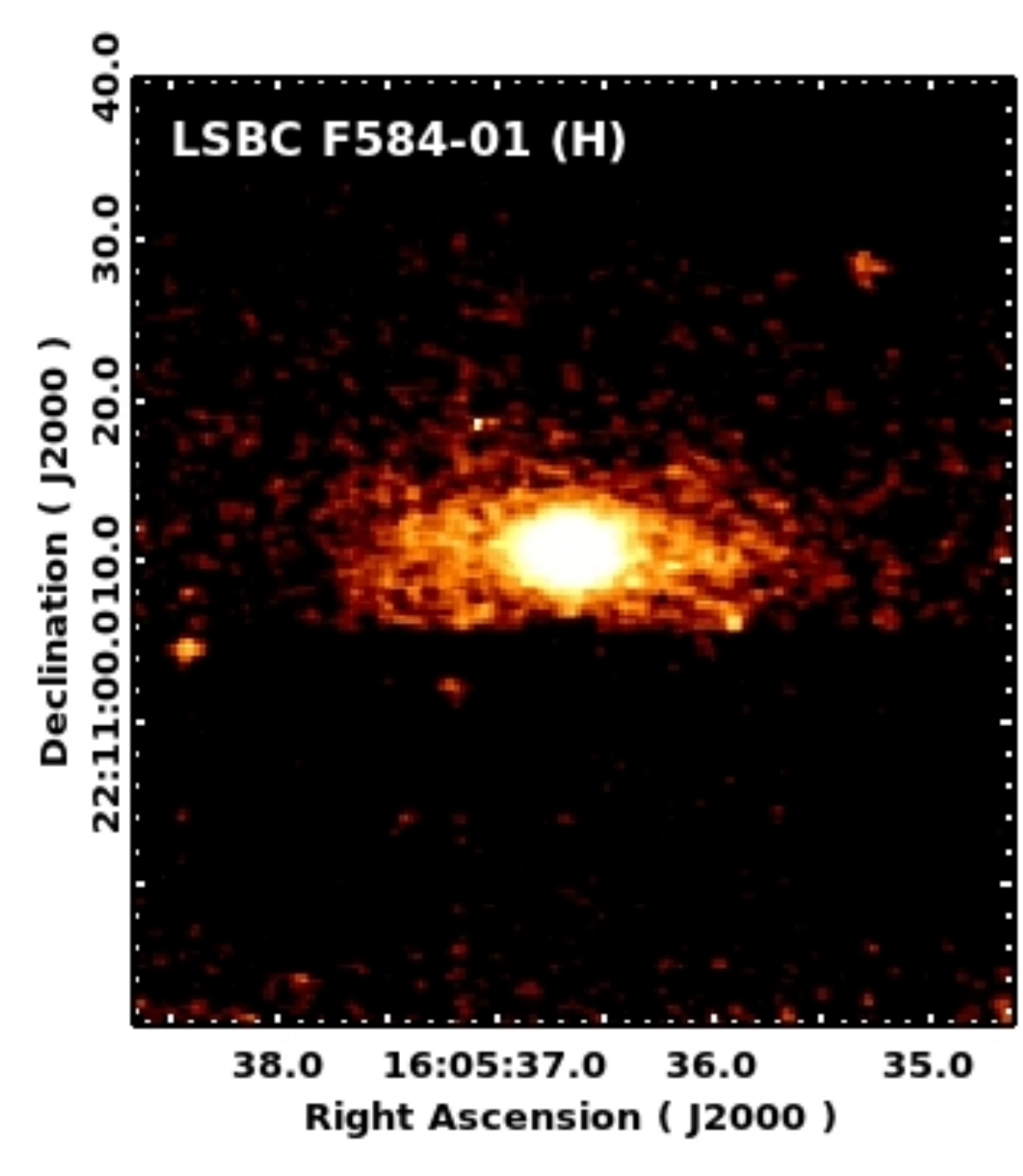}\includegraphics[trim = 50mm 17mm 65mm 8mm, clip,scale=0.42]{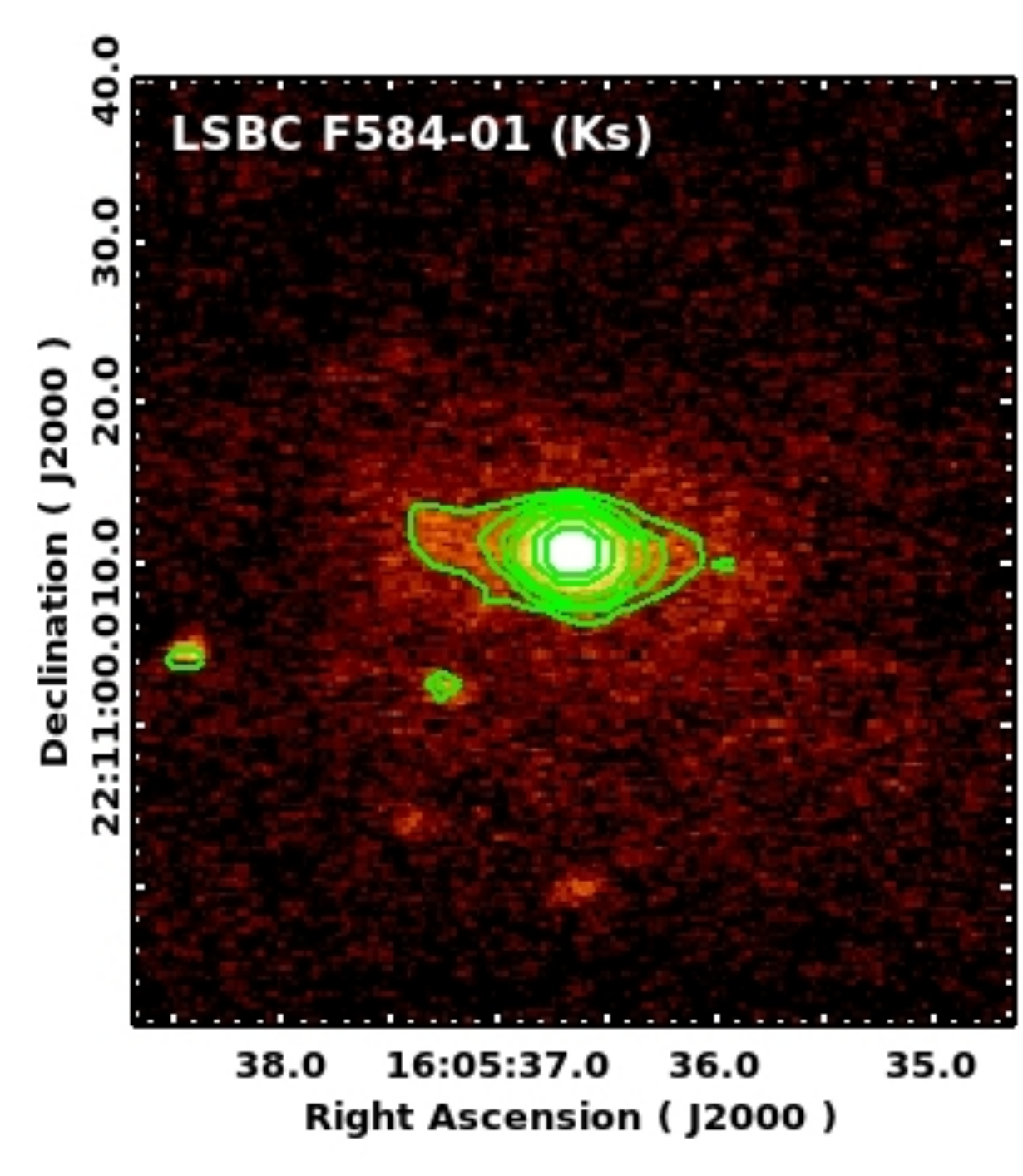}
\includegraphics[trim = 50mm 17mm 65mm 8mm, clip,scale=0.42]{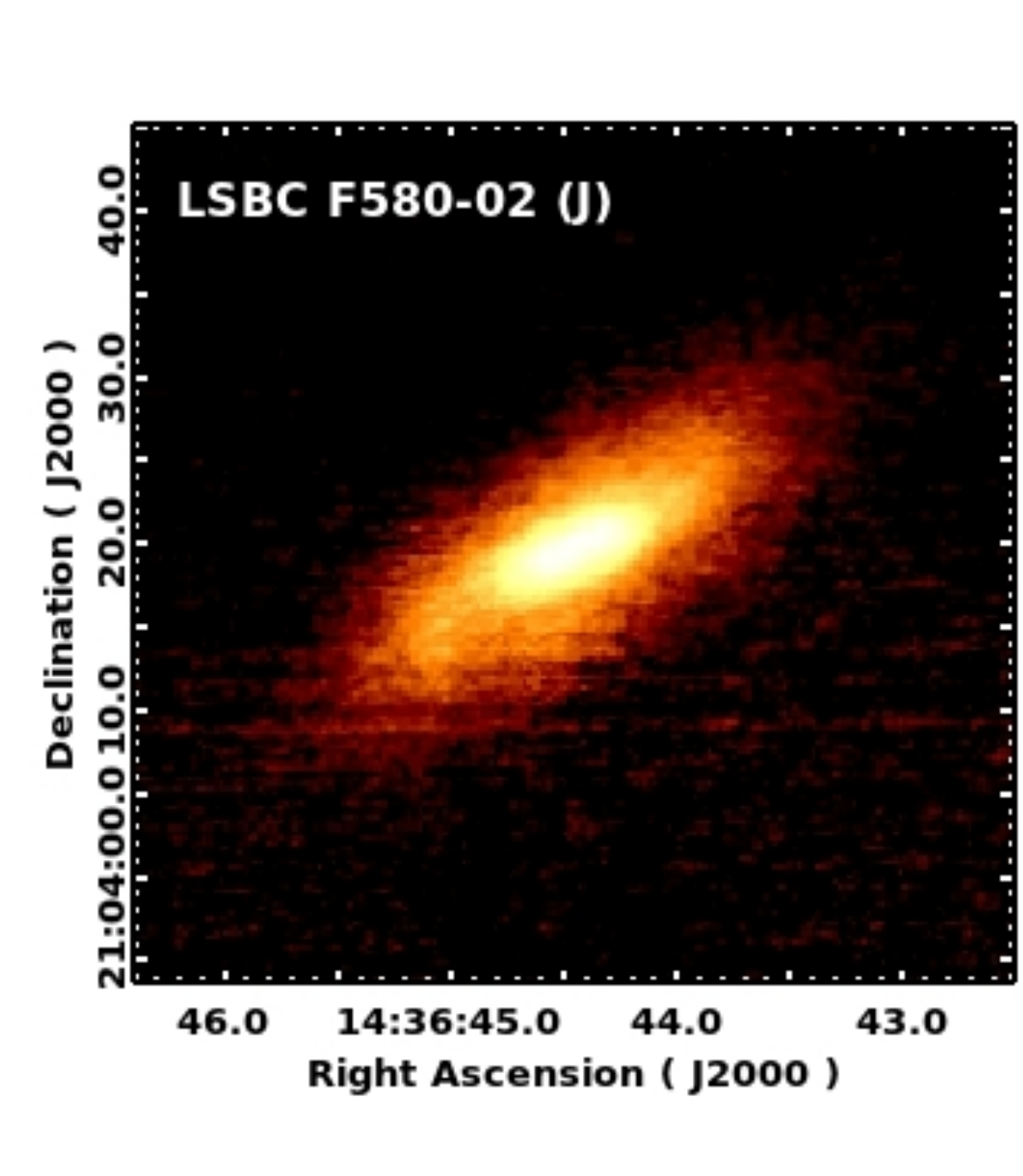}\includegraphics[trim = 50mm 17mm 65mm 8mm, clip,scale=0.42]{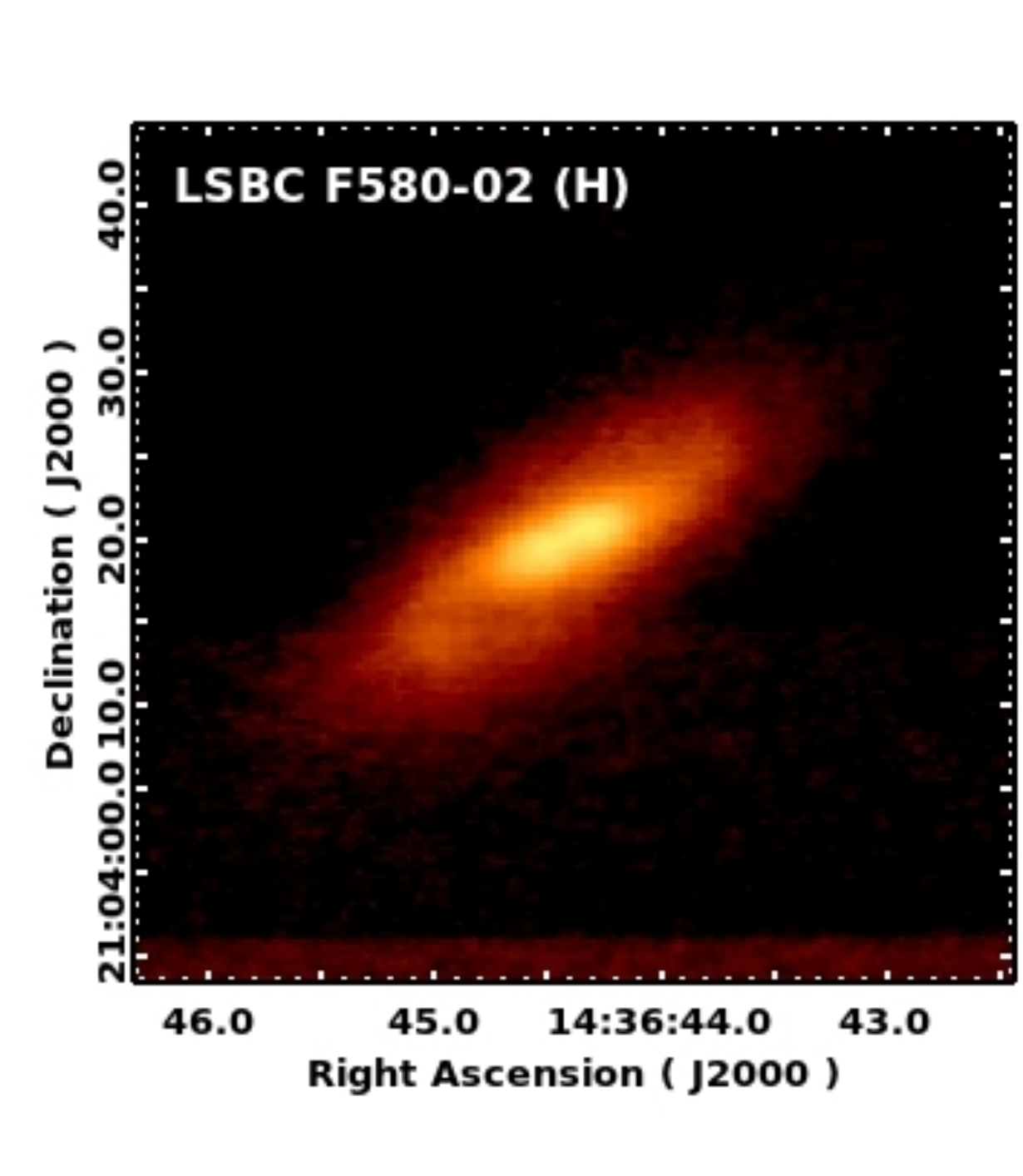}\includegraphics[trim = 50mm 17mm 65mm 8mm, clip,scale=0.42]{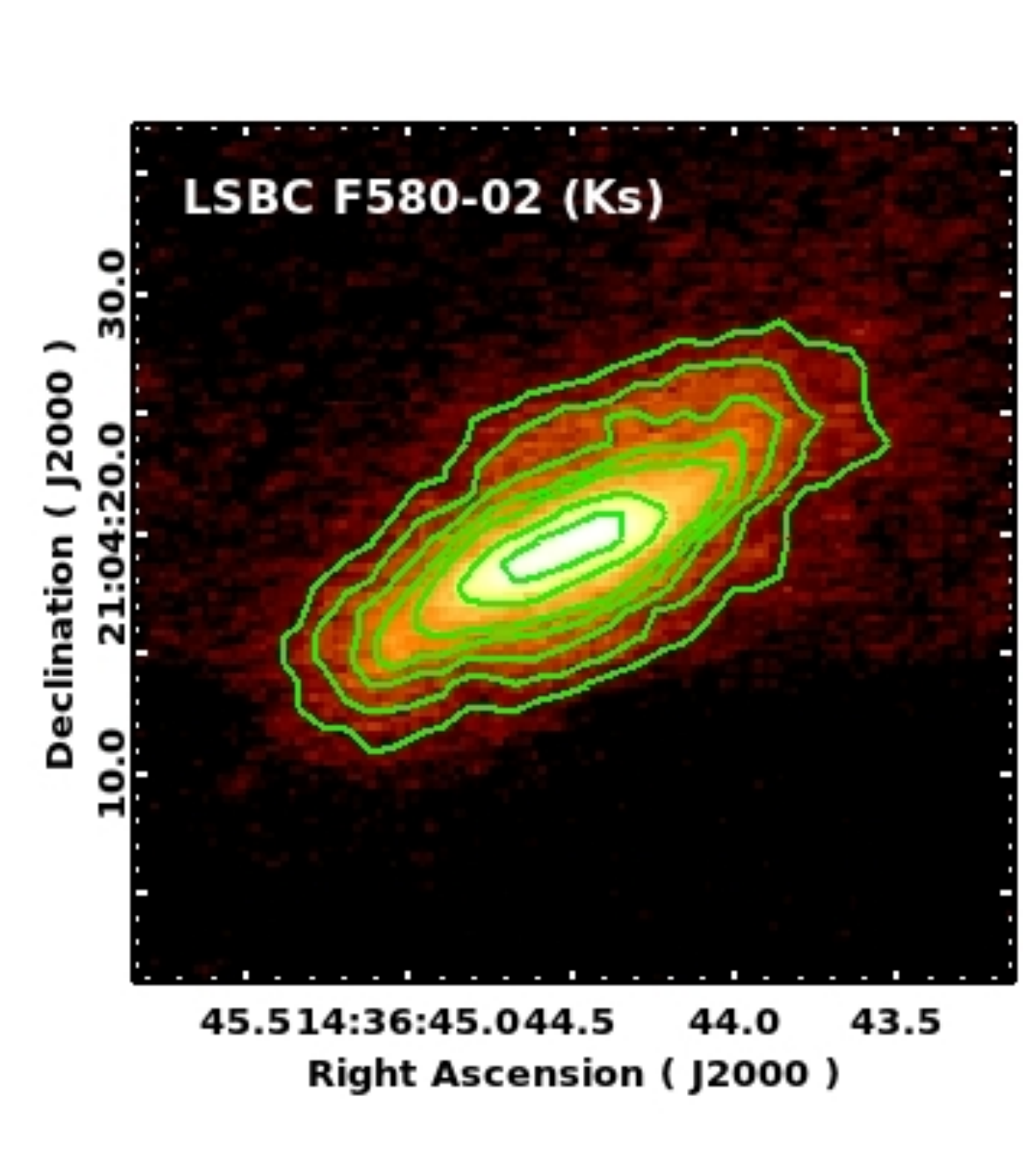}
\end{figure*}

\begin{figure*}
\includegraphics[trim = 50mm 17mm 65mm 8mm, clip,scale=0.42]{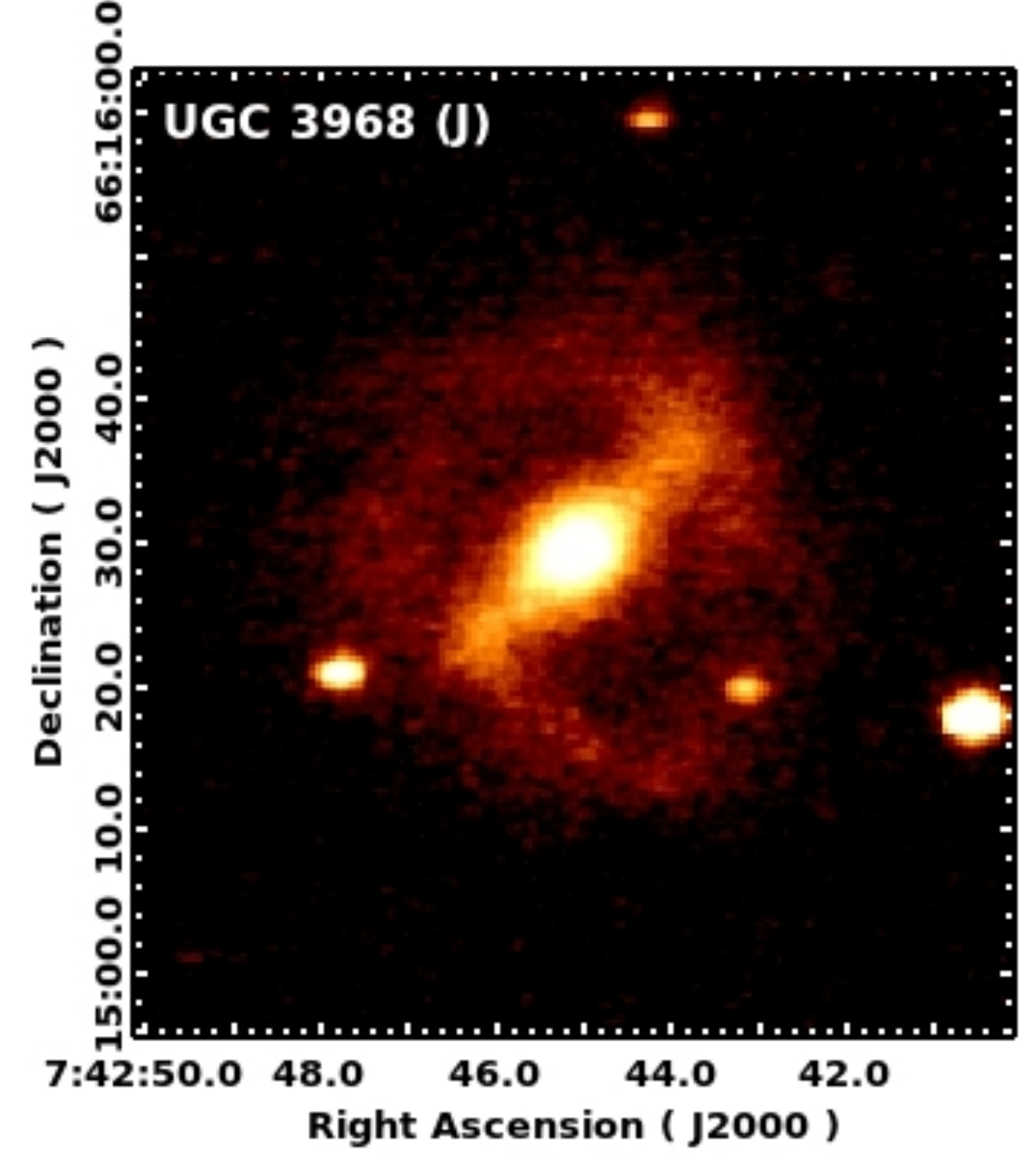}\includegraphics[trim = 50mm 17mm 65mm 8mm, clip,scale=0.42]{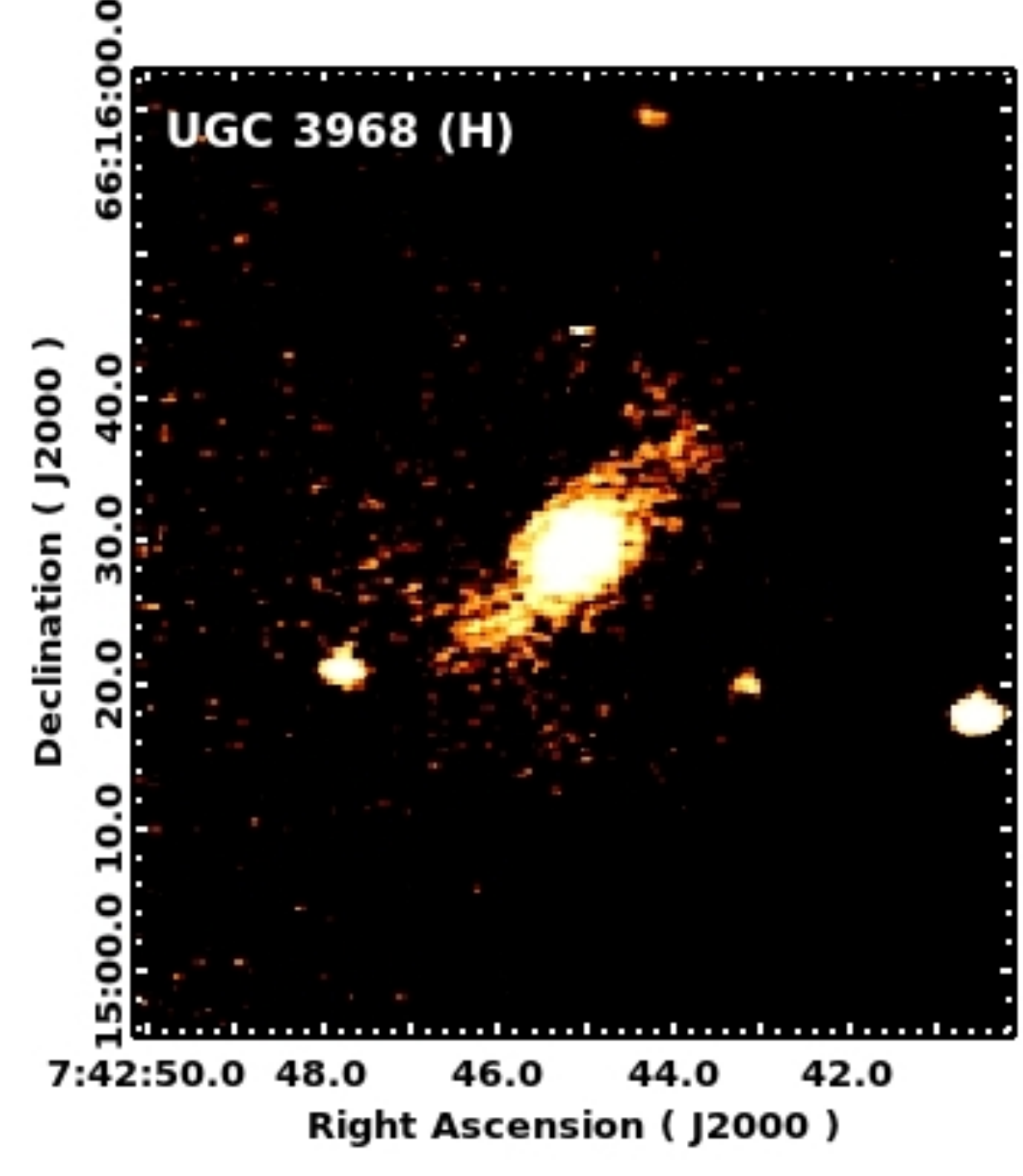}\includegraphics[trim = 50mm 17mm 65mm 8mm, clip,scale=0.42]{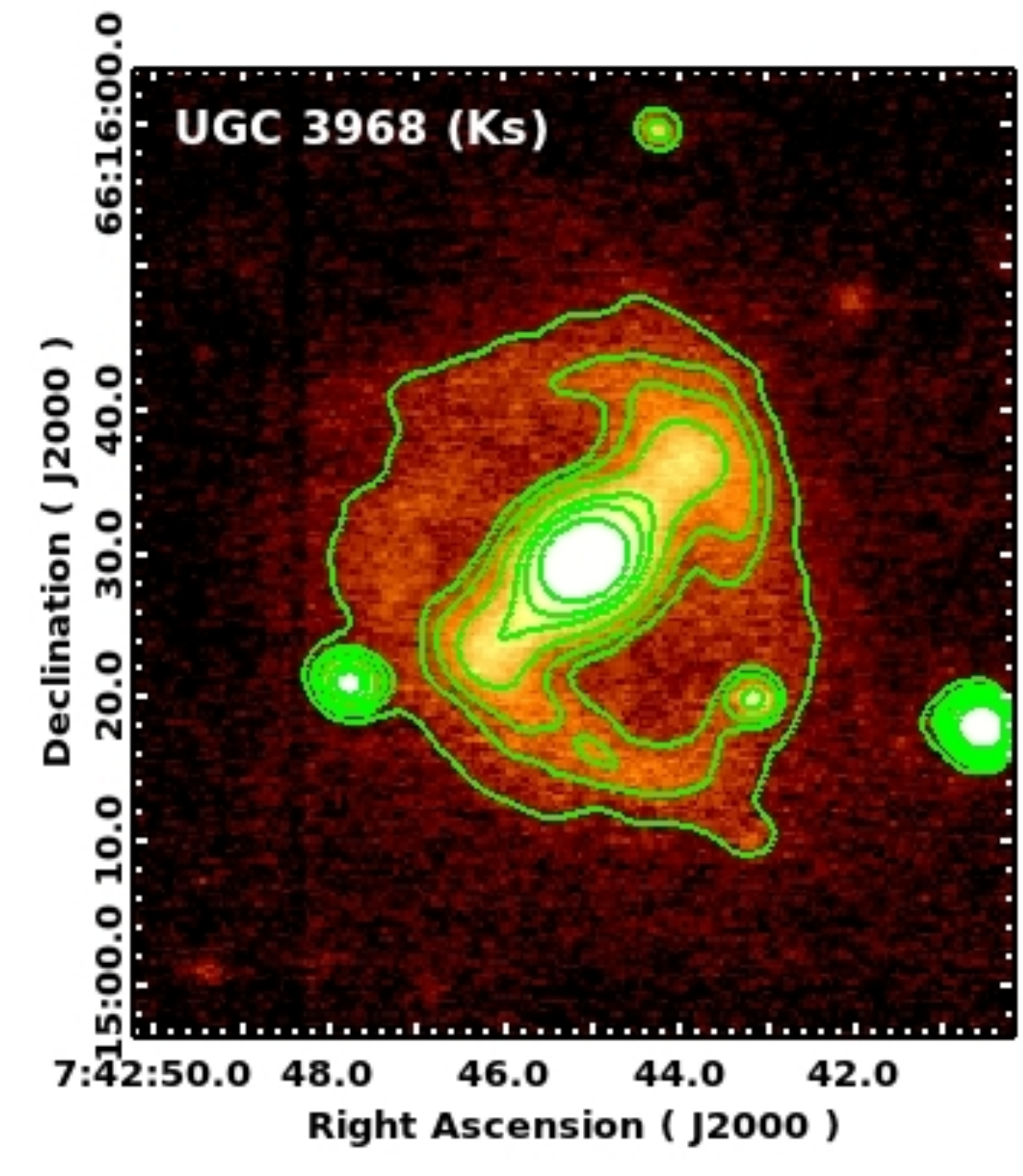}

\includegraphics[trim = 50mm 17mm 65mm 8mm, clip,scale=0.42]{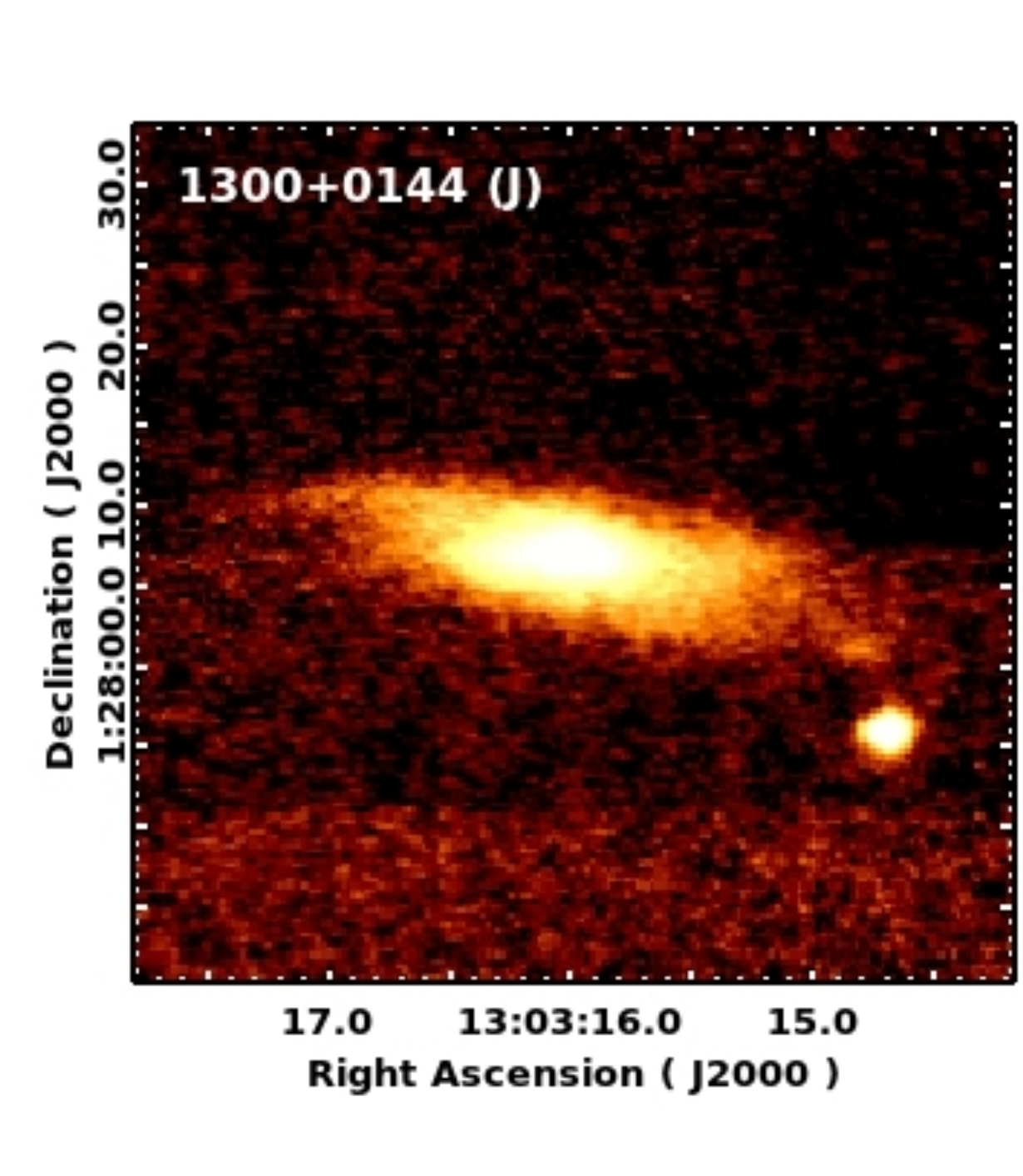}\includegraphics[trim = 50mm 17mm 65mm 8mm, clip,scale=0.42]{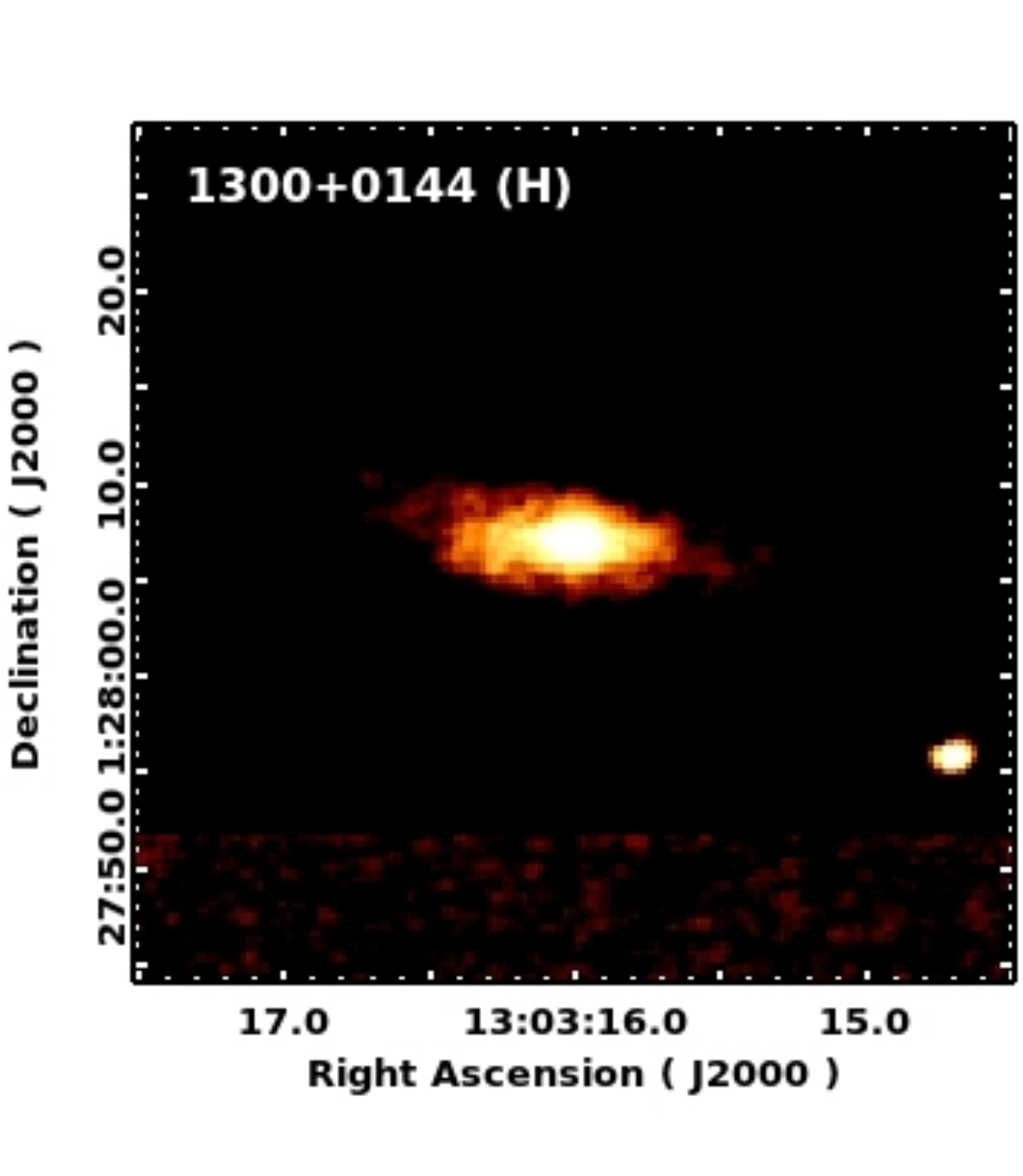}\includegraphics[trim = 50mm 17mm 65mm 8mm, clip,scale=0.42]{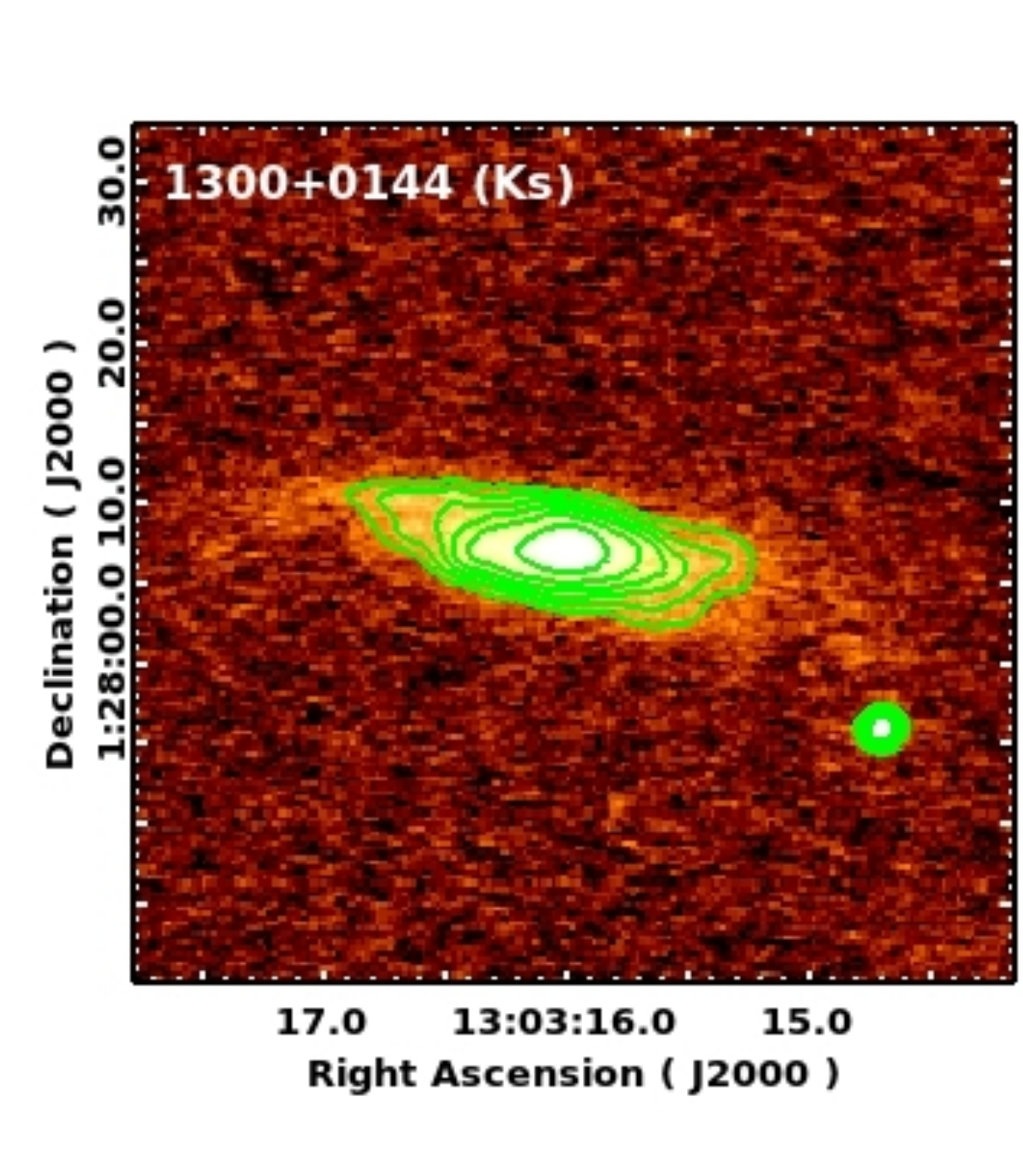}
\includegraphics[trim = 50mm 17mm 65mm 8mm, clip,scale=0.42]{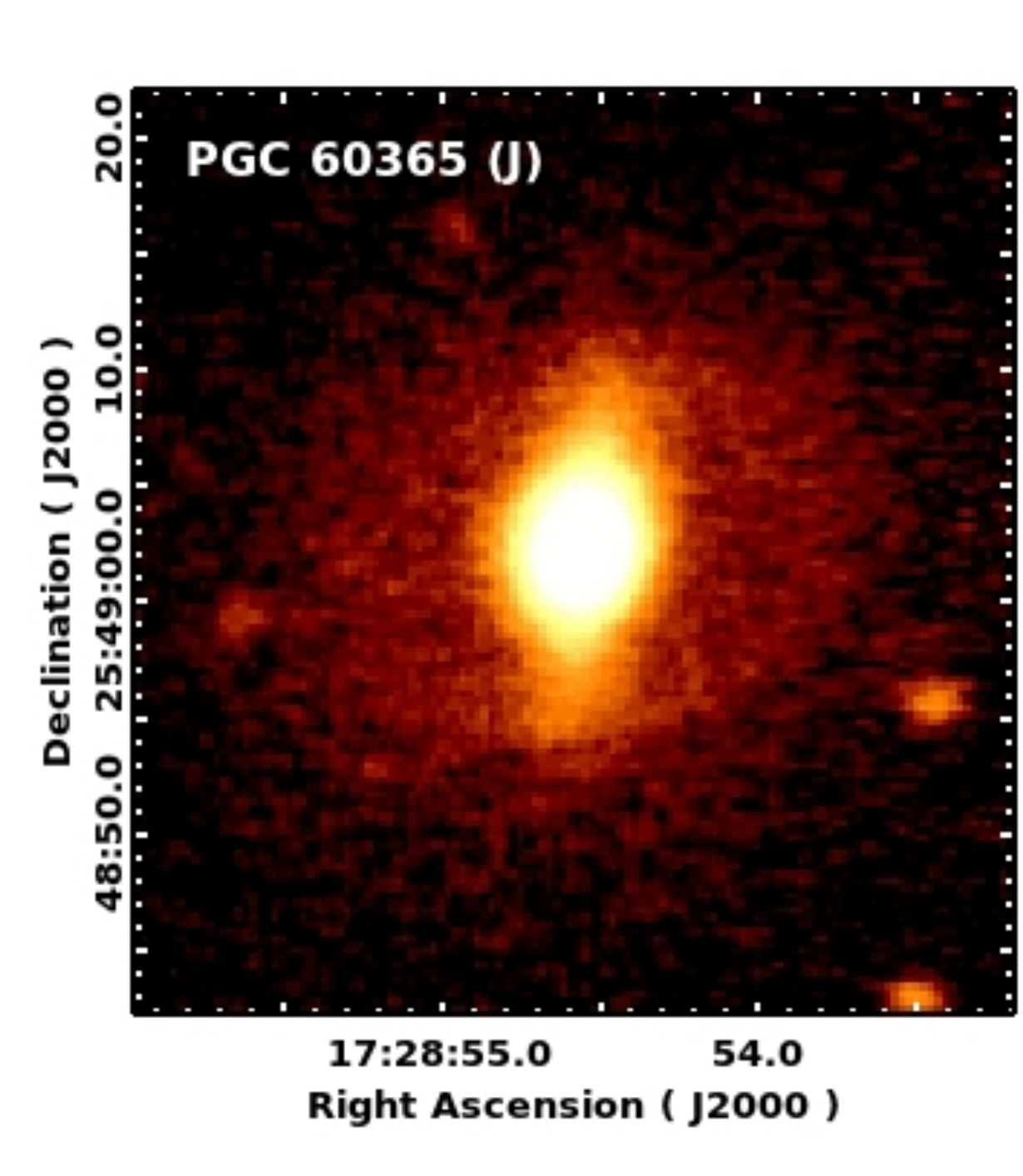}\includegraphics[trim = 50mm 17mm 65mm 8mm, clip,scale=0.42]{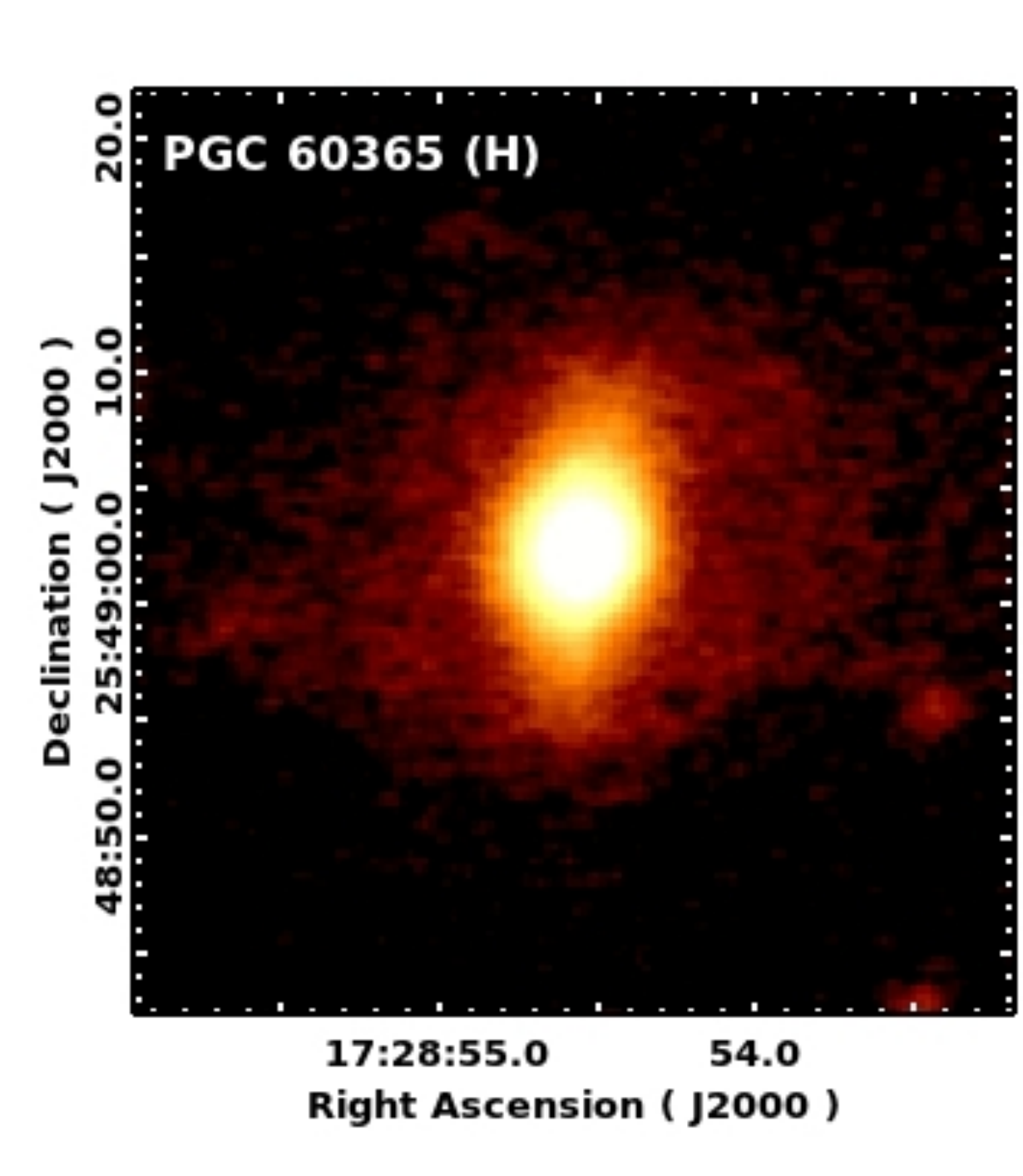}\includegraphics[trim = 50mm 17mm 65mm 8mm, clip,scale=0.42]{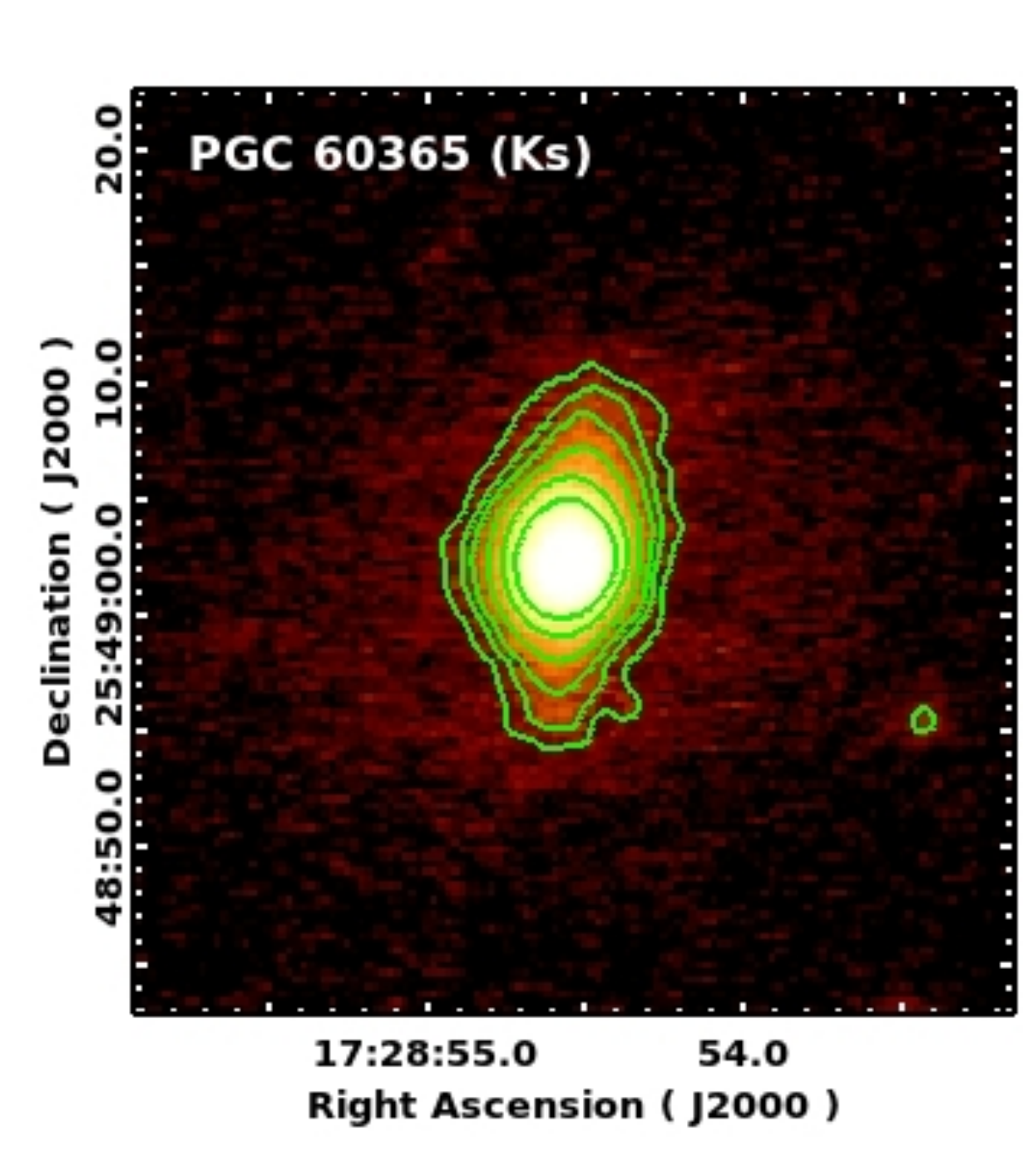}
\includegraphics[trim = 50mm 17mm 65mm 8mm, clip,scale=0.42]{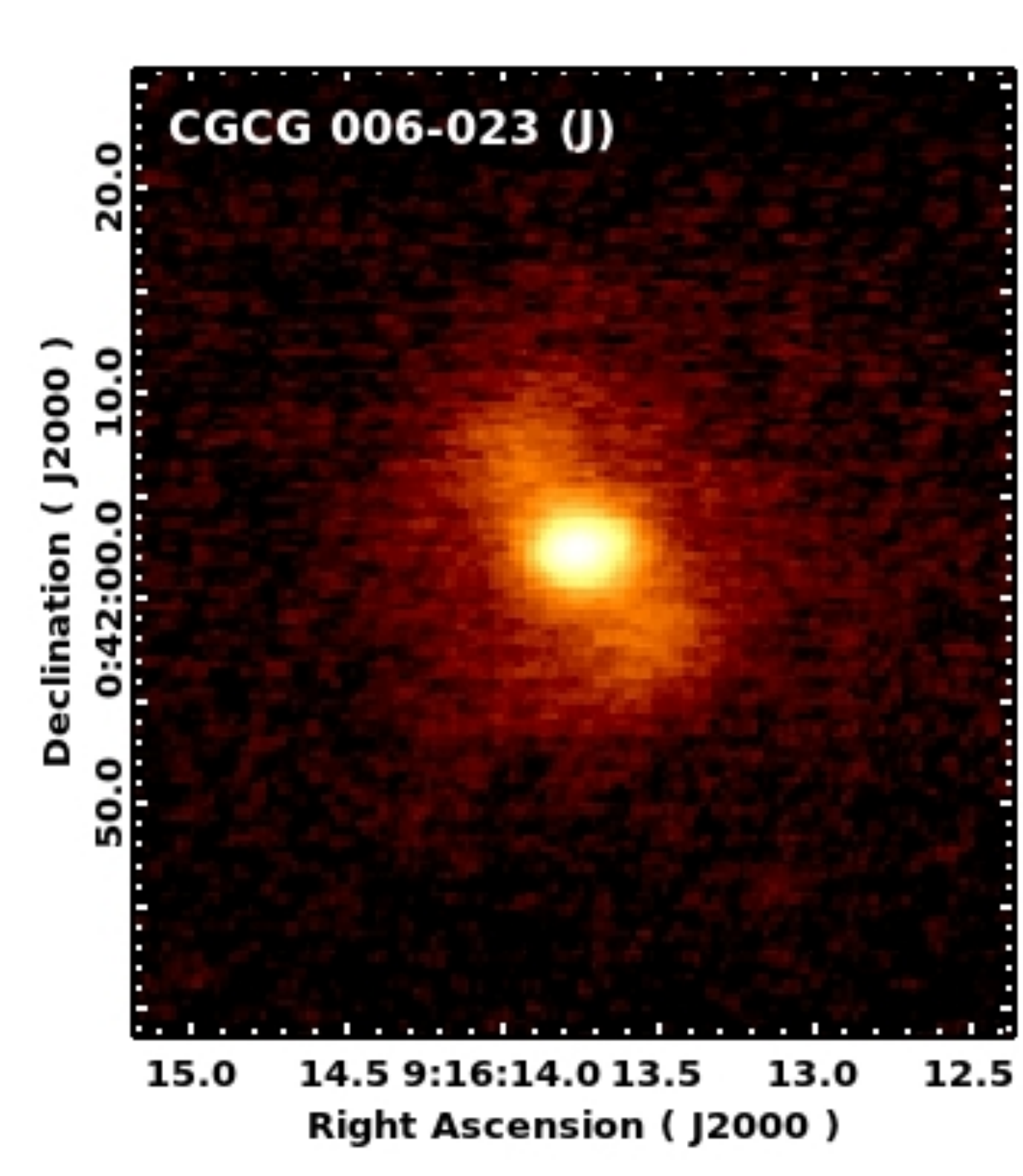}\includegraphics[trim = 50mm 17mm 65mm 8mm, clip,scale=0.42]{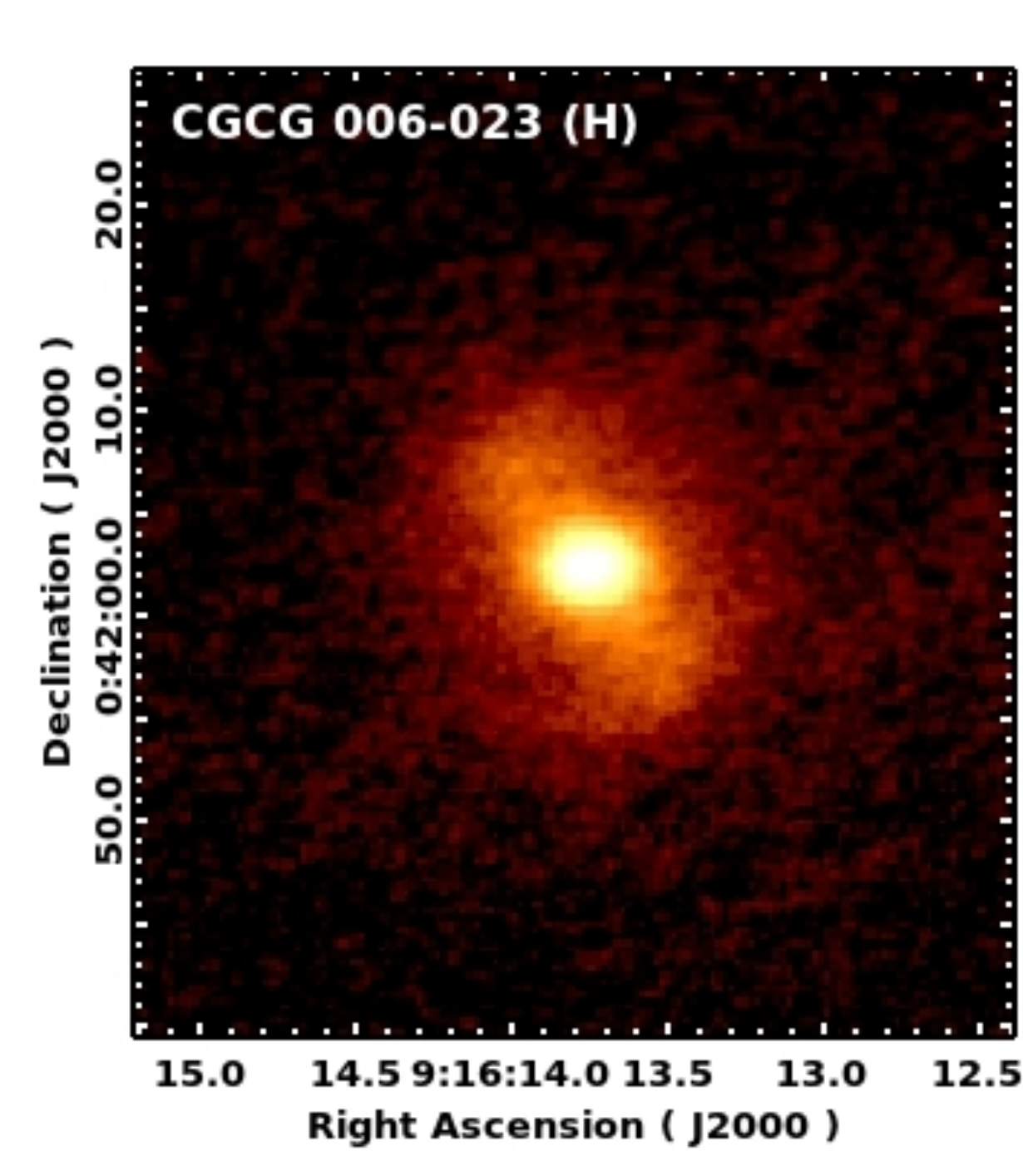}\includegraphics[trim = 50mm 17mm 65mm 8mm, clip,scale=0.42]{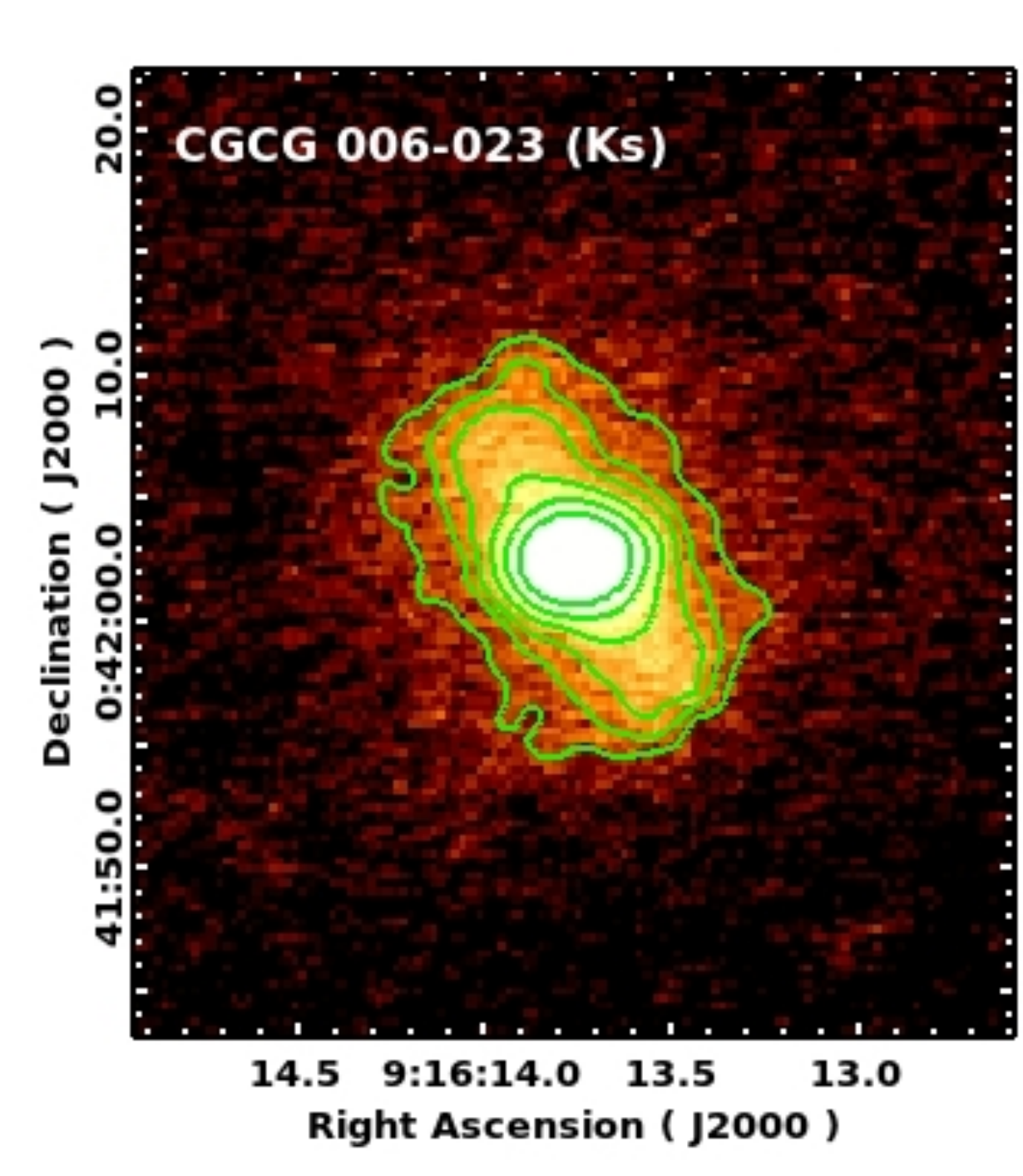}
\end{figure*}

\begin{figure*}
\includegraphics[trim = 50mm 17mm 65mm 8mm, clip,scale=0.42]{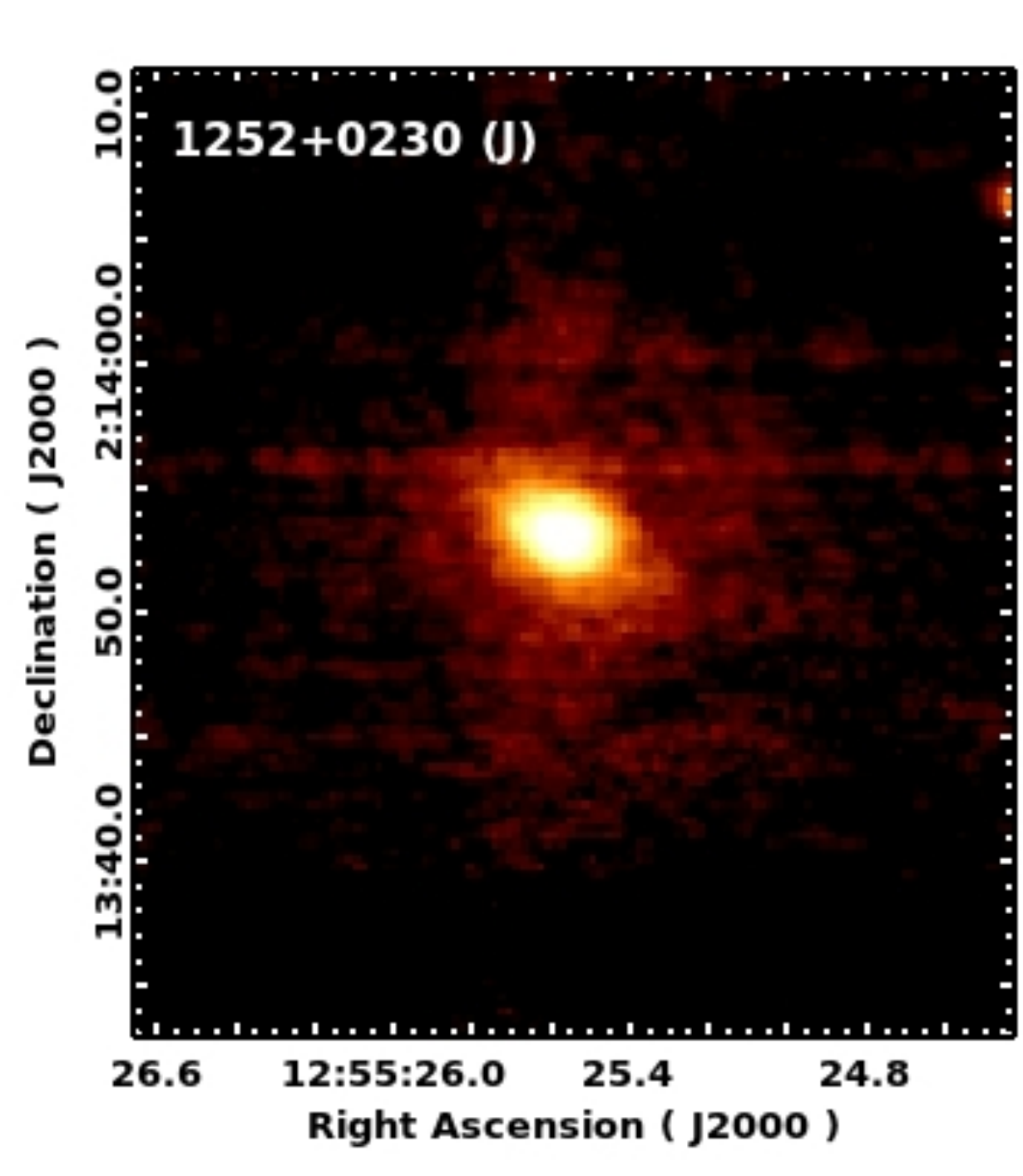}\includegraphics[trim = 50mm 17mm 65mm 8mm, clip,scale=0.42]{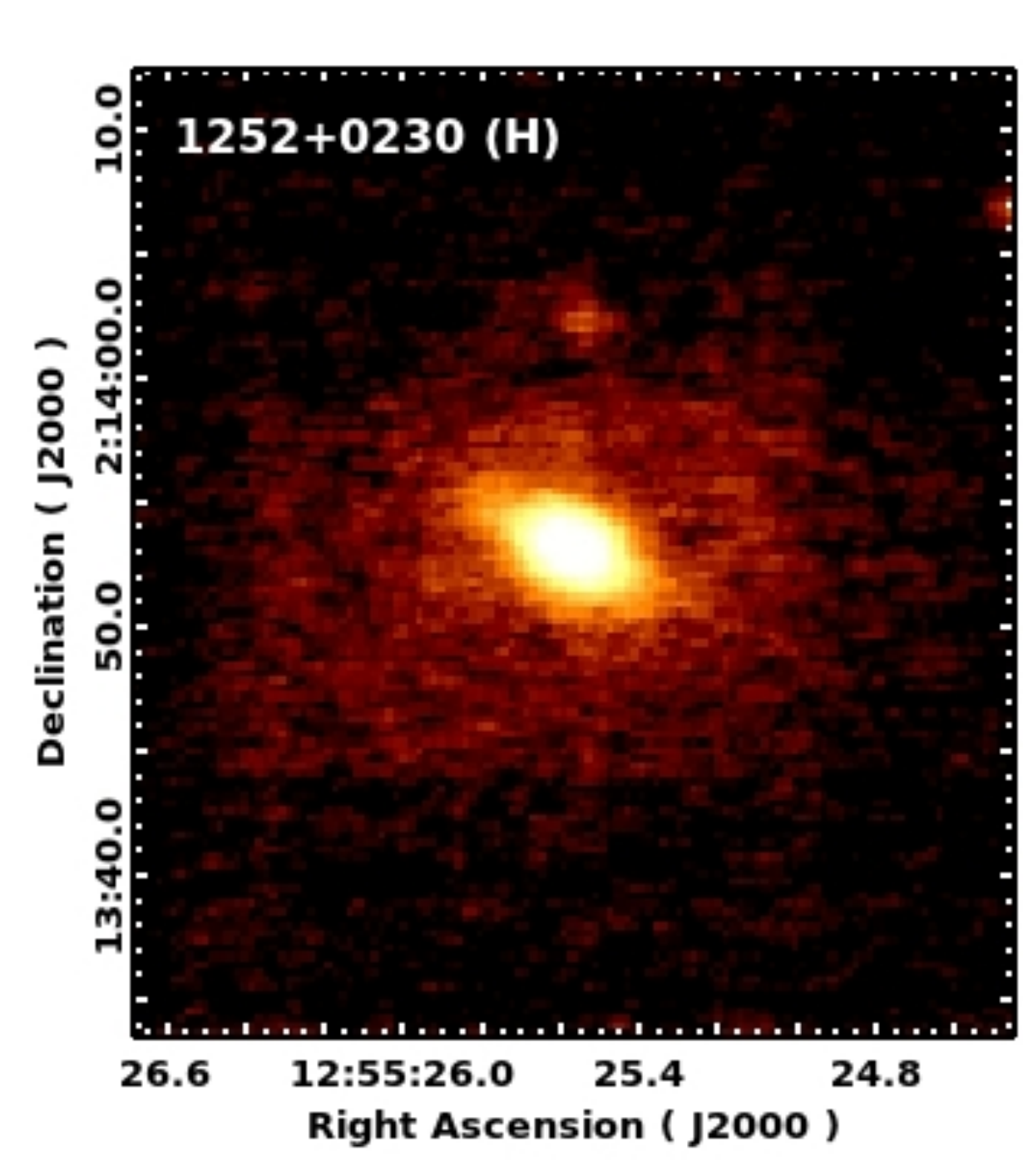}\includegraphics[trim = 50mm 17mm 65mm 8mm, clip,scale=0.42]{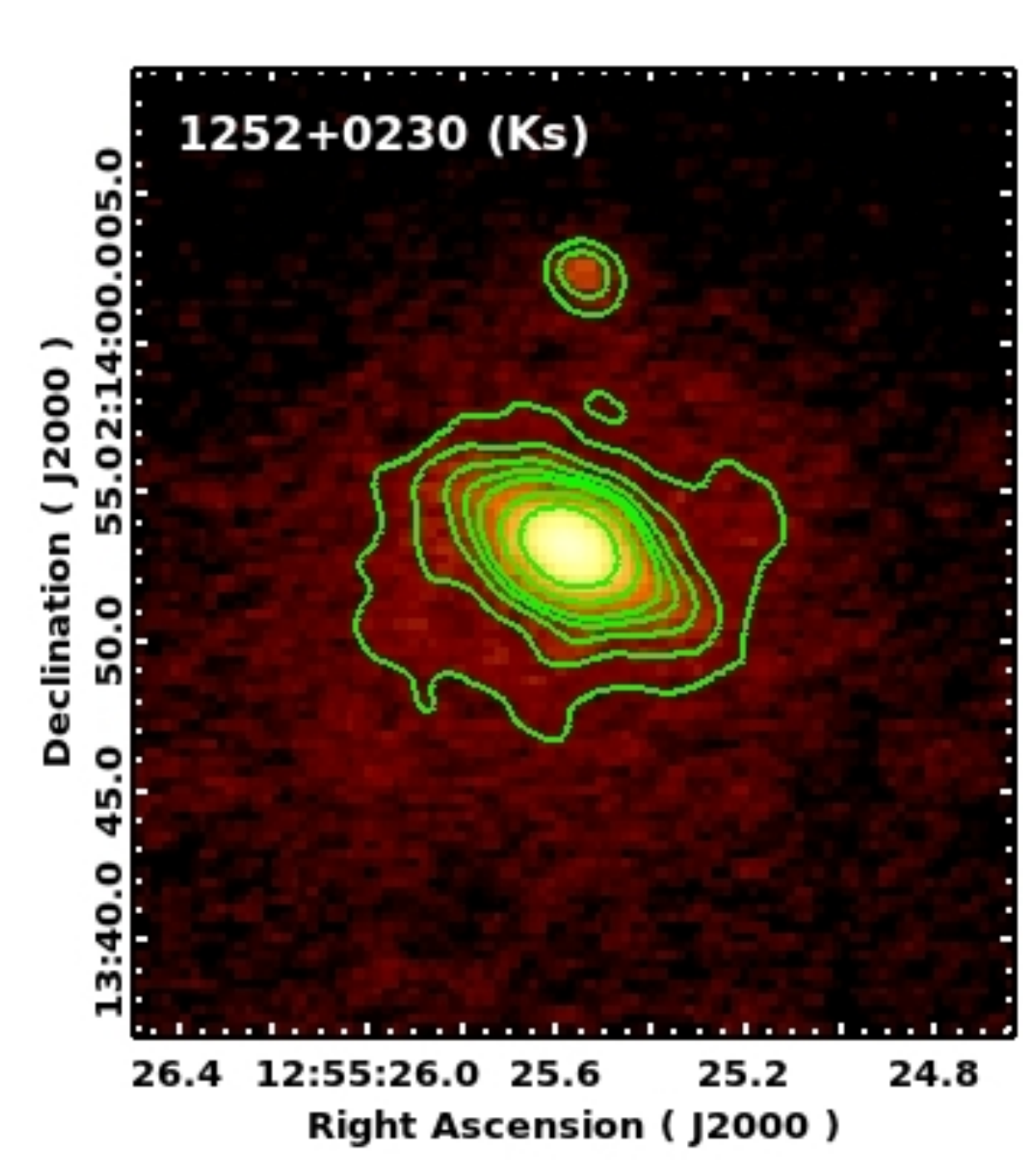}

\includegraphics[trim = 50mm 25mm 65mm 15mm, clip,scale=0.45]{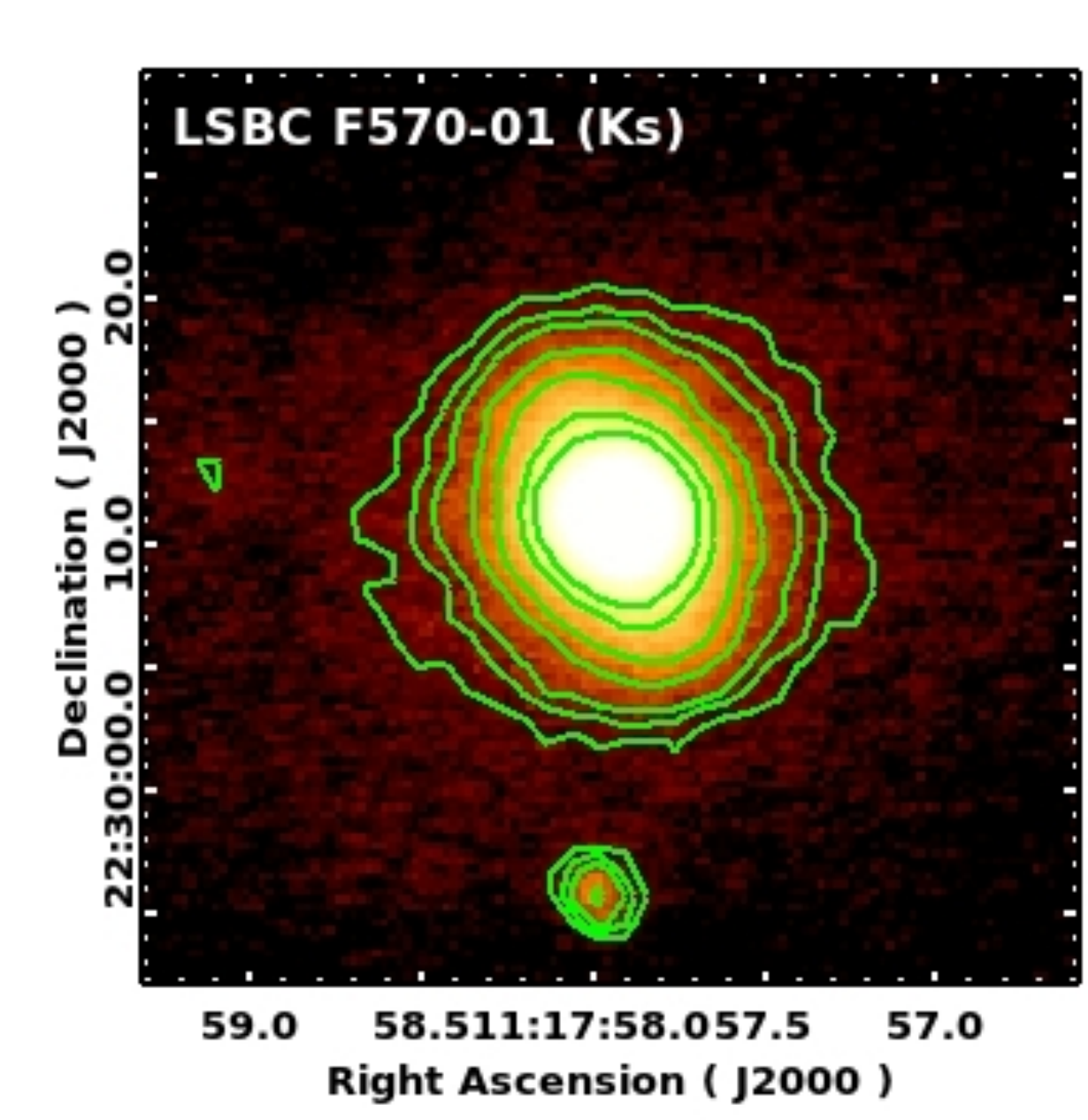}\includegraphics[trim = 50mm 17mm 60mm 8mm, clip,scale=0.4]{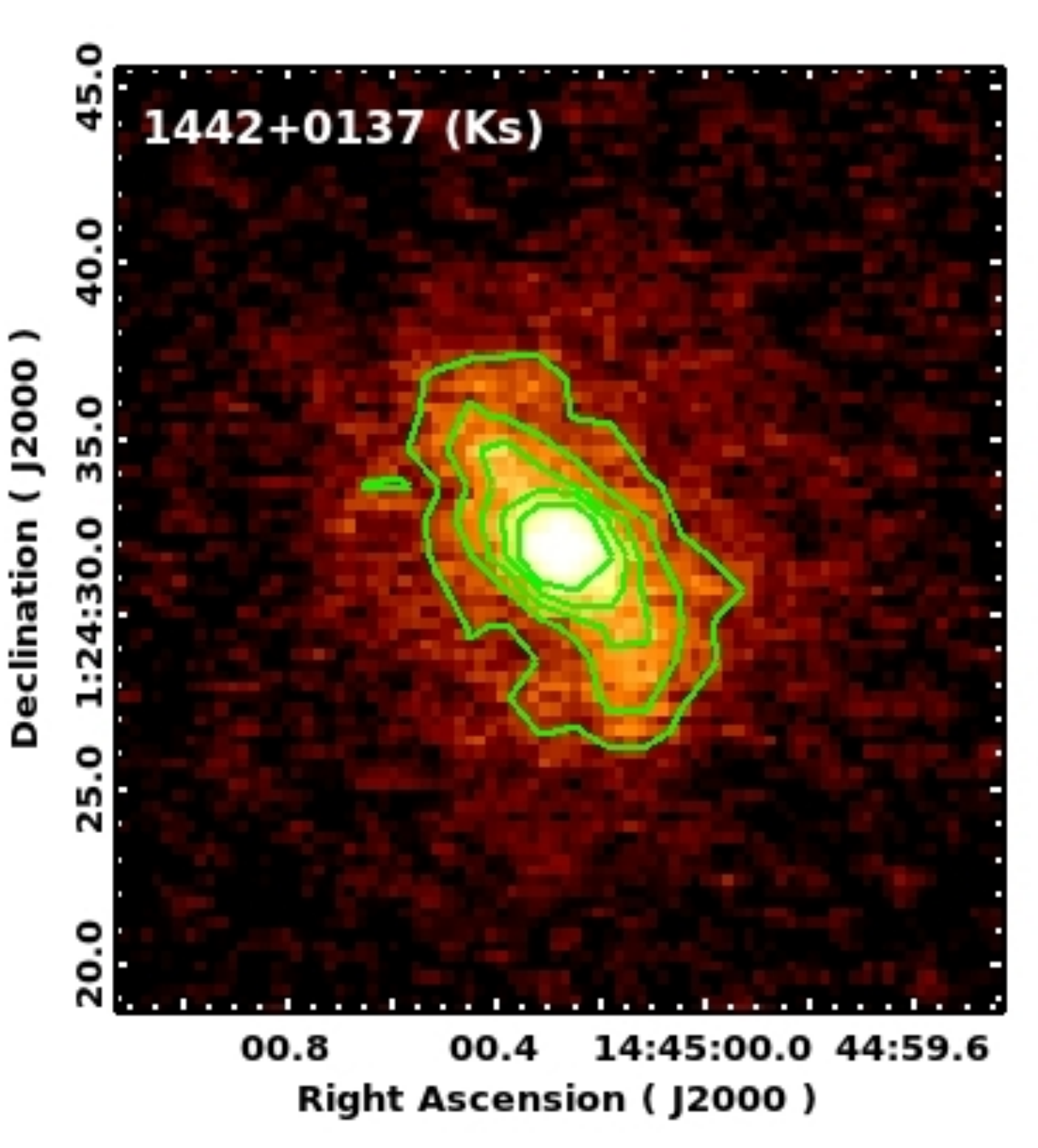}\includegraphics[trim = 50mm 17mm 65mm 8mm, clip,scale=0.4]{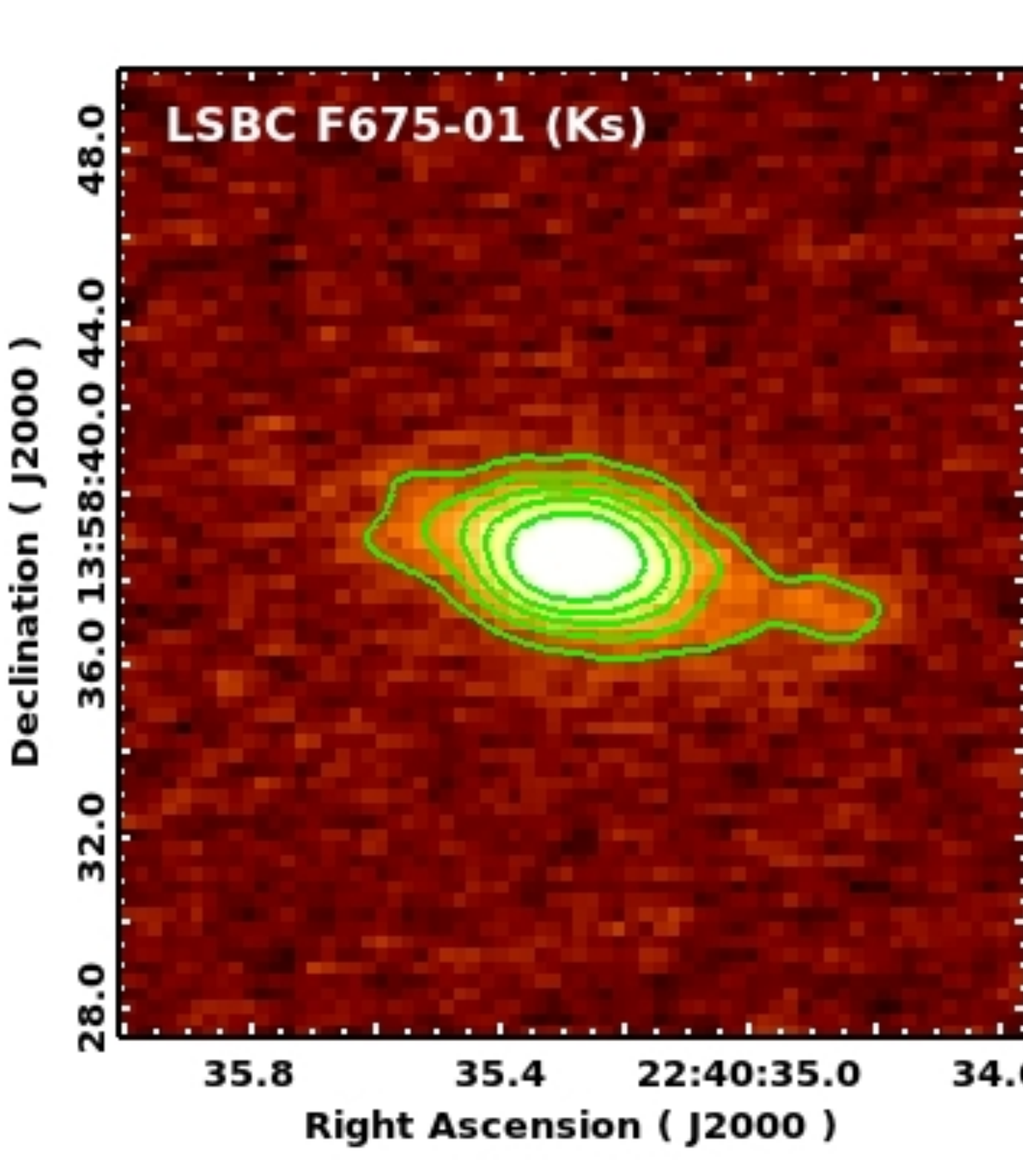}
\includegraphics[trim = 50mm 17mm 60mm 8mm, clip,scale=0.36]{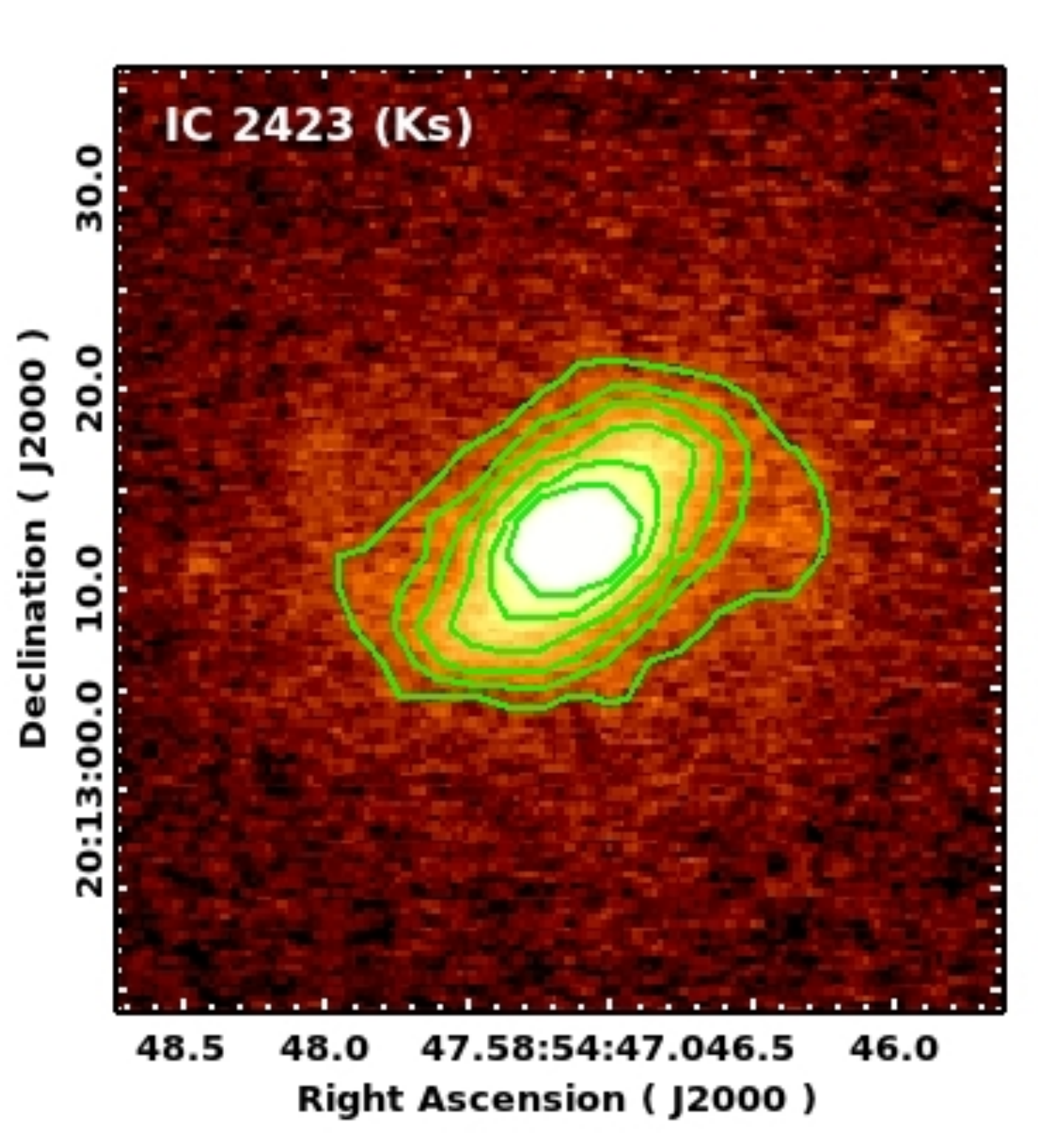}
\\

\end{figure*}

\begin{figure*}
\caption{The J-K$_s$ images of Low Surface Brightness galaxies}

\includegraphics[scale=0.26]{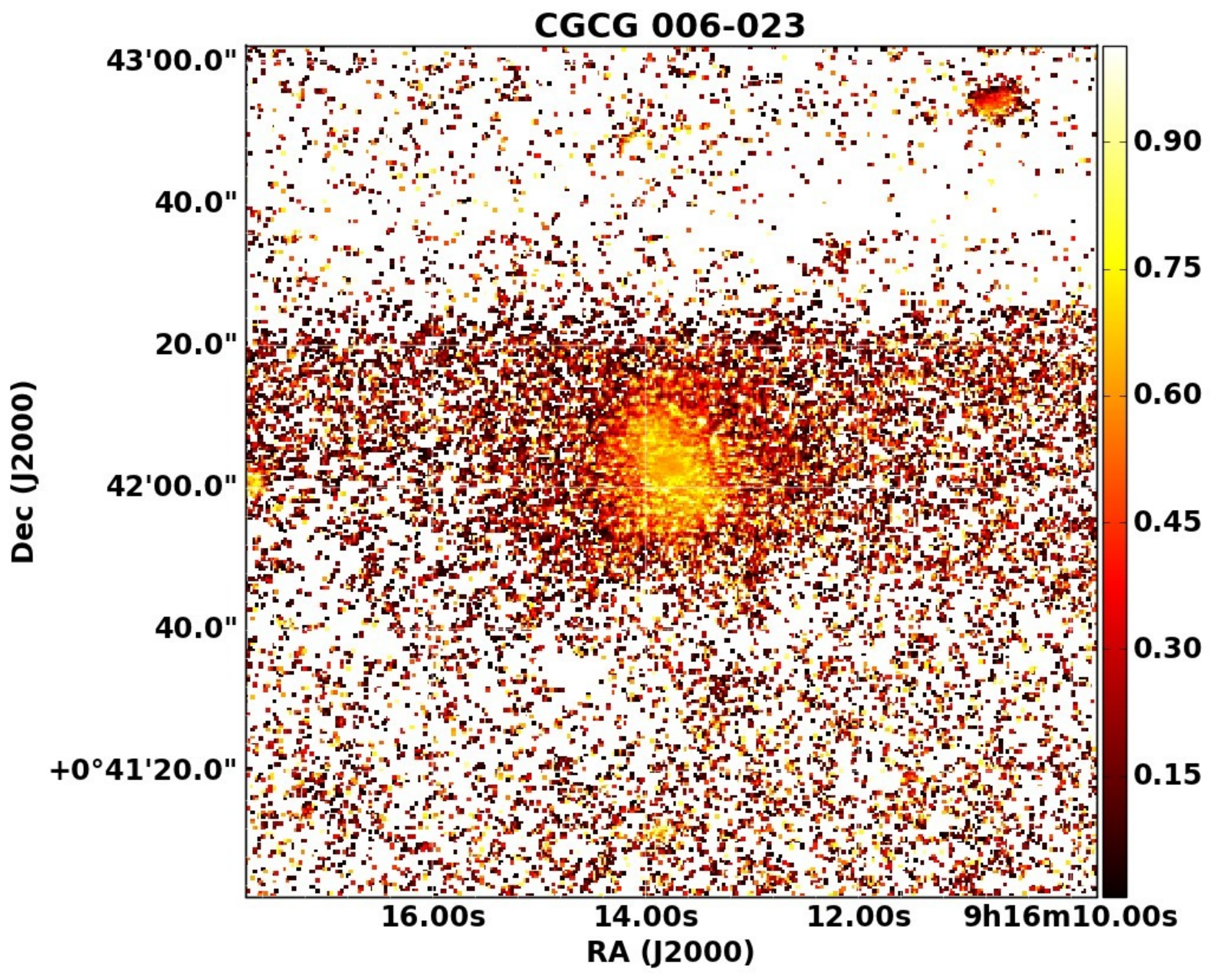}\includegraphics[scale=0.26]{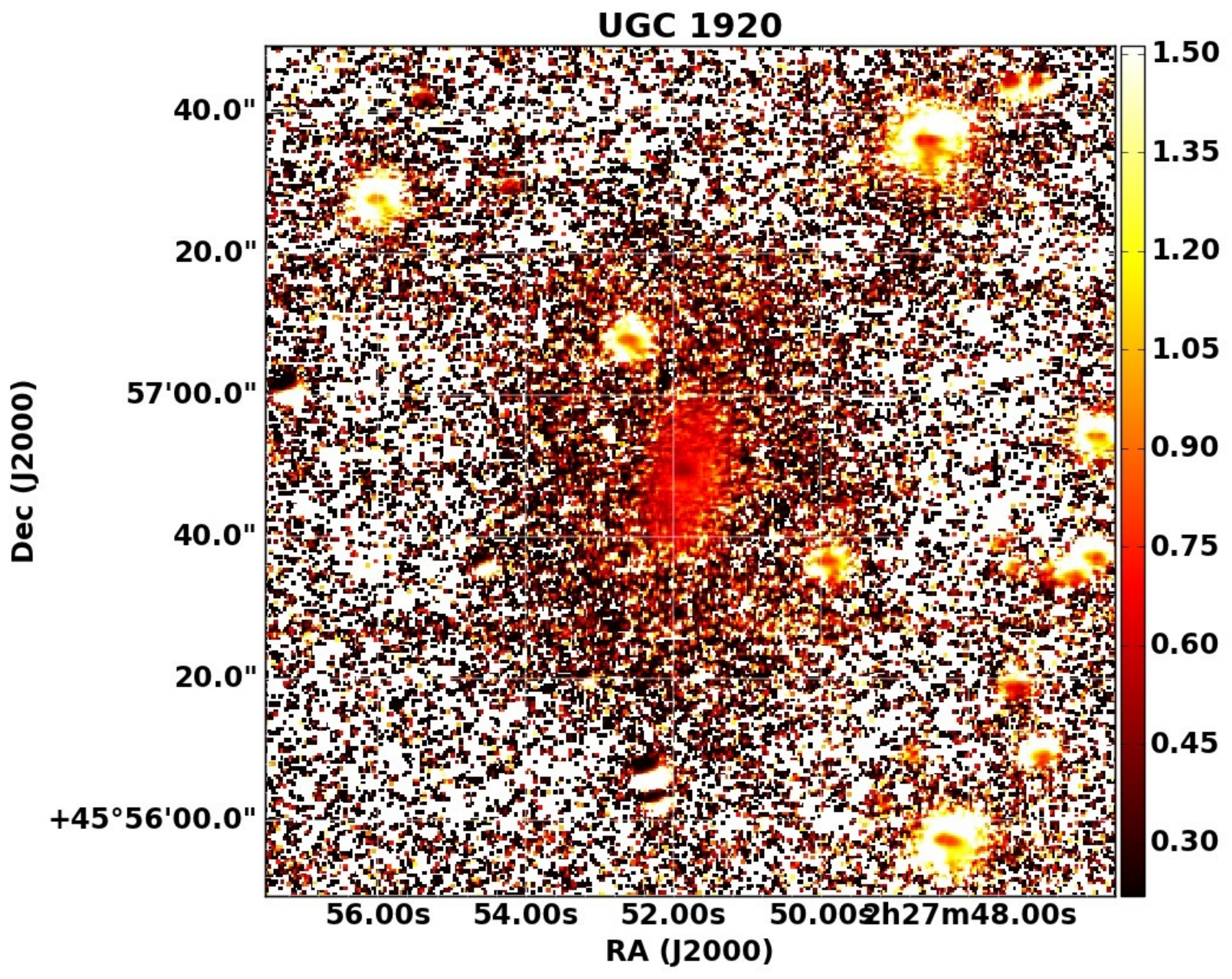}
\includegraphics[scale=0.26]{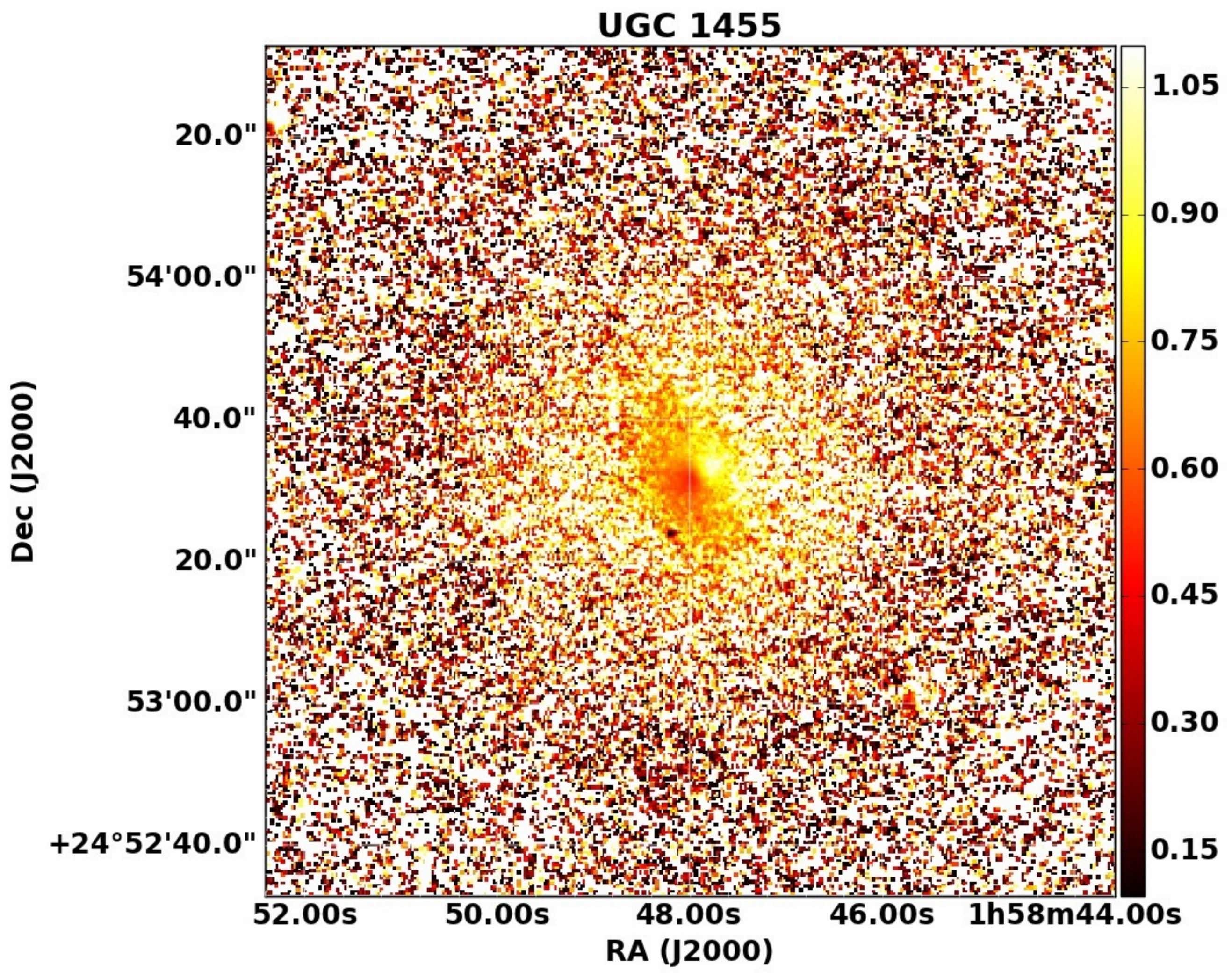}\includegraphics[scale=0.26]{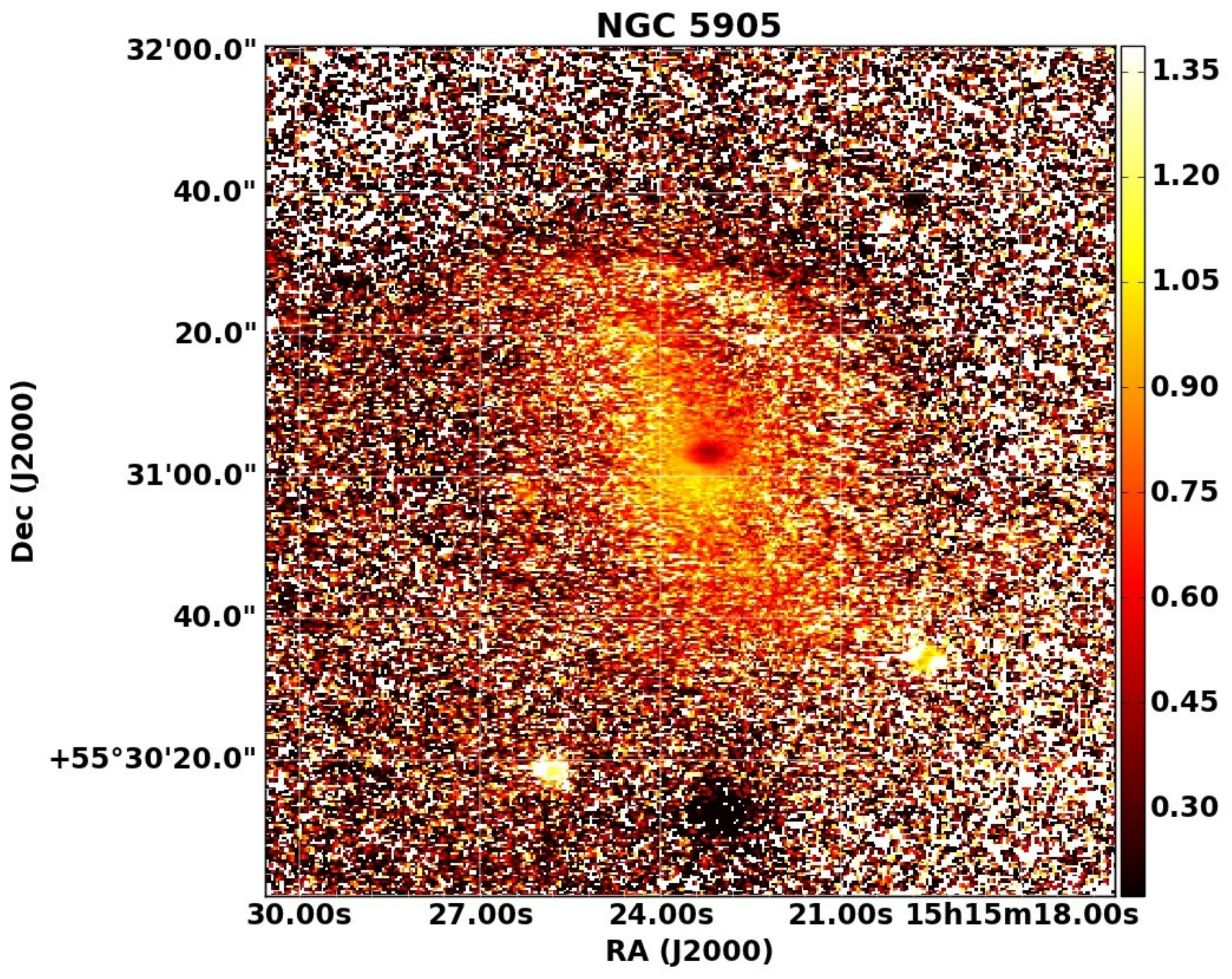}
\includegraphics[scale=0.26]{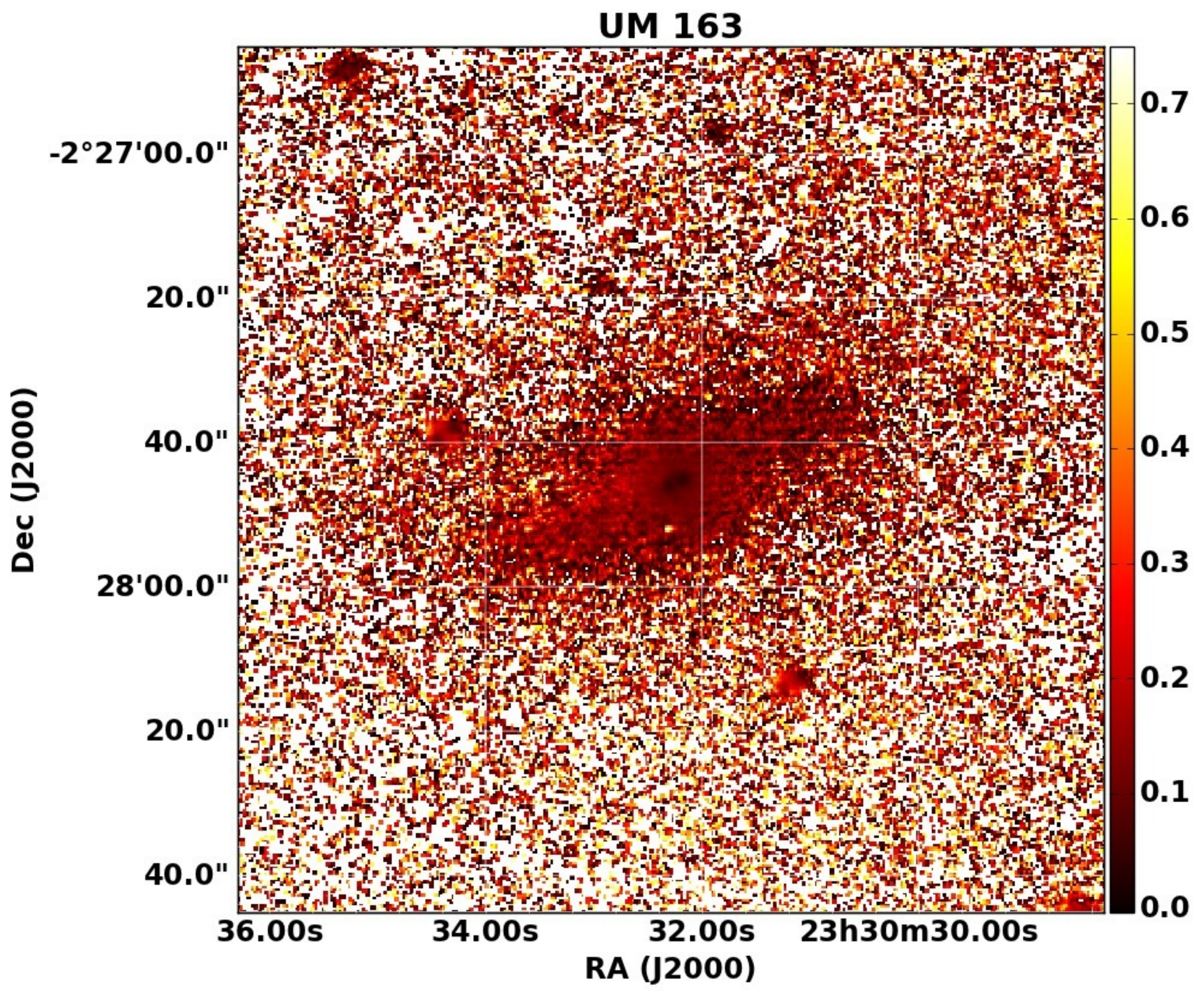}\includegraphics[scale=0.26]{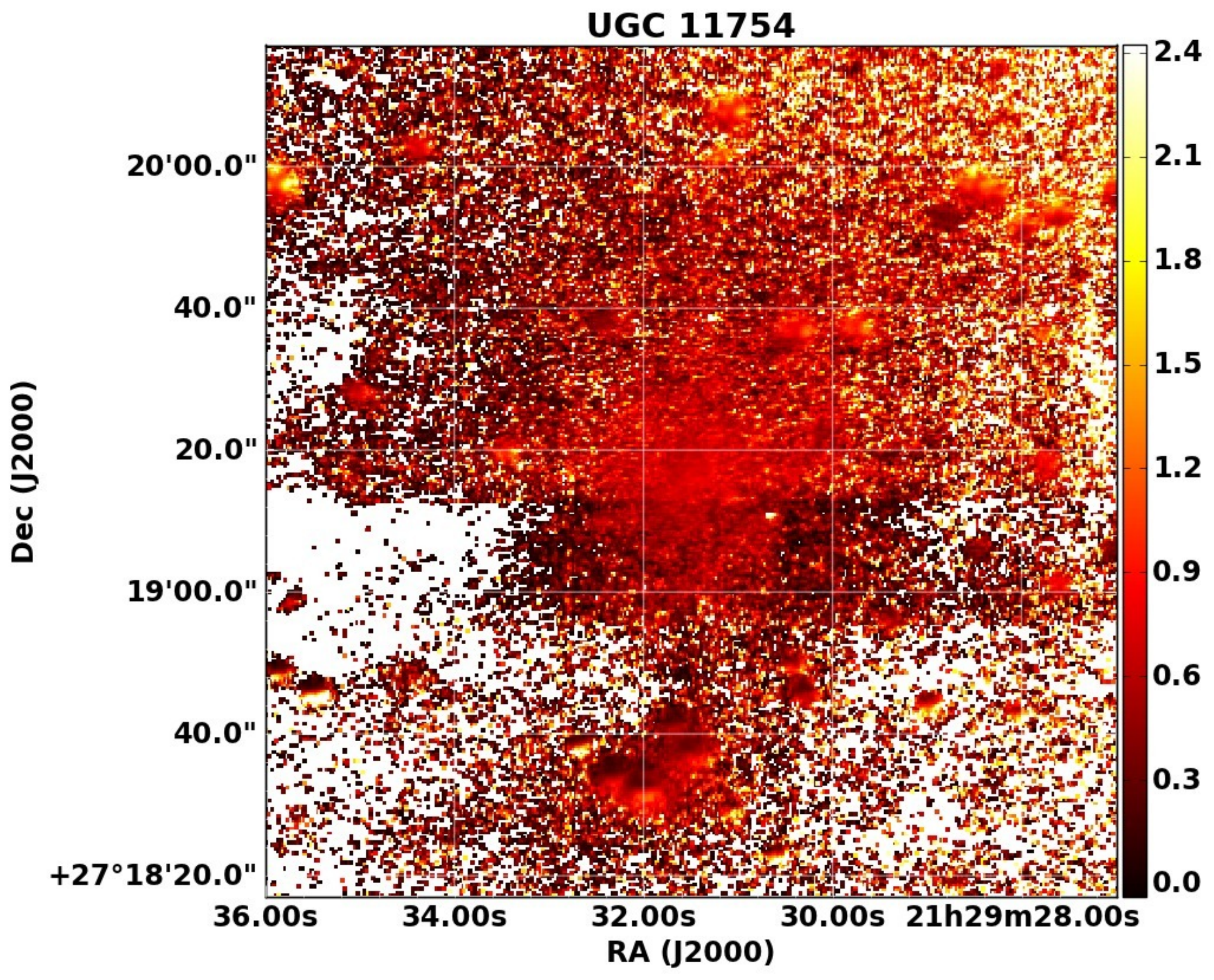}
\includegraphics[scale=0.26]{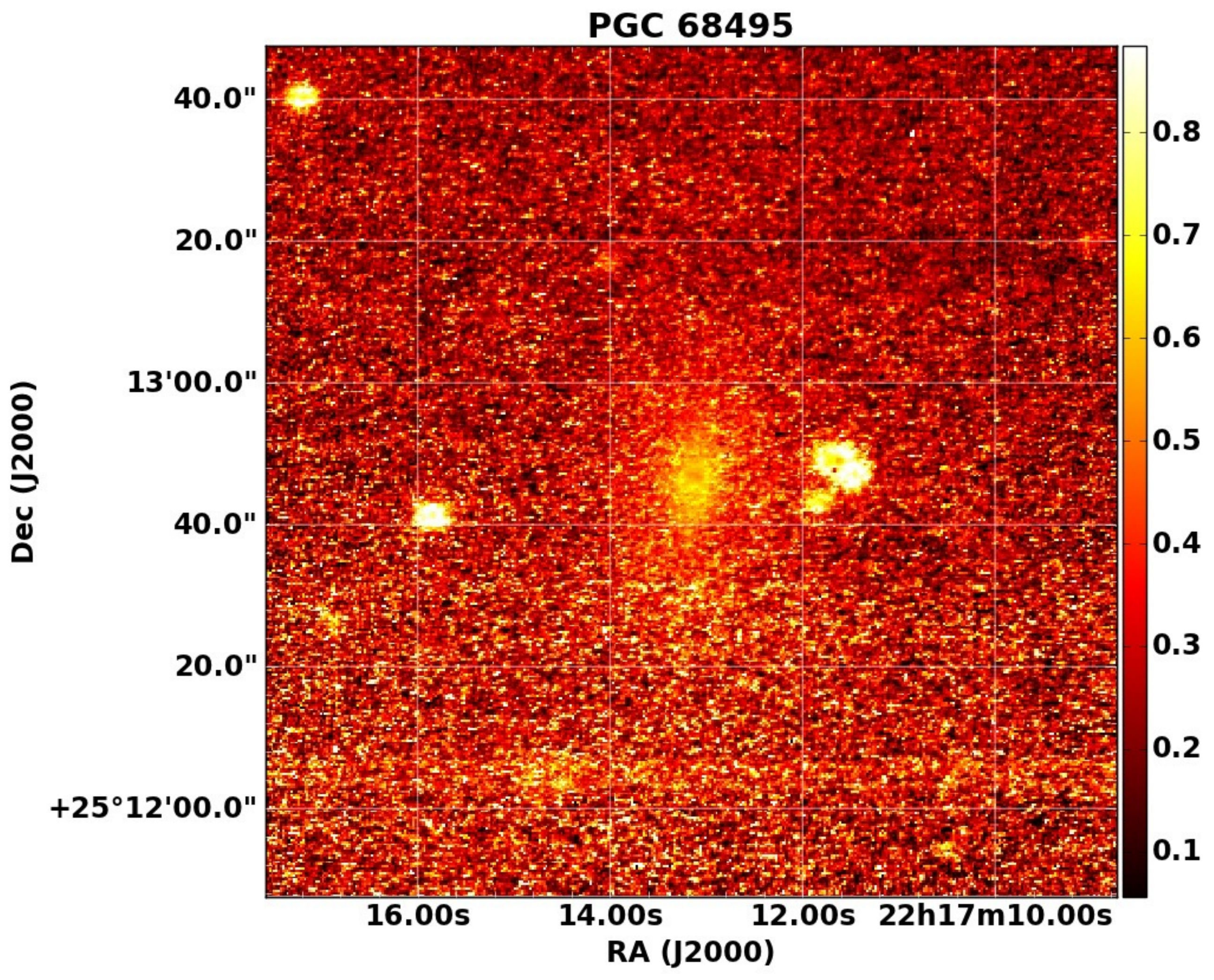}\includegraphics[scale=0.26]{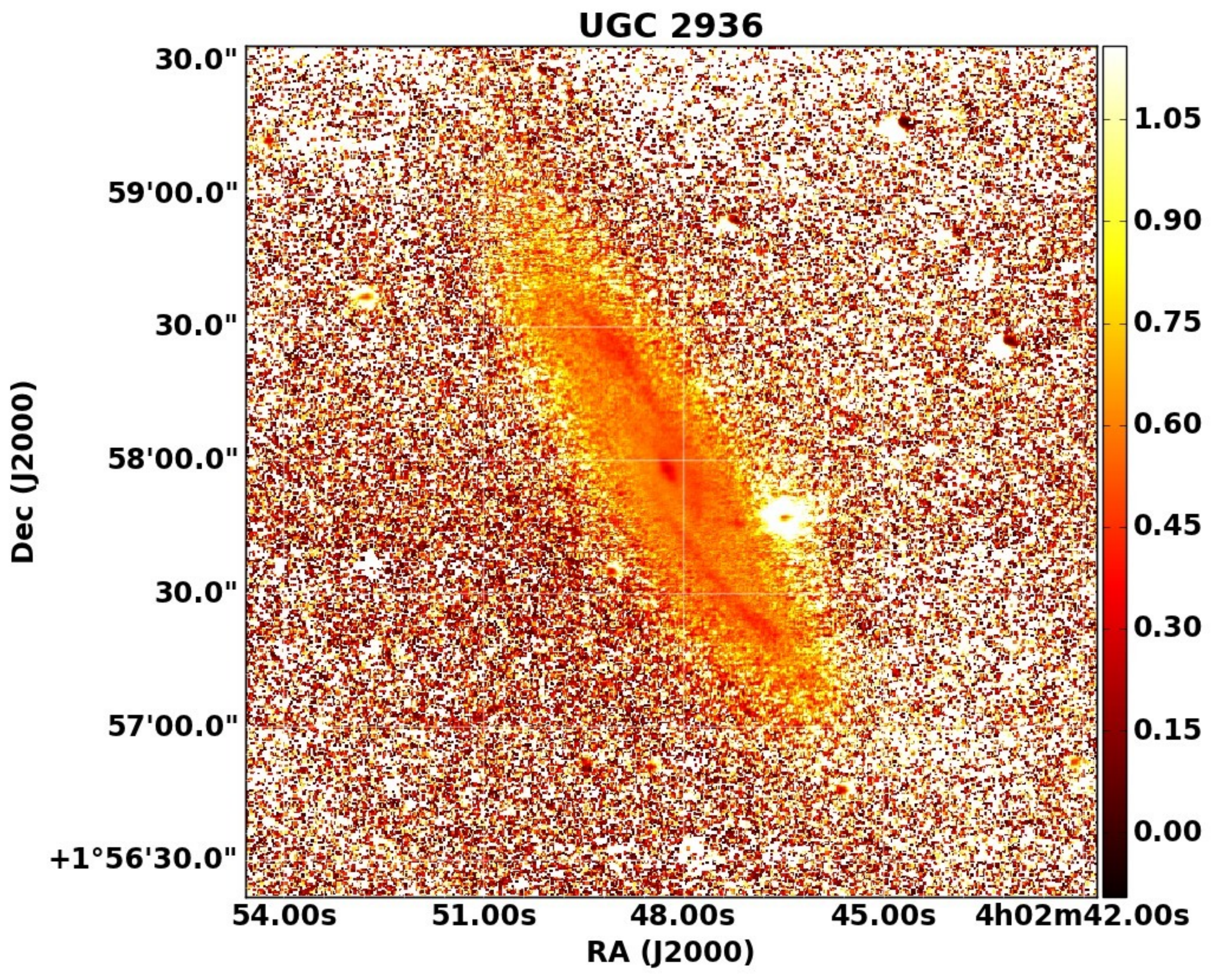}
\end{figure*}
\begin{figure*}

\includegraphics[scale=0.28]{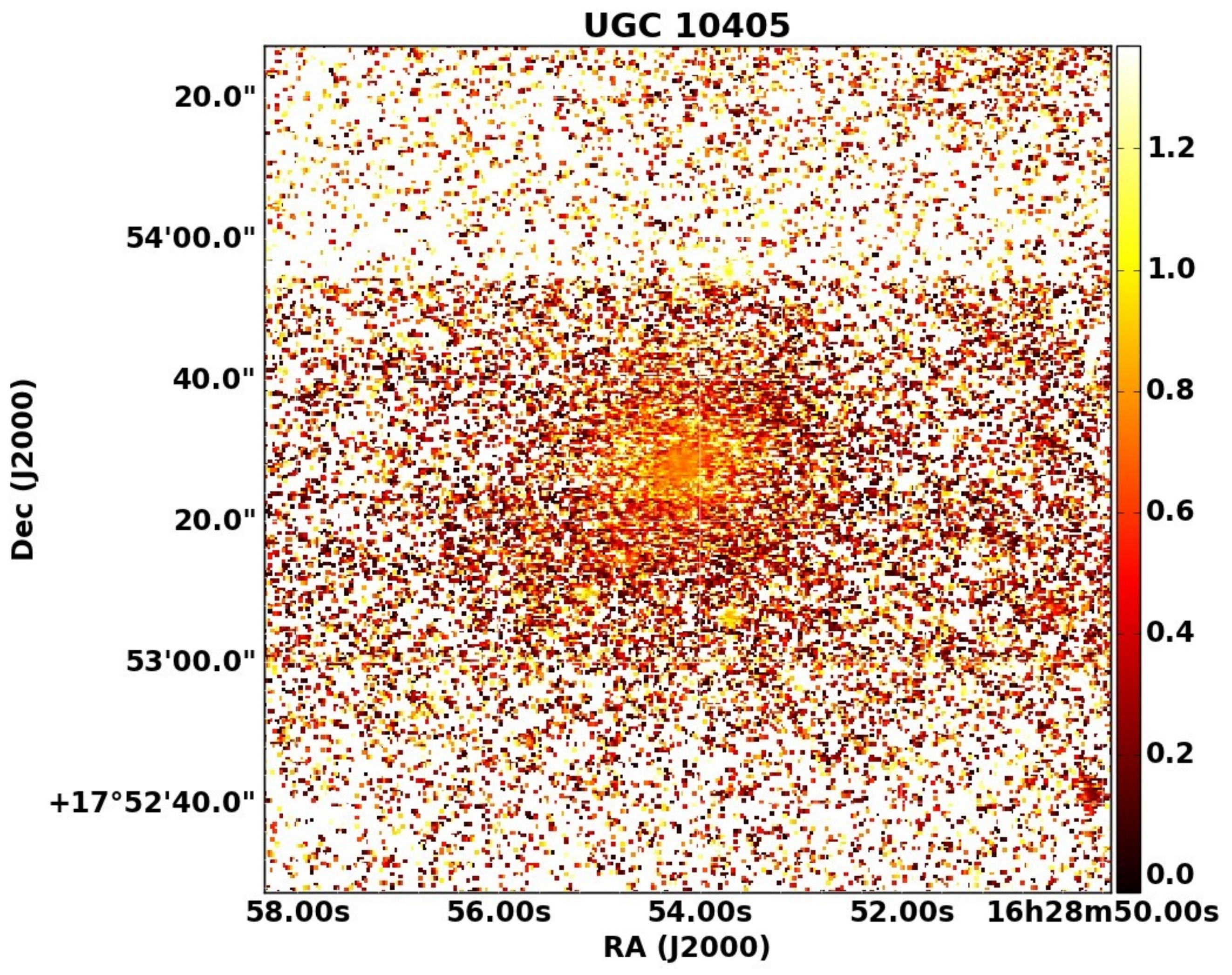}\includegraphics[scale=0.28]{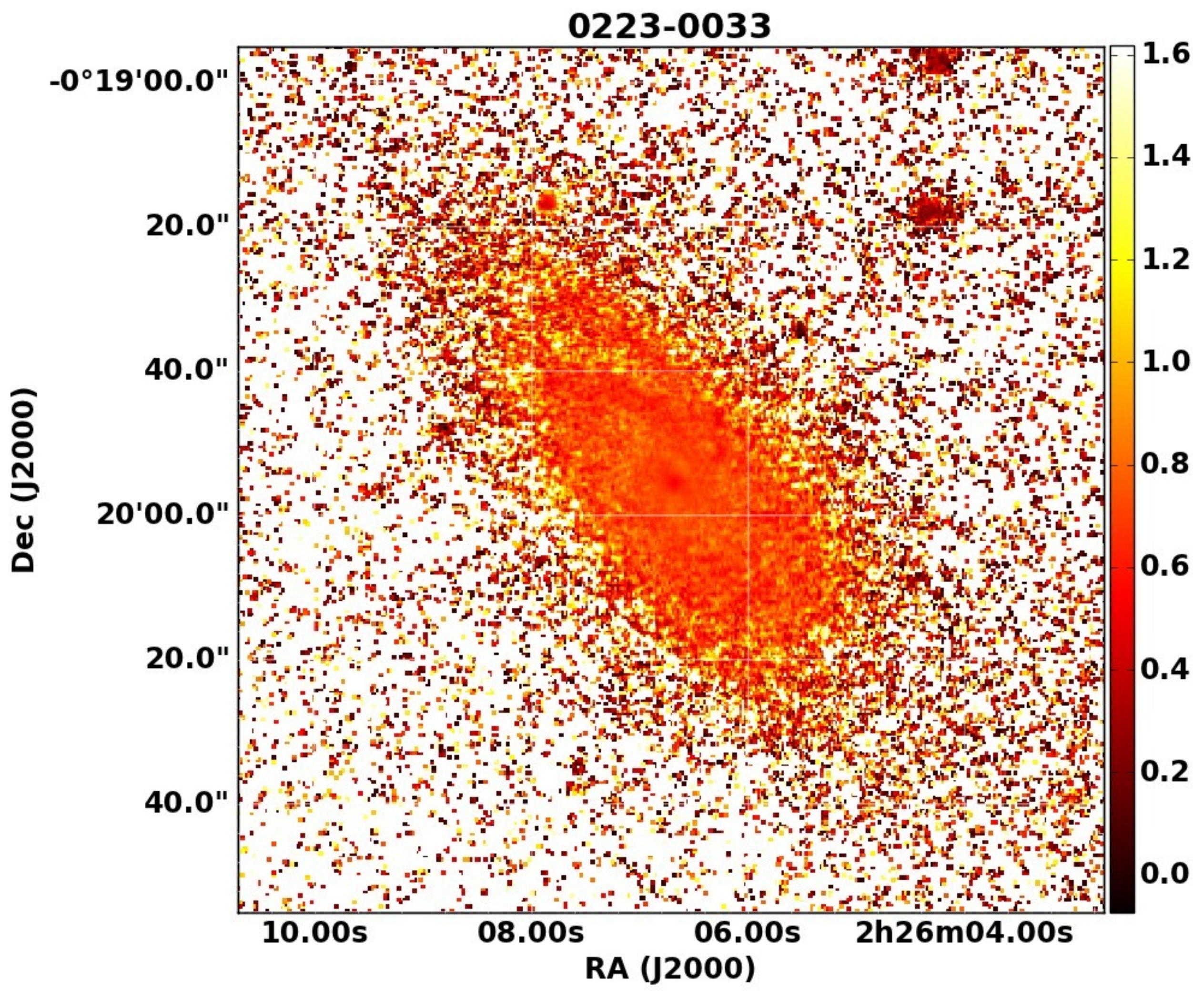}
\includegraphics[scale=0.28]{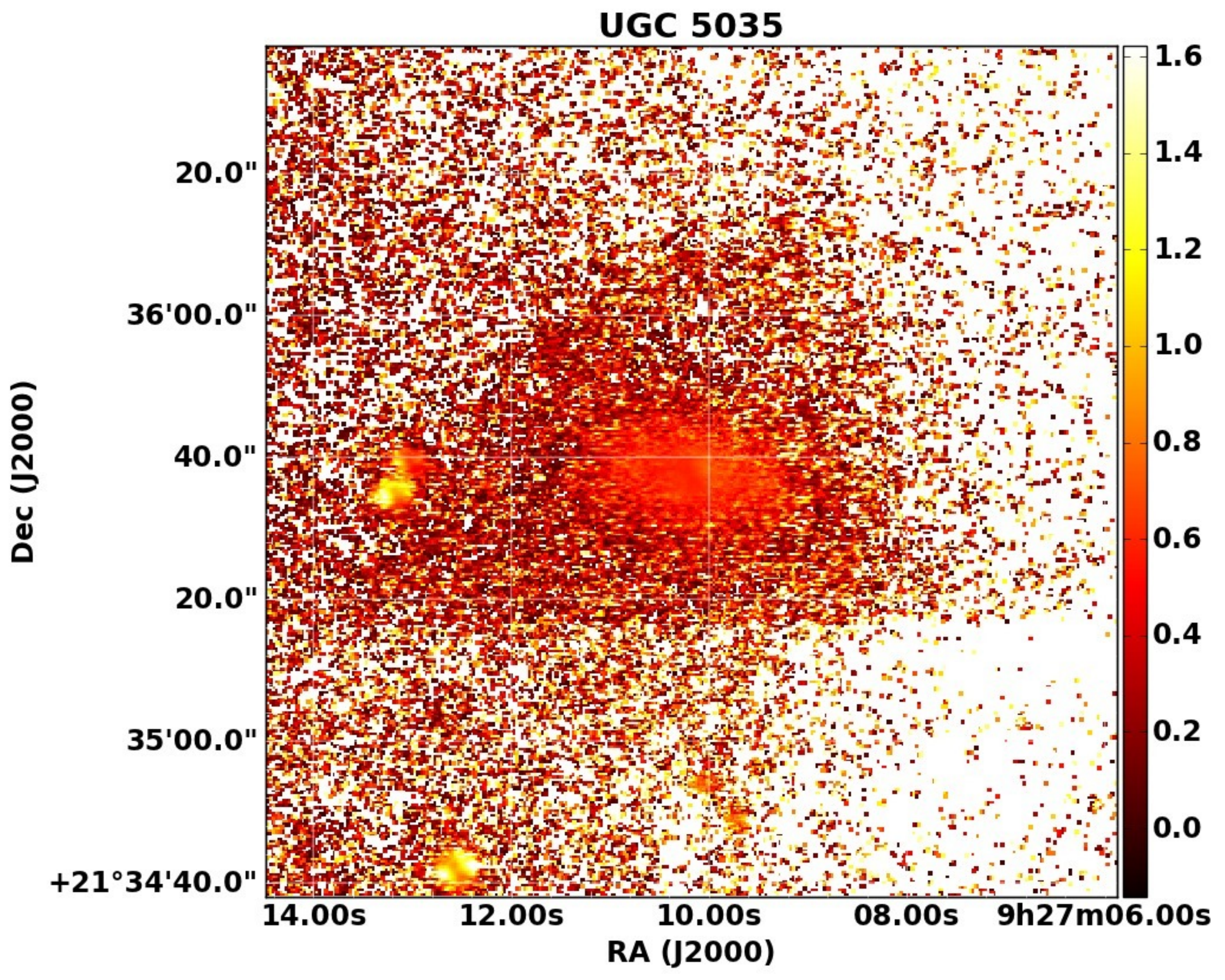}\includegraphics[scale=0.28]{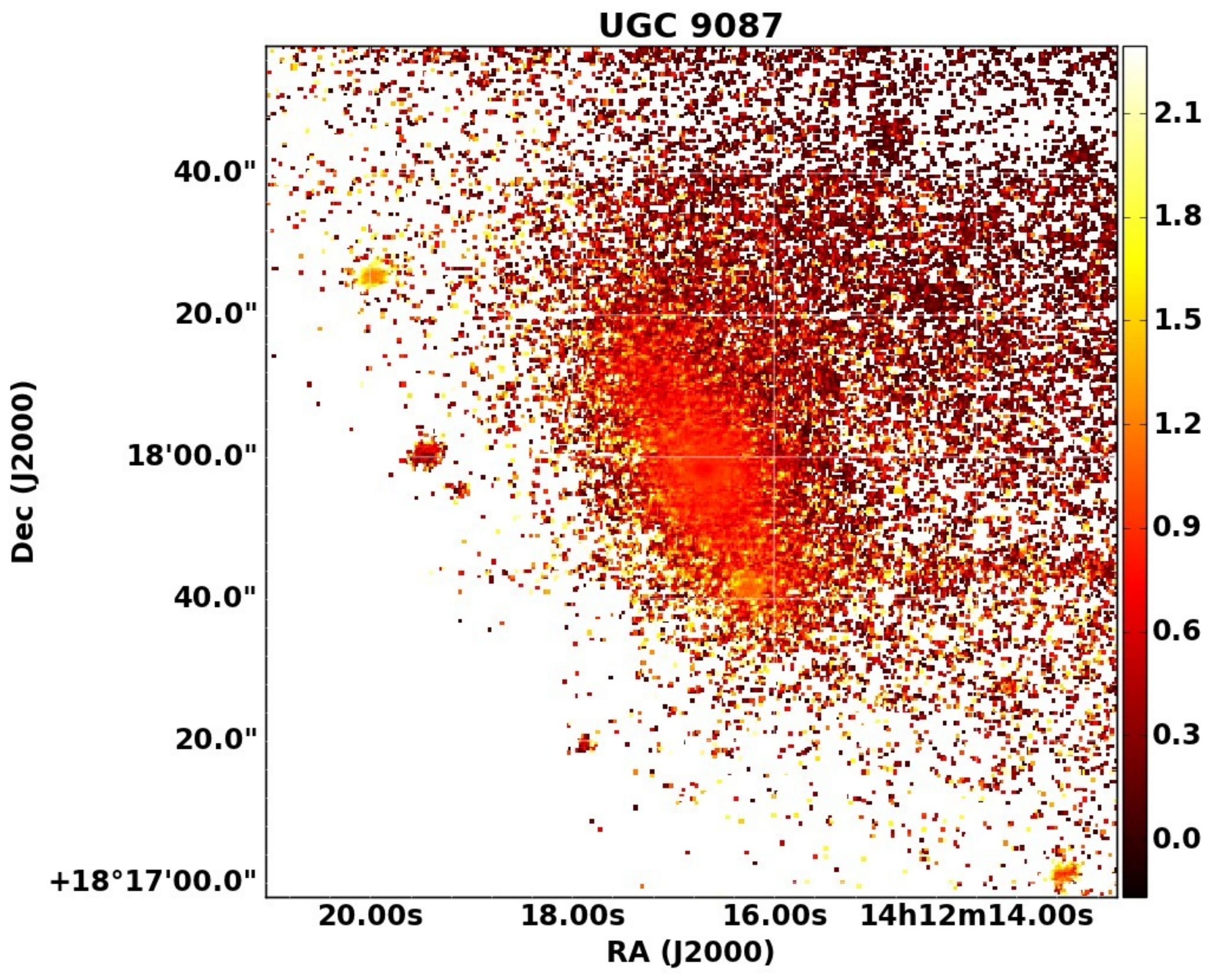}
\includegraphics[scale=0.28]{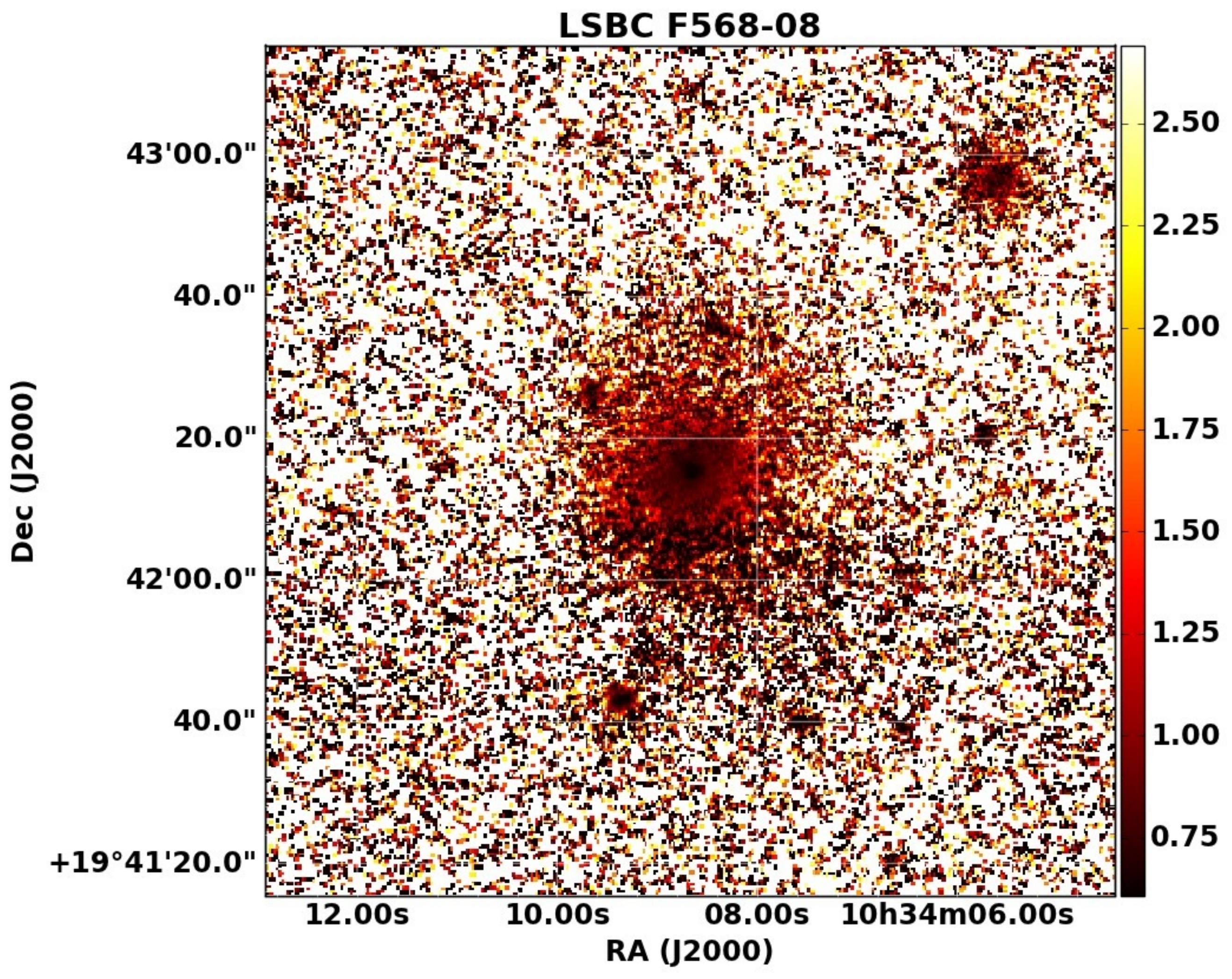}\includegraphics[scale=0.28]{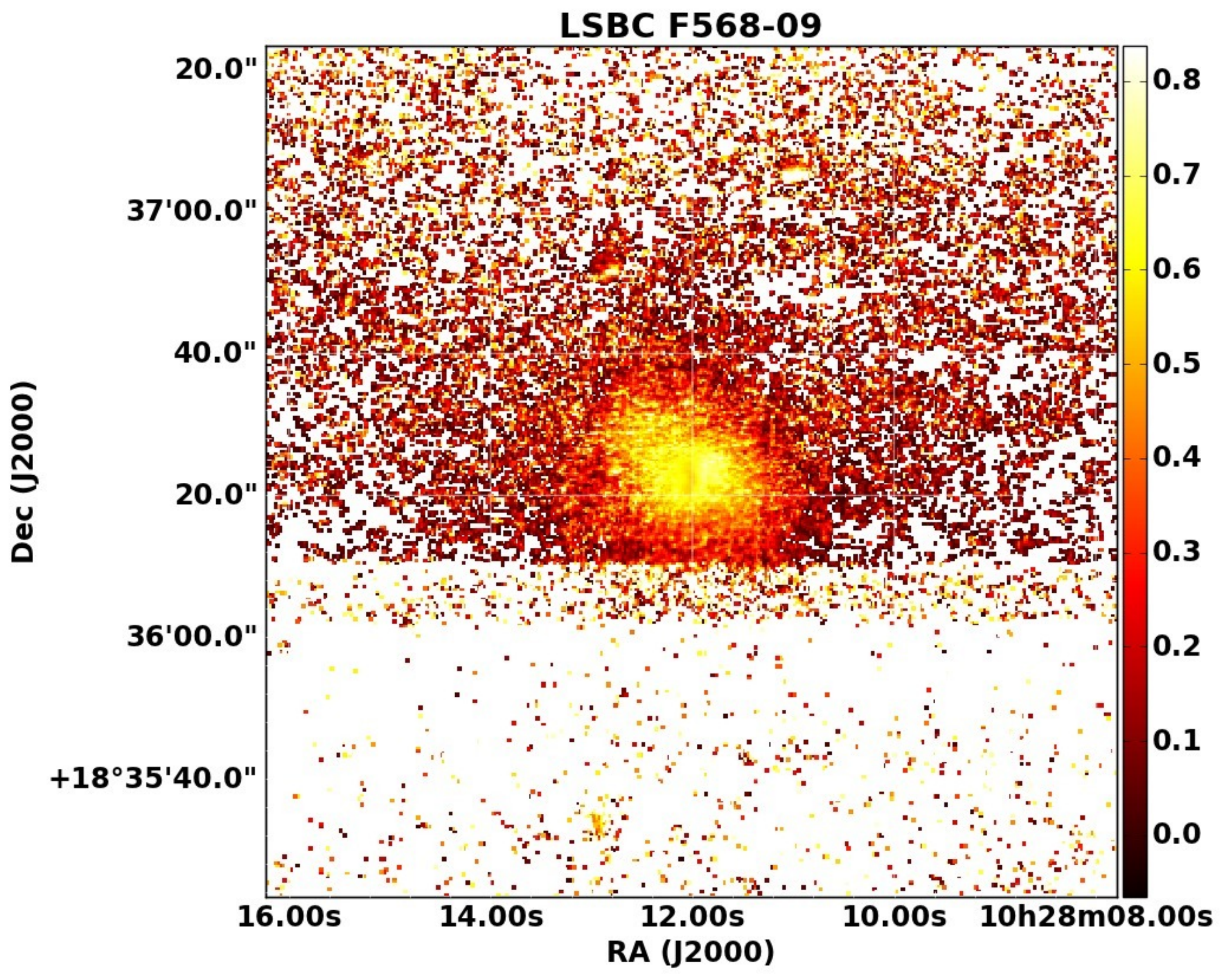}
\includegraphics[scale=0.28]{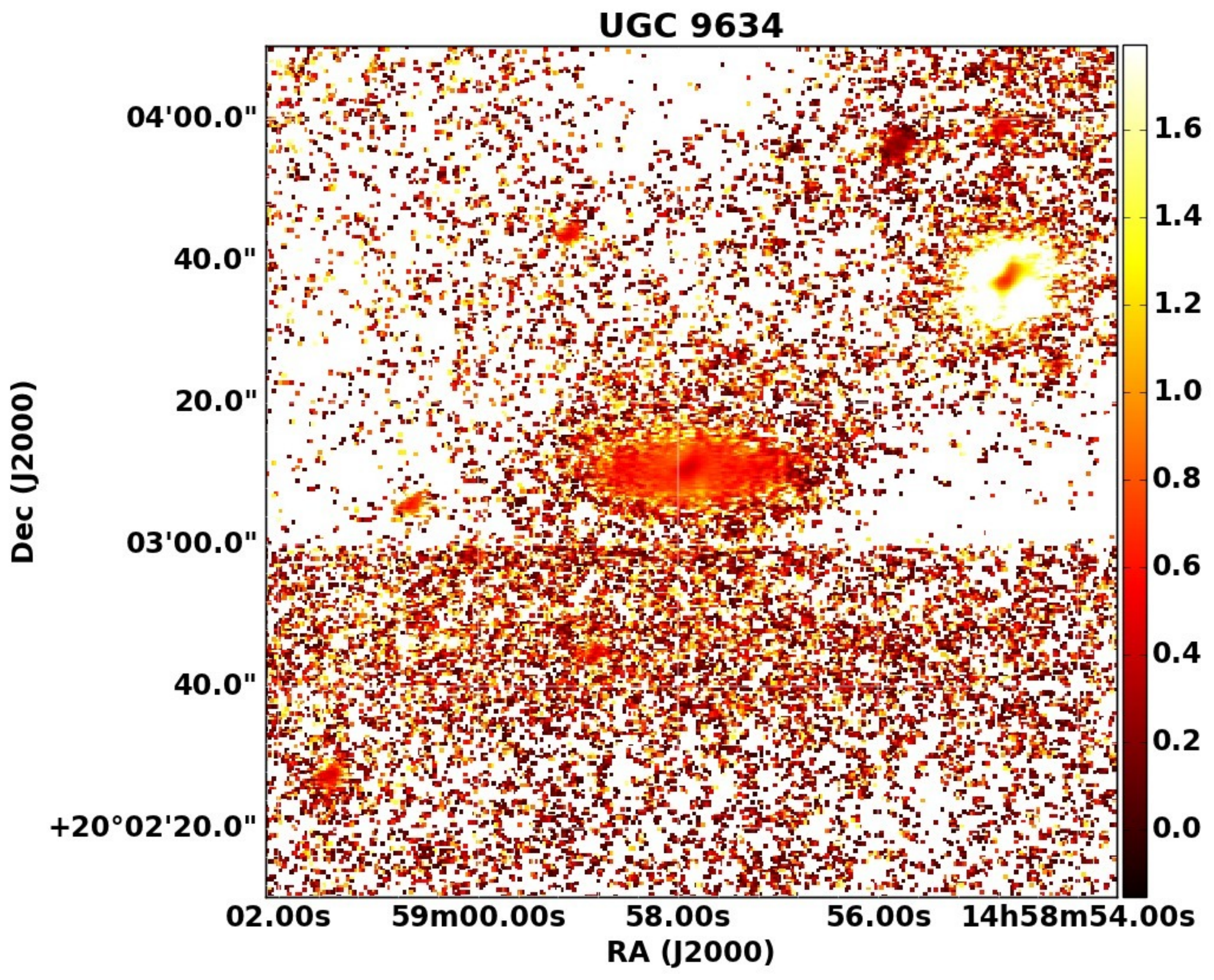}\includegraphics[scale=0.28]{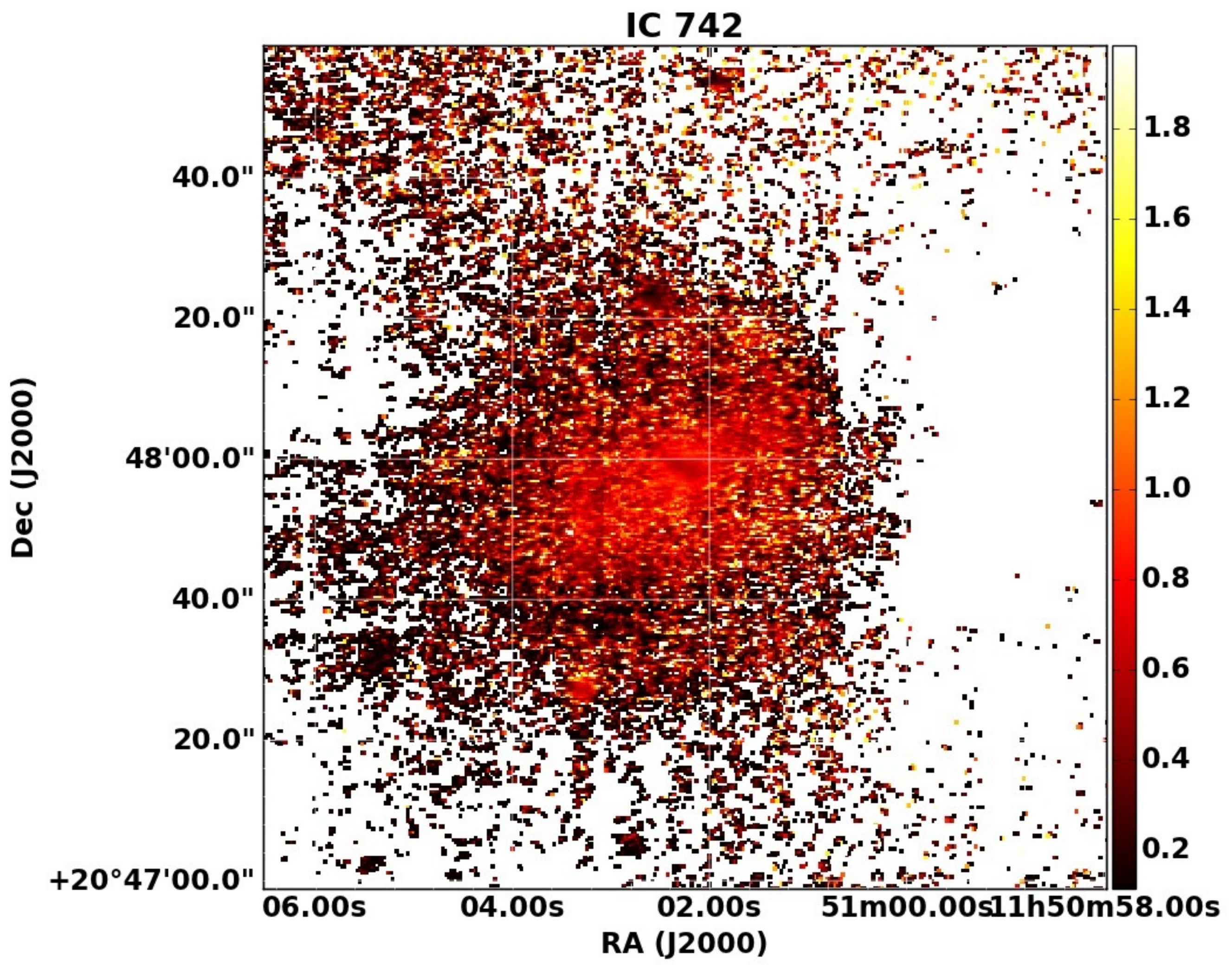}
\end{figure*}
\begin{figure*}

\includegraphics[scale=0.28]{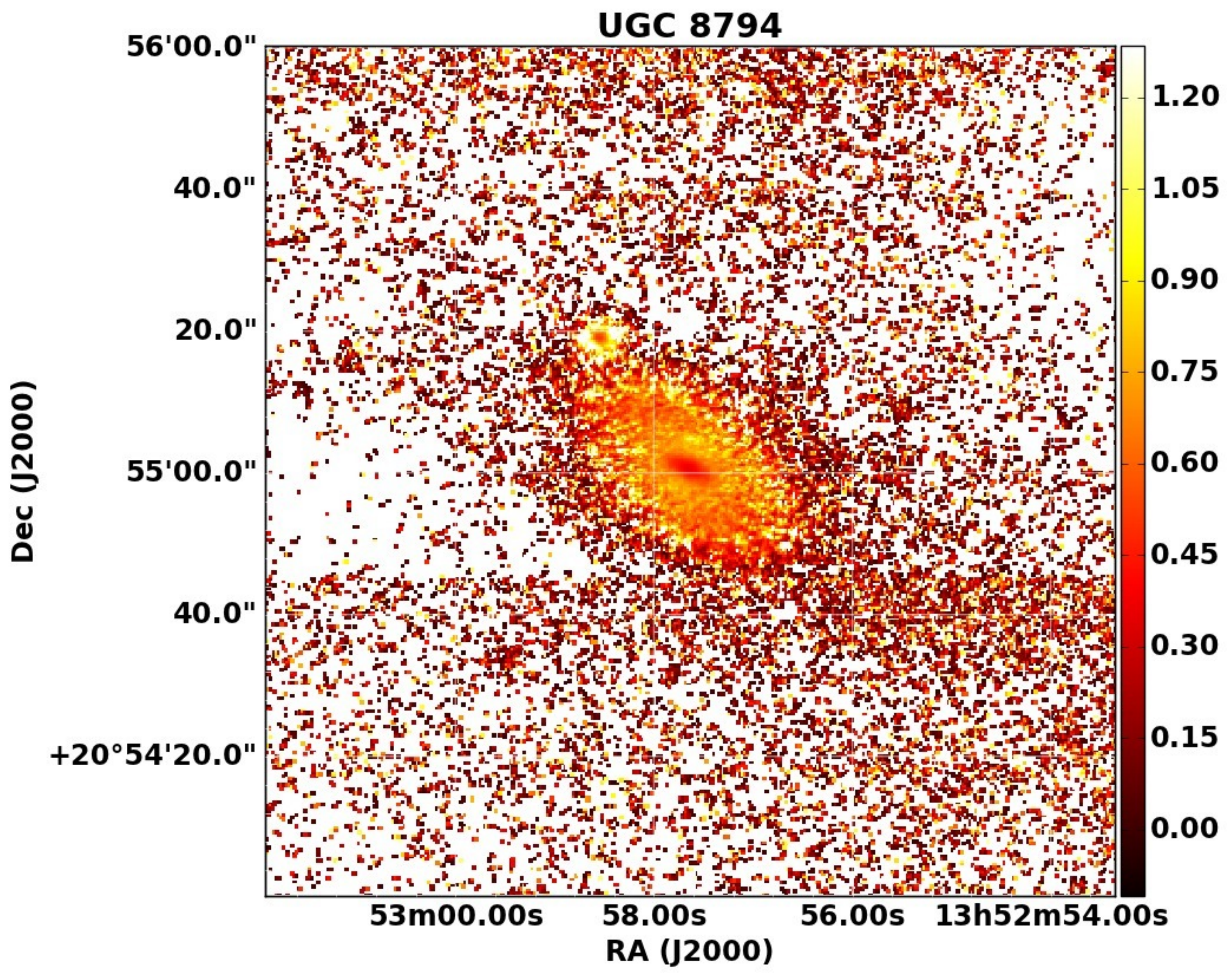}\includegraphics[scale=0.28]{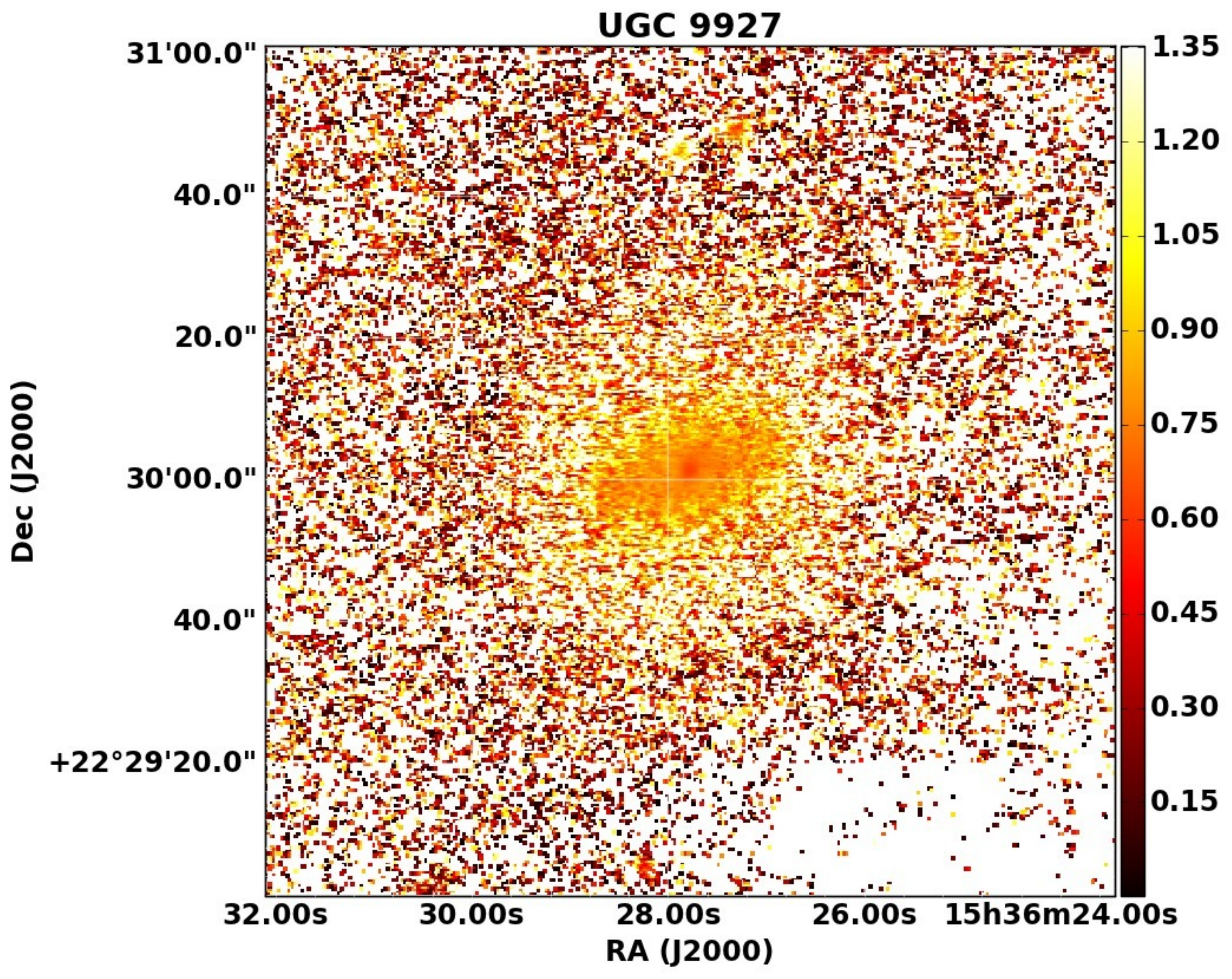}
\includegraphics[scale=0.28]{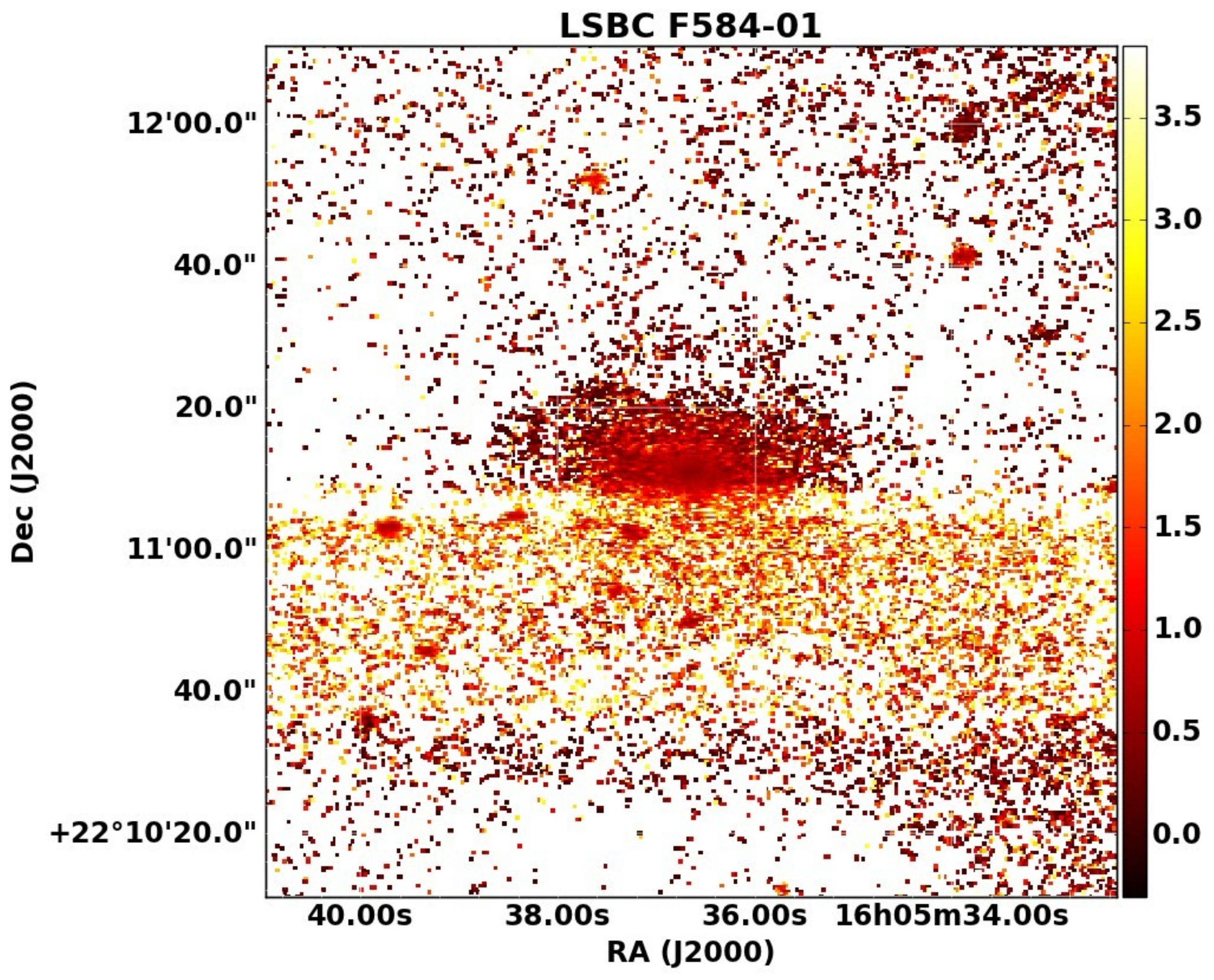}\includegraphics[scale=0.28]{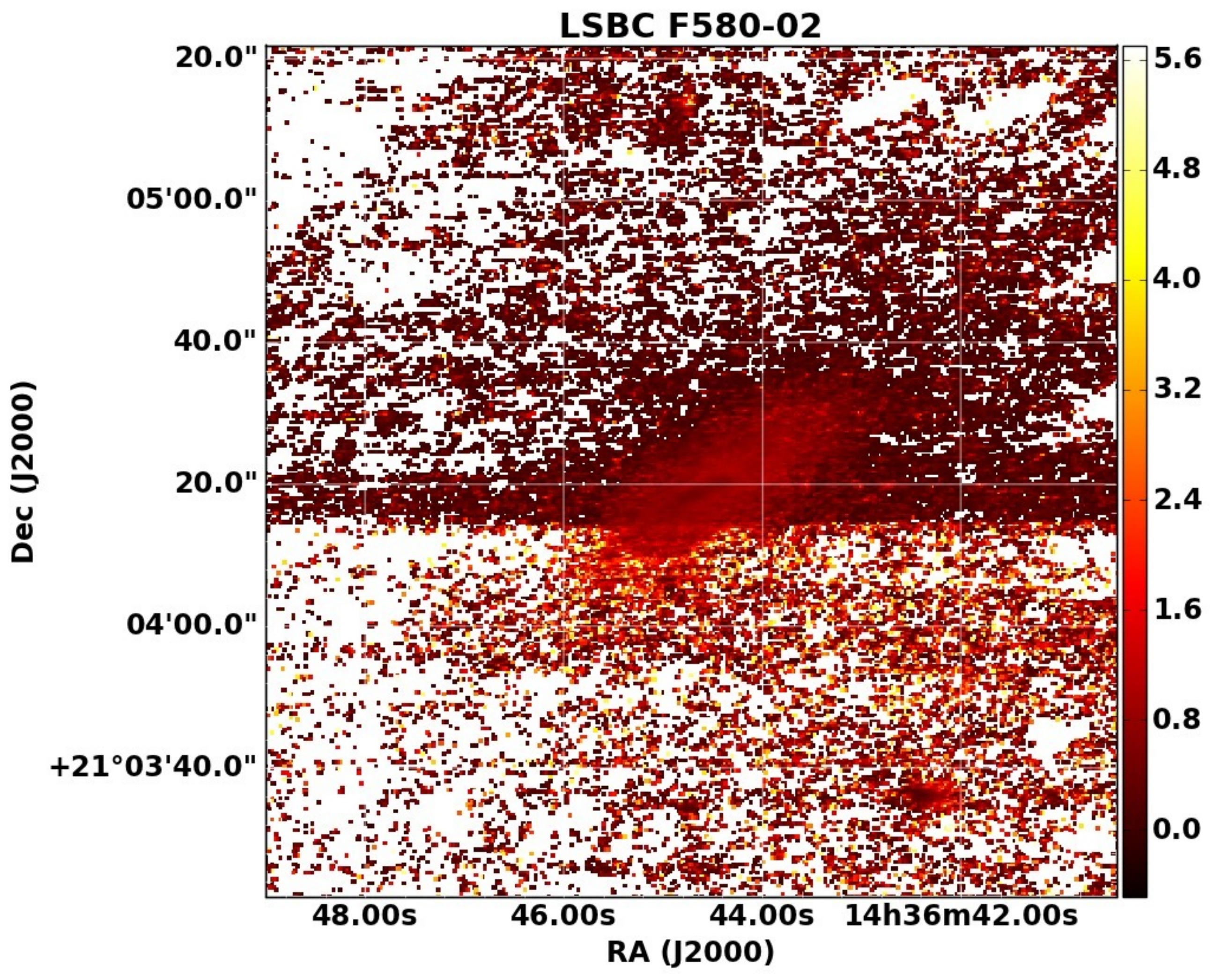}
\includegraphics[scale=0.28]{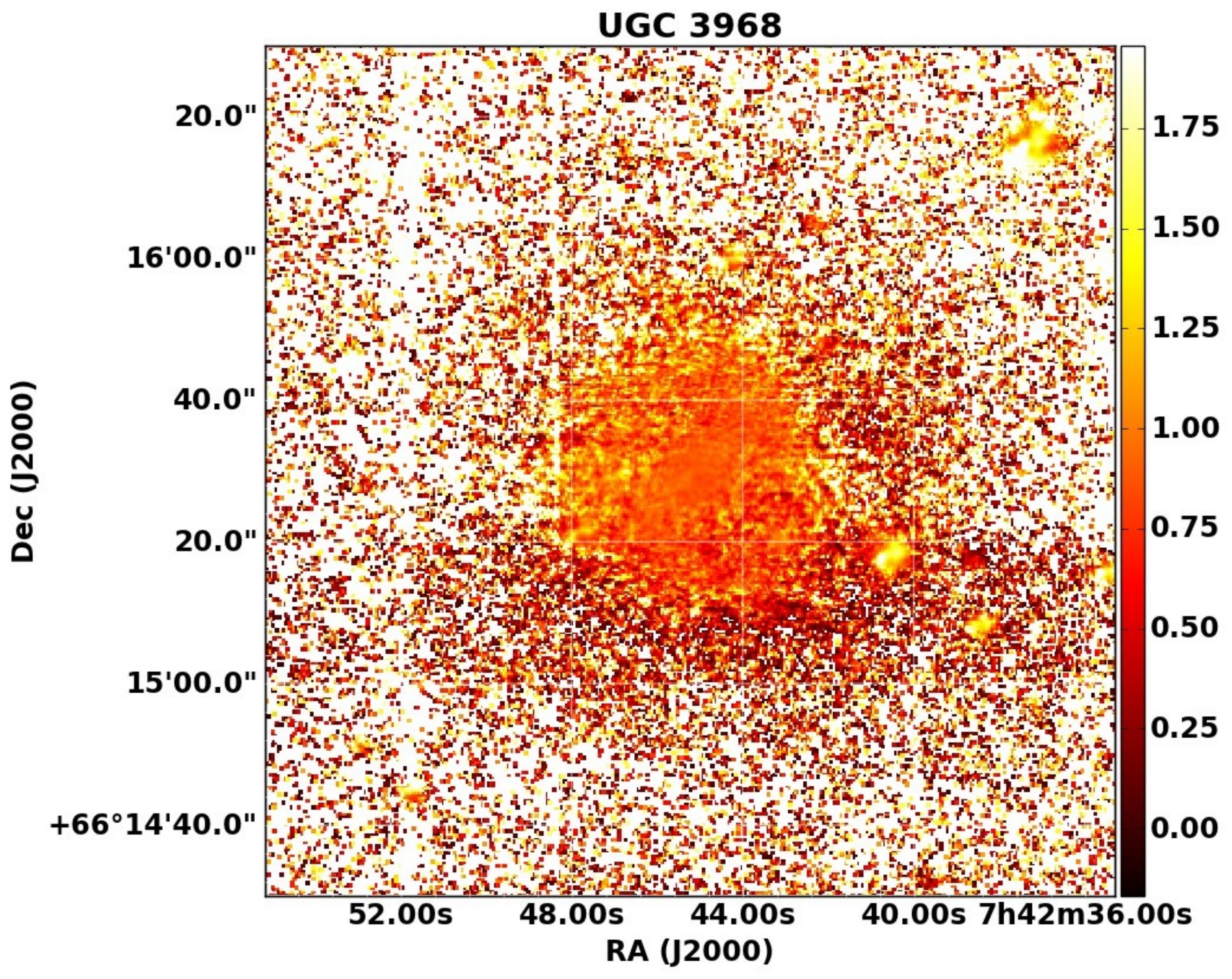}\includegraphics[scale=0.28]{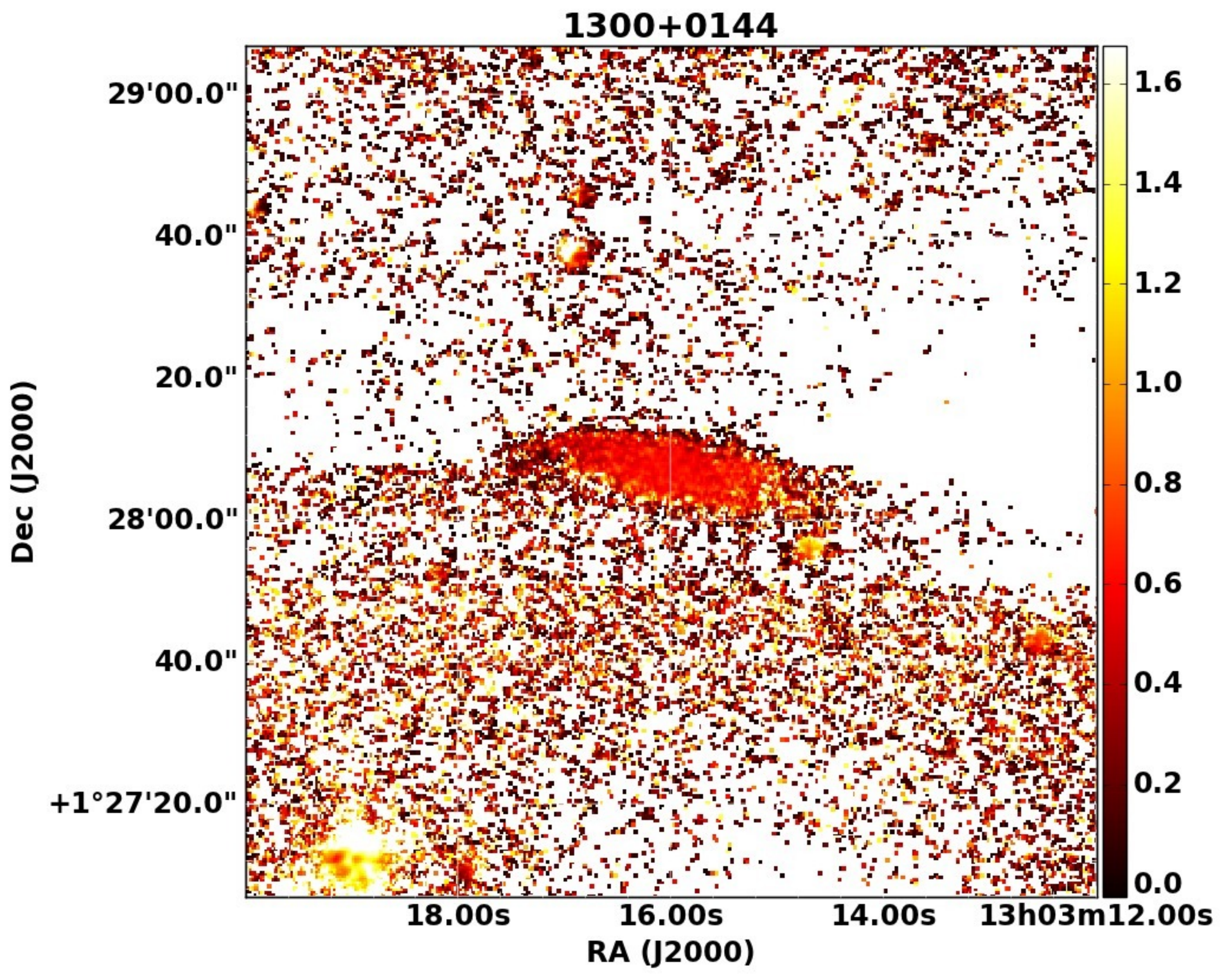}
\includegraphics[scale=0.28]{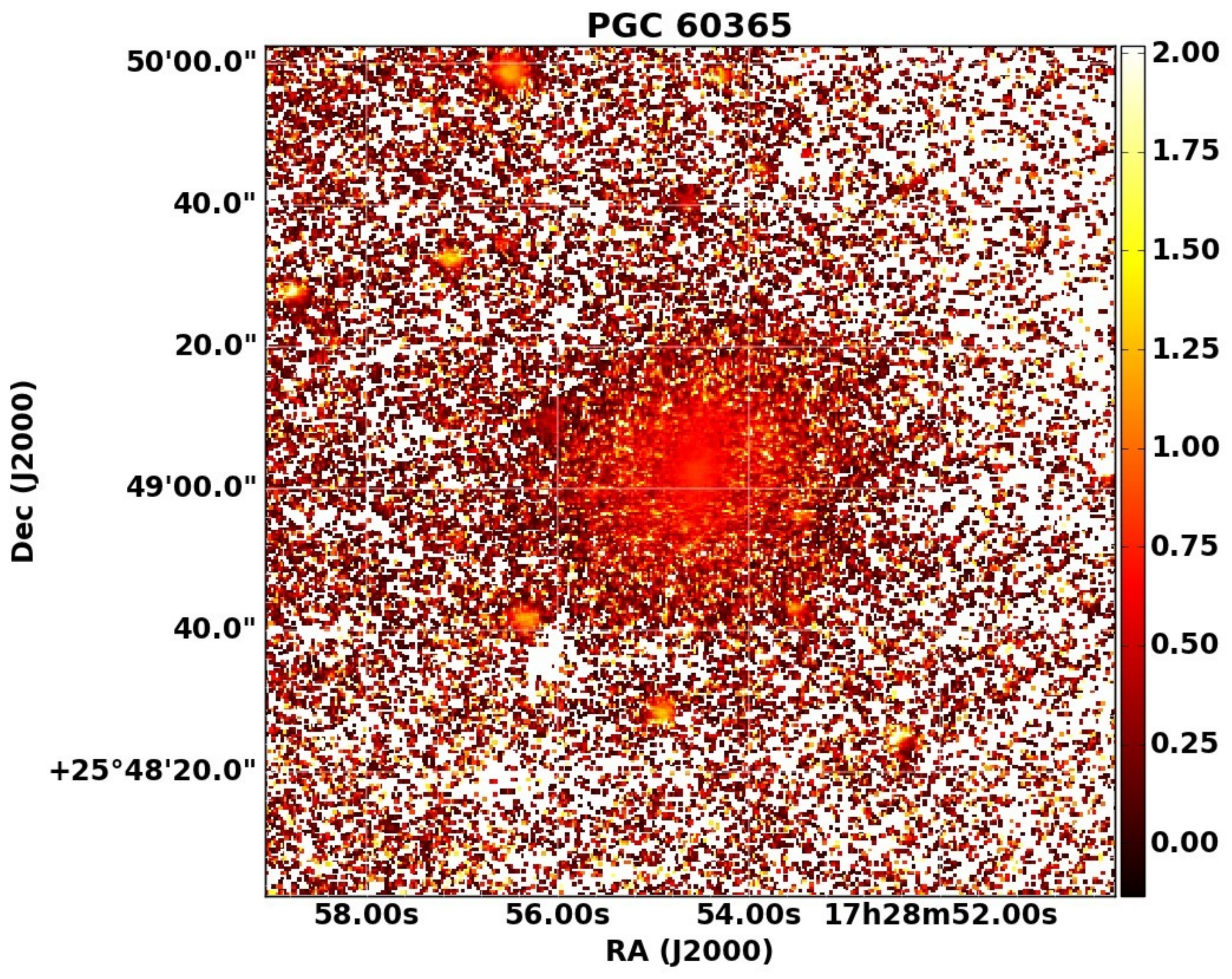}\includegraphics[scale=0.28]{CGCG006-023_JK_WCS_added_mar6.pdf}
\end{figure*}
\begin{figure*}

\includegraphics[scale=0.28]{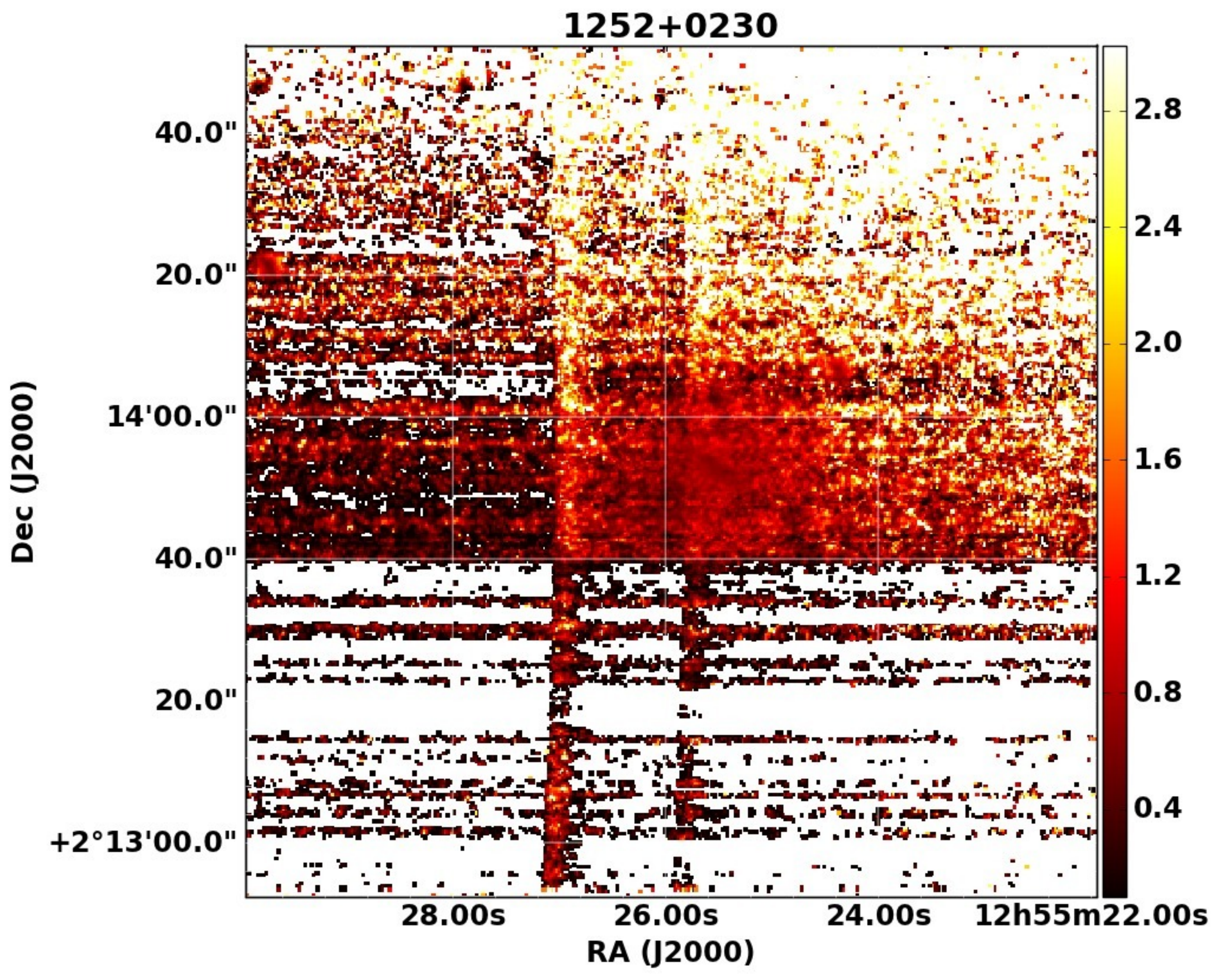}

\end{figure*}


\begin{figure*}
\caption{ The R-band images of galaxies UM 163(left) and UGC 11754(right). Logscale is used in both of the images.}
\label{f:b-band}
\includegraphics[trim = 0mm 18mm 28mm 0mm, clip,scale=0.45]{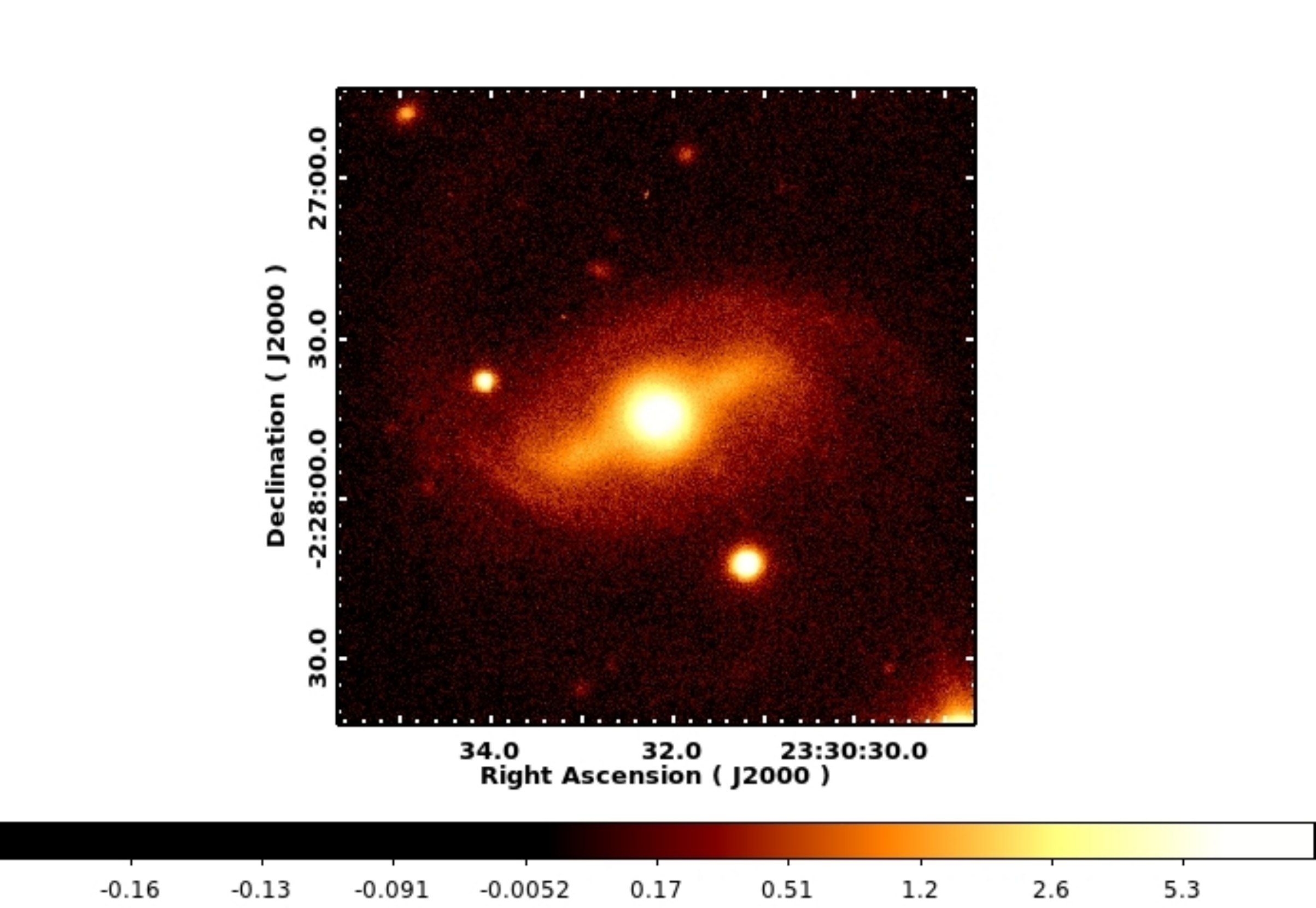}\includegraphics[trim = 18mm 18mm 0mm 0mm, clip,scale=0.45]{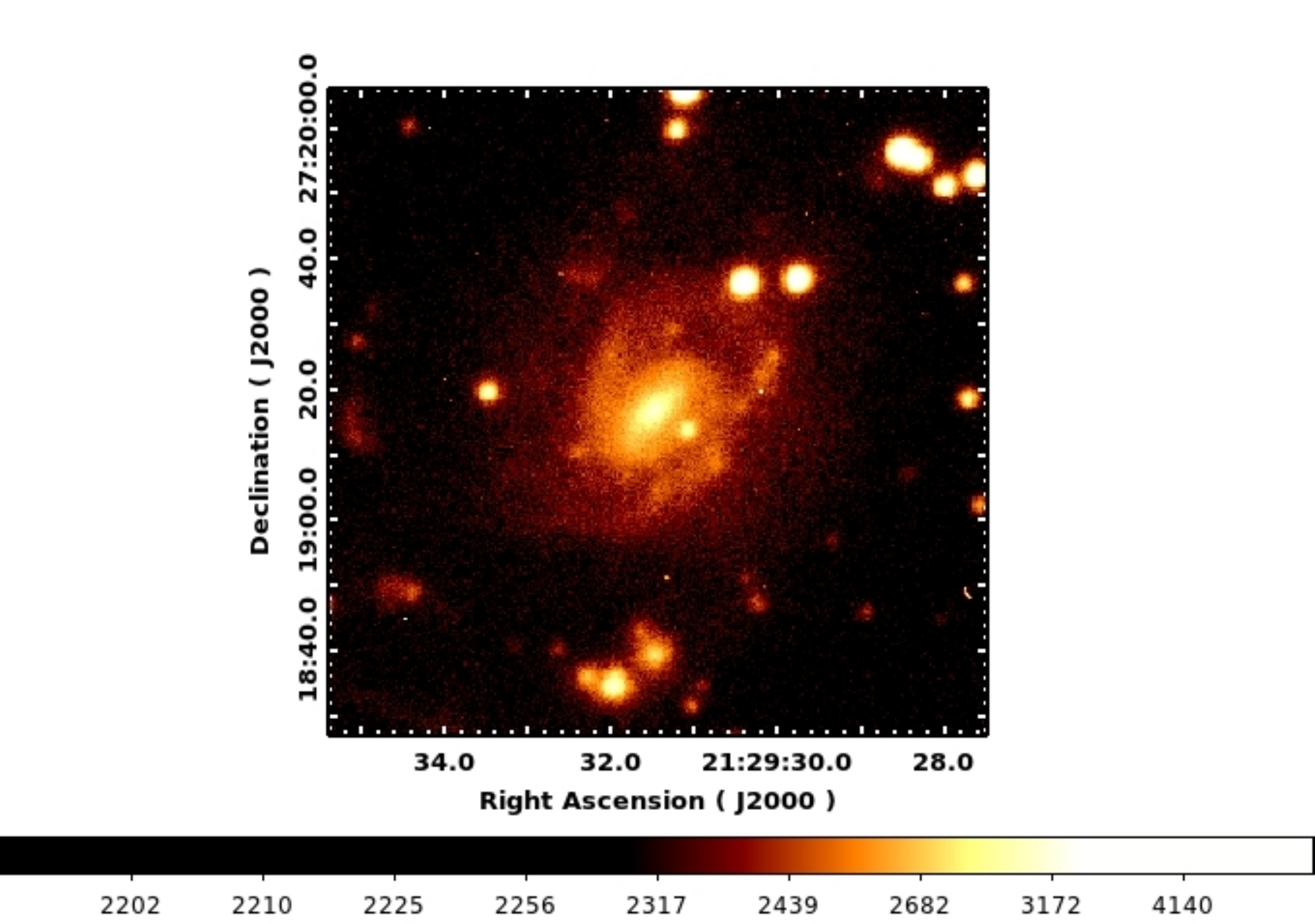}
\end{figure*}

\begin{center}
\begin{table*}
\label{tab:barlength}
\centering

\caption{The bar parameters of LSB galaxies from ellipse fit.The parameters a \& b are the semi major and semi minor axis of the galaxy
in RC3 or SDSS r-band depending upon which one is covering the galaxy disk at greater distance. The angle of
inclination is given as $i$. L$_{obs}$,e$_{obs}$ and PA$_{bar}$ is the projected bar semi major axis length, ellipcity and bar angle obtained 
from ELLIPSE fit. PA$_{gal}$ is the position angle of the galaxy. The parameters b/a,$i$ and PA$_{gal}$ are taken from NED.
( the position angles are taken from north to eastwards given in degrees). The difference between PA$_{bar}$ and PA$_{gal}$ is taken as $\alpha$.
 L$_{dep}$ is the deprojected bar semi major axis length calculated in arcsec. 
 The e$_{dep}$ is the deprojected ellipcity. Using the scale given in table 1, L$_{dep}$ is converted to physical units (kpc) given as L$_{bar}$.
For galaxies 0223-0033, UGC 2936, UGC 9634, UGC 8794 and 1300+0144 are having high inclination angles that are greater than 60$^{\circ}$ so we are avoiding 
those galaxies for the plots even though listed in the table.}

\begin{tabular}{l|l|l|l|l|l|l|l|l|l|l|l}	
\hline
Galaxy		&	b/a   &  $i$    &       L$_{obs}$ & e$_{obs}$ &	PA$_{bar}$ & PA$_{gal}$	 &  $\alpha$ &	L$_{dep}$ & L$_{bar}$ & e$_{dep}$	\\	
 name		&	      & ( $^{\circ}$)&	(arcsec)  &           & ( $^{\circ}$)   &   ( $^{\circ}$)  & ( $^{\circ}$)  &  (arcsec)  &   (kpc)   & 		\\
\hline

CGCG381-048	&  0.78 &  38  &  11.39$\pm$1.14 & 0.38 $\pm$0.03  & 67 $\pm$3.1  &   42   & 25 & 12.00 $\pm$1.21 & 3.83  $\pm$0.39  & 0.32$\pm$0.04 \\
UGC1920	        &  0.72 &  44  &  9.41 $\pm$0.94 & 0.50 $\pm$0.02  & 166$\pm$1.3  &   10   & 24 & 10.11 $\pm$1.01 & 3.93  $\pm$0.39  & 0.42$\pm$0.02 \\
UGC1455	        &  1.00 &  0   &  16.68$\pm$1.67 & 0.44 $\pm$0.03  & 28 $\pm$3.0  &   132  & 76 & 16.68 $\pm$1.67 & 5.27  $\pm$0.53  & 0.44$\pm$0.03 \\
NGC5905	        &  0.66 &  49  &  25.67$\pm$2.57 & 0.54 $\pm$0.03  & 21 $\pm$2.0  &   135  & 66 & 37.21 $\pm$3.73 & 8.41  $\pm$0.84  & 0.67$\pm$0.02 \\
UM163		&  0.69 &  46  &  22.20$\pm$2.22 & 0.41 $\pm$0.04  & 115$\pm$3.5  &   99   & 16 & 23.08 $\pm$2.34 & 14.24 $\pm$1.44  & 0.26$\pm$0.05 \\
UGC11754	&  0.89 &  27  &  8.56 $\pm$0.86 & 0.47 $\pm$0.01  & 145$\pm$2.3  &   157  & 12 & 8.61  $\pm$0.86 & 2.53  $\pm$0.25  & 0.42$\pm$0.02 \\
PGC68495	&  0.65 &  50  &  7.08 $\pm$0.71 & 0.56 $\pm$0.01  & 174$\pm$0.6  &   117  & 57 & 10.00 $\pm$1.00 & 7.79  $\pm$0.78  & 0.66$\pm$0.01 \\
UGC2936	        &  0.27 &  74  &  10.35$\pm$1.04 & 0.43 $\pm$0.01  & 26 $\pm$0.7  &   30   & 4  & 10.66 $\pm$1.07 & 2.58  $\pm$0.26  & 0.52$\pm$0.01 \\
UGC10405	&  0.74 &  42  &  5.32 $\pm$0.53 & 0.44 $\pm$0.02  & 136$\pm$2.1  &   21   & 65 & 6.86  $\pm$0.69 & 4.77  $\pm$0.48  & 0.56$\pm$0.02 \\
0223-0033	&  0.46 &  63  &  7.78 $\pm$0.78 & 0.26 $\pm$0.03  & 176$\pm$3.4  &   39   & 43 & 12.98 $\pm$1.40 & 5.19  $\pm$0.56  & 0.57$\pm$0.02 \\
UGC5035	        &  0.84 &  33  &  13.78$\pm$1.38 & 0.44 $\pm$0.01  & 77 $\pm$1.3  &   170  & 87 & 16.43 $\pm$1.64 & 11.86 $\pm$1.19  & 0.53$\pm$0.01 \\
UGC9087	        &  0.72 &  44  &  19.15$\pm$1.91 & 0.54 $\pm$0.02  & 28.5$\pm$2.0 &   17 & 11.5 & 19.50 $\pm$1.95 & 6.81  $\pm$0.68  & 0.39$\pm$0.03 \\
LSBCF568-08	&  0.78 &  39  &  7.07 $\pm$0.71 & 0.20 $\pm$0.01  & 121$\pm$1.6  &   63   & 58 & 8.58  $\pm$0.86 & 5.76  $\pm$0.58  & 0.33$\pm$0.01 \\
LSBCF568-09	&  0.96 &  16  &  12.53$\pm$1.25 & 0.39 $\pm$0.01  & 60 $\pm$1.1  &   13   & 47 & 12.80 $\pm$1.28 & 6.90  $\pm$0.69  & 0.40$\pm$0.01 \\
UGC9634         &  0.47 &  62  &  10.36$\pm$1.04 & 0.63 $\pm$0.01  & 89 $\pm$0.5  &   100  & 11 & 10.98 $\pm$1.10 & 9.00  $\pm$0.90  & 0.35$\pm$0.01 \\
IC742           &  0.95 &  17  &  16.68$\pm$1.67 & 0.64 $\pm$0.01  & 117.6$\pm$0.7&   0    &62.4& 17.28 $\pm$1.73 & 7.53  $\pm$0.75  & 0.65$\pm$0.01 \\
UGC8794         &  0.31 &  72  &  11.39$\pm$1.14 & 0.48 $\pm$0.01  & 42.1$\pm$0.8 &   68   &25.9& 19.05 $\pm$1.93 & 10.78 $\pm$1.09  & 0.63$\pm$0.01 \\
UGC9927         &  0.88 &  28  &  13.78$\pm$1.38 & 0.42 $\pm$0.01  & 109$\pm$0.8  &   4    &75  & 15.49 $\pm$1.55 & 4.49  $\pm$0.45  & 0.48$\pm$0.01 \\
LSBCF584-01     &  0.75 &  41  &  9.41 $\pm$0.94 & 0.60 $\pm$0.02  & 87$\pm$1.7   &   76   & 11 & 9.54  $\pm$0.96 & 7.31  $\pm$0.73  & 0.48$\pm$0.03 \\
LSBCF580-02     &  0.8  &  37  &  4.83 $\pm$0.48 & 0.57 $\pm$0.03  & 115$\pm$2.0  &   130  & 15 & 4.92  $\pm$0.49 & ***              & 0.48$\pm$0.04 \\
UGC3968         &  0.85 &  32  &  16.68$\pm$1.67 & 0.63 $\pm$0.02  & 136$\pm$1.1  &   26   & 70 & 19.34 $\pm$1.93 & 8.53  $\pm$0.85  & 0.68$\pm$0.01 \\   
1300+0144       &  0.24 &  76  &  7.07 $\pm$0.71 & 0.63 $\pm$0.01  & 82.6$\pm$0.7 &   82   & 0.6& 7.08  $\pm$0.71 & 5.63  $\pm$0.56  & 0.35$\pm$0.02 \\
PGC60365        &  0.71 &  44  &  9.41 $\pm$0.94 & 0.41 $\pm$0.02  & 169$\pm$1.9  &   120  & 49 & 11.64 $\pm$1.17 & ***              & 0.49$\pm$0.02 \\
CGCG006-023     &  0.63 &  51  &  8.56 $\pm$0.86 & 0.37 $\pm$0.01  & 30$\pm$1.3   &   109  & 79 & 13.44 $\pm$1.34 & 10.02 $\pm$1.00  & 0.60$\pm$0.01 \\
1252+0230       &  0.88 &  33  &  7.07 $\pm$0.71 & 0.29 $\pm$0.03  & 88.3$\pm$3.1 &   99   &10.7& 7.12  $\pm$0.71 & 6.57  $\pm$0.66  & 0.18$\pm$0.03 \\
LSBCF570-01     &  0.60 &  53  &  7.78 $\pm$0.78 & 0.17 $\pm$0.01  & 29$\pm$1.4   &   108  & 79 & 12.77 $\pm$1.28 & 6.65  $\pm$0.67  & 0.50$\pm$0.00 \\
1442+0137       &  0.9  &  26  &  6.43 $\pm$0.64 & 0.42 $\pm$0.02  & 31 $\pm$1.8  &   8    & 23 & 6.55  $\pm$0.65 & 4.33  $\pm$0.43  & 0.38$\pm$0.02 \\
LSBCF675-01     &  0.75 &  41  &  6.43 $\pm$0.64 & 0.57 $\pm$0.02  & 79 $\pm$1.8  &   72   & 7  & 6.47  $\pm$0.65 & 4.39  $\pm$0.44  & 0.44$\pm$0.03 \\
IC2423          &  0.81 &  36  &  9.41 $\pm$0.94 & 0.44 $\pm$0.01  & 128 $\pm$1.1 &   100  & 28 & 9.95  $\pm$0.99 & 5.97  $\pm$0.60  & 0.39$\pm$0.01 \\

\hline
\end{tabular}
\end{table*}
\end{center}

\begin{center}
\begin{table*}
\label{tab:barlength}
\centering
\caption{The model magnitudes are taken from SDSS database for G-band and R-Band, which are used to determine the B-V color of the galaxy.
The f$\nu$ represent the total flux in the R-band. For the calculations we have used the luminosity distance taken form NED. Lr, M/L and M$_{stellar}$
denotes the R-band luminosity, the M/L ratio of the galaxy for that particular band and the stellar mass respectively.}
\begin{tabular}{l|c|c|c|c|c|c|c}
\hline
Galaxy		&  g-r & B-V &    f$\nu$                   & Distance& Lr	         & M/L & M$_{stellar}$\\
		&      &     & (ergs cm$^{-2}$ s$^{-1}$ Hz)& (Mpc)   & (ergs s$^{-1}$ Hz)&     &(10$^{10}$M$\odot$)\\
\hline                                                                                                                                     
CGCG 381-048   &    0.629$\pm$0.003  &    0.836   &  8.47E-12   & 68.0  &  4.43E+42 &      2.78  &   0.62$\pm$0.01\\
UGC 1920       &    0.862$\pm$0.003  &     1.065  &  7.22E-12   & 83.4  &  5.68E+42 &      5.66  &   1.61$\pm$0.02\\
UGC 1455       &    0.889$\pm$0.002  &     1.091  &  2.60E-11   & 67.3  &  1.33E+43 &      6.14  &   4.09$\pm$0.03\\
NGC 5905       &    0.823$\pm$0.002  &     1.027  &  4.27E-11   & 47.8  &  1.10E+43 &      5.02  &   2.77$\pm$0.02\\
UM 163         &    0.728$\pm$0.003  &    0.933   &  1.25E-11   & 136.0 &  2.61E+43 &      3.76  &   4.89$\pm$0.04\\
UGC 11754      &    0.385$\pm$0.004  &    0.597   &  8.48E-12   & 62.5  &  3.75E+42 &      1.33  &   0.25$\pm$0.003\\
PGC 68495      &    0.633$\pm$0.004  &    0.840   &  5.40E-12   & 174.0 &  1.85E+43 &      2.82  &   2.61$\pm$0.03\\
UGC 2936       &    ***	             &     ***    &   ***       & 51.2  &    ***    &      ***   &   ***           \\
UGC 10405      &    0.651$\pm$0.004  &    0.858   &  7.60E-12   & 154.0 &  2.04E+43 &      2.98  &   3.03$\pm$0.03\\
0223-0033      &    0.990$\pm$0.003  &     1.191  &  2.03E-11   & 86.0  &  1.70E+43 &      8.35  &   7.11$\pm$0.07\\
UGC 5035       &    0.800$\pm$0.003  &     1.004  &  7.82E-12   & 161.0 &  2.29E+43 &      4.68  &   5.37$\pm$0.05\\
UGC 9087       &    0.939$\pm$0.003  &     1.140  &  1.41E-11   & 74.5  &  8.87E+42 &      7.15  &   3.17$\pm$0.03\\
LSBC F568-8    &    0.843$\pm$0.003  &     1.047  &  8.25E-12   & 148.0 &  2.04E+43 &      5.35  &   5.46$\pm$0.05\\
LSBC F568-9    &    0.682$\pm$0.003  &    0.888   &  6.97E-12   & 117.0 &  1.08E+43 &      3.27  &   1.76$\pm$0.02\\
UGC 9634       &    0.824$\pm$0.004  &     1.027  &  4.25E-12   & 184.0 &  1.63E+43 &      5.03  &   4.09$\pm$0.05\\
IC 742         &    0.934$\pm$0.003  &     1.135  &  8.69E-12   & 94.1  &  8.70E+42 &      7.04  &   3.06$\pm$0.03\\
UGC 8794       &    0.703$\pm$0.003  &    0.909   &  9.25E-12   & 124.0 &  1.61E+43 &      3.49  &   2.80$\pm$0.03\\
UGC 9927       &    0.772$\pm$0.003  &    0.976   &  1.52E-11   & 61.7  &  6.54E+42 &      4.30  &   1.41$\pm$0.01\\
LSBC F584-01   &    0.843$\pm$0.004  &     1.046  &  4.88E-12   & 171.0 &  1.61E+43 &      5.33  &   4.31$\pm$0.05\\
LSBC F580-2    &    0.608$\pm$0.004  &    0.816   &  5.52E-12   &  ***  &   ***     &      ***   &   ***          \\
UGC 3968       &    0.722$\pm$0.003  &    0.928   &  1.17E-11   & 95.2  &  1.20E+43 &      3.69  &   2.21$\pm$0.02\\
1300+0144      &    0.730$\pm$0.005  &    0.935   &  2.99E-12   & 178.0 &  1.07E+43 &      3.78  &   2.03$\pm$0.03\\
PGC 60365      &    0.827$\pm$0.003  &     1.030  &  5.73E-12   &  ***  &    ***    &      ***   &   ***          \\
CGCG 006-023   &    0.548$\pm$0.004  &    0.757   &  4.21E-12   & 166.0 &  1.31E+43 &      2.18  &   1.43$\pm$0.02\\
1252+0230      &    0.596$\pm$0.005  &    0.804   &  3.42E-12   & 209.0 &  1.69E+43 &      2.52  &   2.13$\pm$0.04\\
LSBC F570-01   &    0.812$\pm$0.003  &     1.016  &  8.93E-12   & 113.0 &  1.29E+43 &      4.86  &   3.13$\pm$0.03\\
1442+0137      &    0.587$\pm$0.005  &    0.795   &  2.74E-12   & 146.0 &  6.61E+42 &      2.45  &   0.81$\pm$0.01\\
LSBC F675-01   &    0.881$\pm$0.006  &     1.083  &  1.24E-12   & 150.0 &  3.15E+42 &      5.99  &   0.94$\pm$0.02\\
IC 2423        &    0.609$\pm$0.003  &    0.817   &  1.35E-11   & 132.0 &  2.65E+43 &      2.62  &   3.47$\pm$0.03\\

\hline
\end{tabular}
\end{table*}
\end{center}

\begin{center}
\begin{table*}
\label{tab:HI}
\caption{The J and K$_s$ values are taken from 2MASS extended object catalogue. J and K$_s$ denotes the magnitudes in J and K$_s$ bands.
The HI values are taken from HYPERLEDA archival datadase. The columns m21, S$_{\nu}$, M$_{HI}$ are representing the HI magnitude, flux and the 
neutral hydrogen mass, that corresponds to the gaseous component of the galaxy.}
\begin{tabular}{l|c|c|c|c|c}
\hline

 Galaxy	        &	J  &      K$_s$     &     m21    & S$_{\nu}$ (Jy Km s$^{-1}$) & M$_{HI}$(10$^9$M$\odot$)\\
 \hline
  		&	             &                    &		   &       	                  &     	                  \\ 
 CGCG 381-048   &   12.180$\pm$0.038 &  11.353 $\pm$0.068 & 16.20$\pm$0.15 &$ 3.03\substack{+0.45\\-0.39}$&$3.30\substack{+0.49\\-0.43}$ \\
 		&	             &                    &		   &                              &                     	  \\ 
 UGC 1920       &   12.273$\pm$0.038 &  11.081 $\pm$0.051 & 15.56$\pm$0.12 &$ 5.45\substack{+0.64\\-0.57}$&$8.94\substack{+1.04\\-0.94}$  \\
  		&		     &                    &		   &       	                  &     	                  \\ 
 UGC 1455       &   10.516$\pm$0.023 &  9.502  $\pm$0.040 & 15.05$\pm$0.17 &$ 8.71\substack{+1.26\\-1.48}$&$9.31\substack{+1.58\\-1.35}$  \\
  		&	             &                    &		   &       	                  &                               \\
 NGC 5905       &   10.467$\pm$0.017 &  9.514  $\pm$0.027 & 13.61$\pm$0.12 &$ 32.8\substack{+3.43\\-3.83}$&$17.7\substack{+2.07\\-1.85}$  \\
 		&	             &                    &		   &       	                  &                               \\	
 UM 163         &   11.357$\pm$0.034 &  10.327 $\pm$0.054 & 16.37$\pm$0.09 &$ 2.58\substack{+0.21\\-0.22}$&$11.3\substack{+0.97\\-0.90}$  \\
 		&	             &                    &	           &       	                  &                      	  \\	
 UGC 11754      &   12.823$\pm$0.075 &  11.853 $\pm$0.106 & 15.06$\pm$0.13 &$ 8.63\substack{+0.97\\-1.10}$&$7.96\substack{+1.01\\-0.90}$  \\
 		&		     &                    &		   &       	                  &     	                  \\	
 PGC 68495      &   12.581$\pm$0.070 &  11.436 $\pm$0.111 & 16.38$\pm$0.19 &$ 2.56\substack{+0.41\\-0.49}$&$18.3\substack{+3.50\\-2.93}$  \\
 		&		     &                    &		   &       	                  &     	                  \\	
 UGC 2936       &   10.160$\pm$0.026 &  8.914  $\pm$0.029 & 14.97$\pm$0.14 &$ 9.38\substack{+1.13\\-1.29}$&$5.80\substack{+0.80\\-0.70}$  \\
 		&		     &                    &		   &       	                  &     	                  \\	
 UGC 10405      &   ***	             &   ***	          & 15.57$\pm$0.13 &$ 5.40\substack{+0.61\\-0.69}$&$30.20\substack{+3.84\\-3.41}$ \\
 		&		     &                    &		   &       	                  &     	                  \\	
 0223-0033      &   11.059$\pm$0.023 &  10.115 $\pm$0.045 & 15.52$\pm$0.09 &$ 5.65\substack{+0.45\\-0.49}$&$9.86\substack{+0.85\\-0.78}$  \\
 		&		     &                    &	   	   &       	                  &     	                  \\	
 UGC 5035       &   12.168$\pm$0.040 &  11.144 $\pm$0.049 &  ***           & ***                          &     ***                       \\
 		&		     &                    &		   &       	                  &     	                  \\	
 UGC 9087       &   11.673$\pm$0.027 &  10.811 $\pm$0.046 &  ***           & ***                          &     ***       		  \\
 		&		     &                    &	           &       	                  &                      	  \\	
 LSBC F568-8    &   12.232$\pm$0.061 &  11.177 $\pm$0.072 &  ***           & ***                          &     ***                       \\
 		&		     &                    &		   &       	                  &     	                  \\	
 LSBC F568-9    &   **	             &   **	          & 16.89$\pm$0.09 &$ 1.60\substack{+0.13\\-0.14}$&$5.17\substack{+0.44\\-0.41}$  \\ 
 		&		     &                    &		   &       	                  &                     	  \\	
 UGC 9634       &   12.761$\pm$0.044 &  12.068 $\pm$0.084 & 15.84$\pm$0.22 &$ 4.22\substack{+0.77\\-0.95}$&$33.6\substack{+7.55\\-6.17}$  \\
 		&		     &                    &		   &       	                  &     	                  \\	
 IC 742         &   11.654$\pm$0.029 &  11.199 $\pm$0.082 & 17.58$\pm$0.25 &$ 0.85\substack{+0.17\\-0.22}$&$1.77\substack{+0.46\\-0.36}$  \\
 		&		     &                    &		   &       	                  &     	                  \\	
 UGC 8794       &   11.893$\pm$0.028 &  10.893 $\pm$0.039 & 16.41$\pm$0.15 &$ 2.49\substack{+0.32\\-0.37}$&$9.03\substack{+1.34\\-1.17}$  \\
 		&		     &                    &		   &       	                  &     	                  \\	
 UGC 9927       &   11.559$\pm$0.031 &  10.825 $\pm$0.034 &  ***           & ***                          &     ***                       \\  
 		&		     &                    &		   &       	                  &     	                  \\	
 LSBC F584-01   &   12.551$\pm$0.041 &  11.844 $\pm$0.079 & 17.26$\pm$0.16 &$ 1.14\substack{+0.16\\-0.18}$&$7.85\substack{+1.25\\-1.08}$  \\
 		&		     &                    &		   &       	                  &     	                  \\	
 LSBC F580-2    &   **	             &   **	          &  ***           & ***                          &     ***                       \\
 		&		     &                    &		   &       	                  &     	                  \\	
 UGC 3968       &   12.100$\pm$0.039 &  11.073 $\pm$0.059 & 15.79$\pm$0.12 &$ 4.41\substack{+0.46\\-0.51}$&$9.42\substack{+1.10\\-0.99}$  \\
 		&		     &                    &		   &       	                  &     	                  \\	
 1300+0144      &   13.044$\pm$0.058 &  12.158 $\pm$0.106 & 16.92$\pm$0.27 &$ 1.56\substack{+0.34\\-0.44}$&$11.63\substack{+3.28\\-2.56}$ \\
 		&		     &                    &		   &       	                  &     	                  \\	
 PGC 60365      &   12.473$\pm$0.047 &  11.472 $\pm$0.080 &  ***           & ***                          &     ***                       \\
 		&		     &                    &		   &       	                  &     	                  \\	
 CGCG 006-023   &   13.286$\pm$0.068 &  12.378 $\pm$0.116 &  ***           & ***                          &     ***                       \\
 		&		     &                    &		   &       	                  &     	                  \\	
 1252+0230      &   13.698$\pm$0.083 &  12.752 $\pm$0.145 &  ***           & ***                          &     ***                       \\
 		&		     &                    &		   &       	                  &     	                  \\	
 LSBC F570-01   &   12.116$\pm$0.031 &  11.235 $\pm$0.039 &  ***           & ***                          &     ***                       \\ 
 		&	             &                    &		   &       	                  &     	                  \\	
 1442+0137      &   13.737$\pm$0.080 &  12.770 $\pm$0.139 &  ***           & ***                          &     ***                       \\
 		&		     &                    &		   &       	                  &     	                  \\	
 LSBC F675-01   &   13.954$\pm$0.066 &  13.294 $\pm$0.148 & 16.91$\pm$0.42 & $1.57\substack{+0.53\\-0.74}$&$8.34\substack{+3.94\\-2.68}$  \\
 		&		     &                    &		   &       	                  &     	                  \\	
 IC 2423        &   11.811$\pm$0.028 &  10.866 $\pm$0.044 & 16.47$\pm$0.42 & $2.36\substack{+0.76\\-1.11}$&$9.68\substack{+4.57\\-3.11}$  \\
 		&		     &                    &		   &                              &     	                  \\
 \hline
\end{tabular}
\end{table*}
\end{center}


\begin{figure*}
\centering
\caption{The fitted elliptical isophotes are overlaid on the Ks band image of galaxies UGC 3968, PGC60365 and UGC11754. Logscale is used in all images. The case
of galaxy UGC 3968 is shown detailed with how the parameters like surface brightness, ellipcity, position angle and b4 parameter
changes with the semi major axis}
\includegraphics[trim =0mm 17mm 5mm 0mm, clip,scale=0.44]{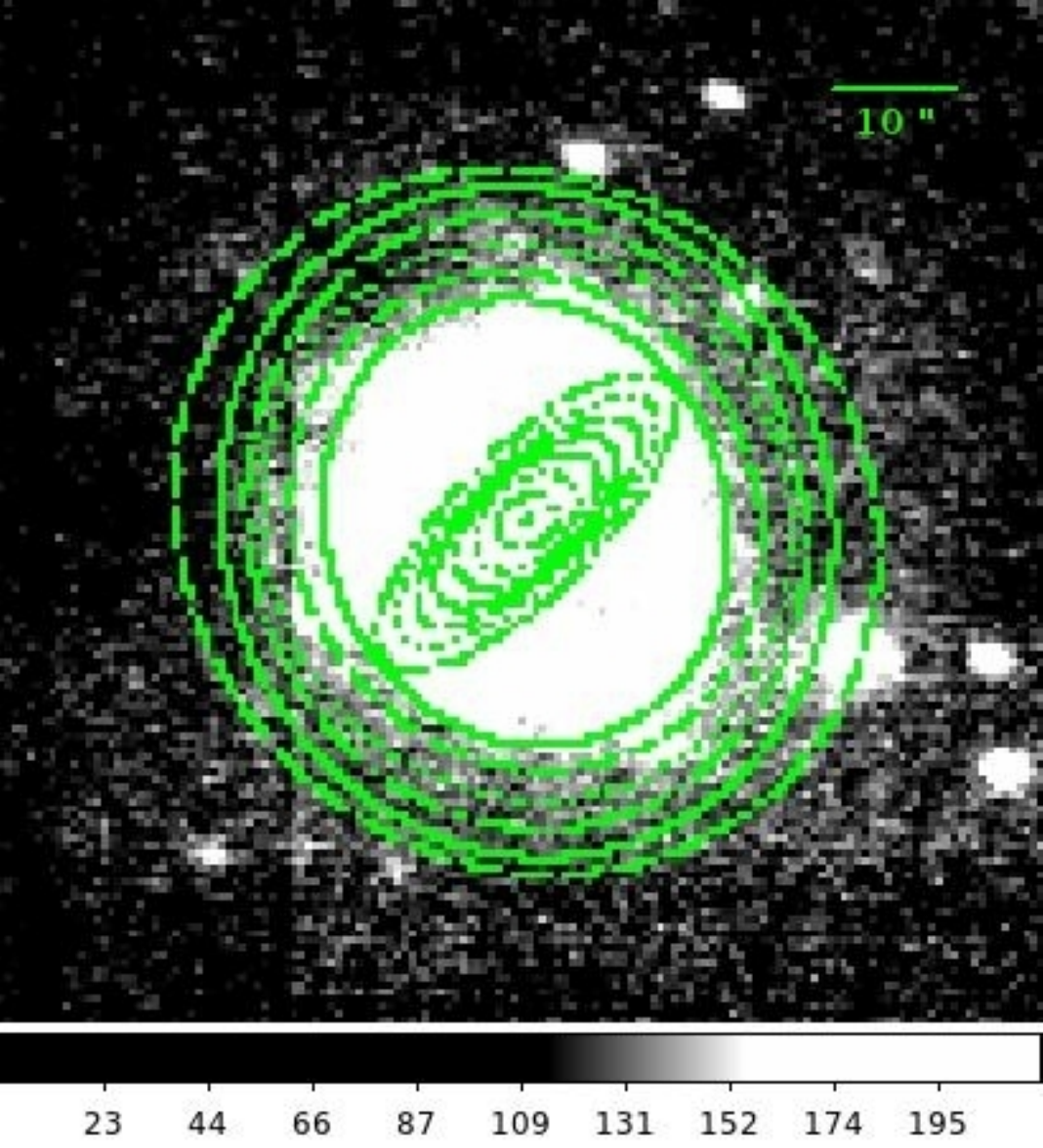}
\includegraphics[trim = 0mm 17mm 5mm 0mm, clip,scale=0.44]{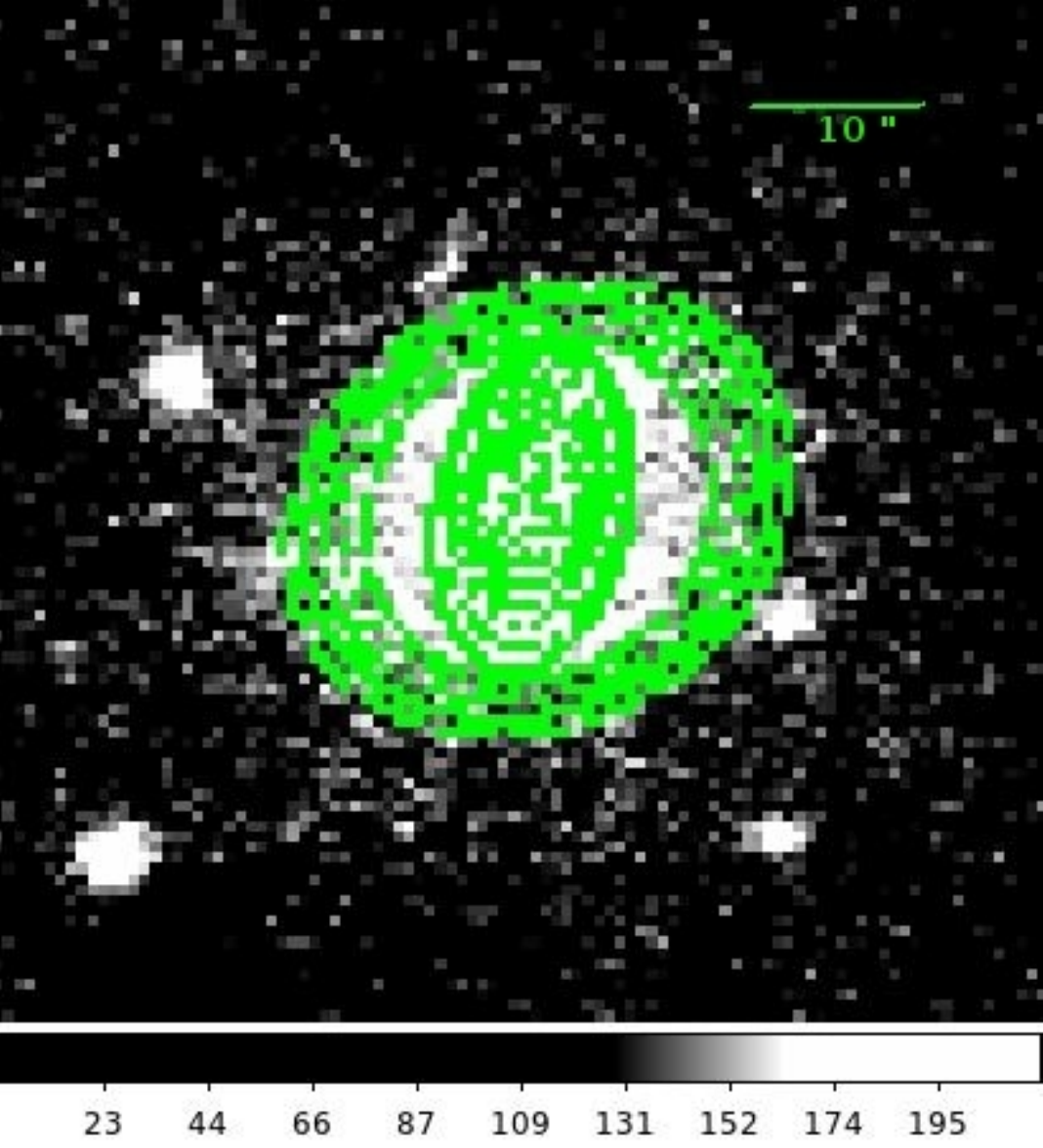}
\includegraphics[trim = 0mm 4mm 0mm 0mm, clip,scale=.47]{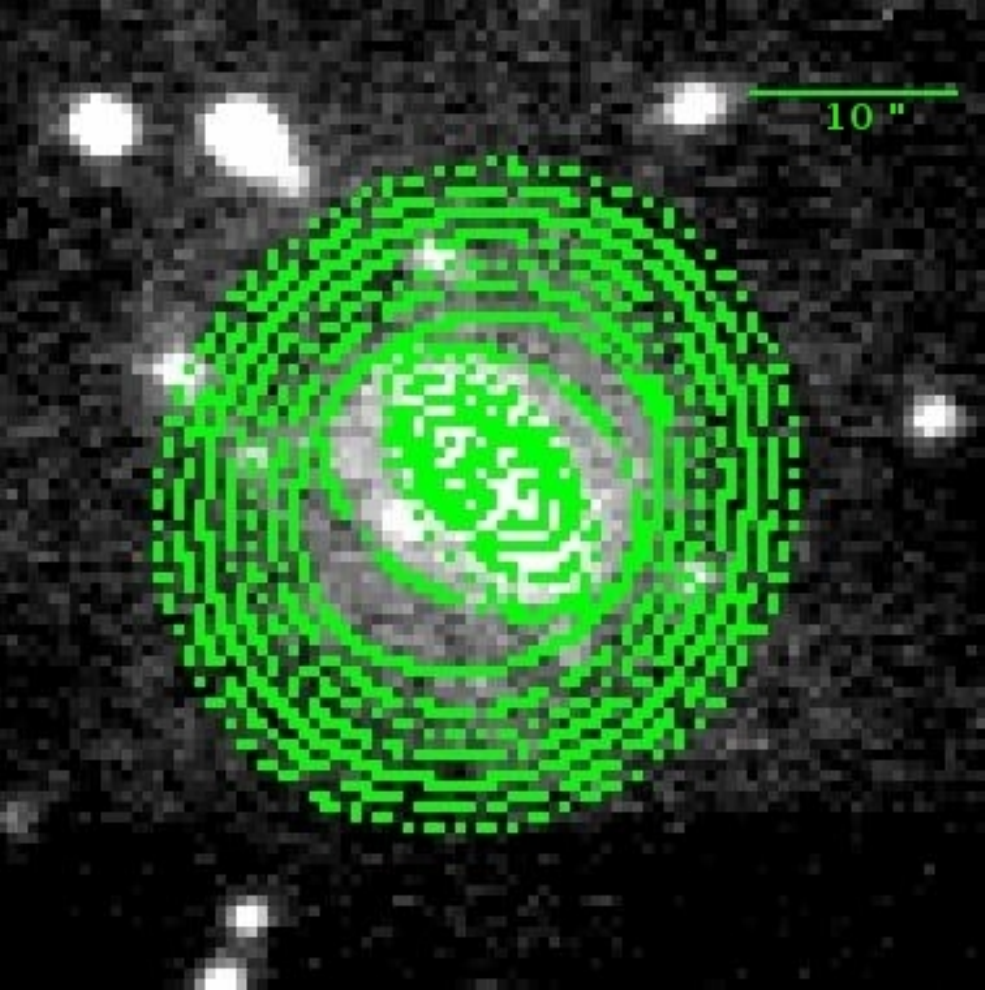}

\vspace{2cm}

\includegraphics[scale=0.5]{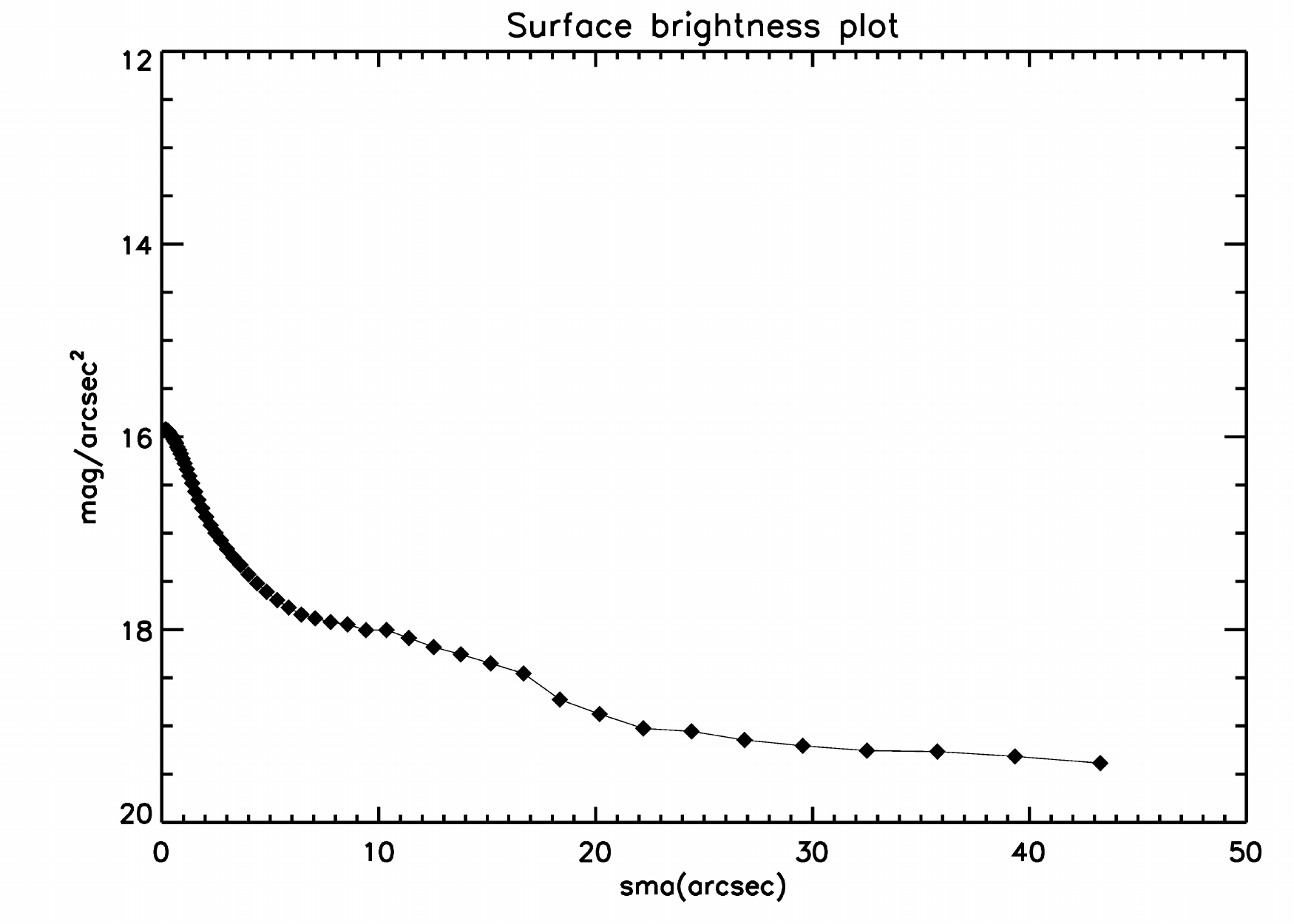}\includegraphics[scale=0.5]{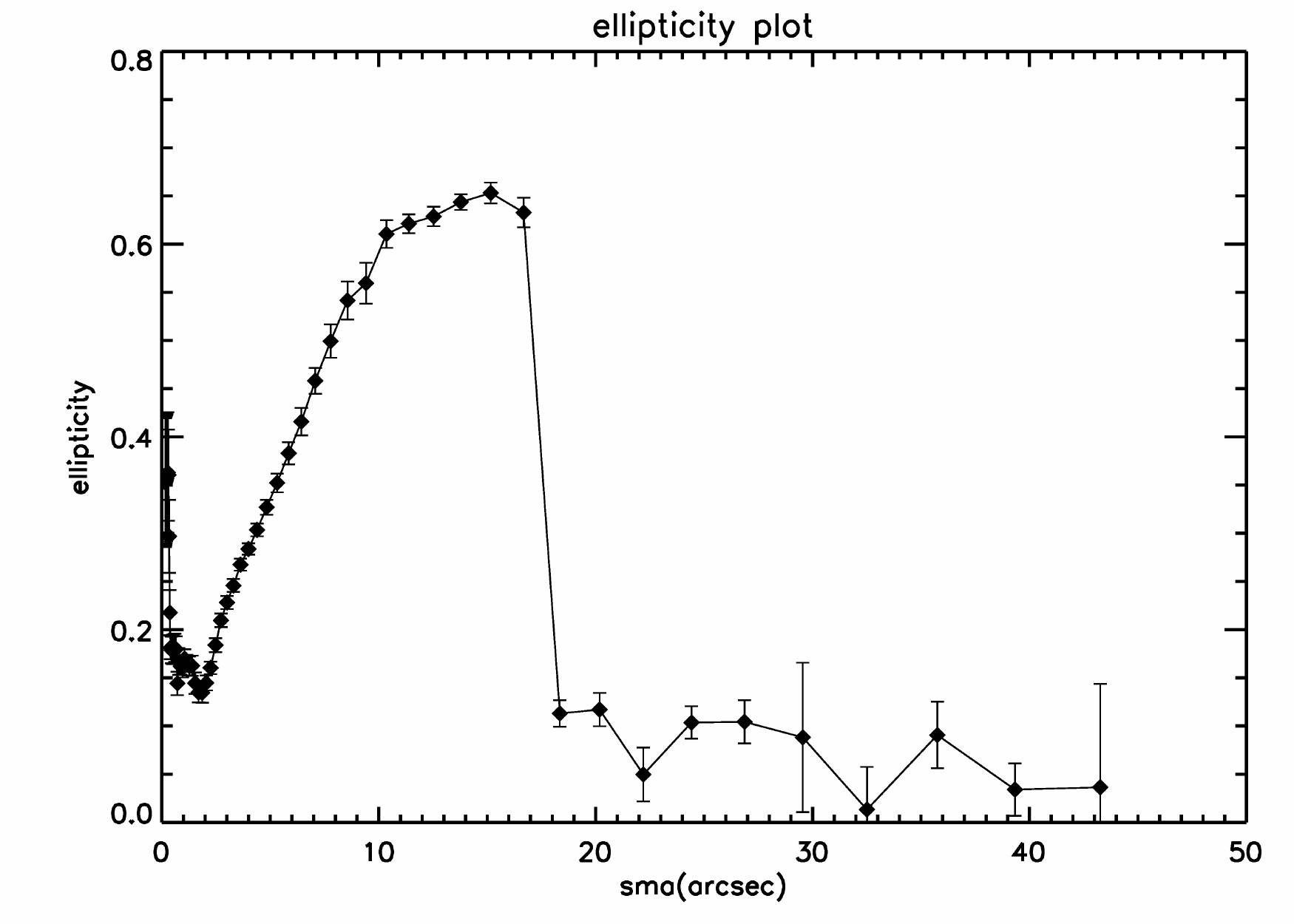}

\includegraphics[scale=0.5]{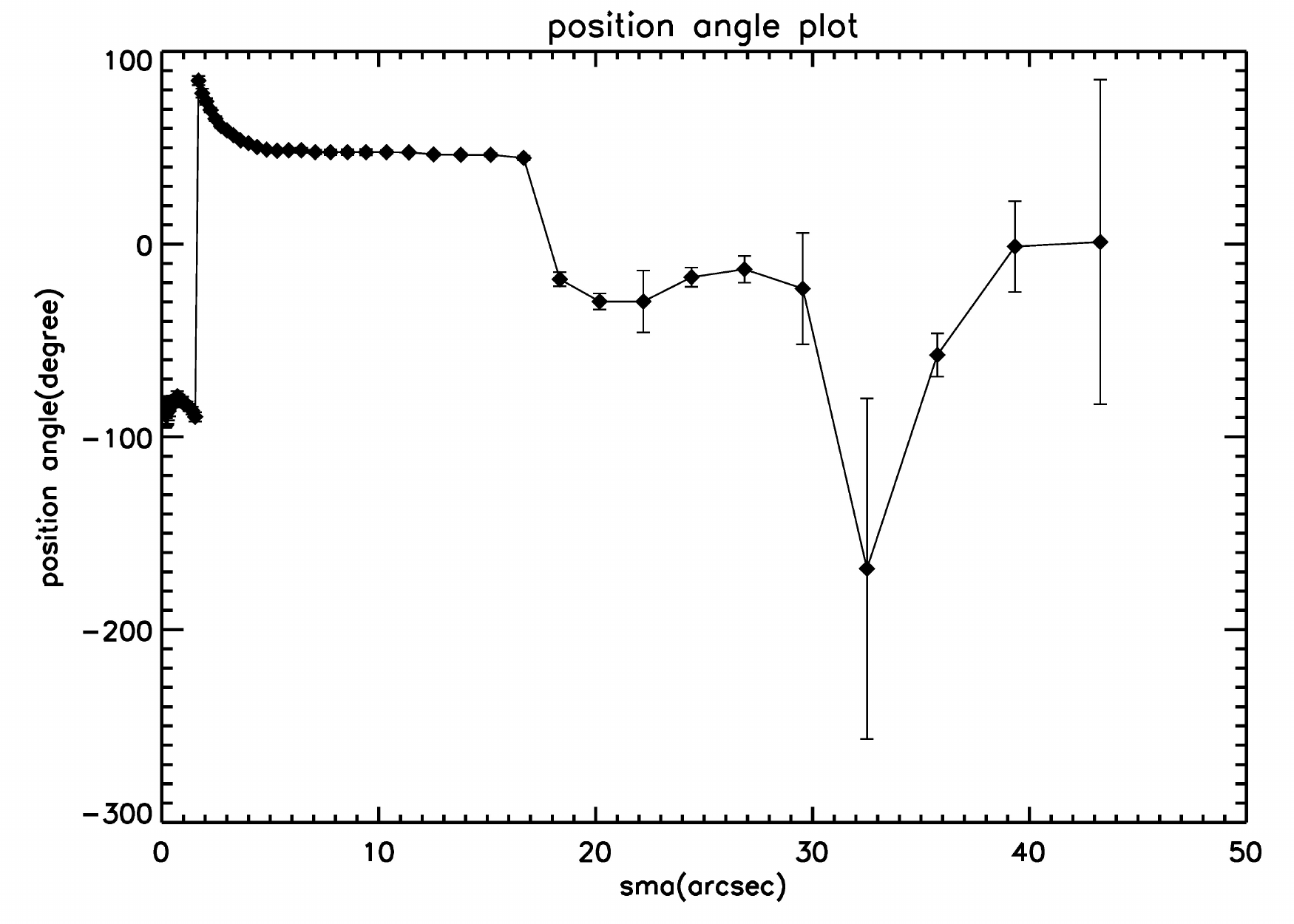}\includegraphics[scale=0.5]{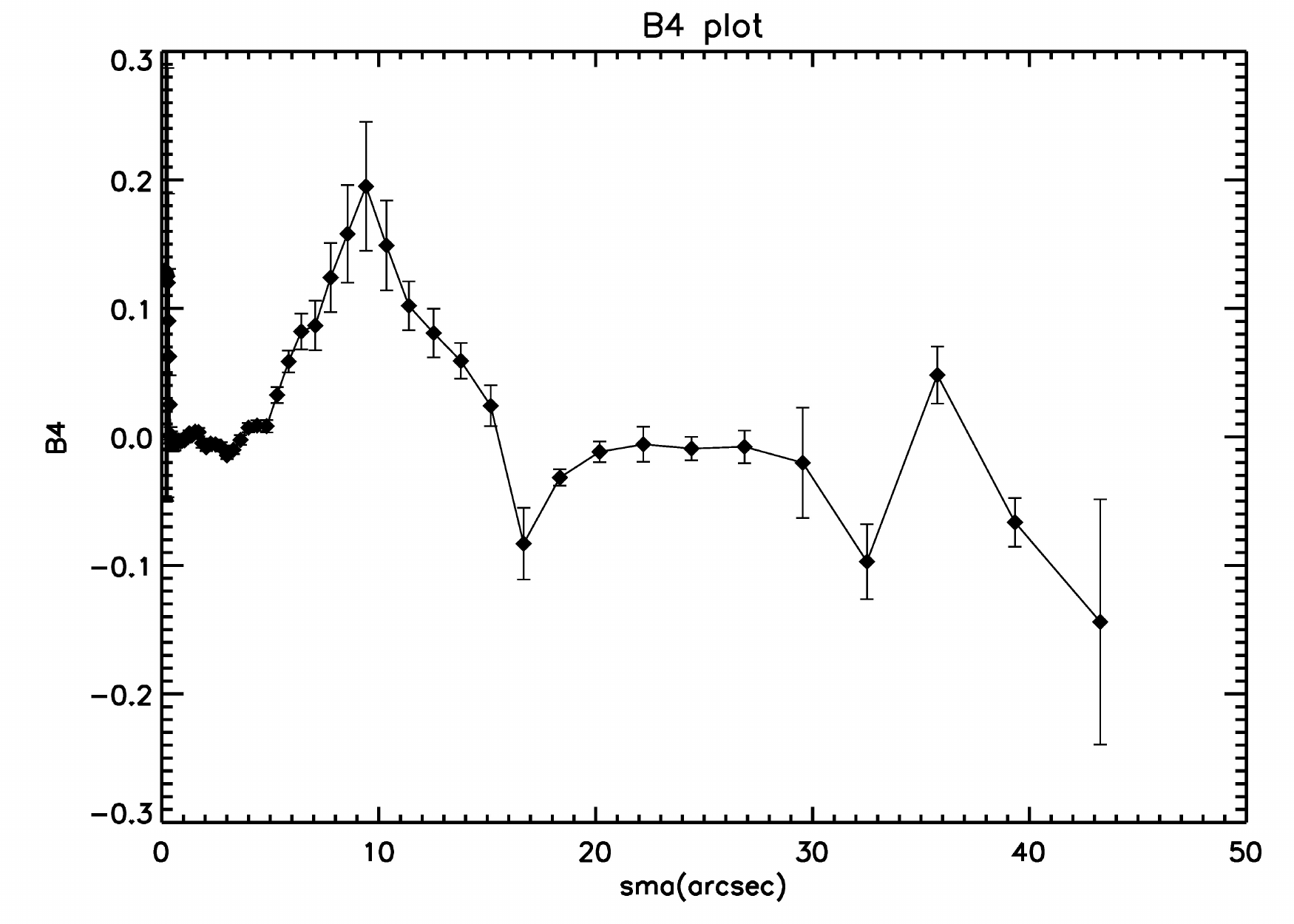}

\end{figure*}

\begin{figure*}
\centering
\caption{The histogram of deprojected barlength and deprojected ellipticities are shown below. The third figure show how the fraction of
barlength to that of D$_{25}$ is distributed}
\includegraphics[scale=0.5]{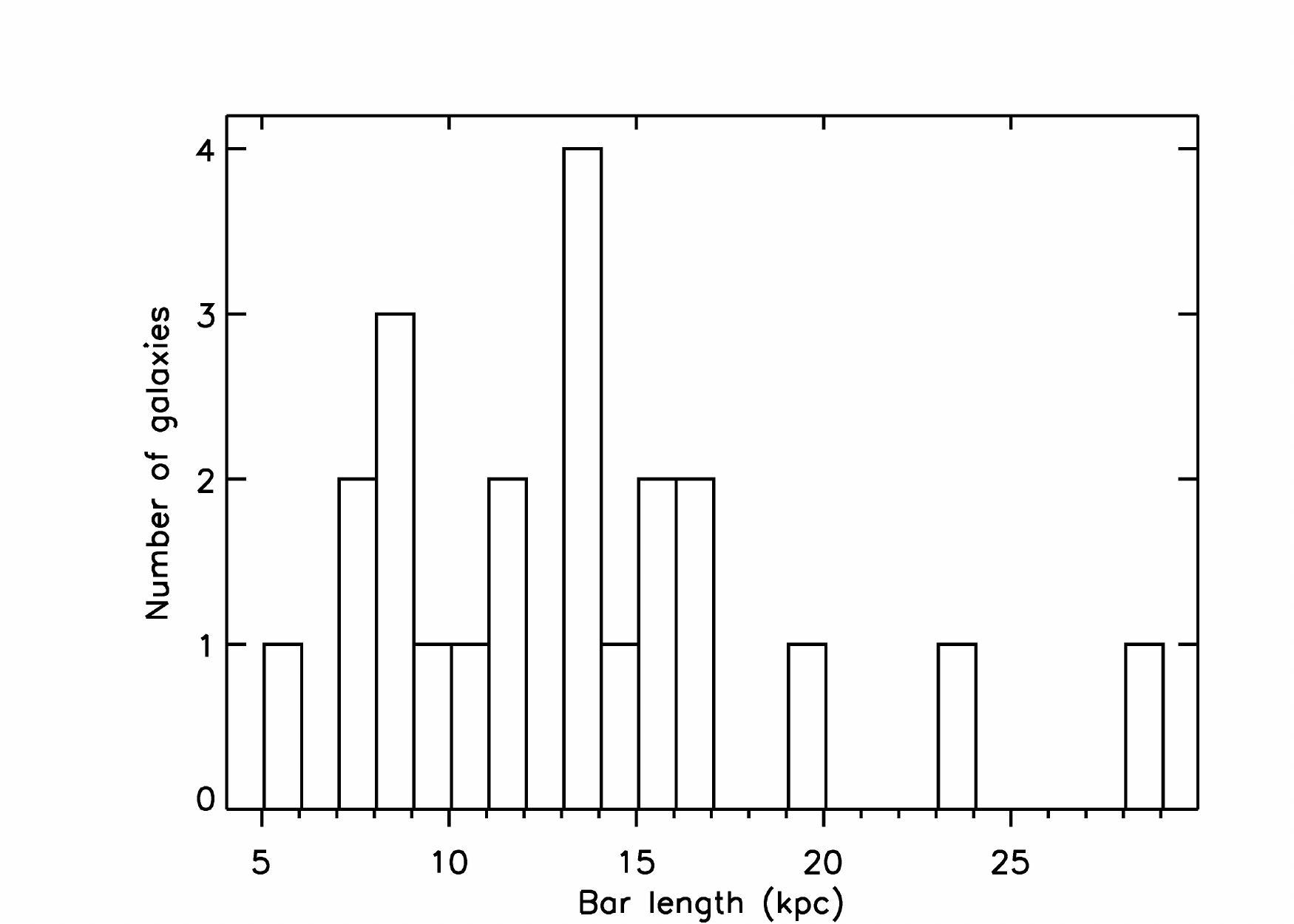}
\includegraphics[scale=0.5]{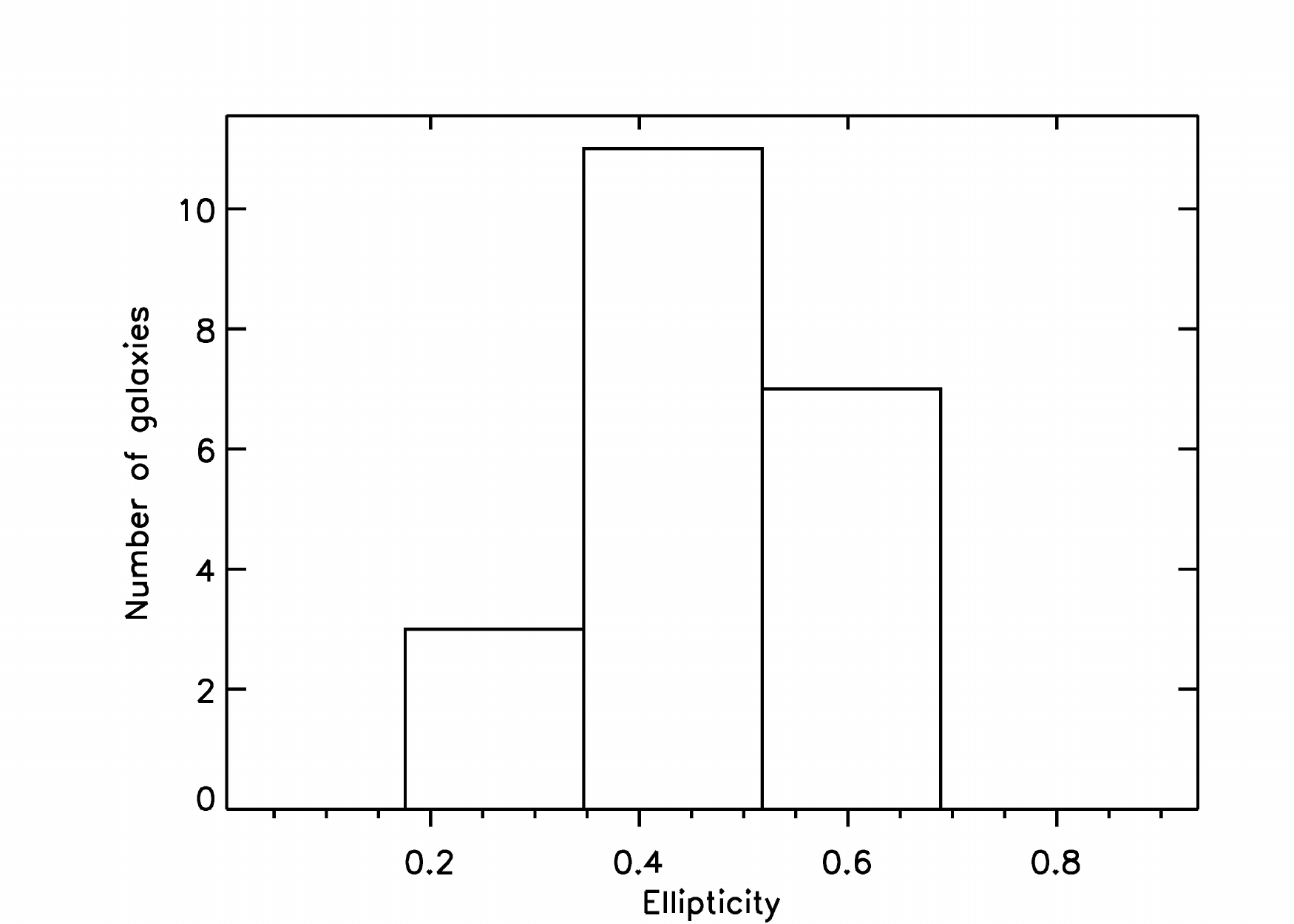}
\includegraphics[scale=0.5]{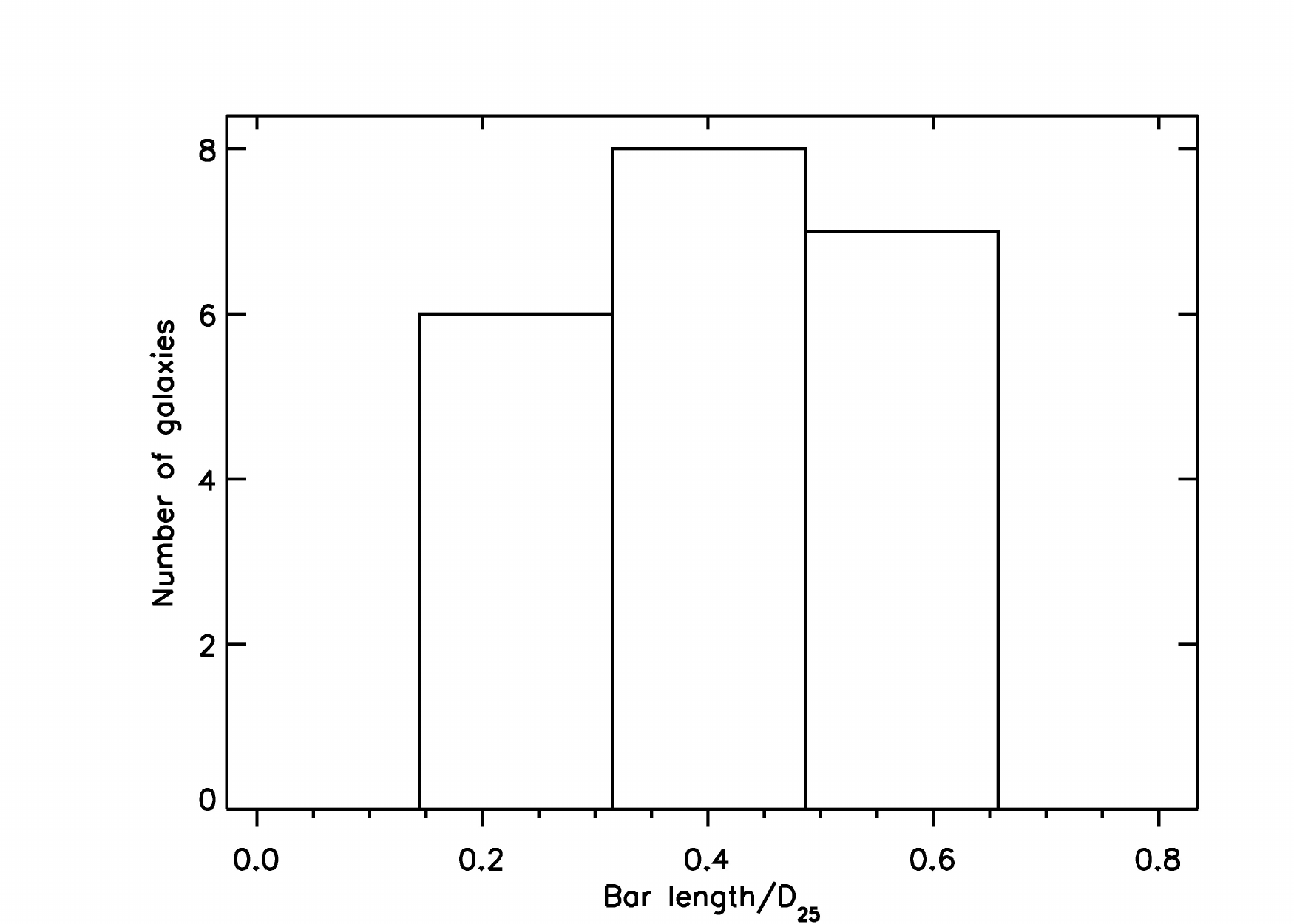}

\end{figure*}

\begin{figure*}
 \centering
 \caption{The ellipticity is plotted against the barlength to D$_{25}$ parameter. The points are very scattered and not showing any 
 correlation}
 \includegraphics{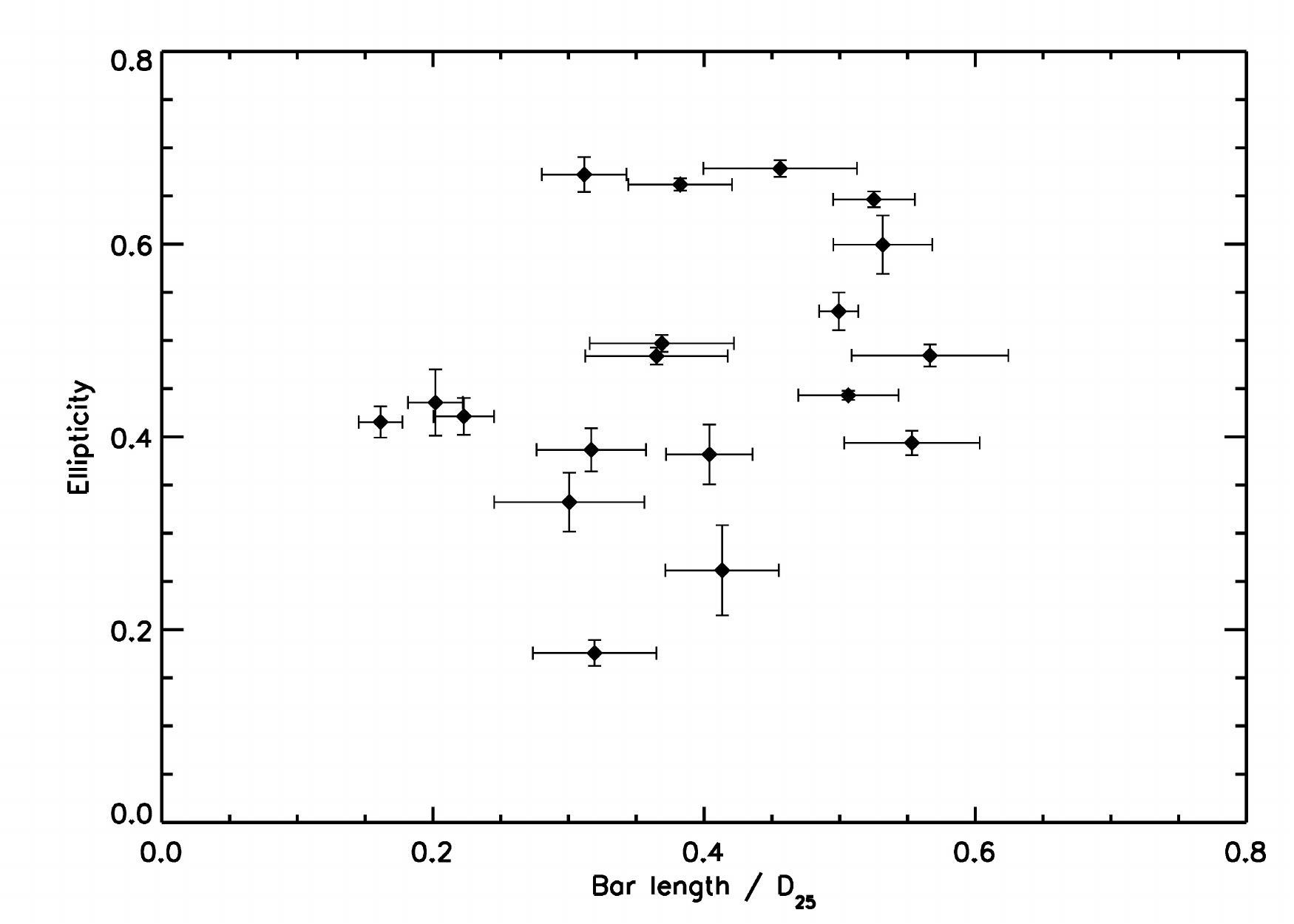}
 \end{figure*}

 
\begin{figure*}
   \centering
   \caption{The bar parameters like barlength to D$_{25}$ and ellipcity are plotted against the gas mass fraction of the galaxy.
 barlength to D$_{25}$  and ellipcity  are not showing any correlation with the M$_{HI}$ to (M$_{HI}$+M$_{stellar}$),barlength to D$_{25}$
 to the total baryonic mass also scattered in nature.}
  \includegraphics[scale=0.9]{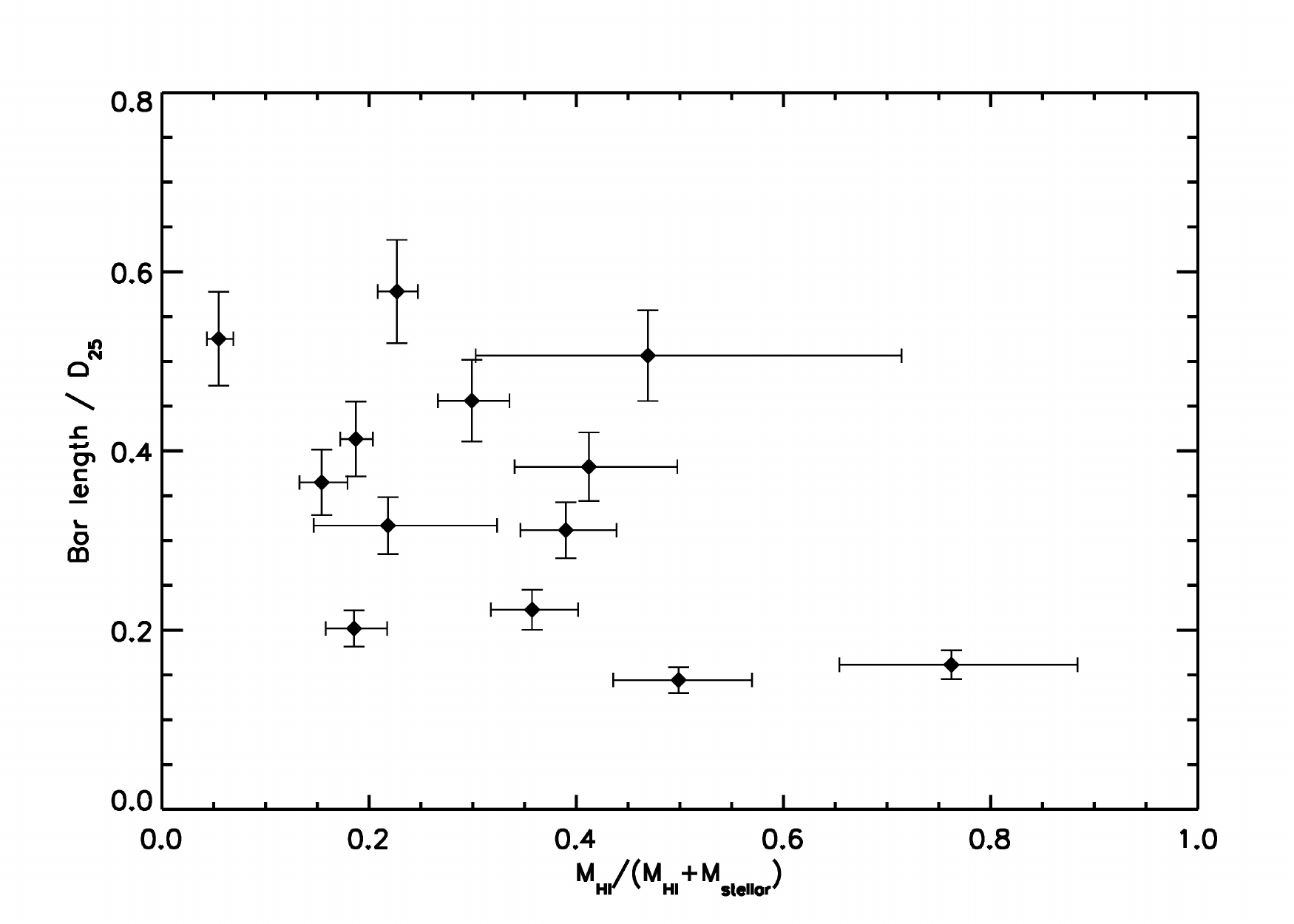}
  \includegraphics[scale=0.9]{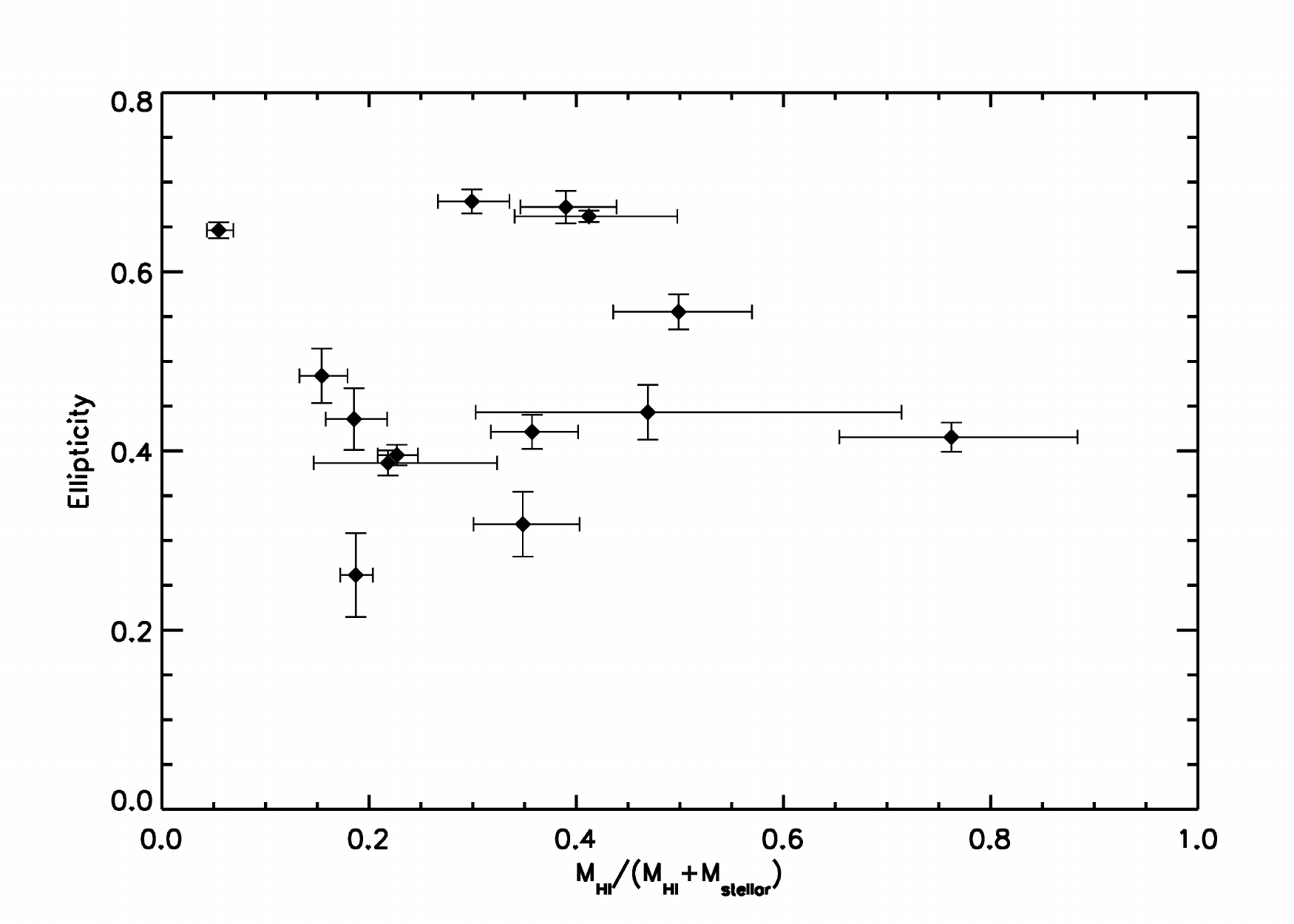}
  \end{figure*}
  \clearpage
  
  \begin{figure*}
  
  \includegraphics[scale=0.9]{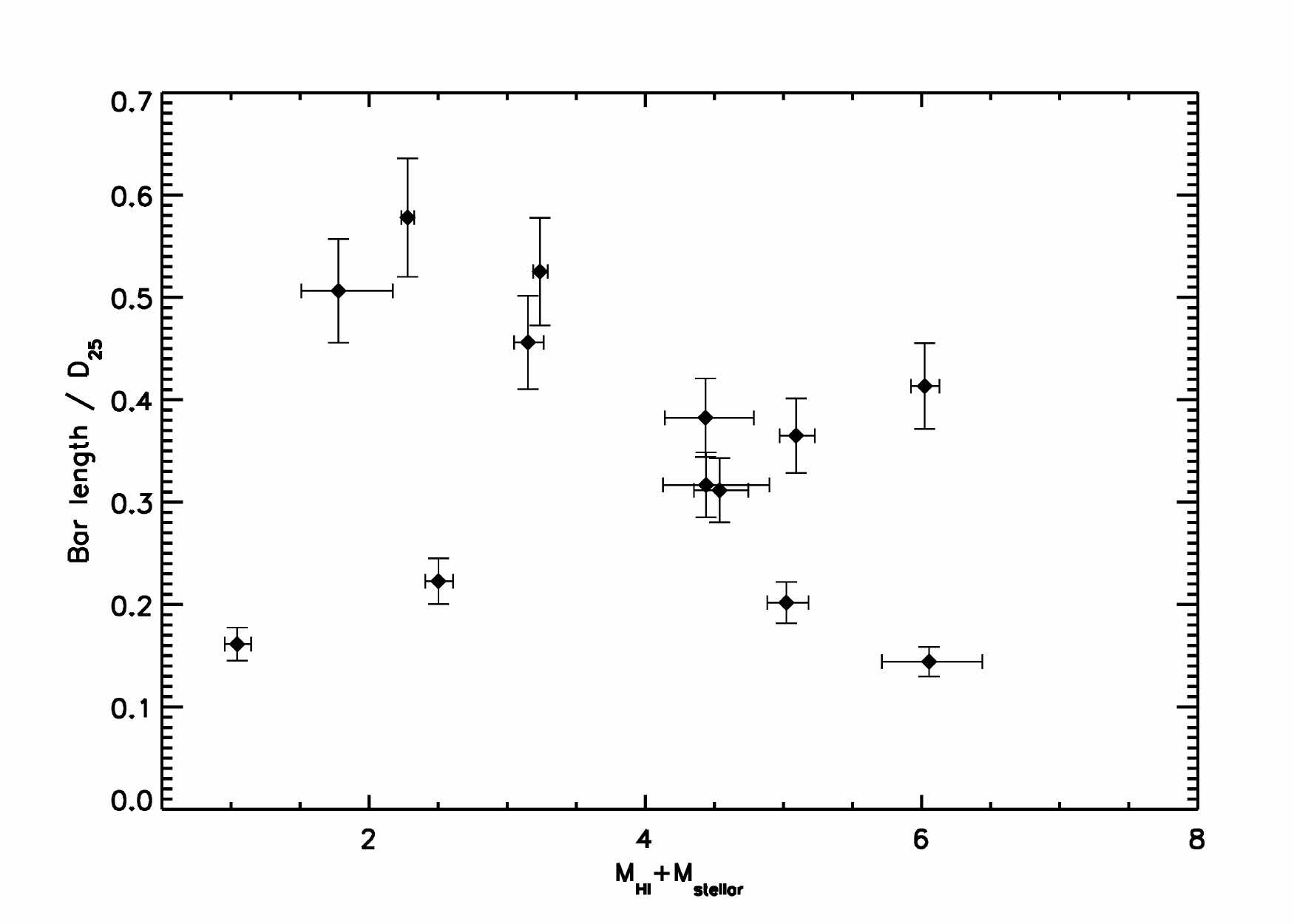}
\end{figure*}

\clearpage

\begin{figure*}
\caption{To check the variation of J-K$_s$ with the bar parameters, we plotted theJ-K$_s$ against the ratio of barlength to D$_{25 }$ in the 
first plot and with ellipcity in the second plot. The ratio of barlength to D$_{25 }$ and ellipcity are showing a weak relation with the J-K$_s$ 
color. The fitting does not include the errors}
\includegraphics[scale=0.9]{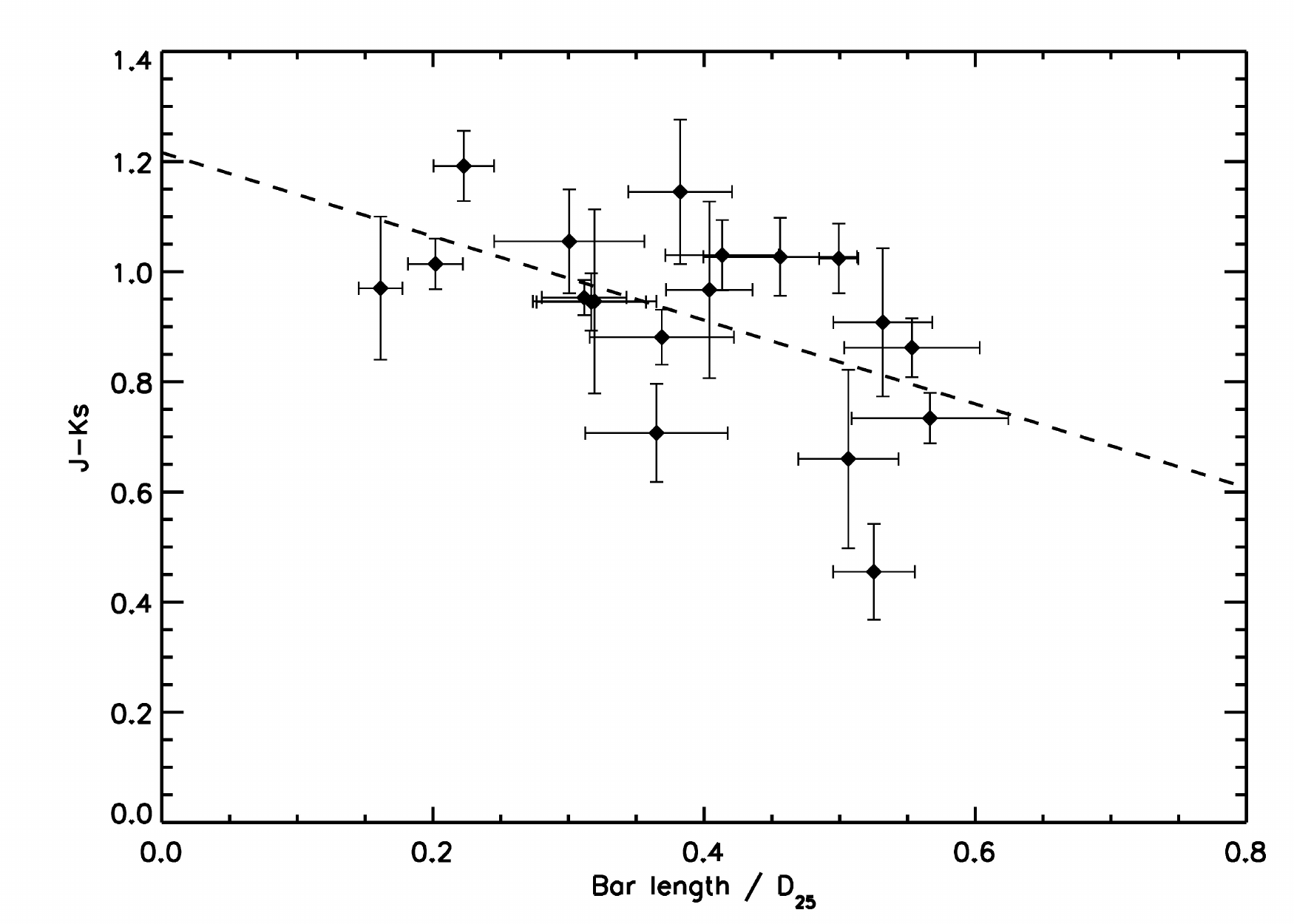}
\includegraphics[scale=0.9]{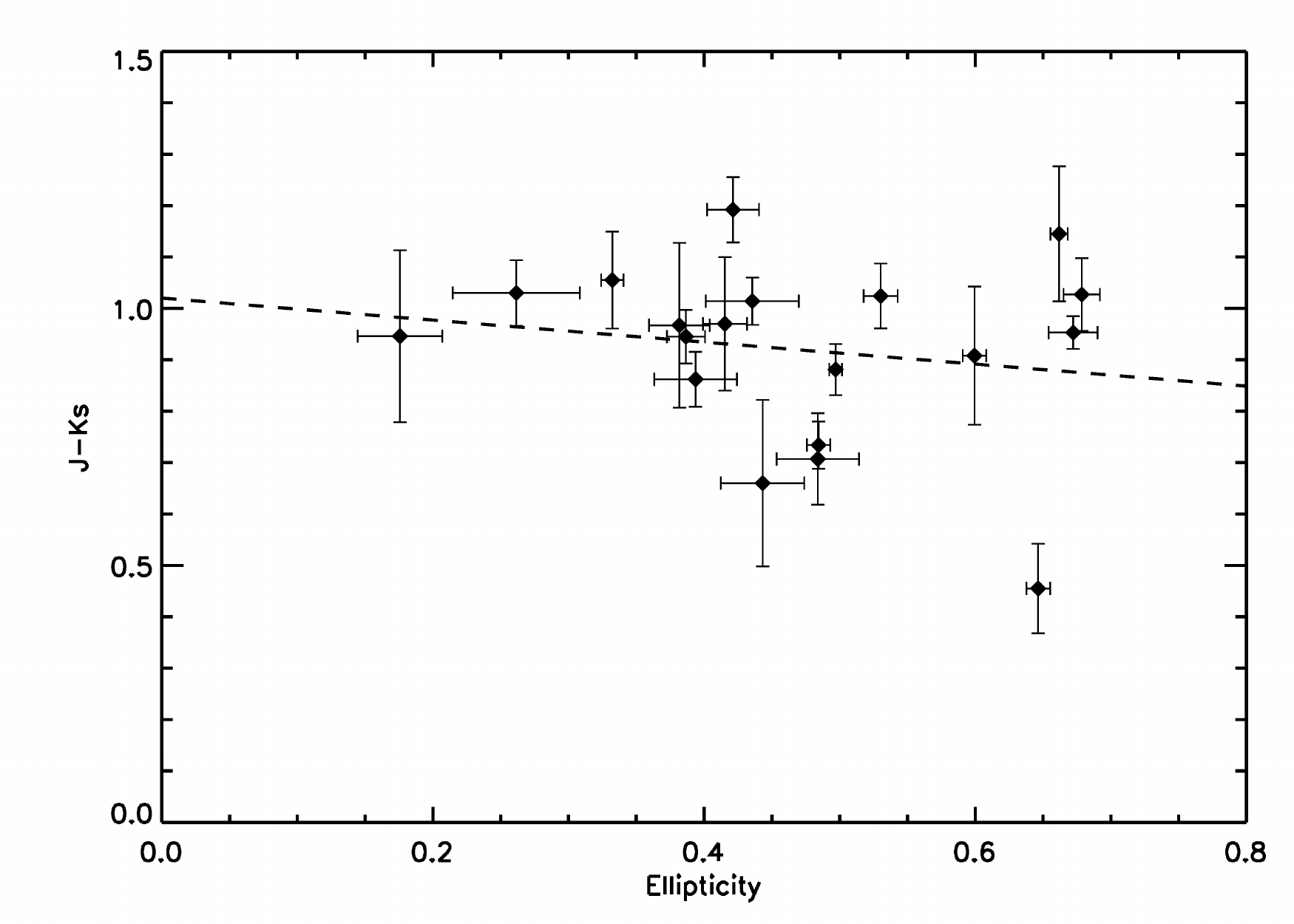}
\end{figure*}
   
\begin{figure*}
   \centering
\caption{The J-K$_s$ color is plotted against the ratio of the gaseous mass to that of the stellar mass of the galaxy}
\includegraphics{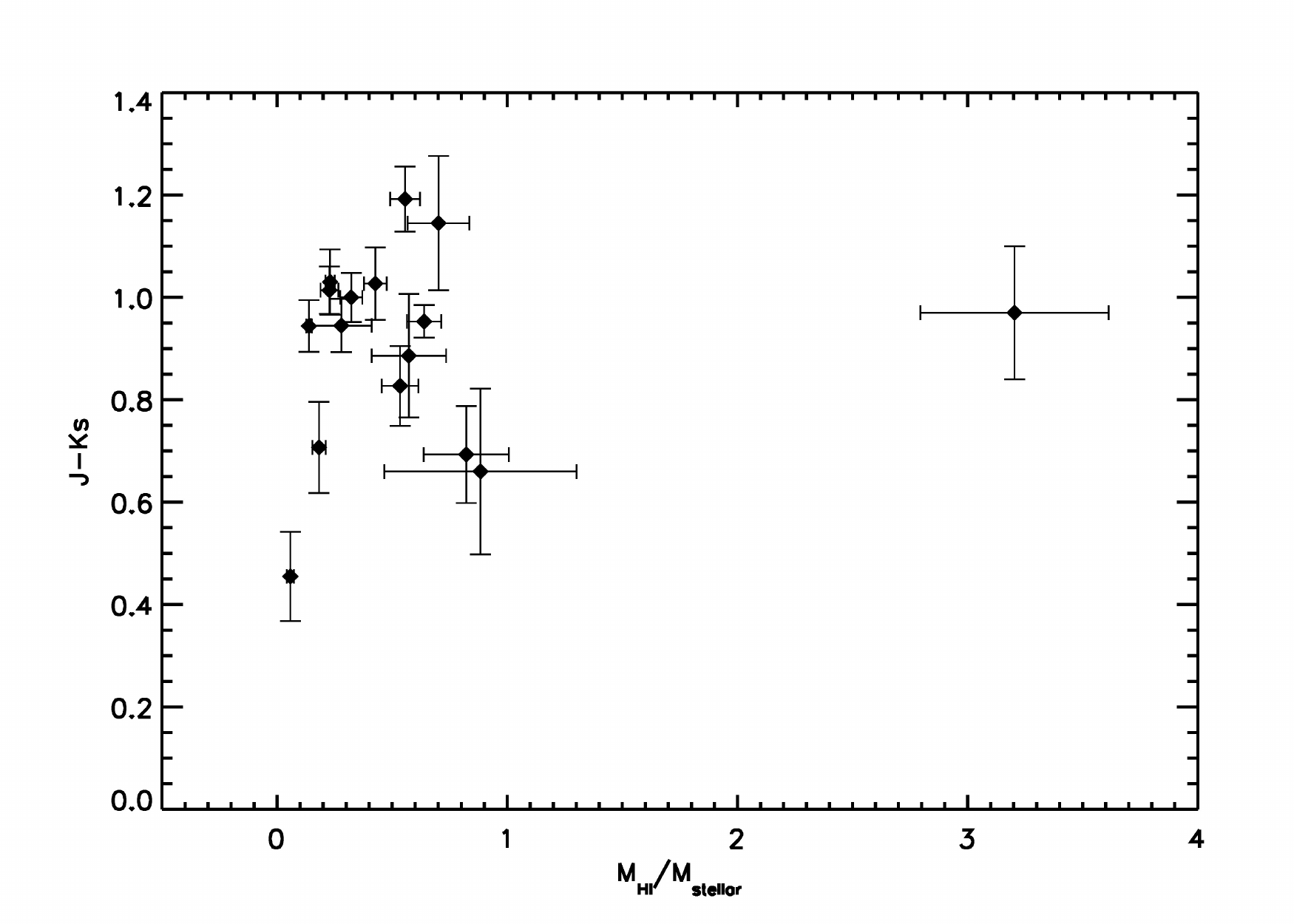}
\end{figure*}
\clearpage

\end{document}